\documentclass[useAMS,usenatbib,usegraphicx]{mn2e}

\usepackage{color}
\usepackage{amssymb}
\usepackage{savesym}
\usepackage{amsmath}

\newcommand{\aap}{A\&A}
\newcommand{\mnras}{MNRAS}

\newcommand{\na}{NewA}
\newcommand{\apjl}{ApJL}

\newcommand{\aj}{AJ}
\newcommand{\apj}{ApJ}
\newcommand{\nat}{Nature}
\newcommand{\ssi}{_i}
\newcommand{\ssj}{_j}
\newcommand{\ssij}{_{ij}}
\newcommand{\raw}{}
\newcommand{\refone}{}
\newcommand{\reftwo}{}

\def\be{\begin{equation}}
\def\ee{\end{equation}}
\def\ds{\displaystyle}
\def\rd{{\rm d}}

\title[Convergence of AMR and SPH simulations]{Convergence of AMR and SPH simulations - I. Hydrodynamical resolution and convergence tests}

\author[Hubber, Falle \& Goodwin]{D. A. Hubber$^{1,2,3}$\thanks{E:mail:dhubber@usm.lmu.de}, S. A. E. G. Falle$^{4}$, S. P. Goodwin$^{1}$ \\
$^{1}$Department of Physics and Astronomy, University of Sheffield, 
Hicks Building, Hounsfield Road, Sheffield, S3 7RH, UK \\
$^{2}$School of Physics and Astronomy, University of Leeds, 
Leeds, LS2 9JT, UK \\
$^{3}$Excellence Cluster Universe, Boltzmannstr. 2, 85748 Garching, Germany \\
$^{4}$Department of Applied Mathematics, University of Leeds, Leeds, LS2 9JT, UK}

\begin{document}
           
\date{February 14th, 2013}

\pagerange{\pageref{firstpage}--\pageref{lastpage}} \pubyear{2013}

\label{firstpage}

\maketitle

\begin{abstract}
We compare  the results  for a set  of hydrodynamical  tests performed
with  the  Adaptive  Mesh  Refinement  (AMR) finite  volume  code,  
{\small MG} and the Smoothed Particle Hydrodynamics (SPH) code,
{\small SEREN}.   The test suite includes  shock tube tests,
with and  without cooling,  the non-linear thin-shell  instability and
the Kelvin-Helmholtz instability.  The  main conclusions are : (i) the
two methods converge  in the limit of high  resolution and accuracy in
most  cases.  All  tests show  good agreement  when  numerical effects
(e.g.  discontinuities  in SPH) are  properly treated.  
(ii)  Both  methods  can  capture
adiabatic shocks and well-resolved  cooling shocks perfectly well with
standard prescriptions.  However, they both have problems when dealing
with under-resolved cooling shocks,  or strictly isothermal shocks, at
high Mach numbers.   The finite volume code only  works well at
1st  order  and  even  then requires  some  additional  artificial
viscosity.   SPH requires  either  a larger  value  of the  artificial
viscosity parameter,  $\alpha_{_{\rm AV}}$, or a modified  form of the
standard  artificial viscosity  term using  the harmonic  mean  of the
density,   rather  than   the  arithmetic   mean.   
(iii)  Some SPH
simulations require  larger kernels  to increase neighbour  number and
reduce particle noise in order to achieve agreement with finite volume
simulations (e.g. the Kelvin-Helmholtz instability).  However, this is
partly due to the need to  reduce noise that can corrupt the growth of
small-scale perturbations (e.g. the Kelvin-Helmholtz instability).  In
contrast,  instabilities seeded  from large-scale  perturbations (e.g.
the non-linear thin shell  instability) do not require more neighbours
and hence work  well with standard SPH formulations  and converge with
the  finite  volume  simulations.   
(iv)   For  purely
hydrodynamical problems,  SPH simulations  take an order  of magnitude
longer  to  run  than   finite  volume  simulations  when  running  at
equivalent resolutions,  i.e.  when  they both resolve  the underlying
physics to  the same  degree.  This requires  about 2-3 times  as many
particles as the number of cells.
\end{abstract}


\section{Introduction} \label{S:INTRO}

The advent of computers has provided a powerful new weapon in the
scientific arsenal: the numerical experiment with computer
simulations.  The aim of a computer simulation is to evolve a given set
of initial conditions according to some physical mathematical
prescription (e.g. solving a set of differential equations).
The numerical solution involves solving a discrete form of the
original mathematical prescription, which can introduce errors into the 
solution depending on the chosen algorithm.  Given that the goal of a 
numerical experiment is to arrive at the
`correct' answer\footnote{Note that the `correct' answer in a
  numerical experiment is properly evolving the initial conditions
  with the input physics.  This may, or may not, match the real
  world.}, it is crucial to understand what problems and inaccuracies it can introduce to the computed solution.

A particular physical problem of great interest in many areas of
science, and in particular astrophysics, is that of hydrodynamics.  
This is the time
evolution of complex fluid systems (liquid or gas) governed by a set
of differential equations, such as the Euler Fluid Equations.  
Hydrodynamics involves numerous complex
physical processes such as turbulence, shocks, shearing, and
instabilities which are not amenable to
an analytic approach except in the most trivial set-ups.  
We are interested in particular in astrophysical problems involving 
a compressible, self-gravitating fluid.

This is the first in a series of papers in which we will closely study
the  performance  and  convergence  of two  very  different  numerical
methods  used  in astrophysics;  Upwind  Finite  Volume combined  with
Adaptive  Mesh Refinement \citep[AMR;][]{AMR1984,AMR1989}  and Smoothed
Particle Hydrodynamics  \citep[SPH;][]{Lucy1977,GM1977}.  Both attempt
to solve the fluid equations,  but use very different algorithms, each
with its  own advantages and  disadvantages (e.g.  grid  vs. particles
and Eulerian vs. Lagrangian).

The reasons for a detailed comparison of AMR and SPH are fourfold; 
firstly, AMR and SPH are the two most popular methods for solving the
fluid equations, especially in astrophysics, and so a full
understanding of their strengths and weaknesses is vital.

Secondly, do both methods converge on the same answer at high enough resolution and accuracy?  And how much resolution is required to achieve convergence?
If both methods give the same results when applied to the same problem, 
this gives us great confidence that this is a `correct' result, as it
is unlikely two such different methods
would both produce the same error.  This is particularly important as
the main purpose of a numerical experiment is to examine situations in
which we do not know the result a priori.

Thirdly, we need to know in what ways do the methodologies diverge at
lower-than optimum  resolution.  In most systems that are
simulated (especially in astrophysics) there is some element of 
sub-resolution physics.  We can 
never simulate every molecule in a fluid and so there will be some
processes that are well modeled and some which will not be resolved
by the limited scope of that simulation.  What problems are introduced
by poor resolution?

The fourth reason is to help educate us in understanding which numerical schemes are appropriate for particular problems.  Aside from differences in particular implementations, there are often several ways of modeling some process even within a particular paradigm.  As well as wishing to understand whether mesh or particle schemes are better in a given situation, simulators need to better understand the subtleties within each method in order to better judge which options should be selected for a particular problem.

\subsection{Previous studies}

Various comparisons  between particular  aspects of finite  volume and
particle   codes   have  been   made   in   recent   years.

\citet{Frenk1999}  conducted  a  comparison  simulation  involving  12
different SPH,  static grid and moving  grid codes of a  single set of
initial conditions which represent the formation of an isolated galaxy
cluster  in a  cold-dark  matter dominated  Universe.  The  comparison
showed that the  major features of the galaxy  cluster were reproduced
in  all  codes, especially  large-scale  features  which are  strongly
dependent on the dark matter  gravity.  The comparison did reveal some
discrepancies  between  particle and  mesh  methods,  most
noticeably  in  the  distributions  of the  temperature  and  specific
entropy  profiles, the  origin of  which has  been debated  by various
authors subsequently \citep[e.g.][]{Mitchell2009}.

\citet{Agertz2007} considered the Kelvin-Helmholtz instability and the so-called `blob'-test, to demonstrate that SPH could not, in its most basic form, model mixing processes as well as finite volume codes.  However, \cite{Price2008} suggests that this is due to the discretisation of the SPH equations resulting in artificial surface terms that can be mitigated against by the use of appropriate dissipation terms.  \cite{Price2008} also then demonstrates that using an appropriate artificial conductivity term allows SPH to quite easily model the Kelvin-Helmholtz instability.  Other authors \cite[e.g.][]{GSPH2010,OSPH} have shown that modifications to SPH can allow mixing without extra dissipation.

\cite{Tasker2008} performed a suite of tests using two SPH codes and two
finite volume AMR codes for simple problems with analytical or semi-analytical solutions.  They suggest that to achieve similar levels of resolution and therefore similar results, that one particle is required per grid cell in regions of interest, e.g. high density regions of shocks.

{\refone 
\cite{Commercon2008} performed a comparison study of the two methods by modeling the fragmentation of a rotating prestellar core with initial conditions similar to the Boss-Bodenheimer test \citep{BBSIT1979}.  They found that broad agreement between the two methods could be achieved given sufficient resolution, i.e. when the local Jeans length/mass is sufficiently resolved.  In both cases, they found that insufficient resolution could lead to significant angular momentum errors.

\cite{Kitsionas2009} performed a comparison study of isothermal turbulence using four mesh codes and three SPH codes.  They found generally good agreement between the various implementations for similar levels of resolutions, and that the effect of low resolution in the simulations was dependent on the individual implementations.  They also found the SPH codes to be more dissipative requiring more advanced artificial viscosity switches to reduce this problem.

\cite{Federrath2010} performed a comparison of SPH and AMR via the formation of sink particles in various problems, including turbulent fragmenting prestellar cores.  They found good agreement between the gas properties and the sinks that formed from each simulation, including the total numbers formed and their mass accretion properties.

\cite{AREPO} compared both SPH and AMR simulations to his new finite volume tessellation code, {\small AREPO}.  \cite{AREPO} demonstrates that the new method is capable of giving improved results over fixed-mesh codes in problems with high advection velocities due to its Galiliean invariance.

}

\subsection{Our study}
This is  the first paper in  a series comparing finite  volume AMR and
SPH codes.  In this paper,  we consider a set of purely hydrodynamical
problems, ignoring self-gravity which we will cover in future papers.
In Section \ref{S:NUMERICALMETHODS}, we discuss the
main features and characteristics of AMR and SPH, and the relative
merits and weaknesses of each method.  In Section \ref{S:TESTS}, we
introduce our first suite of tests, describe the initial conditions used, 
perform the tests at various resolutions and describe the results. 
In Section \ref{S:DISCUSSION}, we discuss our results 
 and their practical implications with regards 
to how AMR and SPH perform relative to each other.

\section{Numerical methods} \label{S:NUMERICALMETHODS}

The two most popular methods used in astrophysical hydrodynamical 
simulations are Adaptive Mesh Refinement Finite Volume Hydrodynamics 
and Smoothed Particle Hydrodynamics.  The fundamental approaches of these 
two methods are very different, one being Eulerian (AMR) and the other 
being Lagrangian (SPH). Although there exists a large number of 
codes that can be considered hybrid Eulerian-Lagrangian, such as 
Particle-in-cell \citep{PIC1983} and {\small AREPO} \citep{AREPO}, 
AMR and SPH represent pure Eulerian and Lagrangian methods and 
therefore allow us to highlight the fundamental differences more clearly.  
We describe here the exact details of our implementations of both methods 
for clarity and for future comparisons with our work.

\subsection{Adaptive mesh refinement (AMR) Finite Volume Code} \label{SS:MG}

We use the AMR code {\small MG} \citep{VanLoo2006} to perform all the finite
volume  
simulations presented in this paper.  This uses an upwind finite volume 
scheme to solve  the standard equations of compressible flow in conservation form:
\begin{equation}
\frac{\partial {\bf U}}{\partial t} + \frac{\partial {\bf F}}{\partial
  x} + \frac{\partial {\bf G}}{\partial y} + \frac{\partial {\bf
  H}}{\partial z} = {\bf S},
\label{cons_eqn}
\end{equation}
\noindent where
\begin{equation}
\begin{array}{l}
{\bf U} = (\rho, \rho v_x, \rho v_y, \rho v_z, e), \\
{\bf F} = (\rho v_x, P + \rho  v_x^2, \rho v_x v_y, \rho v_x v_z, (e +
P) v_x), \\
{\bf G} = (\rho v_y, \rho  v_x v_y, P + \rho v_y^2, \rho v_y v_z, (e +
P) v_y), \\
{\bf H} = (\rho v_z, \rho  v_x v_z, \rho v_y v_z, P + \rho v_z^2, (e +
P) v_z).\\
\end{array}
\label{cons_var_flux}
\end{equation}

\noindent
Here $\rho$ is  the density, $v_x$, $v_y$, $v_z$  are the velocities in
the $x$, $y$, $z$ directions, $P$ is the pressure and 
\begin{equation}
e = \frac{P}{\gamma - 1} + \frac{1}{2} (\rho v_x^2 + \rho v_y^2 + \rho
v_z^2)
\label{tot_en}
\end{equation}
\noindent
is the total  energy per unit volume. ${\bf S}$ is  a vector of source
terms to account for gravity, heating and cooling etc.

The  fluxes are  calculated with  an exact  Riemann solver  and second
order accuracy is achieved by using a first order step to determine the
solution  at  the  half-timestep.   The van  Leer  averaging  function
\citep{VanLeer1977} is used to reduce the scheme to first order at
shocks  and contact discontinuities.   The details  of the  scheme are
described in \cite{Falle1991}.

It has long been known  that upwind schemes suffer from an instability
in certain types of flow  e.g. when a shock propagates nearly parallel
to  the grid  \citep{Quirk1994}.  This  can  be cured  by adding  a
second order  artificial dissipative flux to the  fluxes determined from
the  Riemann  solver.  Here  we  adopt  the  prescription described  in
\cite{Falle1998}  in which  the  viscous momentum  fluxes  in the  $x$
direction are
\begin{equation}
\mu (v_{xl} - v_{xr})
\label{art_vis_vel}
\end{equation}
\noindent
and similarly for  the $y$ and $z$ directions.  Here the suffixes $l$,
$r$  denote the  left and  right states  in the  Riemann  problem. The
coefficient, $\mu$, is given by
\begin{equation}
\mu = \eta \frac{1}{[1/(c_l \rho_l) + 1/(c_r \rho_r)]}
\label{art_vis_vel}
\end{equation}
\noindent
where $c$ is  the sound  speed and $\eta$  is a dimensionless  parameter (in
most  cases $\eta =  0.2$ is  appropriate). The  harmonic mean  of the
densities and sound speed is  used to avoid large viscous fluxes where
there is  a large density contrast  in the Riemann  problem. In smooth
regions, this gives a viscosity of order $\Delta x^2$ i.e. it does not
reduce the order of the scheme.

{\small MG} uses  a hierarchy of grids, $G_0 \cdots  G_N$ such that if
the mesh  spacing is  $\Delta x_n$  on grid $G_n$  then it  is $\Delta
x_n/2$ on  $G_{n+1}$. Grids $G_0$  and $G_1$ cover the  entire domain,
but  finer grids  only exist  where  they are  required for  accuracy.
Refinement in {\small MG} is on a cell-by-cell basis.  The solution is
computed on all  grids and refinement of a cell  on $G_n$ to $G_{n+1}$
occurs whenever the the  difference between the solutions on $G_{n-1}$
and $G_n$  exceeds a given error 
{\reftwo for any of the conserved variables.}
{\refone $G_0$ and  $G_1$ must therefore cover the  entire domain since
they are used to determine  refinement to $G2$. In all the simulations
in this  paper, the error  tolerance was set  to $1\%$.} Each  grid is
integrated at its own timestep.

\begin{table*}
\label{TAB:SHOCKS}
\caption{Mathematical expressions for the post-shock quantities of the density, $\rho_s$, the velocity, $v_s$ and the sound speed squared, $a_s^2$, for isothermal, adiabatic, and strong (i.e. ${\cal M} \gg 1$) adiabatic shocks.}
\begin{tabular}{cccc}
\hline
  Physical quantity & Isothermal & Adiabatic & Adiabatic (${\cal M} \gg 1$) \\ \hline
$\rho_s$ & ${\cal M}^2\,\rho_0$ & 
$\displaystyle{\frac{(\gamma + 1)\,{\cal M}^2}
{ (\gamma - 1)\,{\cal M}^2 + 2 }\,\rho_0}$ 
& $\displaystyle{\frac{(\gamma + 1)}{(\gamma - 1)}\,\rho_0}$ \\ \vspace{0.05cm} \\
$v_s$ & ${\cal M}^{-2}\,v_0$ & 
$\displaystyle{\frac{(\gamma - 1)\,{\cal M}^2 + 2}
{(\gamma + 1)\,{\cal M}^2}\,v_0}$ 
& $\displaystyle{\frac{(\gamma - 1)}{(\gamma + 1)}\,v_0}$ \\ \vspace{0.05cm} \\
$a_s^2$ & $a_0^2$ & $\displaystyle{\frac{\left[(\gamma - 1)\,{\cal M}^2 + 2 \right]\,
\left[ 2\,\gamma\,{\cal M}^2 - (\gamma - 1) \right]}
{(\gamma + 1)^2\,{\cal M}^2}\,a_0^2}$ &
$\displaystyle{\frac{2\,\gamma\,(\gamma - 1)}{(\gamma + 1)^2}\,v_0^2}$
\\ \hline
\end{tabular}
\end{table*}

\subsection{Smoothed Particle Hydrodynamics Code} \label{SS:SPH}

We use the SPH code {\small SEREN} \citep{Hubber2011} to perform all SPH
simulations presented in this paper.  {\small SEREN} uses a conservative form
of SPH \citep{SH2002,PM2007} to integrate all particle properties.
The SPH density of particle $i$, $\rho\ssi$, is computed by 
\begin{equation} \label{EQN:RHO-SPH}
\rho\ssi = \sum \limits_{j=1}^{N}  m\ssj W({\bf r}\ssij,h\ssi)\,.
\end{equation}
where $h\ssi$ is the smoothing length of particle $i$, ${\bf r}\ssij = {\bf r}\ssi - {\bf r}\ssj$, $W({\bf r}\ssij,h\ssi)$ is the smoothing kernel and $m\ssj$ is the mass of particle $j$.  The smoothing length of every SPH particle is constrained by the simple relation
\begin{equation} \label{EQN:HRHO-SPH}
h_i = \eta\,\left( \frac{m_i}{\rho_i} \right)^{1/D}\,
\end{equation}
where $D$ is the dimensionality of the simulation and $\eta$ is a dimensionless parameter that relates the smoothing length to the local particle spacing.  {\refone We use the default value, $\eta = 1.2$, throughout this paper}.  Since $h$ and $\rho$ are inter-dependent, we must iterate $h$ and $\rho$ to achieve consistent values for both quantities \citep[see][for strategies on this computation]{PM2007}.  
Equation \ref{EQN:HRHO-SPH} effectively constrains the smoothing length so each smoothing kernel contains approximately the same total mass/number of neighbouring particles.  {\refone In this paper, we use both the M4 cubic spline and quintic spline kernels.  Expressions for each kernel and derivative quantities are given in \citet{Hubber2011}.}

The SPH momentum equation is 
\begin{equation} \label{EQN:GRADHMOMEQN}
\frac{d{\bf v}\ssi }{dt} = -
\sum \limits_{j=1}^{N}  m\ssj  \left[ 
\frac{P\ssi}{\Omega\ssi \rho\ssi^2} \nabla\ssi W({\bf r}\ssij ,h\ssi) + 
\frac{P\ssj}{\Omega\ssj \rho\ssj^2} \nabla\ssi W({\bf r}\ssij ,h\ssj) \right]
\,,
\end{equation}
where $P\ssi = (\gamma - 1)\,\rho\ssi\,u\ssi$ is the thermal pressure of particle $i$, $u\ssi$ is the specific internal energy of particle $i$, $\nabla\ssi W$ is the gradient of the kernel function, and 
\begin{equation} \label{EQN:OMEGA}
\Omega\ssi = 1 - \frac{\partial h\ssi }{\partial \rho\ssi } 
\sum \limits_{j=1}^{N}  m\ssj  \frac{\partial W}{\partial h}
({\bf r}\ssij , h\ssi )\,.
\end{equation}
$\Omega\ssi$ is a dimensionless correction term that accounts for the spatial variability of $h$ amongst the neighbouring particles.  $\partial h\ssi / \partial \rho_i$ is obtained explicitly from Eqn. (\ref{EQN:HRHO-SPH}) and $\partial W / \partial h$ is obtained by directly differentiating the employed kernel function.  For the thermodynamics, we integrate the specific internal energy, $u$, with an energy equation of the form
\begin{equation} \label{EQN:GRADHENEQN}
\frac{du\ssi }{dt} = 
\frac{P\ssi }{\Omega\ssi  \rho\ssi ^2} \sum \limits_{j=1}^{N}
m\ssj  {\bf v}\ssij \cdot \nabla W\ssij ({\bf r}\ssij , h\ssi )\,,
\end{equation}
where ${\bf v}\ssij = {\bf v}\ssi - {\bf v}\ssj$.

We include dissipation terms following \citet{Monaghan1997} and \citet{Price2008}.
\begin{eqnarray} 
\frac{d{\bf v}\ssi}{dt} &=& 
\sum\limits_{j=1}^{N}\,\frac{m\ssj}{\overline{\rho}\ssij}\,\left\{\alpha_{_{\rm AV}}\,v_{_{\rm SIG}} {\bf v}\ssij\cdot\hat{\bf r}\ssij\right\}\,\overline{\nabla\ssi W}\ssij\,,\label{EQN:MON97ARTVISC} \\
\frac{du\ssi}{dt} &=& -\,\sum\limits_{j=1}^{N}\,\frac{m\ssj}{\overline{\rho}\ssij}\,\frac{\alpha_{_{\rm AV}}\,v_{_{\rm SIG}}({\bf v}\ssij\cdot\hat{\bf r}\ssij)^2}{2} \,\hat{\bf r}\ssij\cdot\overline{\nabla\ssi W}\ssij\,, \nonumber \\
&& +\,\sum\limits_{j=1}^{N}\,\frac{m\ssj}{\overline{\rho}\ssij}\, \alpha_{_{\rm AC}}\,v_{_{\rm SIG}}'(u\ssi - u\ssj) \,\hat{\bf r}\ssij\cdot\overline{\nabla\ssi W}\ssij\,,
\label{EQN:MON97ENERGYDISS}
\end{eqnarray}
where $\alpha_{_{\rm AV}}$ and $\alpha_{_{\rm AC}}$ are user specified coefficients of order unity, $v_{_{\rm SIG}}$ and $v_{_{\rm SIG}}'$ are signal speeds {\refone for artificial viscosity and conductivity respectively}, $\hat{\bf r}\ssij = {\bf r}\ssij/|{\bf r}\ssij|$ {\refone and $\overline{\nabla\ssi W}\ssij = \frac{1}{2}\,\left(\nabla\ssi W({\bf r}\ssij ,h\ssi) + \nabla\ssi W({\bf r}\ssij ,h\ssj) \right)$}.  {\refone For artificial viscosity, we use $v_{_{\rm SIG}} = c\ssi + c\ssj - \beta_{_{\rm AV}}\,{\bf v}\ssij \cdot \hat{\bf r}\ssij$ and $\beta_{_{\rm AV}} = 2\,\alpha_{_{\rm AV}}$.  If using artificial conductivity, we use the signal speeds defined by \citet{Price2008}, $v_{_{\rm SIG}}' = \sqrt{|P\ssi - P\ssj|/\overline{\rho}\ssij}$ and \citet{Wadsley2008}, $v_{_{\rm SIG}}' = |{\bf v}\ssij\cdot\hat{\bf r}\ssij|$.}  We consider two different forms of the mean density, the arithmetic mean, $\overline{\rho} = \frac{1}{2}\left( \rho\ssi + \rho\ssj \right)$, and the harmonic mean, $\overline{\rho} = 2 / \left[ (1/\rho\ssi) + (1/\rho\ssj) \right]$.

We use the Leapfrog kick-drift-kick integration scheme \citep[e.g.][]{Gadget2} to integrate all particle positions and velocities.  All other non-kinematic quantities are integrated in the same way as the velocity (i.e. time derivatives calculated on the full-step).  {\small SEREN} uses hierarchical block timestepping in tandem with the neighbour-timestep constraint \citep{SM2009} to minimise errors between neighbouring particles with large timestep differences.  {\small SEREN} uses a Barnes-Hut octal spatial decomposition tree \citep{BH1986} for efficient neighbour finding.

\section{Tests} \label{S:TESTS}
We have prepared a suite of tests which we will 
use to investigate the performance and relative merits and weaknesses of 
finite volume (uniform grid and AMR) and SPH.  We perform tests of 
(i) adiabatic, isothermal and cooling shocks (section \ref{SS:SHOCKTUBE}), 
(ii) the non-linear thin-shell instability (section \ref{SS:NTSI}), 
and (iii) the Kelvin-Helmholtz instability (section \ref{SS:KHI}). 
Details of each test, the initial conditions, the physics used and any other special additions will be discussed in each section before the results are presented and discussed.

\begin{figure*}
\centerline{
\includegraphics[width=6.2cm]{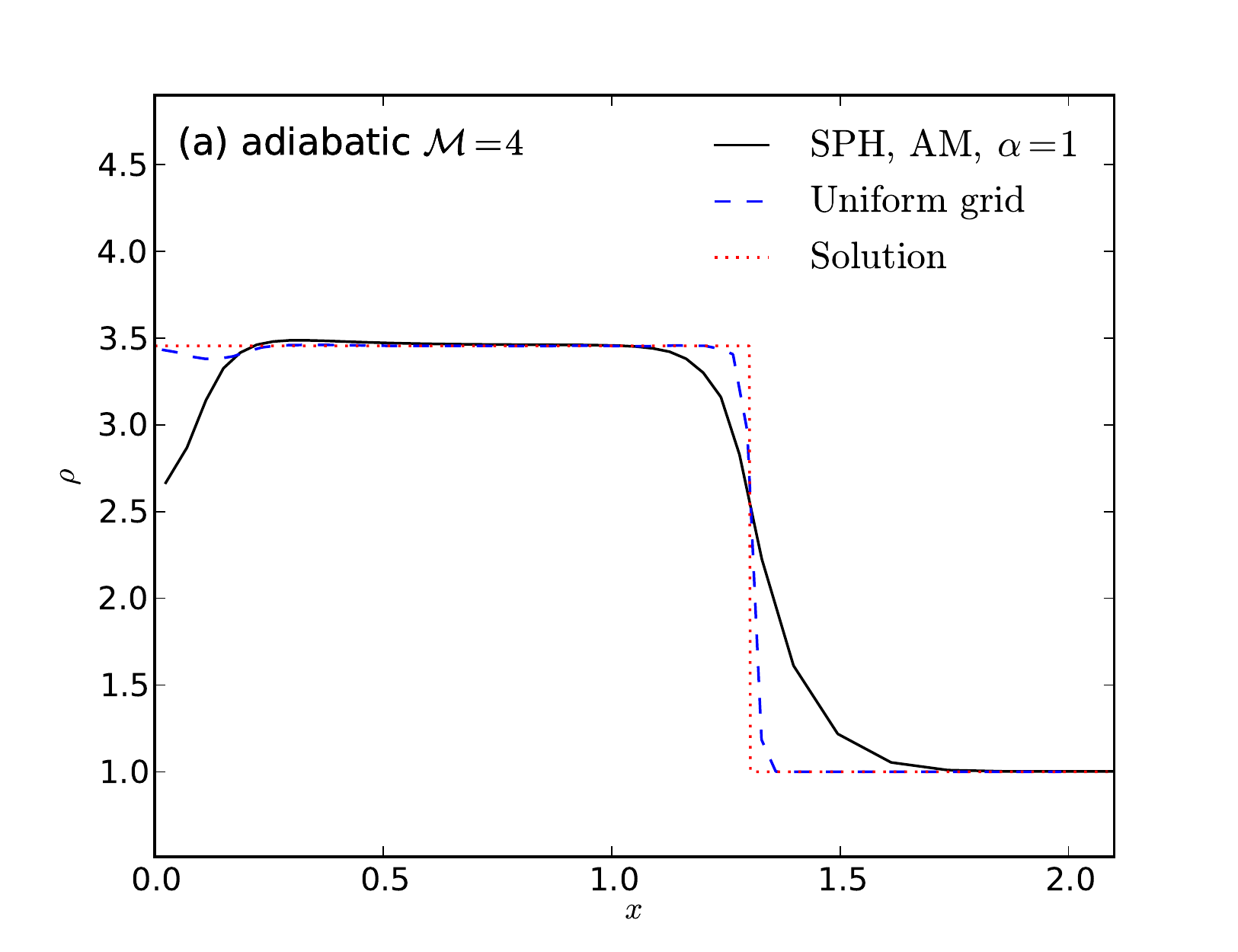}
\includegraphics[width=6.2cm]{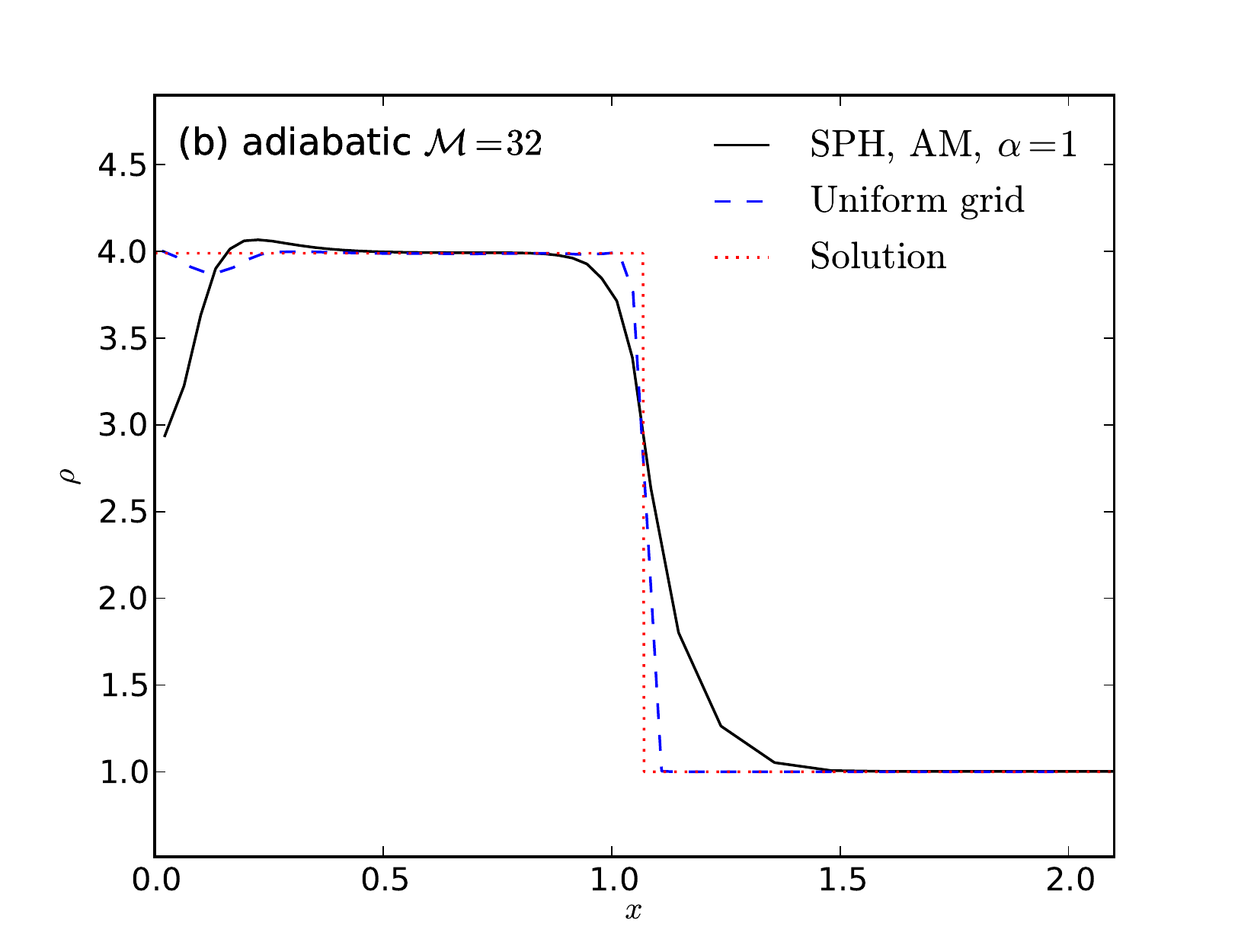}
\includegraphics[width=6.2cm]{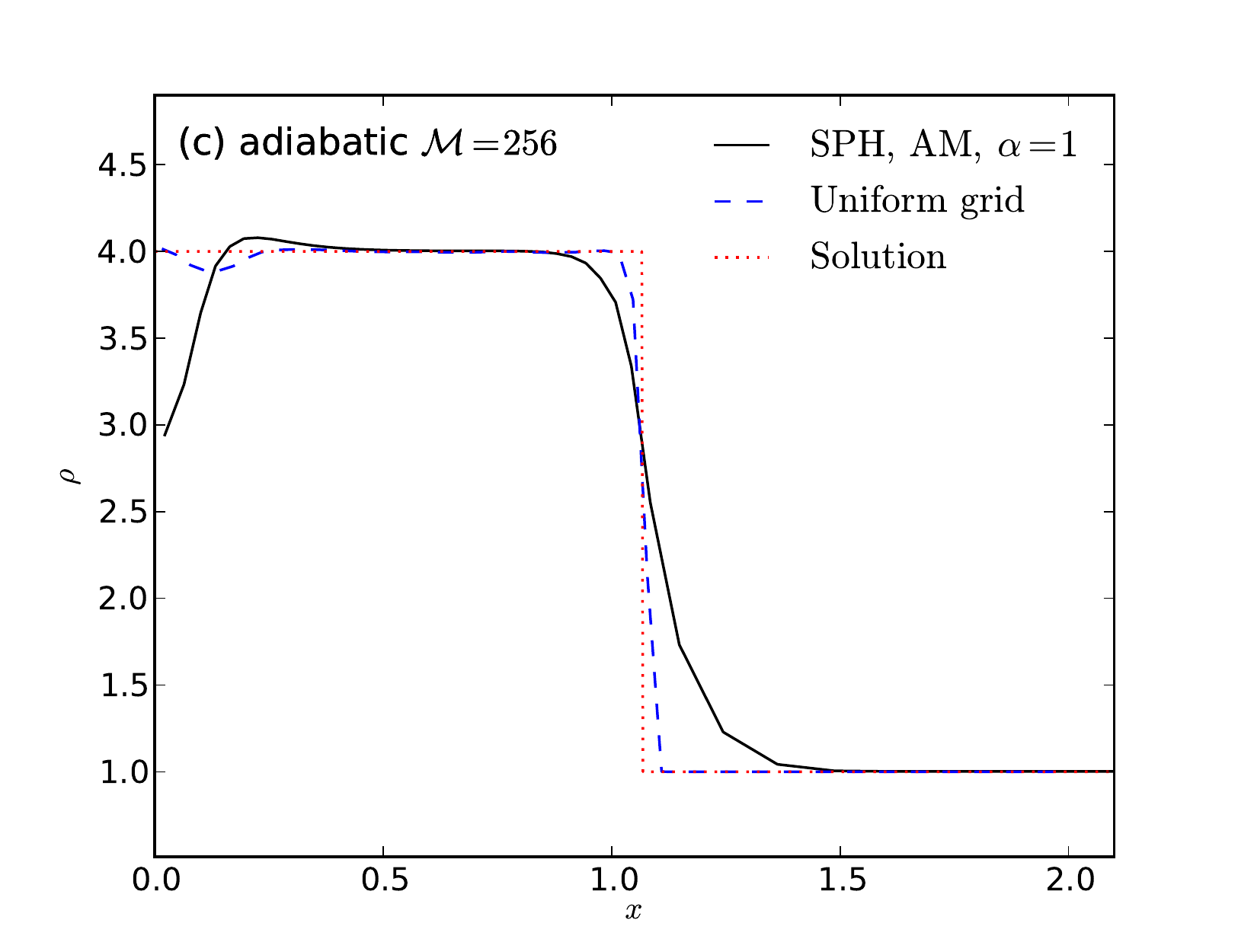}}
\vspace{0.04cm}
\caption{Density profiles of 1-D adiabatic shocks simulated with SPH and uniform grid for shocks with (a) ${\cal M}' = 4$ at $t = 0.8$, (b) ${\cal M}' = 32$ at $t = 0.1$, (c) ${\cal M}' = 256$ at $t = 0.0125$.  All SPH simulations using the arithmetic mean viscosity are performed with $\alpha_{_{\rm AV}} = 1$. Also plotted are the solutions from a Riemann solver.}
\label{FIG:ADSHOCK}
\end{figure*}

\subsection{Shock tube tests} \label{SS:SHOCKTUBE}
A simple, but demanding,  shock tube test is one in which 
uniform-density flows collide supersonically to produce a dense
shock layer.  
Despite  their importance  in  astrophysical simulations,  comparisons
between finite volume and SPH codes in simple shock-capturing problems
have  not  received  as  much  attention  as  other more  complicated
hydrodynamical  processes.  \citet{Tasker2008}  looked  at the  Sod
shock  tube problem,  both parallel  and diagonal  to  the grid.
\citet{Creasey2011} have performed detailed comparisons between
finite  volume and  SPH  in cooling  shocks  and derived  resolution
criteria in both  cases for resolving the cooling  region.  
Comparisons of isothermal shocks using finite volume and SPH
codes have been  made in the context of  driven, isothermal turbulence
\citep{Kitsionas2009,PF2010}.

We consider three types of shock; 
(i) adiabatic shocks, 
(ii) strictly isothermal shocks, and
(iii) cooling shocks.
These three cases cover the most important types of shocks 
modeled in numerical astrophysics.
For isothermal and adiabatic shocks, the solutions for the 
post-shock properties can be obtained via
the Rankine-Hugoniot conditions
\citep[e.g.][See Table \ref{TAB:SHOCKS}]{Shore2007}.  
It is important to note the different behaviour of 
isothermal and adiabatic shocks.  Adiabatic shocks have a
maximum density compression ratio, no matter how high the Mach-number
of the shock is, whereas the sound speed of the post-shock gas can
increase without limit.  Isothermal shocks, however, have a constant
sound speed, due to the imposed isothermal equation of state, but have
no limit on the post-shock density.

For cooling shocks, the initial post-shock state follows that of the
adiabatic shock, but as the shock cools towards the equilibrium
temperature, the post-shock properties tend towards those of the
isothermal shock.  We chose a simple linear cooling law of the form 
\begin{equation} \label{EQN:COOLING}
\frac{du}{dt}_{_{\rm COOL}} = -A \left( T - T_{_{\rm EQ}} \right)
\end{equation}
where $A$ is the cooling rate constant, $u$ is the specific internal
energy of the particle or cell, and $T_{_{\rm EQ}}$ is the equilibrium
temperature and $T = (\gamma - 1)\,u$ in dimensionless units.  
The solution for the shock structure is given in Appendix \ref{A:COOLSHOCK}.
We note that this is a slightly different cooling law to that considered 
by \citet{Creasey2011}.

\subsubsection{Initial conditions} \label{SSS:SHOCK-IC}

We set up two uniform density flows, each with density $\rho_0 = 1$, pressure $P_0 = 1$ and ratio of specific heats $\gamma = 5/3$.  The initial specific internal energy of the gas is $u = P_0 / \rho_0 / (\gamma - 1) = 3/2$ and the temperature is therefore $T = 1$.  We set the equilibrium temperature of the gas equal to the initial temperature of the gas, $T_{_{\rm EQ}} = 1$.  The initial velocity profile of the flows is of the form
\begin{equation}
v_x(x) =  
\begin{cases}
+\,{\cal M}'c_s\,, &\;\;\;  x < 0 \\
-\,{\cal M}'c_s\,, &\;\;\;  x > 0
\end{cases}
\end{equation}
where $\cal M'$ is the ratio of the inflow velocity to the isothermal
sound speed, $c_s = 1$.  
We note that ${\cal M}'$ is {\it not} the Mach number of the shock.
The true Mach number, $\cal M$, is the ratio of the inflow speed
relative to the shock front to the sound speed.   
Formally, the gas extends to infinity in both directions, but in
practice we use a finite box size that is long enough to allow enough
gas to form the shock before we terminate the simulation.  
Also, we only model the gas for $x > 0$ and use mirror boundary 
conditions at $x = 0$ exploiting the symmetry of the problem to reduce the 
computational effort. 

Due to the scale-free nature of the isothermal and adiabatic shock
simulations, there is no benefit in performing a resolution test with
different numbers of grid cells or particles.  However, in the
cooling-shock simulations, there is a typical length scale, i.e. the
size of the cooling region, which we may need to resolve to obtain
convergence.  Therefore, we will perform a convergence test of the
cooling shock with a range of different resolutions.

For the finite volume simulations, we perform simulations of 
(i) adiabatic shocks with ${\cal M}' = 2$, $8$ and $32$, 
(ii) isothermal shocks with ${\cal M}' = 4$, $8$, $16$ and $32$ using 
both 1st and 2nd order, and 
(iii) cooling shocks with ${\cal M}' = 32$ with the cooling parameter 
$A = 256$ and both 1st and 2nd order.  

For the SPH simulations, we perform simulations of 
(i) adiabatic shocks with ${\cal M}' = 2$, $8$ and $32$ using 
artificial viscosity with $\alpha_{_{\rm AV}} = 1$, 
(ii) isothermal shocks with ${\cal M}' = 4$, $8$, $16$ and $32$ using 
$\alpha_{_{\rm AV}} = 1$ and $2$, and 
(iii) cooling shocks with ${\cal M}' = 32$ with the cooling parameter 
$A = 256$, $1024$ and $4096$ using $\alpha_{_{\rm AV}} = 1$ and $2$. 
We perform all simulations using the M4 spline kernel, and using the 
\citet{Monaghan1997} artificial viscosity 
with both the arithmetic mean and harmonic mean density.  
We smooth the initial velocity profile near the flow-interface for 
consistency with the later evolution of the velocity, which will 
itself be naturally smoothed due to the acceleration profile 
being smooth at the shock interface.  
The smoothed velocity is calculated using  
\begin{equation} \label{EQN:SMOOTHV}
{\bf v}'\ssi = \frac{1}{\rho\ssi}\sum \limits_{j=1}^{N} 
{ m\ssj\,{\bf v}\ssj\,W({\bf r}\ssij,h\ssi)}\,.
\end{equation}

\begin{figure*}[width=4.6cm]
\centerline{
\includegraphics[width=4.6cm]{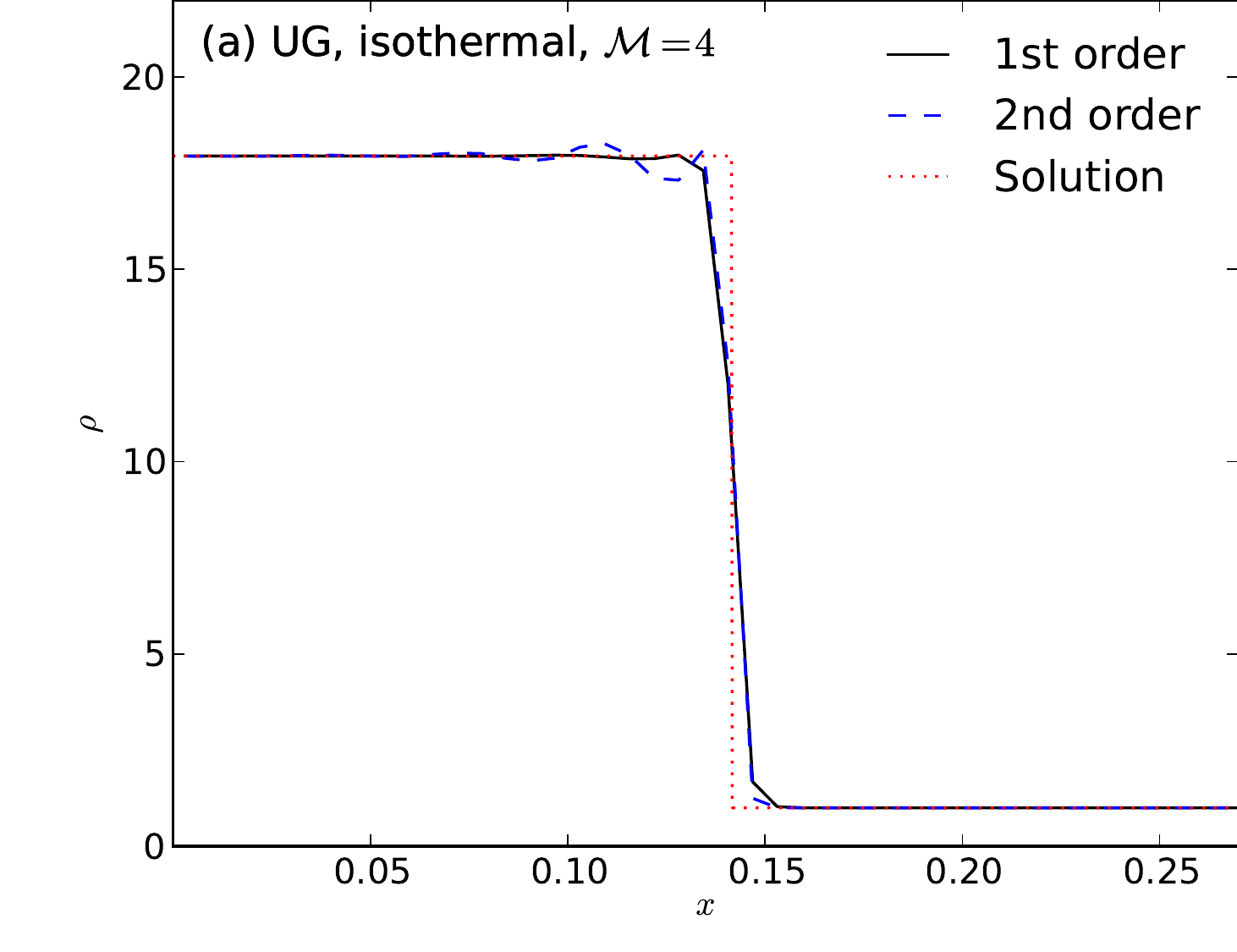}
\includegraphics[width=4.6cm]{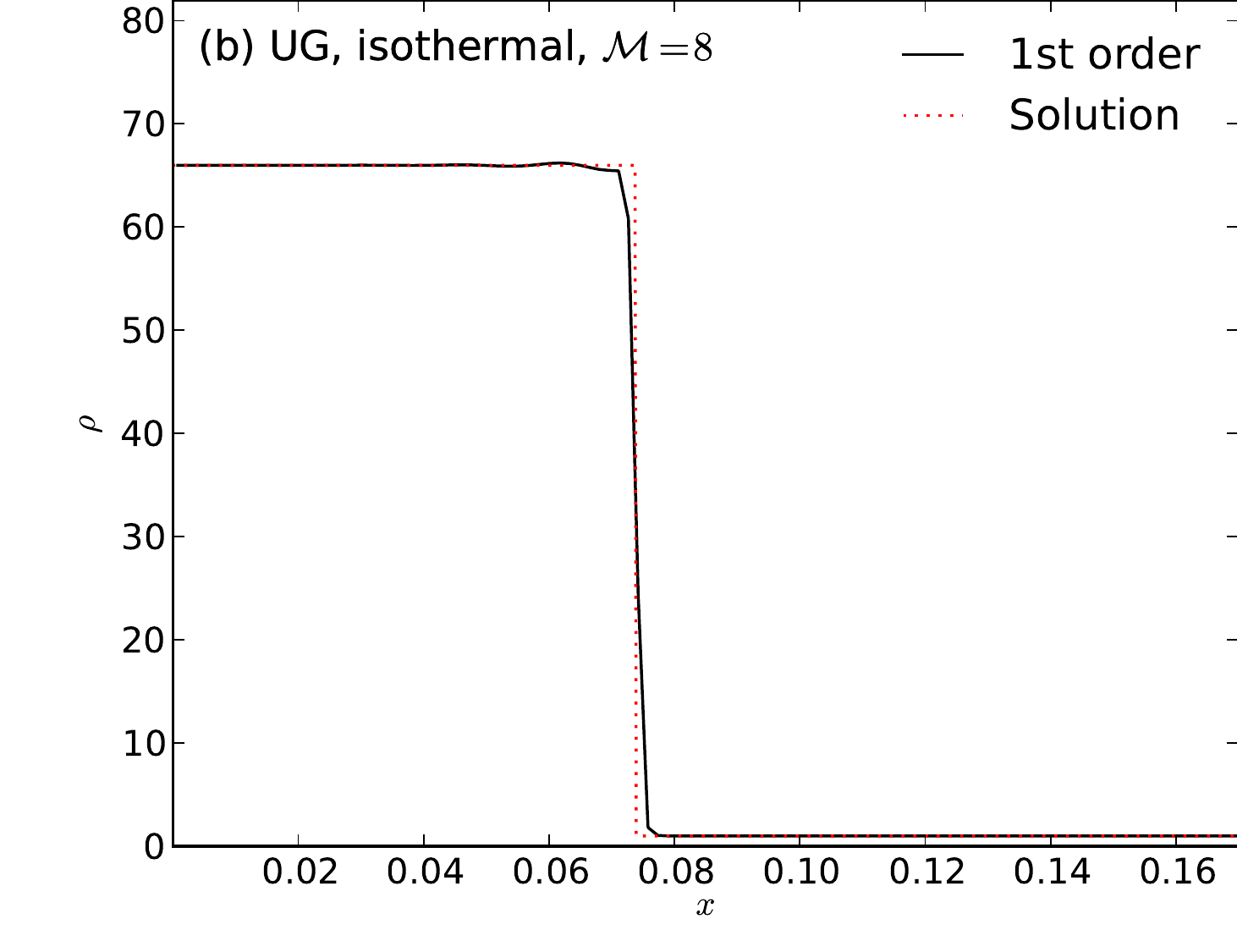}
\includegraphics[width=4.6cm]{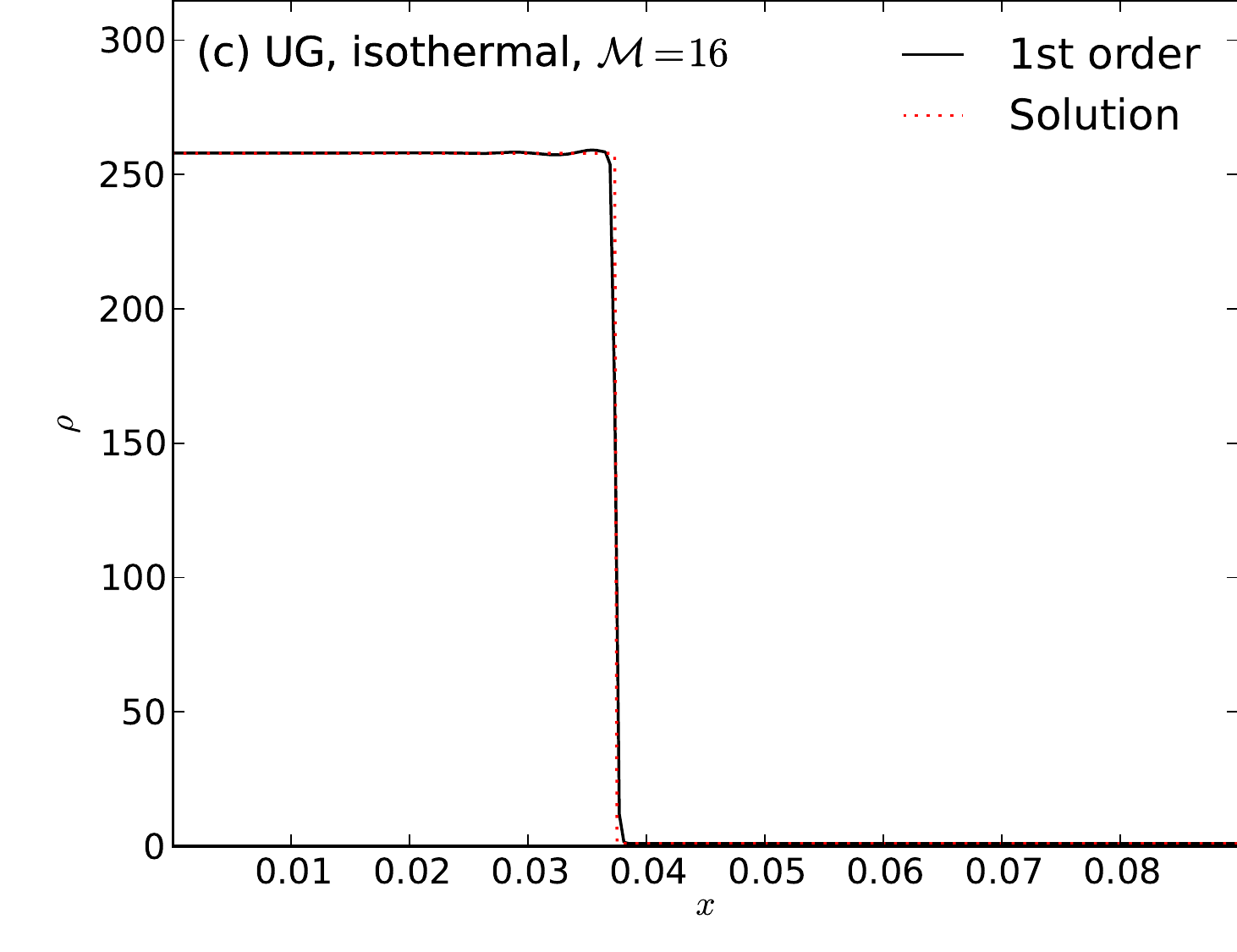}
\includegraphics[width=4.6cm]{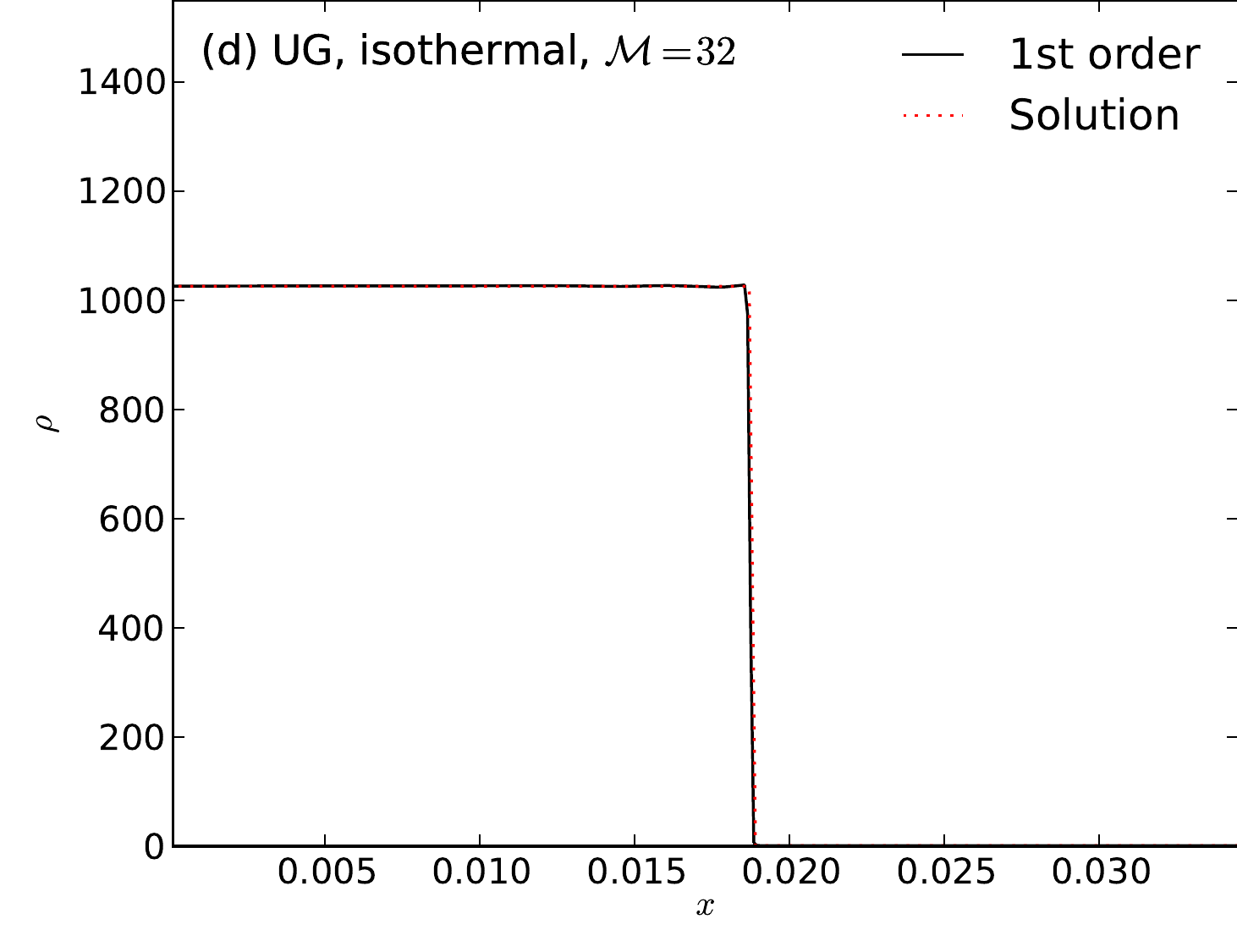}}
\centerline{
\includegraphics[width=4.6cm]{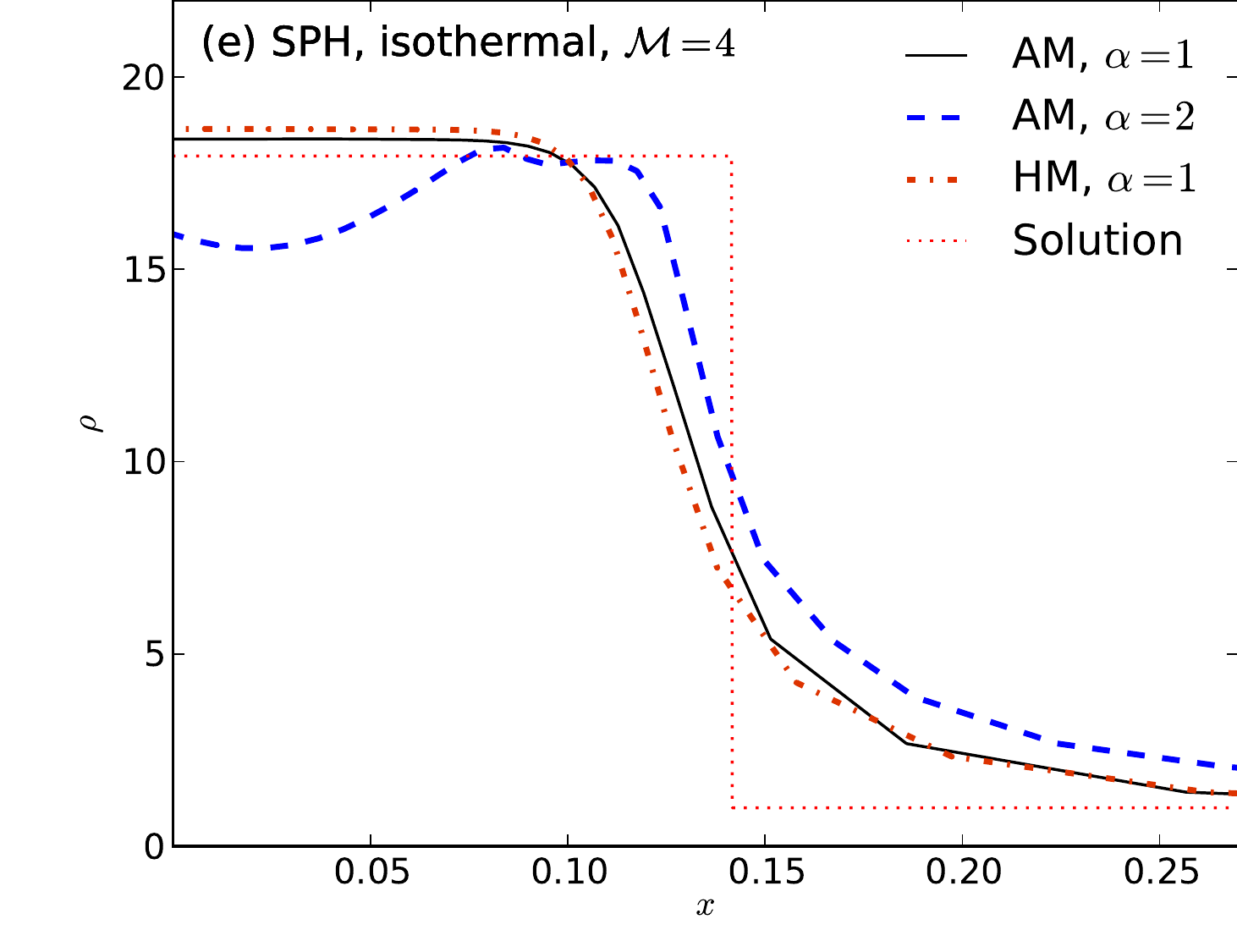}
\includegraphics[width=4.6cm]{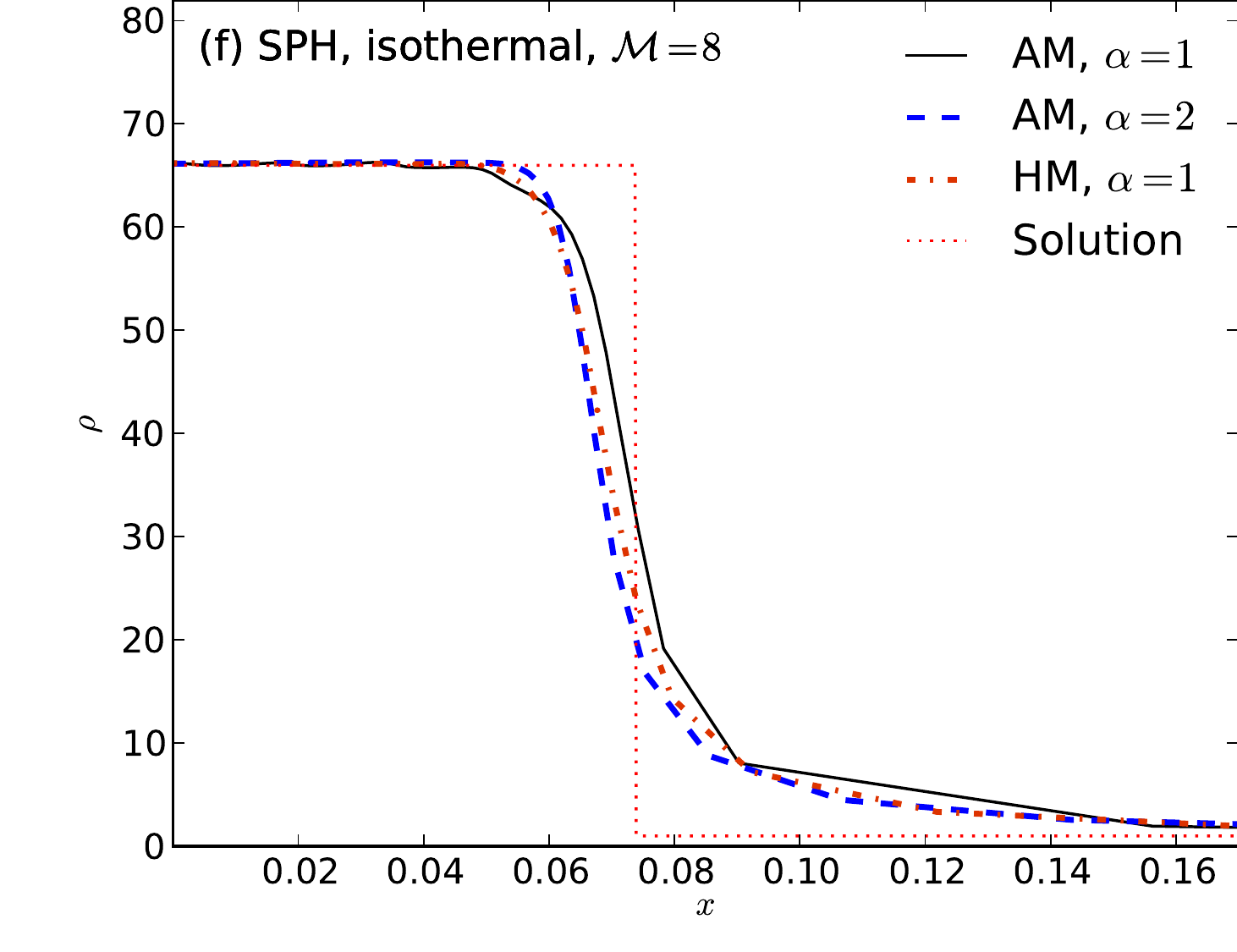}
\includegraphics[width=4.6cm]{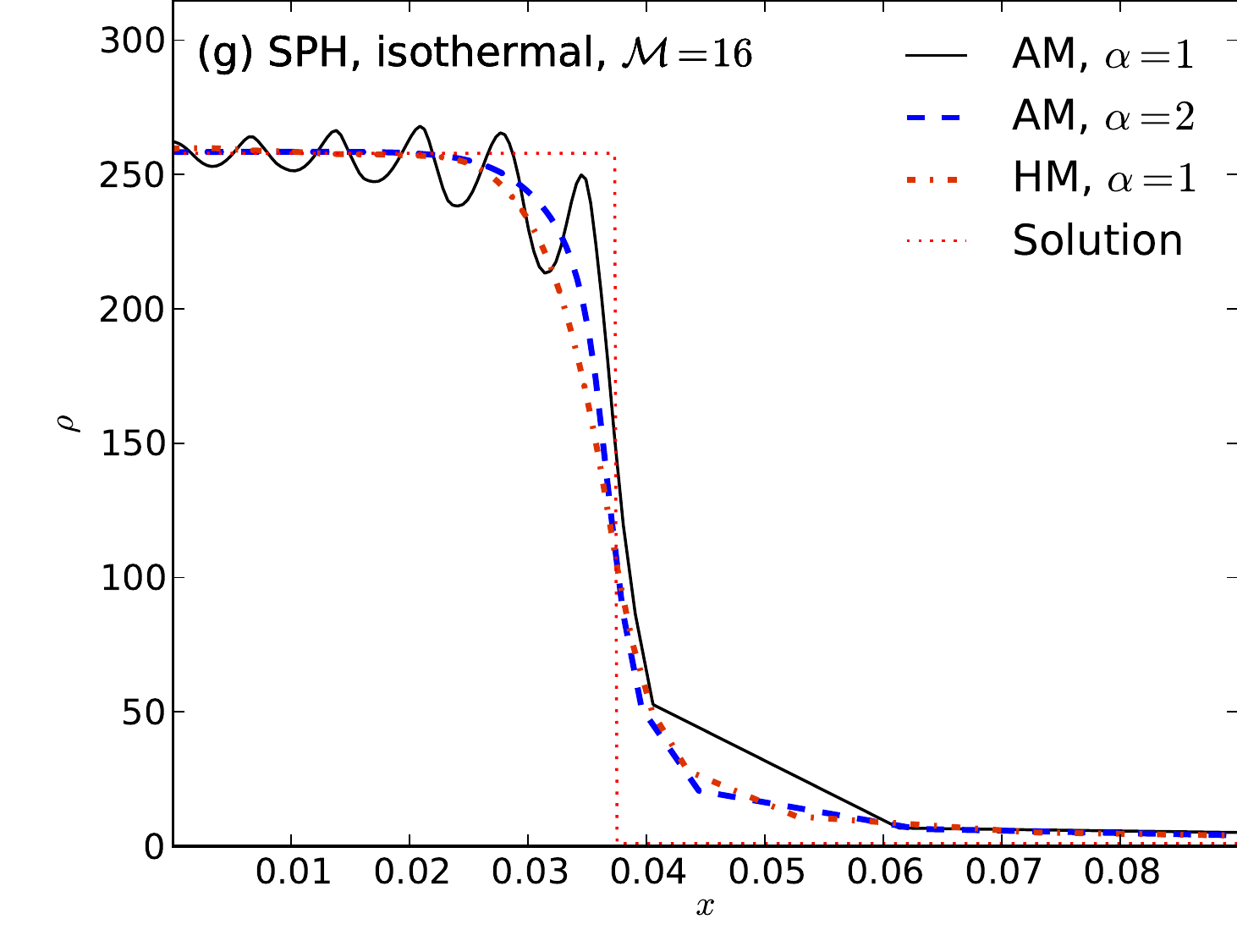}
\includegraphics[width=4.6cm]{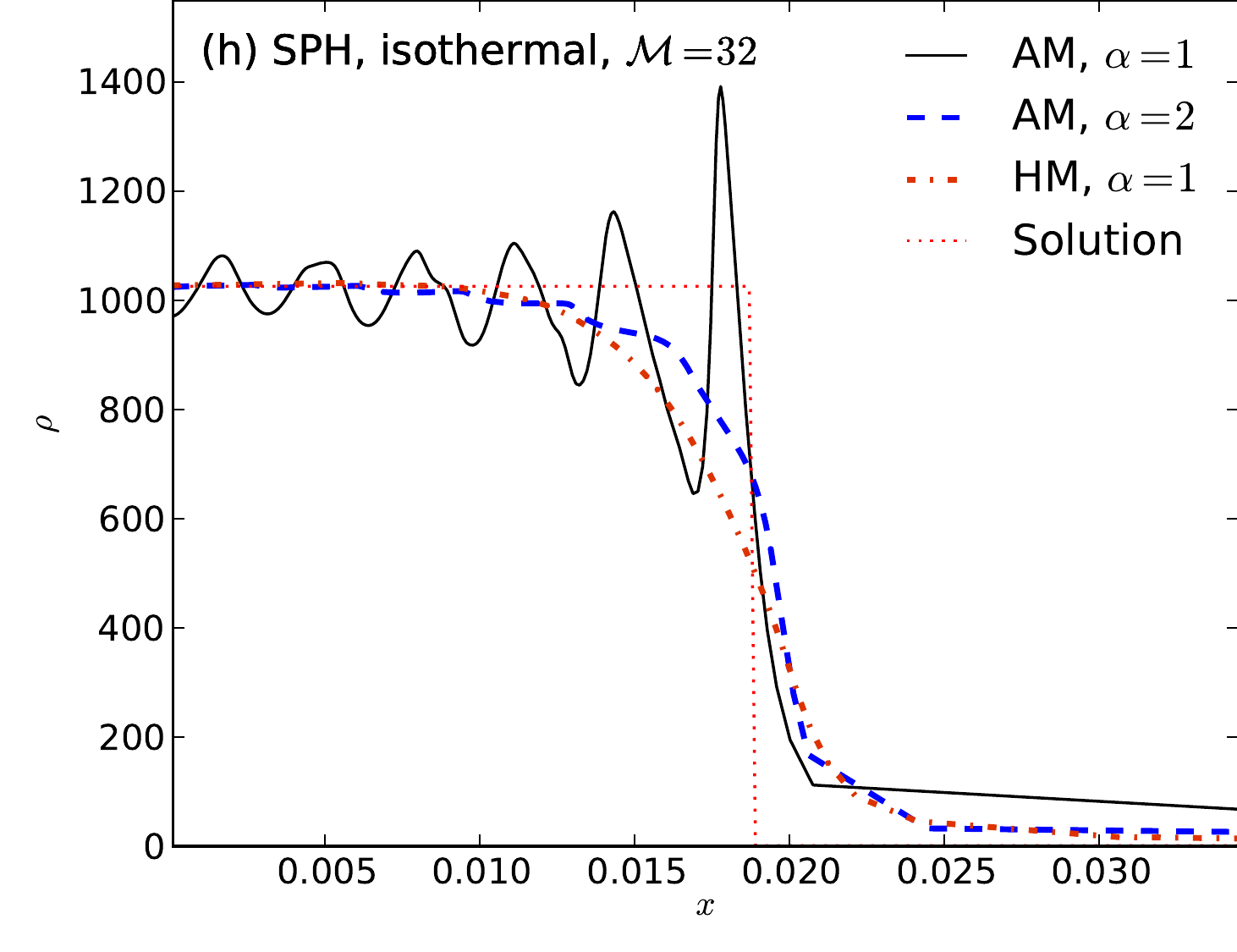}}
\vspace{0.04cm}
\caption{Density profiles of 1-D isothermal shocks simulated with uniform mesh finite volume at a time $t = 0.6$ for (a) ${\cal M}' = 4$, (b) ${\cal M}' = 8$, (b) ${\cal M}' = 16$, (c) ${\cal M}' = 32$, and with SPH for (e) ${\cal M}' = 4$, (f) ${\cal M}' = 8$, (g) ${\cal M}' = 16$, (h) ${\cal M}' = 32$.  The simulations are performed with both a 1st and 2nd order Riemann solver for ${\cal M}' = 4$, but only 1st order at higher values of ${\cal M}'$.  All SPH simulations using the arithmetic mean viscosity are performed with both $\alpha = 1$ and $2$, whereas the harmonic mean simulations are performed only with $\alpha = 1$. Also plotted are the solutions from an exact Riemann solver.}
\label{FIG:ISOSHOCK}
\end{figure*}

\subsubsection{Adiabatic shocks} \label{SSS:ADIABATIC-SHOCK}
We compute adiabatic shocks with both codes using (a) ${\cal M}' = 4$, (b) ${\cal M}' = 32$, and (c) ${\cal M}' = 256$ for fluids with $\gamma = 5/3$.  We use a uniform grid spacing of $\Delta x = 1/32$ for the finite-volume simulations and an initial particle spacing of $\Delta x = 1/8$ for the SPH simulations, which gives similar resolutions in the shocked region.   Figure \ref{FIG:ADSHOCK} shows the density profile of these three cases.   The finite voume simulations accurately capture the shock and describe the correct density profile, with only a small wall-heating effect near $x = 0$.  One benefit of many Finite-Volume codes is the use of a Riemann solver which is designed to model  shocks correctly.  The Rankine-Hugoniot conditions (Table \ref{TAB:SHOCKS}) show that for high-Mach numbers, the maximum compression ratio is $(\gamma + 1)/(\gamma - 1) = 4$ for $\gamma = 5/3$.  Therefore, the size of the shocked-region grows quickly reducing any problem with the initial shock.   Overall, finite-volume codes also have no trouble capturing adiabatic shocks, regardless of the Mach number.

Figure \ref{FIG:ADSHOCK} shows that SPH using the standard \citet{Monaghan1997} artificial viscosity with $\alpha_{_{\rm AV}} = 1$ is capable of capturing adiabatic shocks for all the tested Mach numbers with no sign of any post-shock oscillations.  There is a more prominent wall-heating effect than mesh codes near $x = 0$, but this is the only undesirable numerical artifact with all other features of the shock well-modelled.  We also notice that the commonly used M4 spline kernel is sufficient to capture the shock, even for steep velocity gradients.  

{\refone One noticeable difference between the SPH and mesh simulations (for all shocks modelled) is the larger broadening around the discontinuity in the SPH shocks.  Although artificial dissipation plays a role in smoothing the discontinuity, another reason for this is the large transition in the smoothing length between the pre-shock and the post-shock regions.  This is particularly true in 1D simulations where $h \propto \rho^{-1}$ (in comparison to 3D simulations where the $h \propto \rho^{-1/3}$).  Mesh codes on the other hand can have uniform resolution (for uniform grid codes) on either side of the shock and therefore broaden the shock uniformlly over a few grid cells.}

\begin{figure*}
\centerline{
\includegraphics[width=6.2cm]{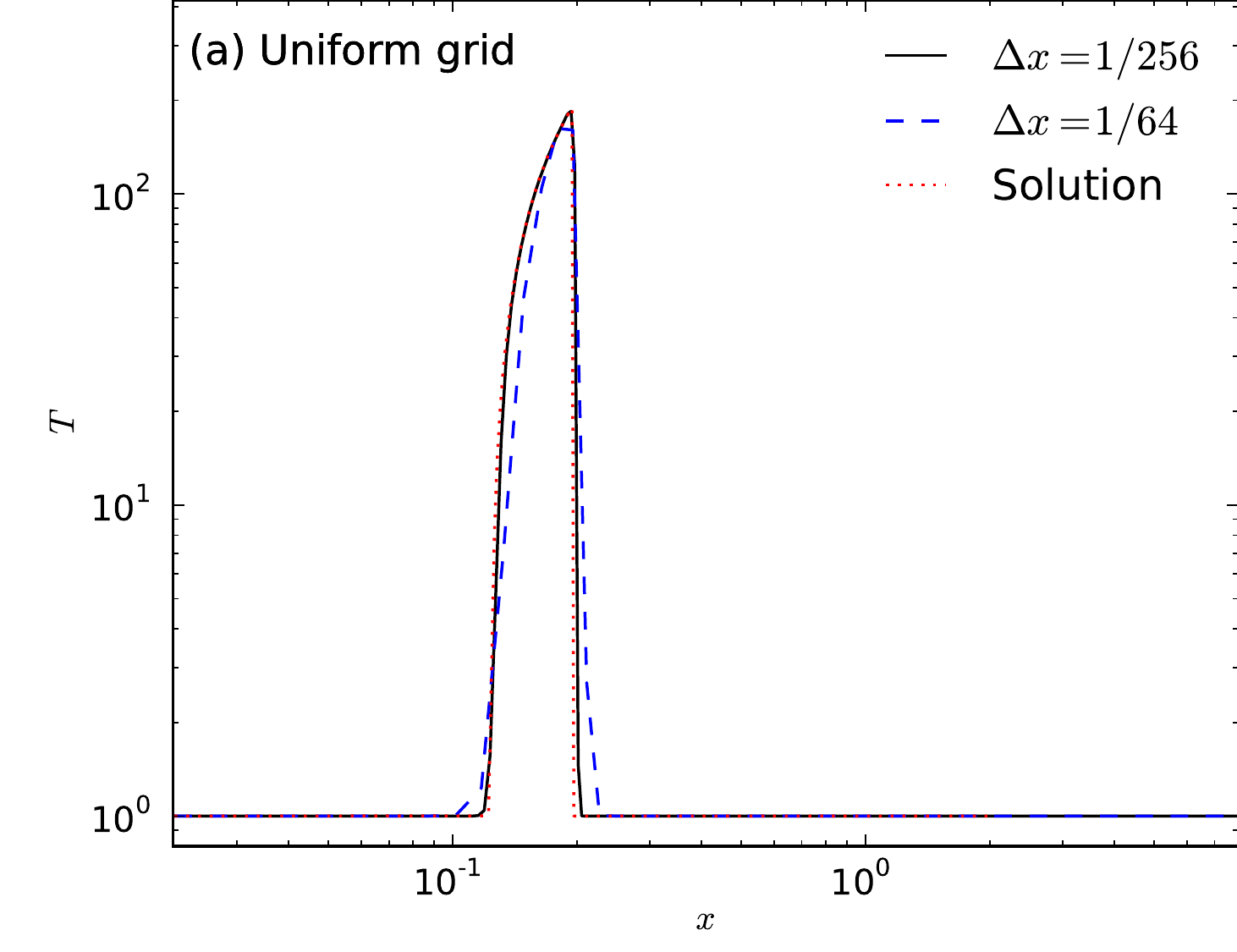}
\includegraphics[width=6.2cm]{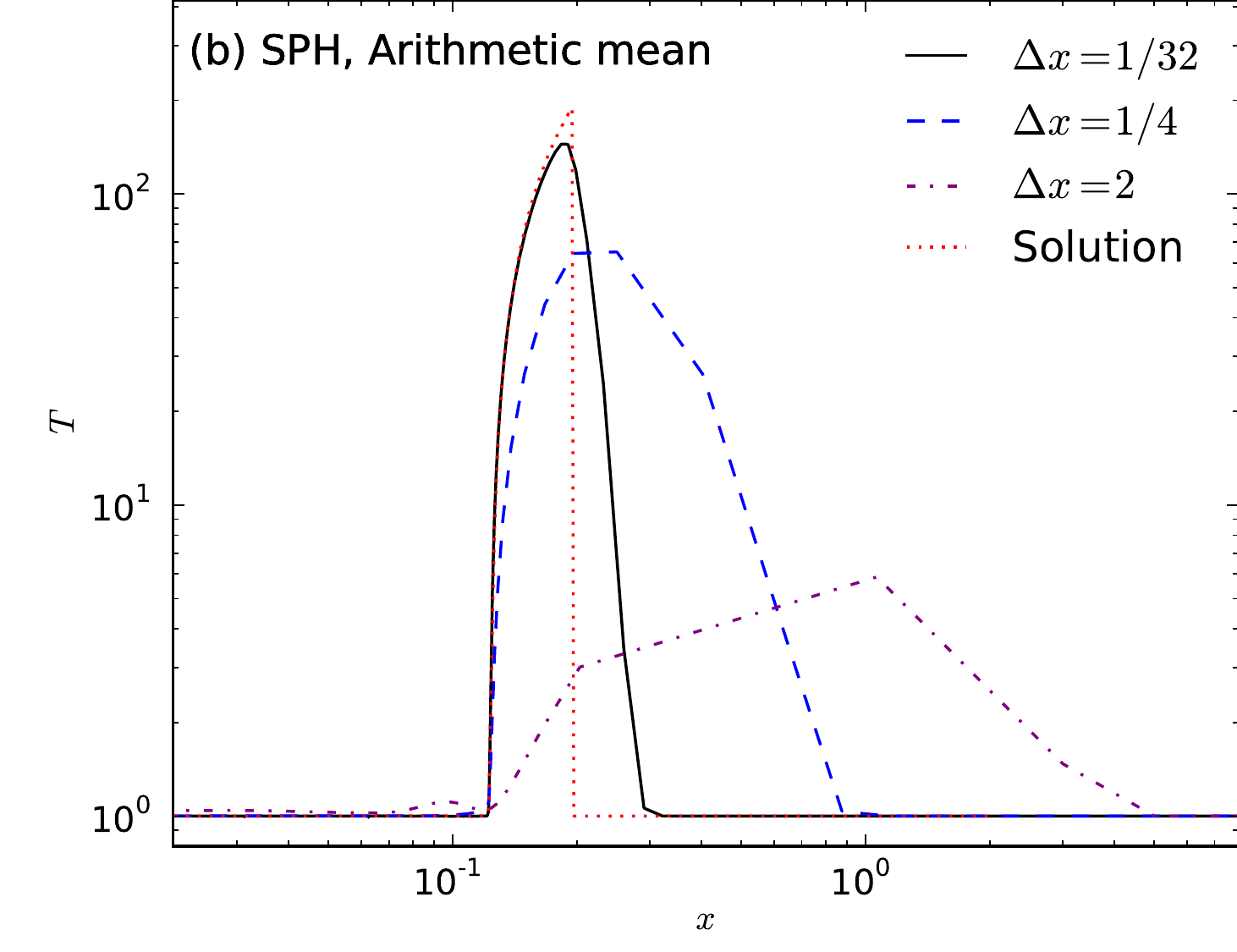}
\includegraphics[width=6.2cm]{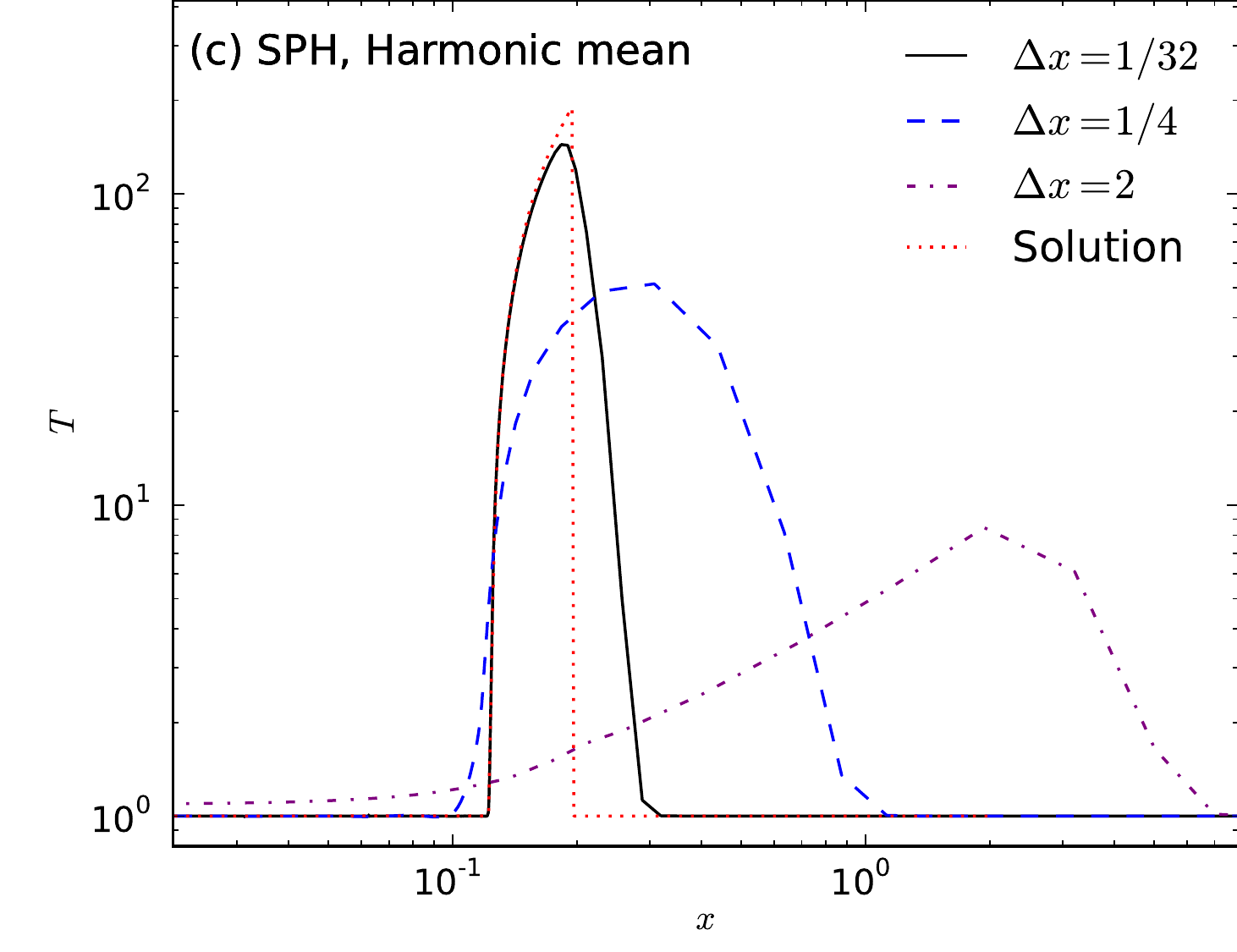}}
\centerline{
\includegraphics[width=6.2cm]{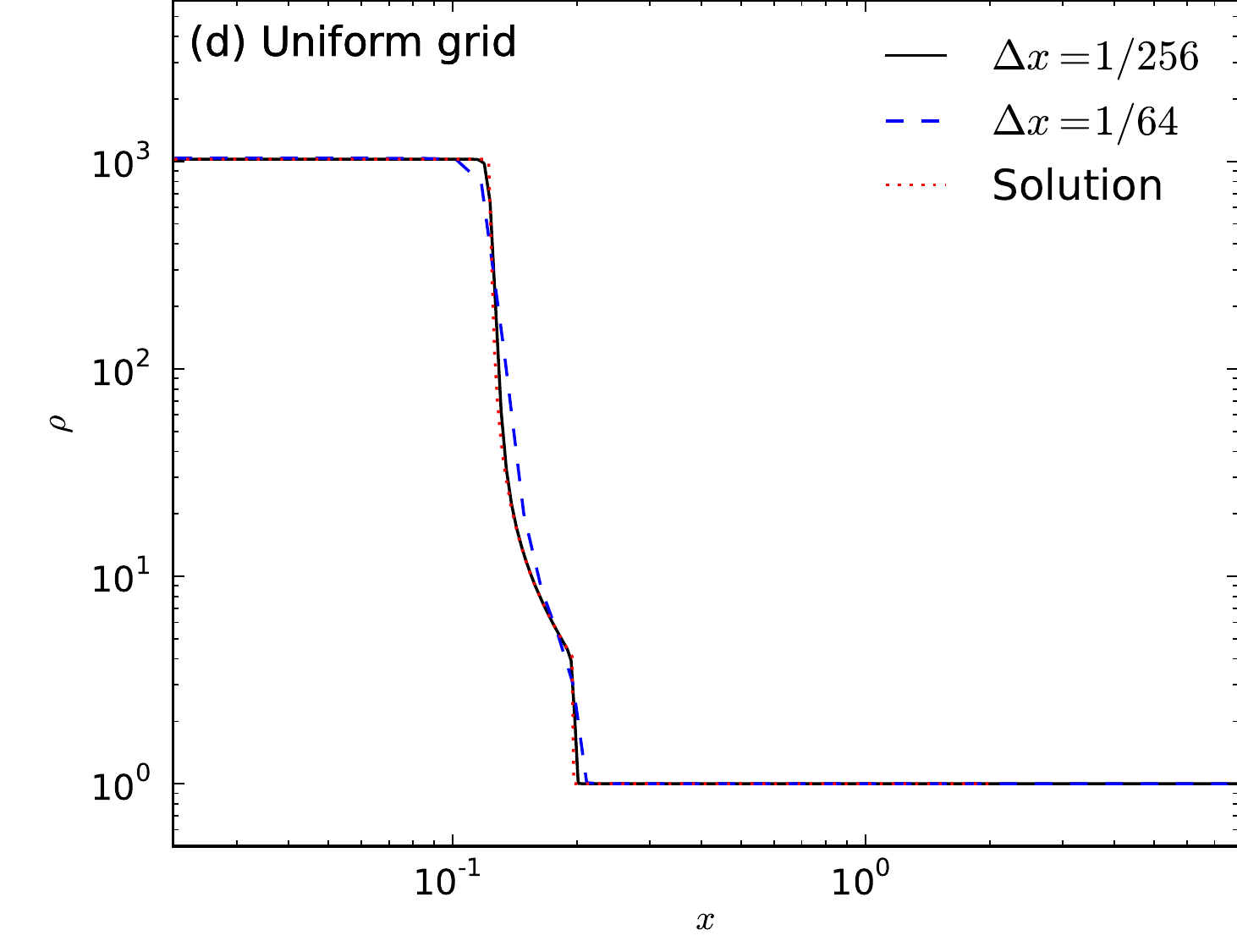}
\includegraphics[width=6.2cm]{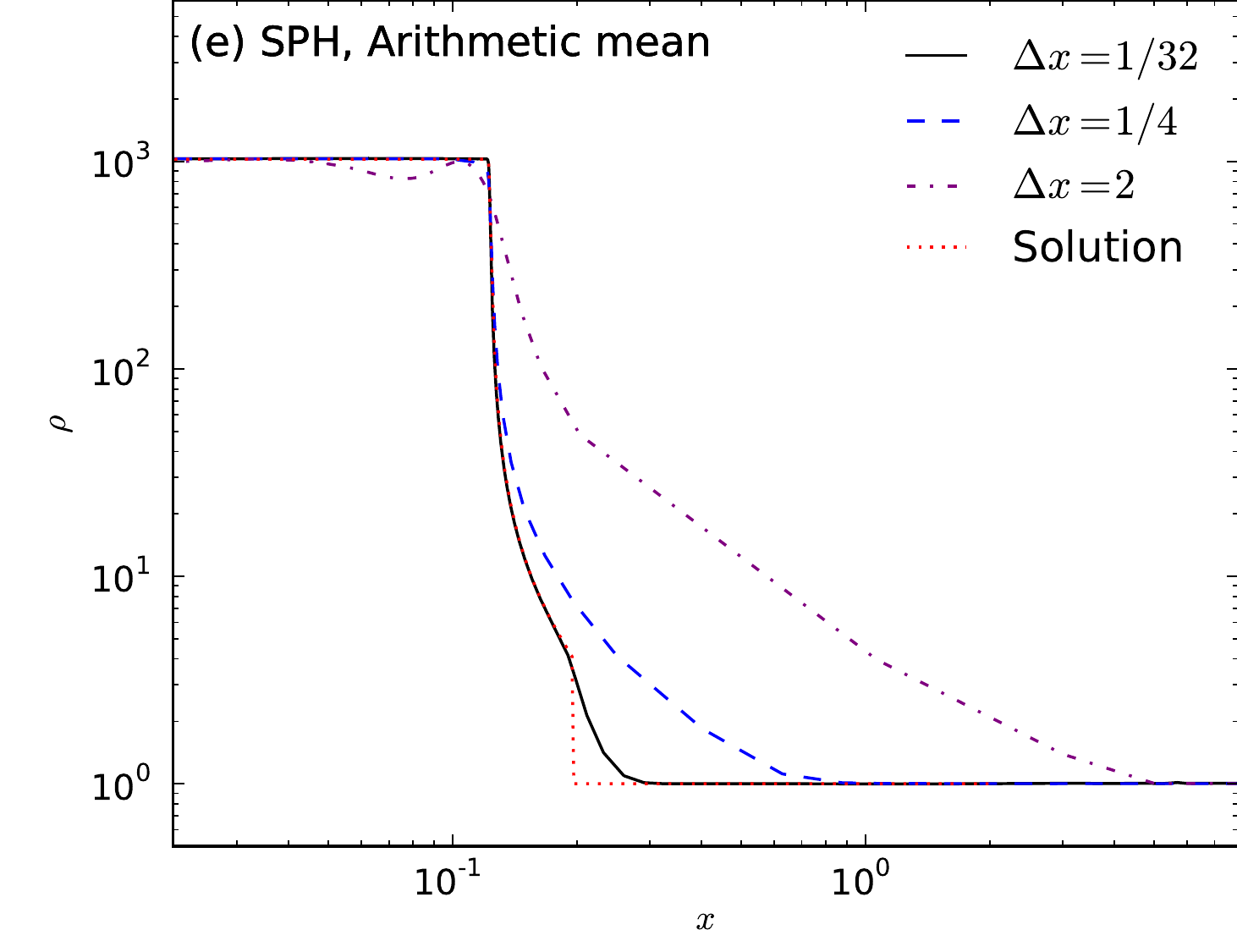}
\includegraphics[width=6.2cm]{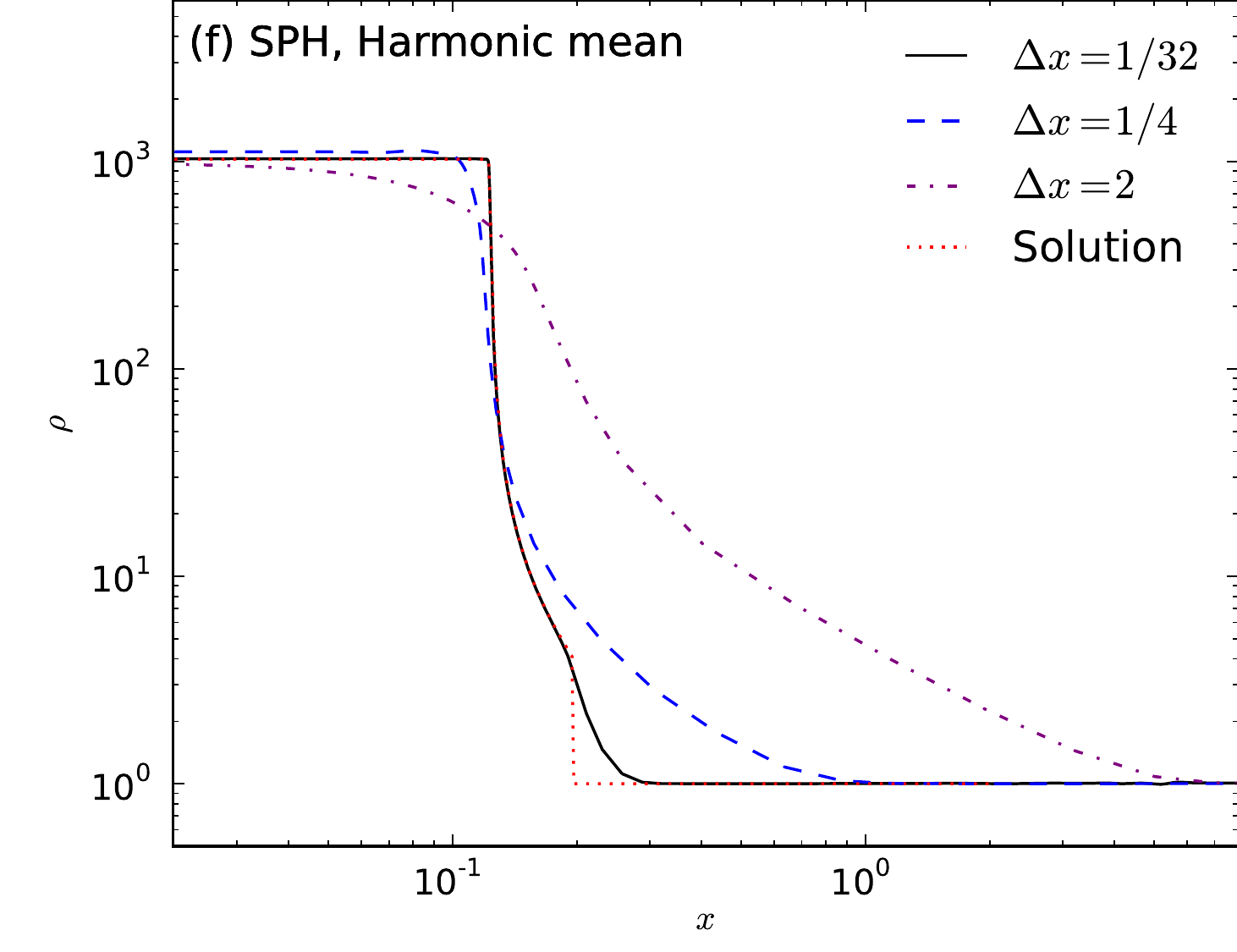}}
\vspace{0.04cm}
\caption{Simulations of cooling shocks with ${\cal M}' = 32$ with uniform grid and SPH at $t = 4$; Shock temperature using {\refone (a) Uniform grid, (b) SPH with the arithmetic mean density, and (c) SPH using the harmonic mean density; Shock density profiles using (d) Uniform grid, (e) SPH with the arithmetic mean density, and (f) SPH using the harmonic mean density.}  The shock solution derived in Appendix \ref{A:COOLSHOCK} is also shown.}
\label{FIG:COOLSHOCK}
\end{figure*}

\subsubsection{Isothermal shocks} \label{SSS:ISOTHERMAL-SHOCK}
We perform  simulations of isothermal shocks using  both finite volume
and SPH with (a) ${\cal M}' = 4$, (b) ${\cal M}' = 8$, (c) ${\cal M}' = 16$, and (d) ${\cal
M}' = 32$.  For the SPH simulations, we use an initial particle spacing of 
$\Delta x = 1/10$.  For the grid simulations, we use $\Delta x = 1/160$, $1/250$, $1/500$ and $1/1000$ for ${\cal M}' = 4$,$8$,$16$ and $32$ respectively in order to match the inner-shock resolution in the SPH code.  Figure \ref{FIG:ISOSHOCK} shows the isothermal simulations for both the grid and SPH simulations.

For upwind finite volume codes one needs to use an isothermal Riemann solver to
capture isothermal shocks:  here we used a Riemann  solver provided by
O'Sullivan (Private  communication).  This works well  for shocks with
${\cal M}'  < 5$ (Fig \ref{FIG:ISOSHOCK}(a)),  but for stronger shocks (Fig \ref{FIG:ISOSHOCK}(b,c,d)) one  needs to go  to 1st
order and add an artificial viscous momentum flux of the form
\begin{equation} \label{EQN:VISCMOMFLUX}
f = \frac{1}{2} \alpha (\rho_l + \rho_r) |v_l - v_r| (v_l - v_r),
\end{equation}
where $\rho_l$, $\rho_r$, $v_l$, $v_r$  are left and  right densities
and velocities in the Riemann  problem and $\alpha$ is a parameter. We
find that $\alpha = 1$ works well even for very strong shocks (${\cal
M}' > 100$).
The reason for this is simply that the shock is moving slowly relative
to the grid and becomes very sharp if second order is used. One can
smear it out using the artificial viscous flux given by (\ref{EQN:VISCMOMFLUX}), but this requires a large value of $\alpha$ and the time-step must be
reduced.  Note this problem is much less severe if the shock is moving
at a reasonable speed relative to the grid.

For the SPH code, we perform simulations using the \citet{Monaghan1997} artificial viscosity with (a) the arithmetic mean of density with $\alpha_{_{\rm AV}} = 1$ and $2$, and (b) the harmonic mean of the density with $\alpha_{_{\rm AV}} = 1$.  In all cases, $\beta_{_{\rm AV}} = 2\,\alpha_{_{\rm AV}}$.  For weaker isothermal shocks (${\cal M'} \leq 8$), standard artificial viscosity with $\alpha_{_{\rm AV}} = 1$ is sufficient to capture the shock (Figure \ref{FIG:ISOSHOCK} (a)) {\refone with no noticeable sign of post-shock oscillations.}  
 Using $\alpha_{_{\rm AV}} = 2$ has little noticeable effect in this simulation, although in principle it can smooth out the discontinuity even further due to larger dissipation.  Using the harmonic mean viscosity yields very similar results to the arithmetic mean case.  For both the ${\cal M}' = 16$ and $32$ isothermal shocks using the arithmetic mean with $\alpha_{_{\rm AV}} = 1$, noticeable post-shock oscillations are present (Figures \ref{FIG:ISOSHOCK} (b)) which suggests that the artificial viscosity prescription is not adequate for capturing shocks.  Increasing the viscosity parameter to $\alpha_{_{\rm AV}} = 2$ allows the shock to be successfully captured.  Alternatively, using the harmonic mean allows the shocks in both to be captured successfully without increasing $\alpha_{_{\rm AV}}$ yielding similar results to the $\alpha = 2$ arithmetic mean case.

The issue of SPH viscosity failing to capture strong isothermal shocks has been suggested in several papers in the literature \citep[e.g.][]{PF2010,Hubber2011}, where {\refone values of the artificial viscosity parameters larger than the canonical values of $\alpha_{_{\rm AV}} = 1$ and $\beta_{_{\rm AV}} = 2$ are} required to capture strong shocks.  Our short study shows that this is only really an issue in strong isothermal shocks and could in part be down to the mathematical form of the mean density in the SPH artificial dissipation equations (Eqns \ref{EQN:MON97ARTVISC} \& \ref{EQN:MON97ENERGYDISS}).  The standard choice of $\rho = \frac{1}{2}\,\left(\rho_i + \rho_j \right)$ is motivated to ensure the added dissipation obeys conservation laws, such as conservation of momentum.  However, this form tends to reduce the effective artificial viscosity in shocks with high compressibility where there is a high-density contrast near the shock surface.  The harmonic mean reverses this by biasing the required viscosity to the lower-density component.  Since it is usually the low-density component of the inflow that must be `slowed-down' at the shock surface, it follows that it may be more prudent to bias the effective viscosity to the lower-density material.  Using the harmonic mean therefore allows standard artificial viscosity to capture highly compressible isothermal shocks.

\begin{figure*}
\centerline{
\includegraphics[width=4.35cm]{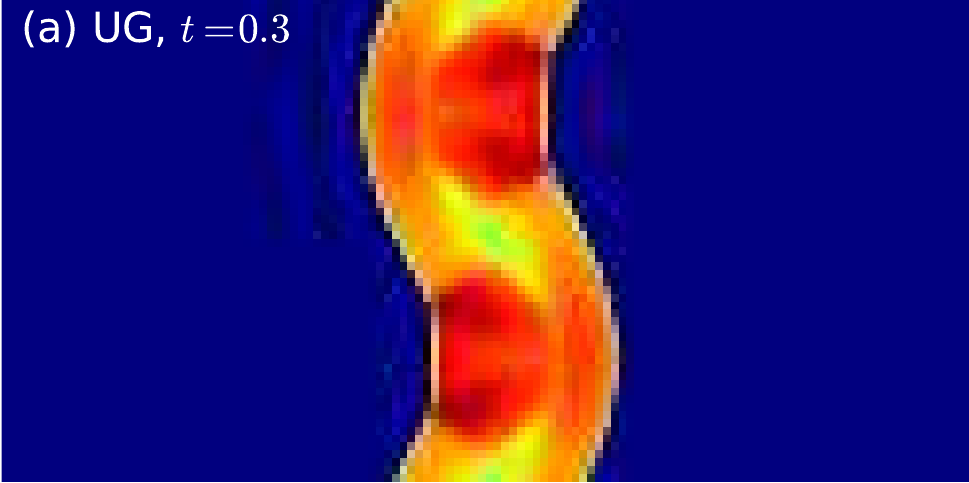}
\includegraphics[width=4.35cm]{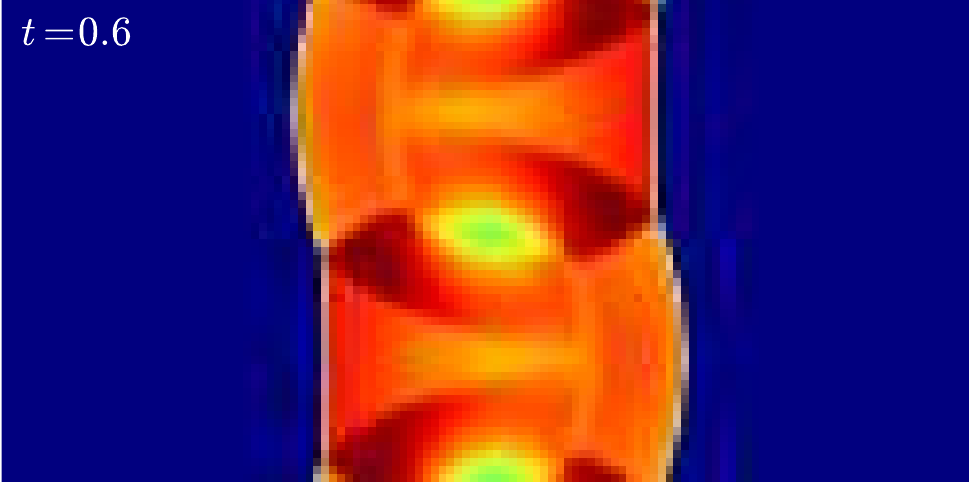}
\includegraphics[width=4.35cm]{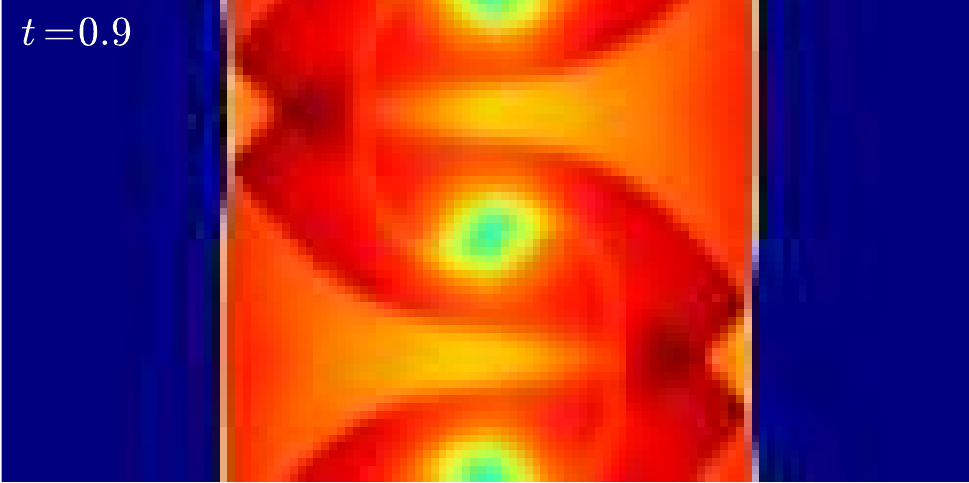}
\includegraphics[width=4.35cm]{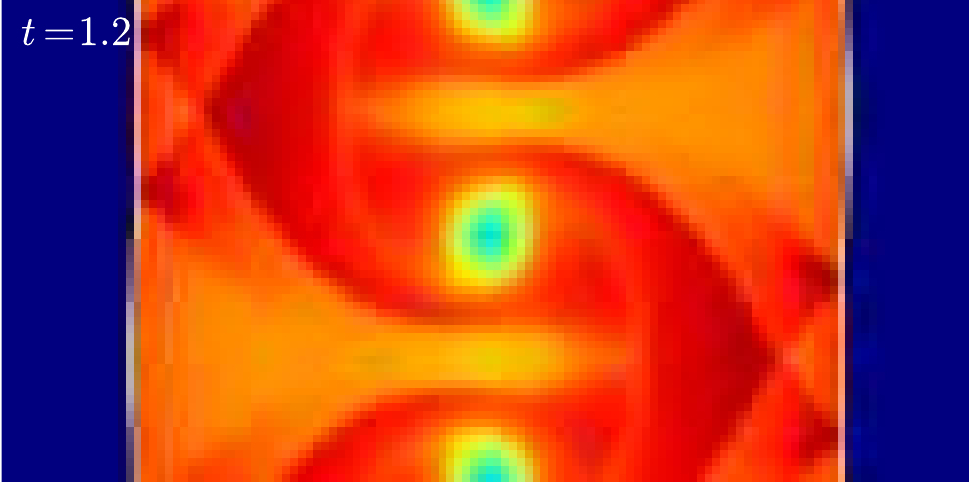}}
\vspace{0.04cm}
\centerline{
\includegraphics[width=4.35cm]{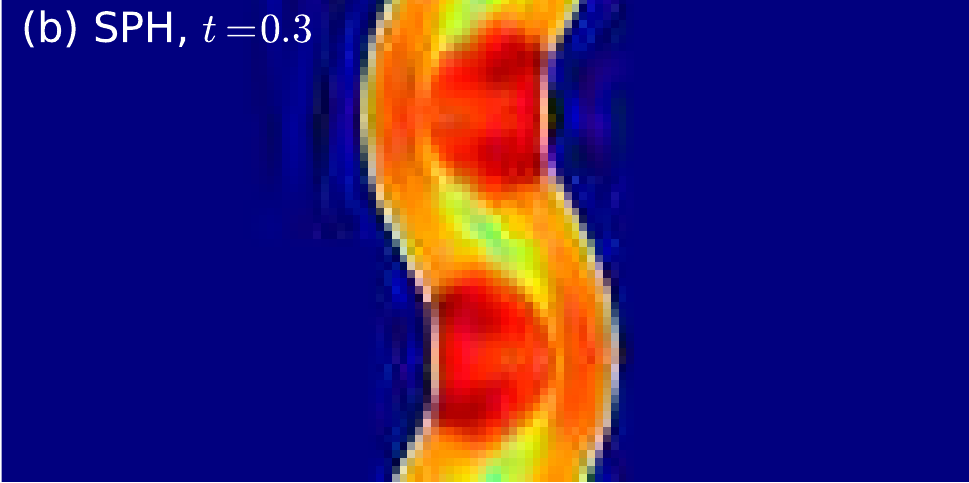}
\includegraphics[width=4.35cm]{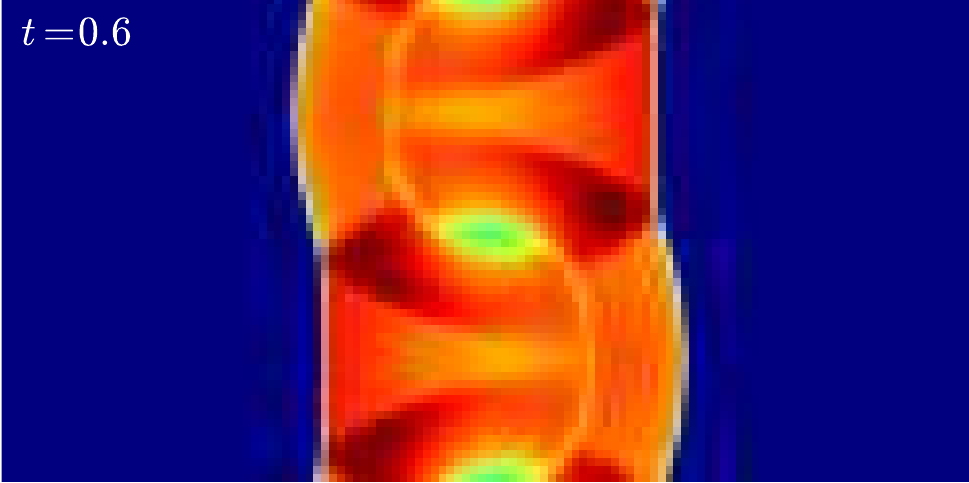}
\includegraphics[width=4.35cm]{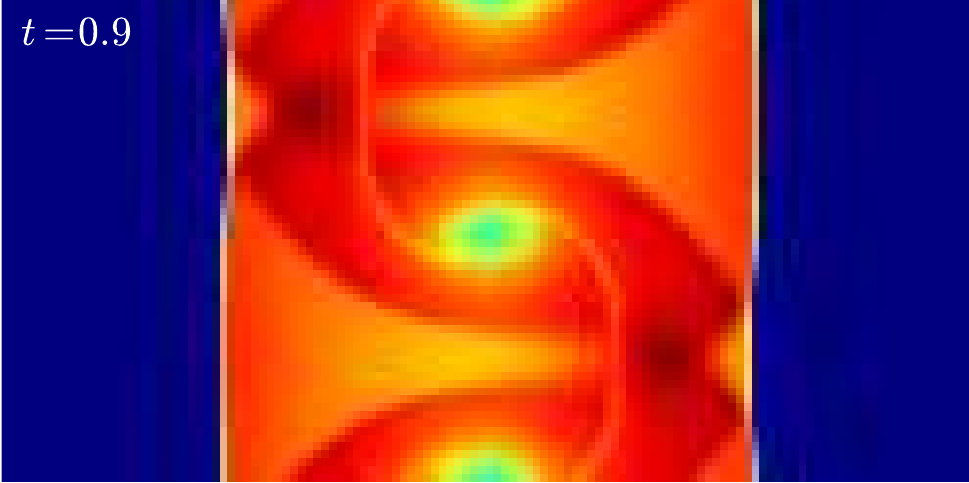}
\includegraphics[width=4.35cm]{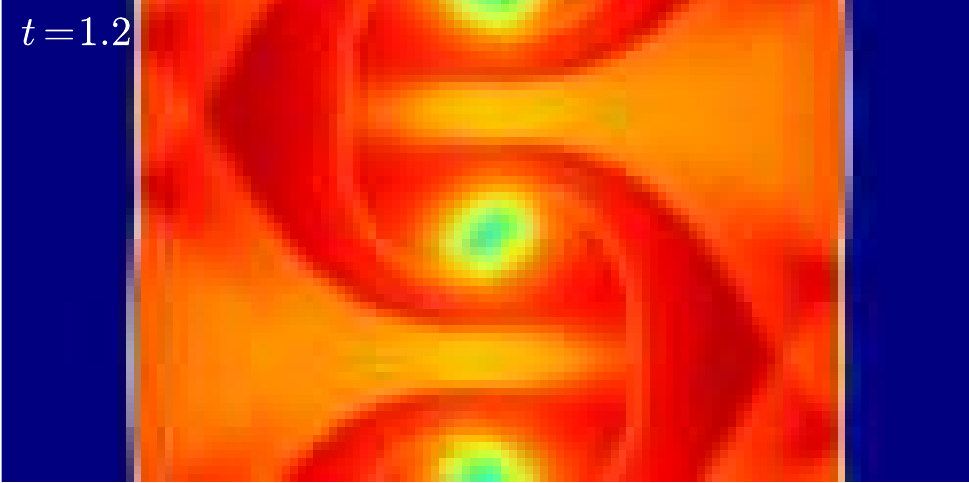}}
\vspace{0.04cm}
\caption{Development of the Non-linear thin shell instability for a ${\cal M}' = 2$ shock with a $\lambda_y = 1$, $A = 0.1$ boundary perturbation for (a) {\small MG} with a $1280 \times 128$ uniform grid, and (b) {\small SEREN} with $640,000$ particles using conservative SPH with the quintic kernel and the \citet{Wadsley2008} artificial conductivity.  The columns from left to right show the instability at times $t = 0.3$, $0.6$, $0.9$ and $1.2$.  Each sub-figure shows the density field (blue : low density - red : high density) in the region $-1 < x < +1$, $0 < y < 1$.}
\label{FIG:NTSI-MACH2-TIME}
\end{figure*}

\subsubsection{Cooling shocks} \label{SSS:COOLING-SHOCK}
Figure \ref{FIG:COOLSHOCK} shows the temperature and density profiles
of cooling shocks for both SPH and finite volume simulations with ${\cal M}' = 32$ and $A = 256$.  We only consider these values because of the overlap with isothermal shocks for very high values of $A$.  The semi-analytical solution (See Appendix \ref{A:COOLSHOCK}) is also plotted for reference.  At the initial shock interface, the shock obeys the adiabatic shock jump conditions reaching a density $\rho \sim 4$ and peak temperature $T \sim 180$.  The post-shock gas then cools according to Eqn. \ref{EQN:COOLING} to the equilibrium temperature, $T_{_{\rm EQ}} = 1$ at a density $\rho \sim 10^3$.  The size of the cooling region for these initial conditions and cooling law is about $\lambda_{_{\rm COOL}} \sim 0.075 \sim 1/13$ (See Fig. \ref{FIG:COOLSHOCK}(a); red dotted line).

For the  finite volume code, we perform simulations at three different resolutions, 
$\Delta x = \frac{1}{256}, \frac{1}{64}$ and $\frac{1}{16}$.  We find from 
our simulations that resolving  the cooling  region by  four or  more grid
cells  seems adequate to  allow the  full shock  to be  captured.  The
temperature profile  of the shock  (Fig. \ref{FIG:COOLSHOCK}(a)) shows
that for  $\lambda_{_{\rm COOL}} >> 4\,\Delta
x$, the  peak temperature of the  shock is correctly  captured and the
width of the cooling  region also matches the semi-analytical solution
(red  dotted line).   For the  lowest resolution  case that  can still
capture the cooling region ($\Delta  x = 1/64$), the cooling region is
broadened a little  but this is not unexpected for  a feature only 5-6
grid   cells    thick.    The   density   profiles    of   the   shock
(Fig. \ref{FIG:COOLSHOCK}(d))  show that the  well-resolved cases also
capture  the correct  density  profiles, with  the just-resolved  case
broadening  the density  profile also.

{\raw 
For SPH simulations, we simulate cooling shocks using both the
arithmetic and harmonic means with $\alpha_{_{\rm AV}} = 1$ at initial
resolutions $\Delta x = \frac{1}{32}, \frac{1}{4}$ and $2$.  Unlike
the finite volume code, the smoothing length and resolution changes as the density of the shock structure evolves.  We note three key resolutions, the pre-shock resolution ($h \sim \Delta x$), the adiabatic-shock resolution $h \sim \frac{1}{4}\Delta x$) and the isothermal-phase resolution ($h \sim \frac{1}{1000}\Delta x$).  For the highest resolution case, the peak temperature and cooling region width (Fig. \ref{FIG:COOLSHOCK}(b)), are well-modeled by the SPH code.  The most notable numerical artifact of reducing the resolution is that the peak temperature is less-well resolved and the shock becomes broader extending into the pre-shock region.  This can also be seen in the density profile (Fig. \ref{FIG:COOLSHOCK}(e)) where the SPH density is higher in the pre-shock regions.  For the lowest resolution case ($\Delta x = 2$), we begin to see evidence of post-shock oscillations in the temperature and density profiles.  At this resolution, we can consider the cooling region as severely under-resolved to the extent that we cannot model the cooling correctly.  If the resolution were decreased further, then the shock becomes more and more like the pure isothermal shock with similar numerical artifacts.  As with the pure isothermal shocks, the harmonic mean variant of the artificial viscosity allows the shock to be captured without significant post-shock oscillations, even when the cooling region is under-resolved (Fig. \ref{FIG:COOLSHOCK}(c) \& (f)).}

These results lead to similar conclusions to those by
\citet{Creasey2011}, who suggest the need for a resolution criteria
for cooling shocks for both finite volume and SPH codes to ensure cooling shocks are modelled correctly.  
In our case, moderate under-resolution of the cooling region leads to a broader shock, but no problematic numerical effects.  More severe under-resolution of the cooling region leads to the same problems as those resulting from purely isothermal shocks as described above.  For both methods, minor alterations to the standard algorithms can reduce these numerical problems.

\subsection{Non-linear thin-shell instability} \label{SS:NTSI}

The non-linear thin-shell instability \cite[hereafter NTSI;][]{NTSI}
occurs when two colliding streams of gas form a shock along a
non-planar boundary. We consider an interface between the two flows as a 
sinusoidal boundary, e.g. $x_{_{\rm BOUNDARY}} \sim
A\,\sin(k\,y)$ where $A$ is the boundary displacement and $k$ is the
wavenumber of the sinusoid.  The evolution of the shock interface can evolve 
due to a number of competing effects \cite[See][for a detailed analysis]{NTSI}, which can decrease or increase the amplitude of the sinusoidal displacement.
If the amplitude of the boundary displacement becomes comparable to, or 
greater than, the thickness of the shock, then this
shape can effectively `funnel' material towards the extrema of the
sinusoid.  This leads to the growth of density enhancements as more
material flows into the shock, as well as a growth in the amplitude of
the boundary displacement, which causes the interface to `bend' more.
For small displacements, the growth rate of the instability is $\sim
c_s\,k\,(A\,k)^{1/2}$ where $c_s$ is the sound speed of the shocked
gas.  The NTSI has only been simulated numerically by a few authors 
\citep[e.g.][]{Blondin1996,Klein1998,Heitsch2007}.

\begin{figure*}
\centerline{
\includegraphics[width=4.35cm]{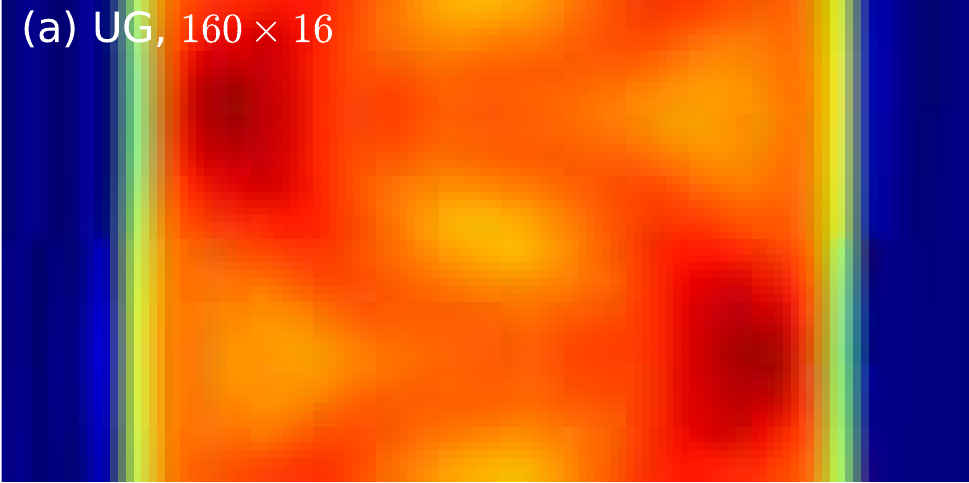}
\includegraphics[width=4.35cm]{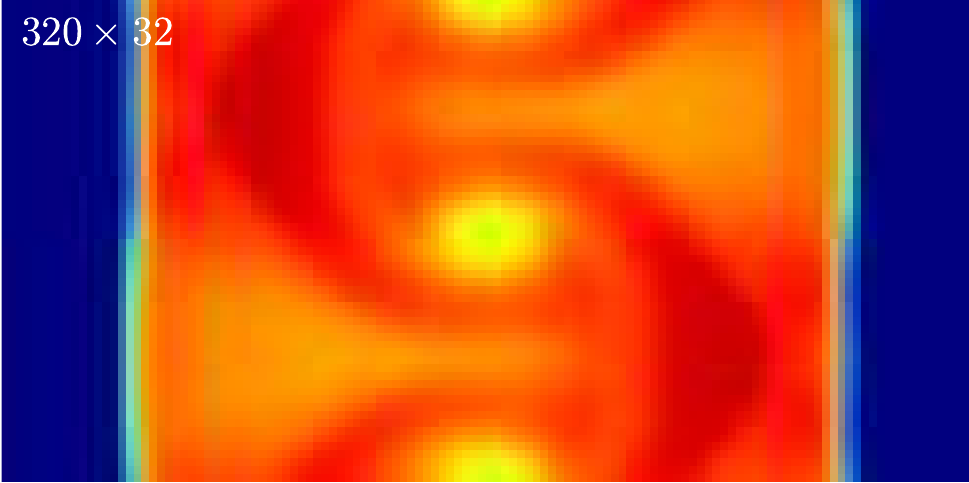}
\includegraphics[width=4.35cm]{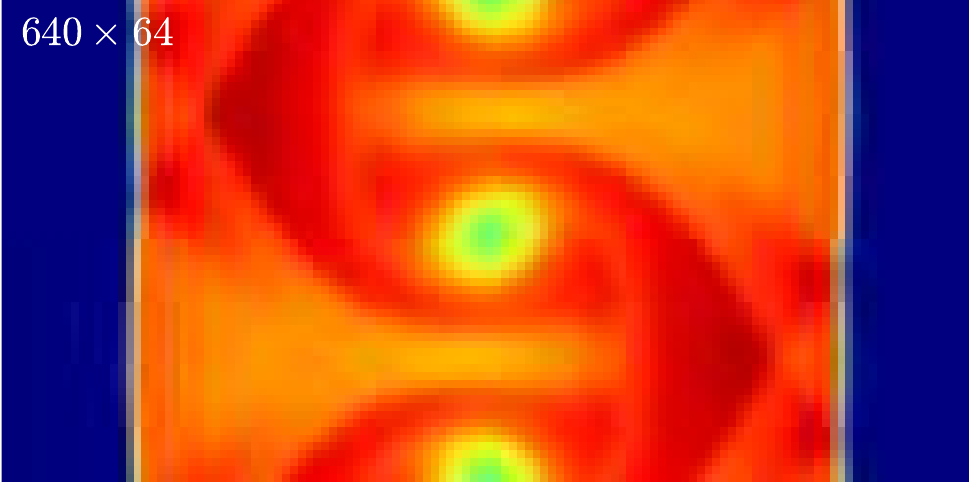}
\includegraphics[width=4.35cm]{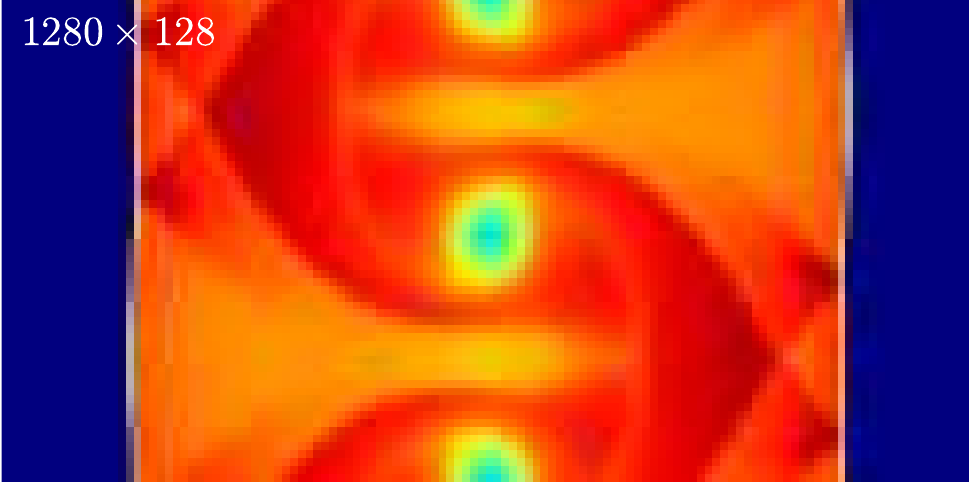}}
\vspace{0.04cm}
\centerline{
\includegraphics[width=4.35cm]{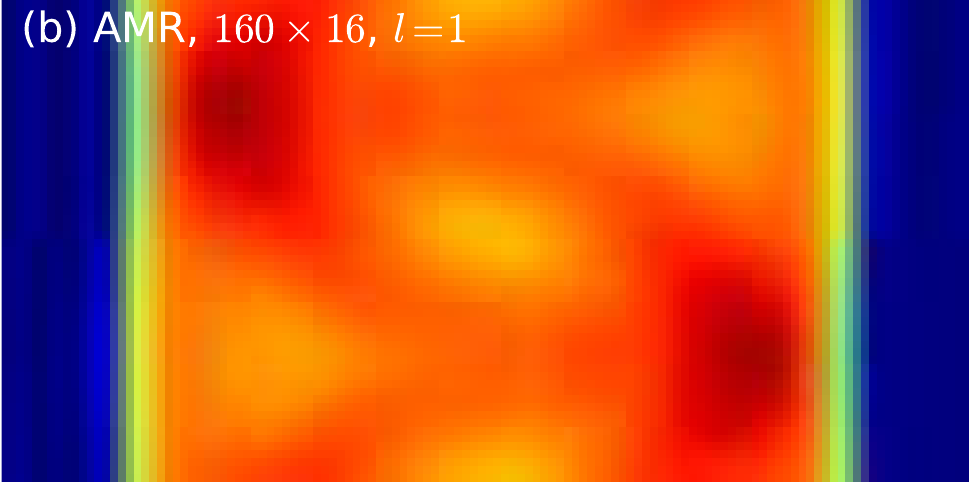}
\includegraphics[width=4.35cm]{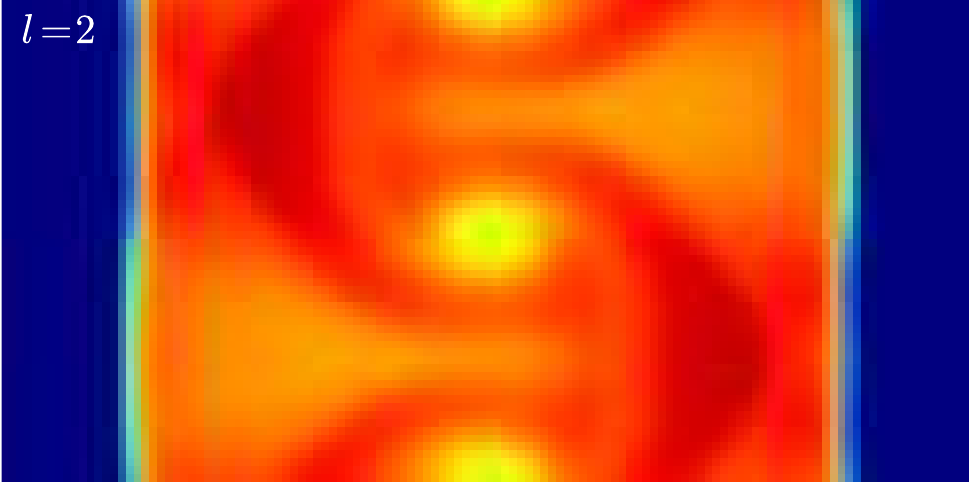}
\includegraphics[width=4.35cm]{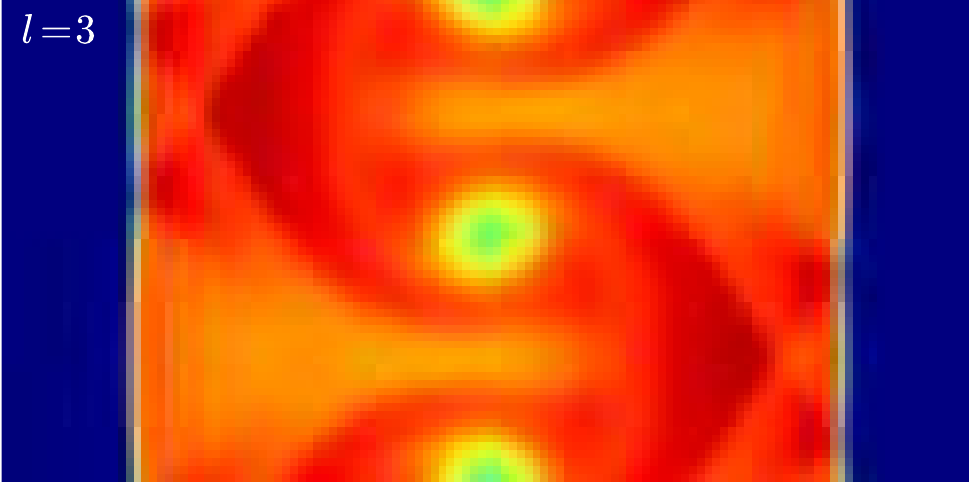}
\includegraphics[width=4.35cm]{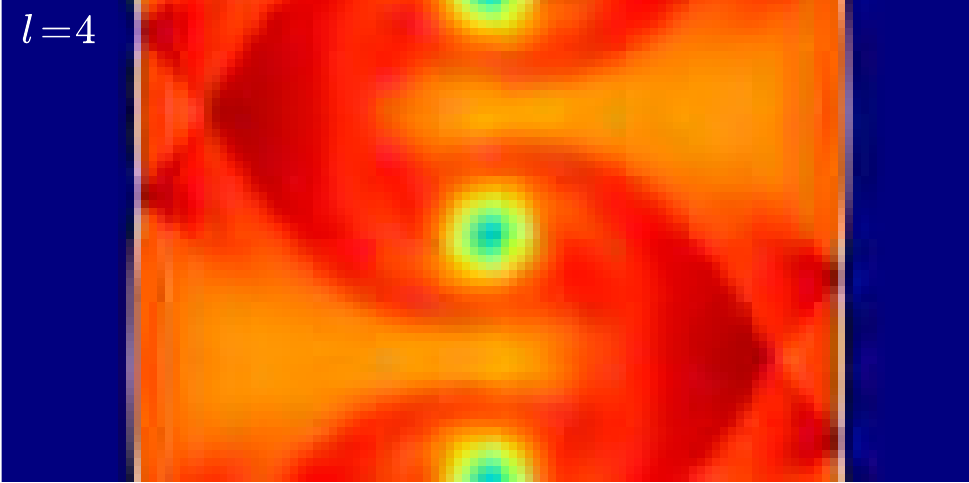}}
\vspace{0.04cm}
\centerline{
\includegraphics[width=4.35cm]{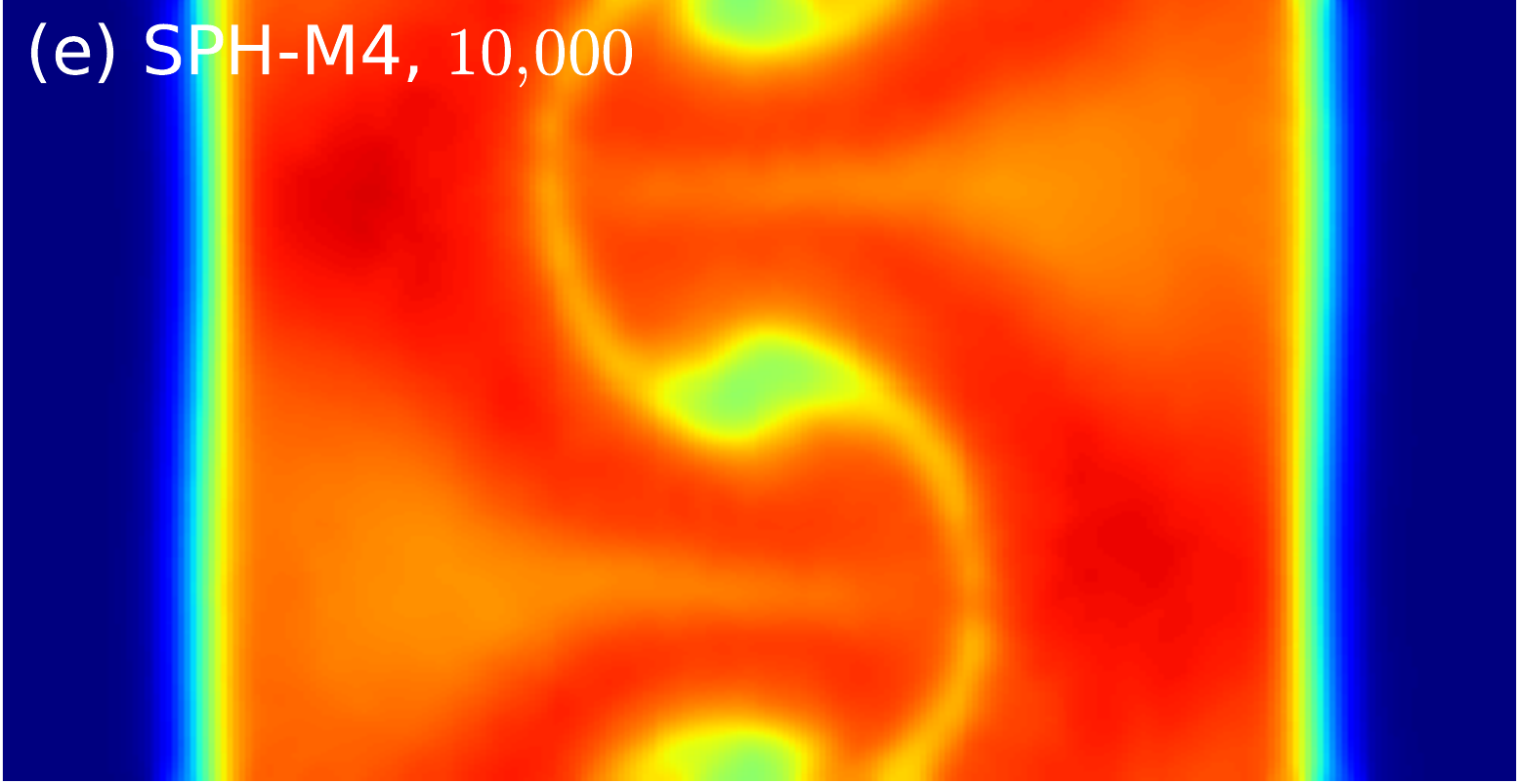}
\includegraphics[width=4.35cm]{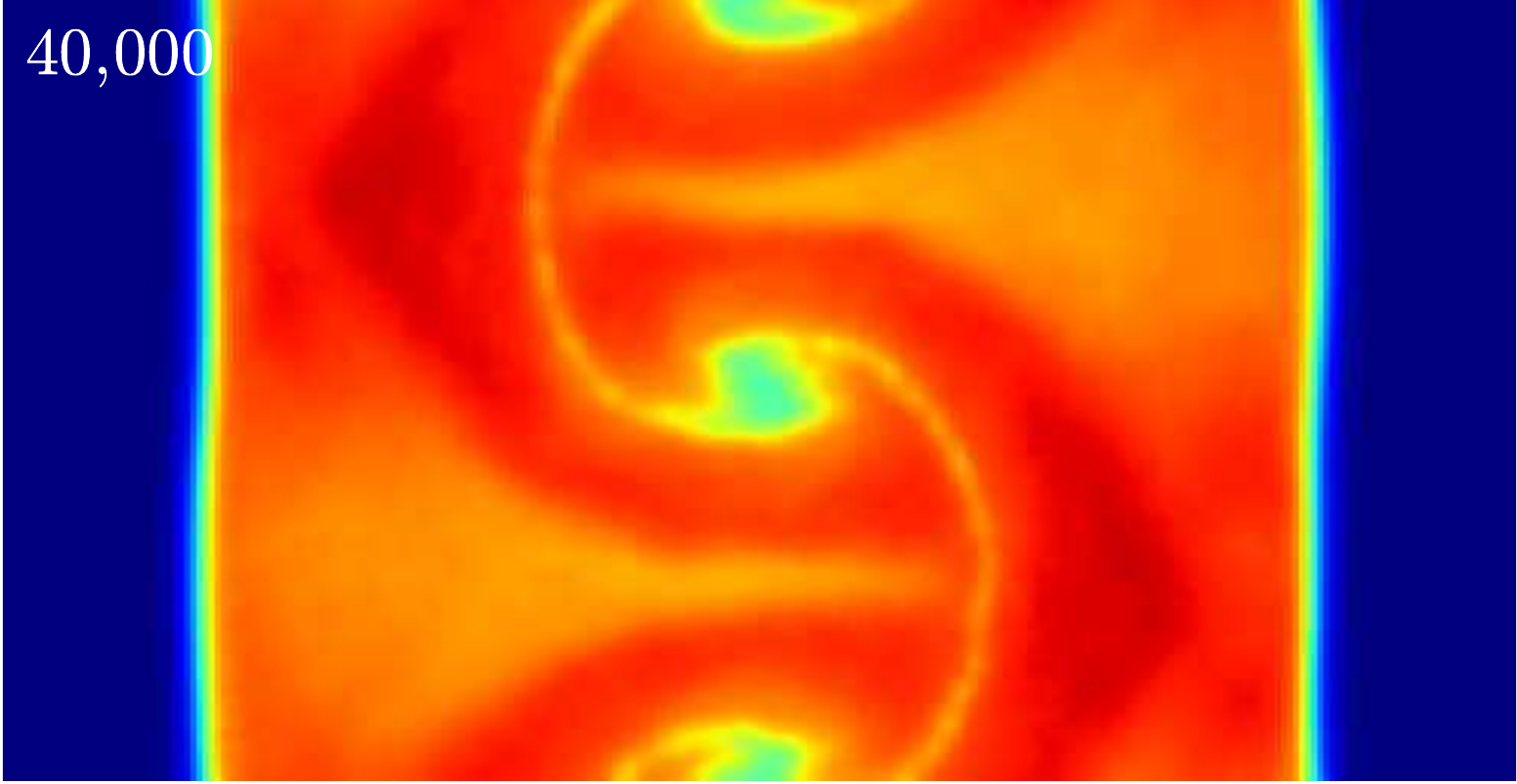}
\includegraphics[width=4.35cm]{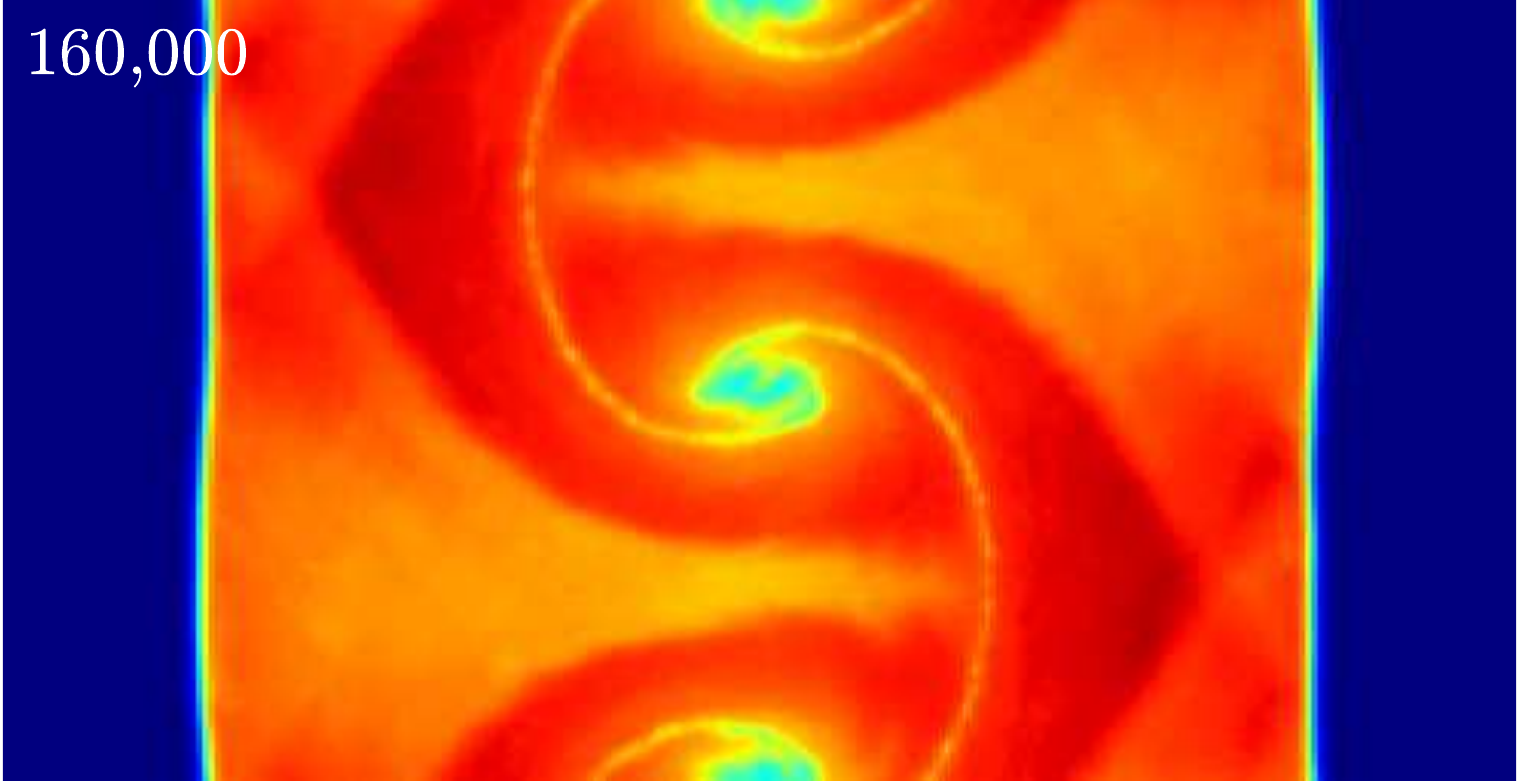}
\includegraphics[width=4.35cm]{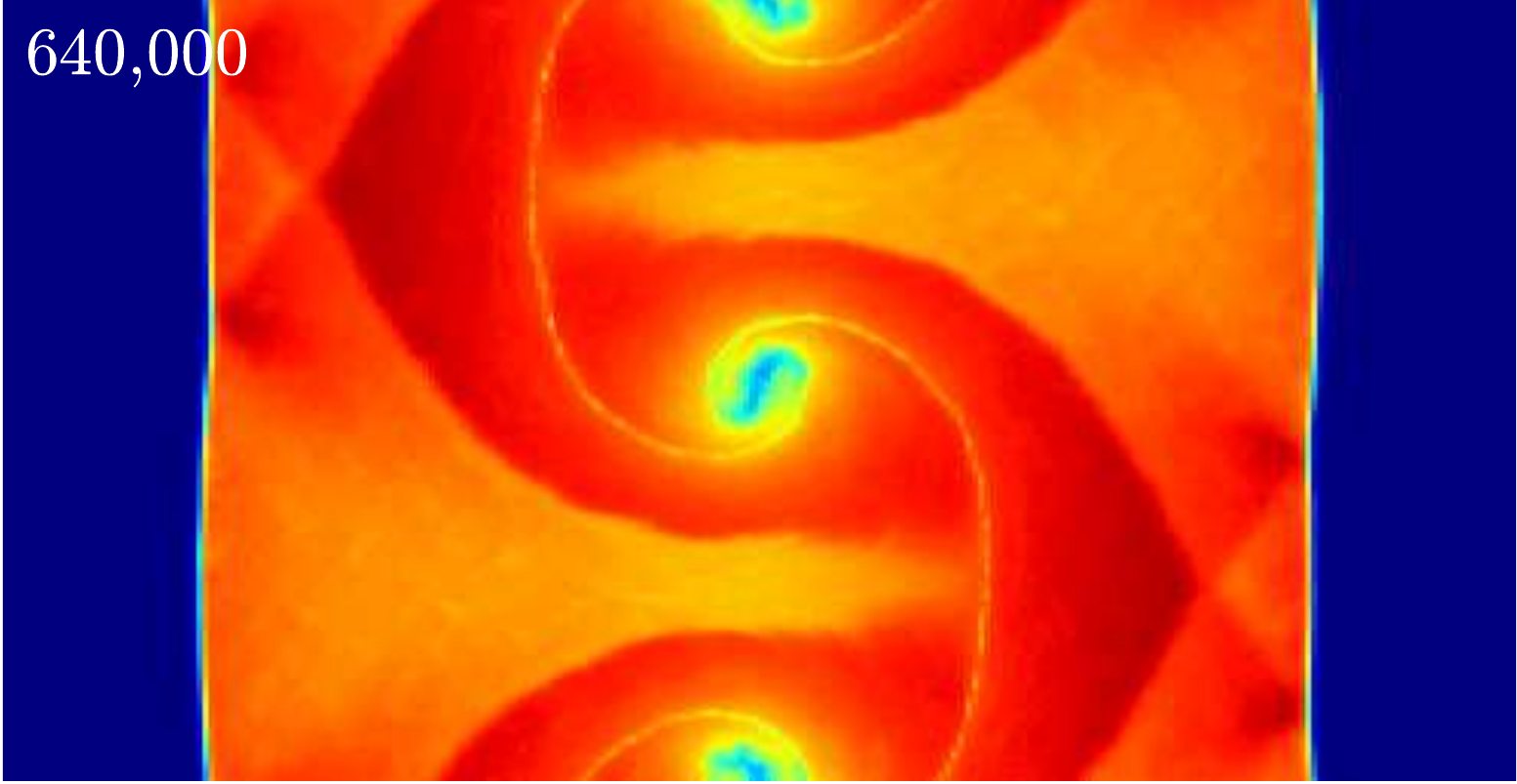}}
\vspace{0.04cm}
\centerline{
\includegraphics[width=4.35cm]{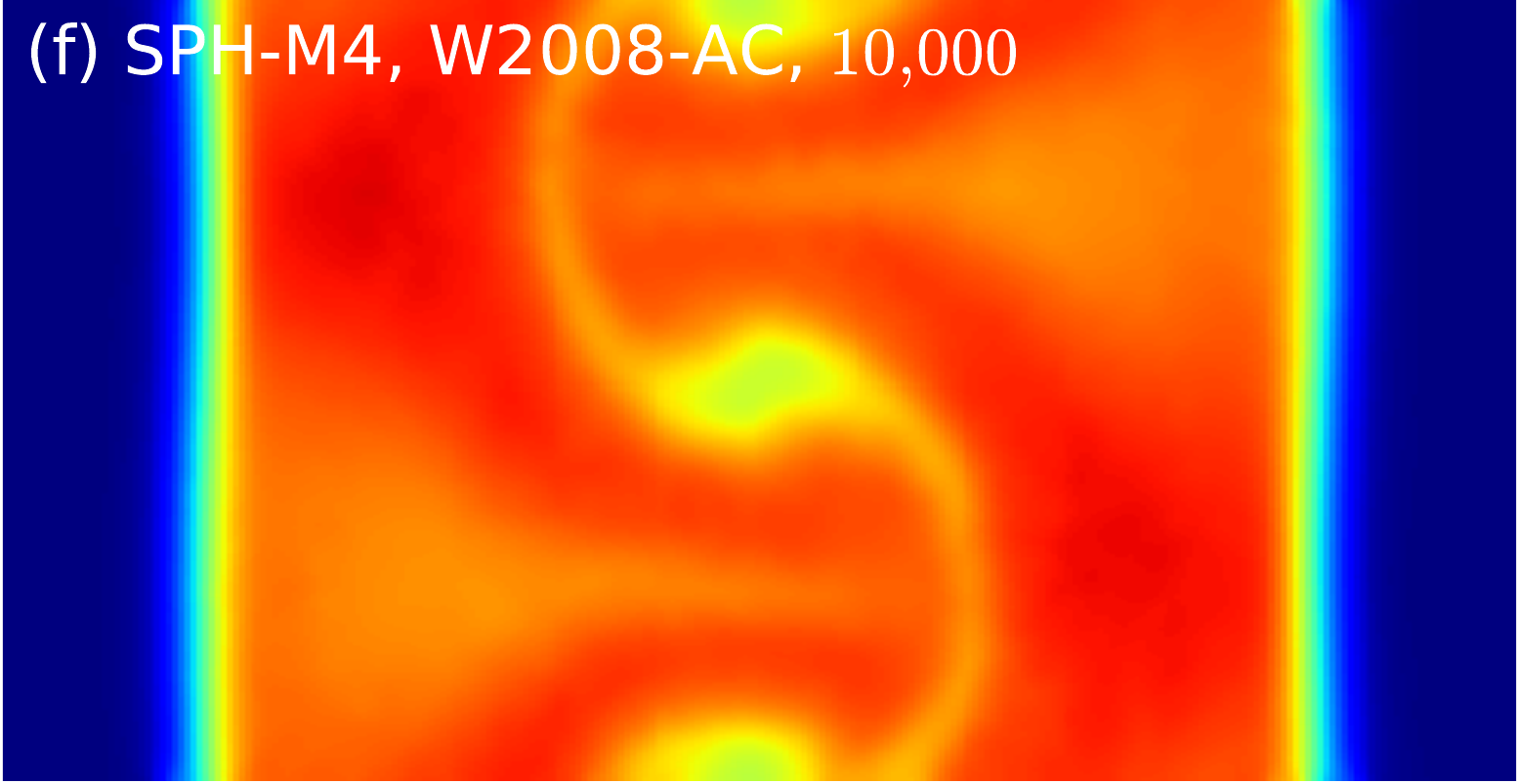}
\includegraphics[width=4.35cm]{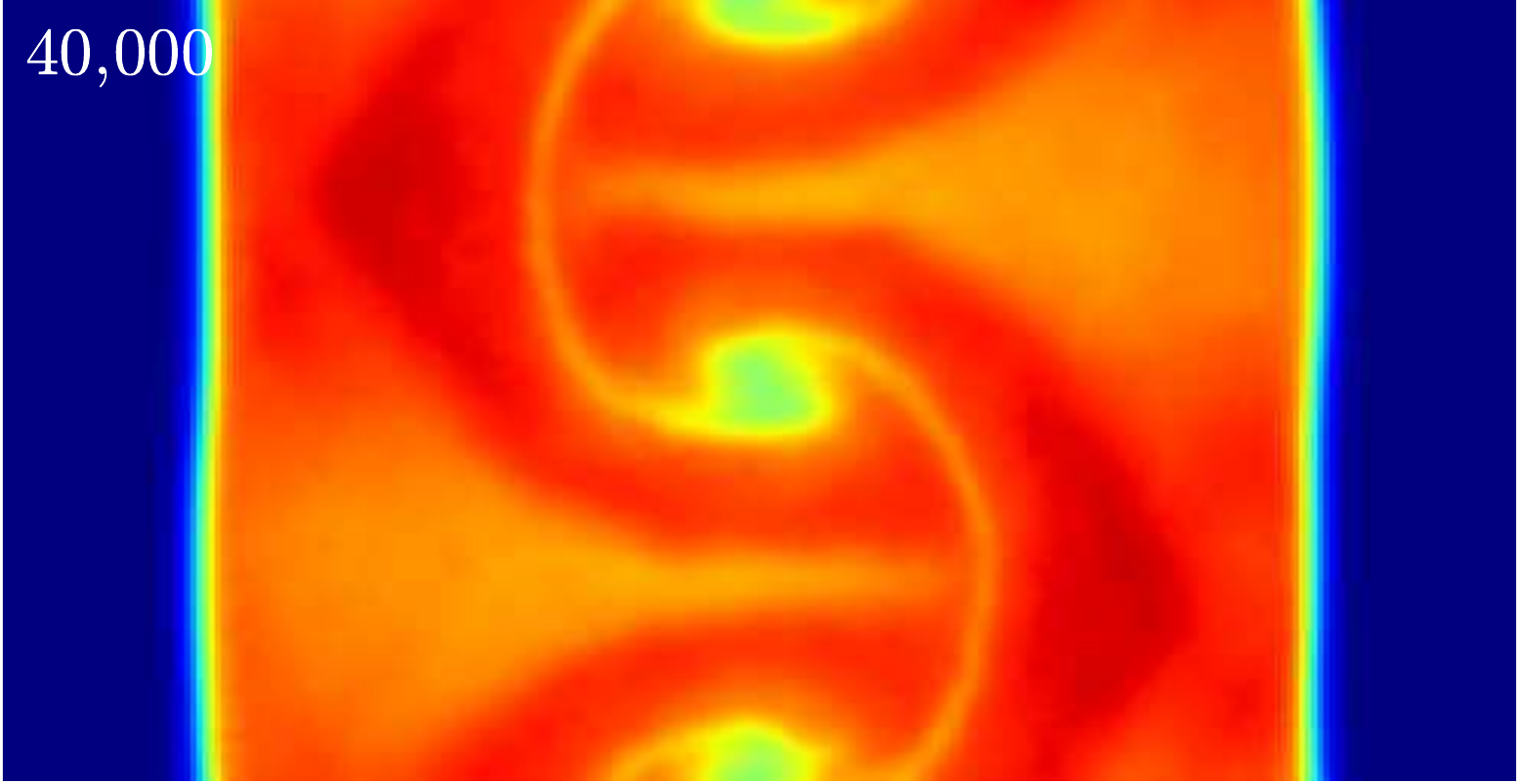}
\includegraphics[width=4.35cm]{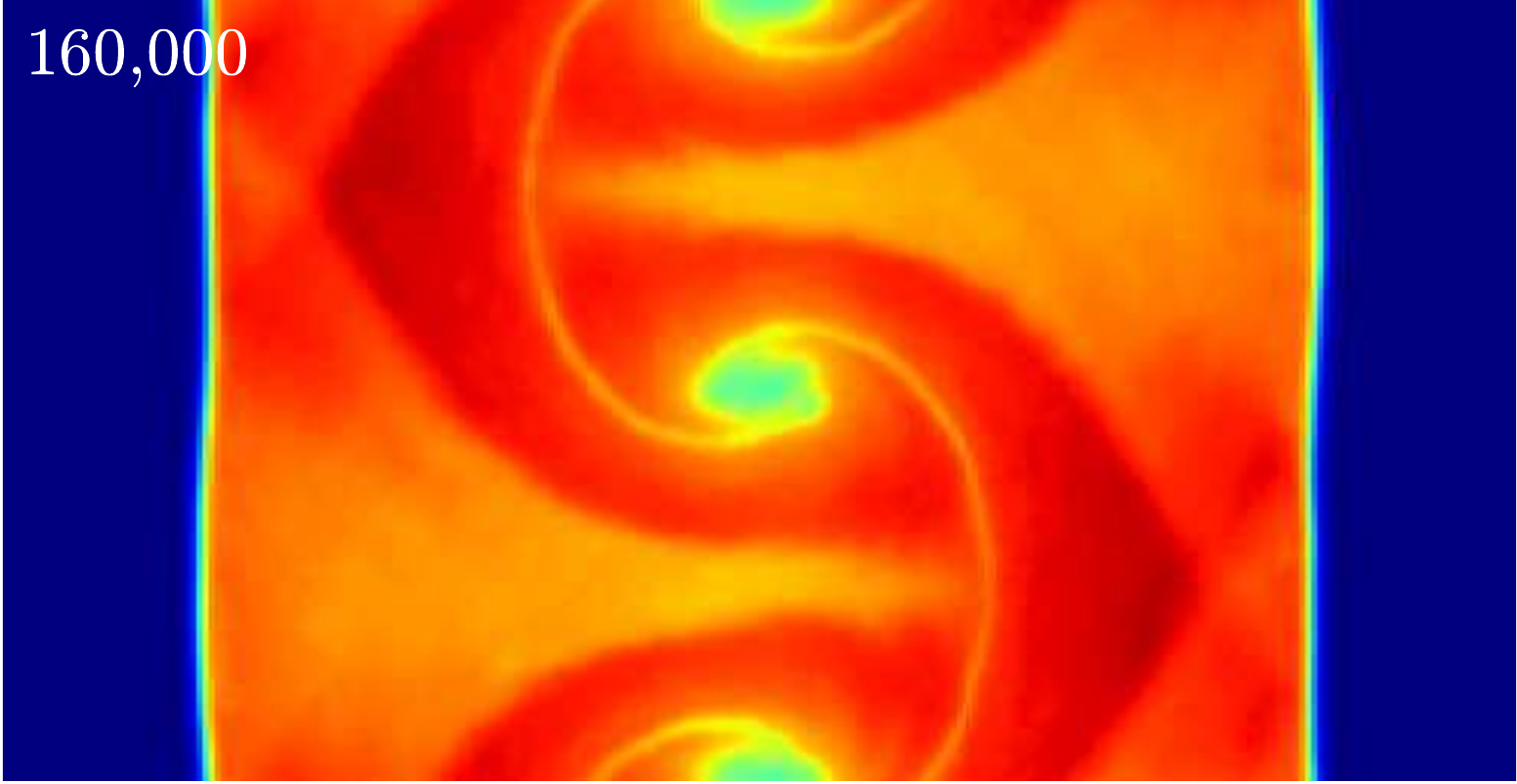}
\includegraphics[width=4.35cm]{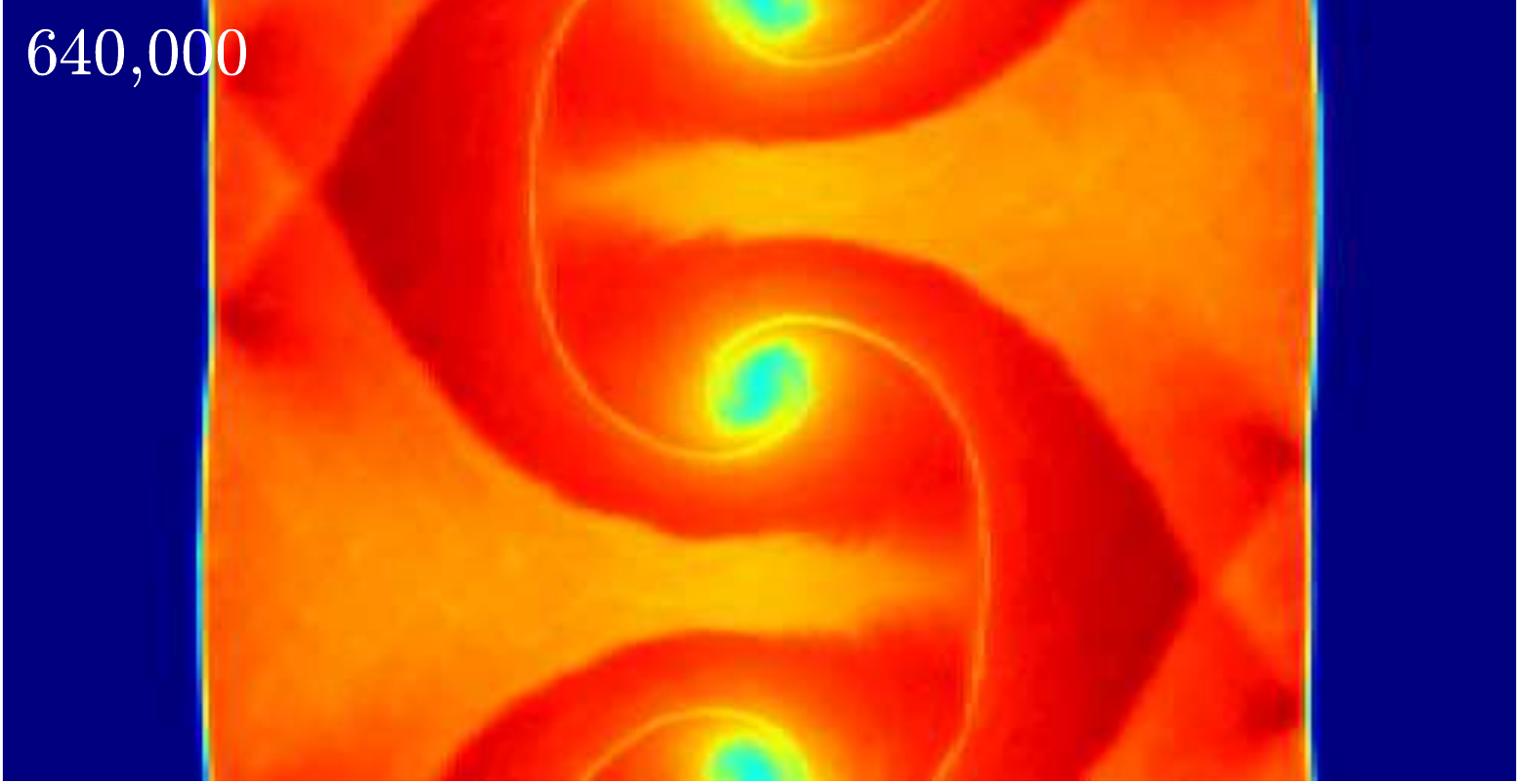}}
\vspace{0.04cm}
\centerline{
\includegraphics[width=4.35cm]{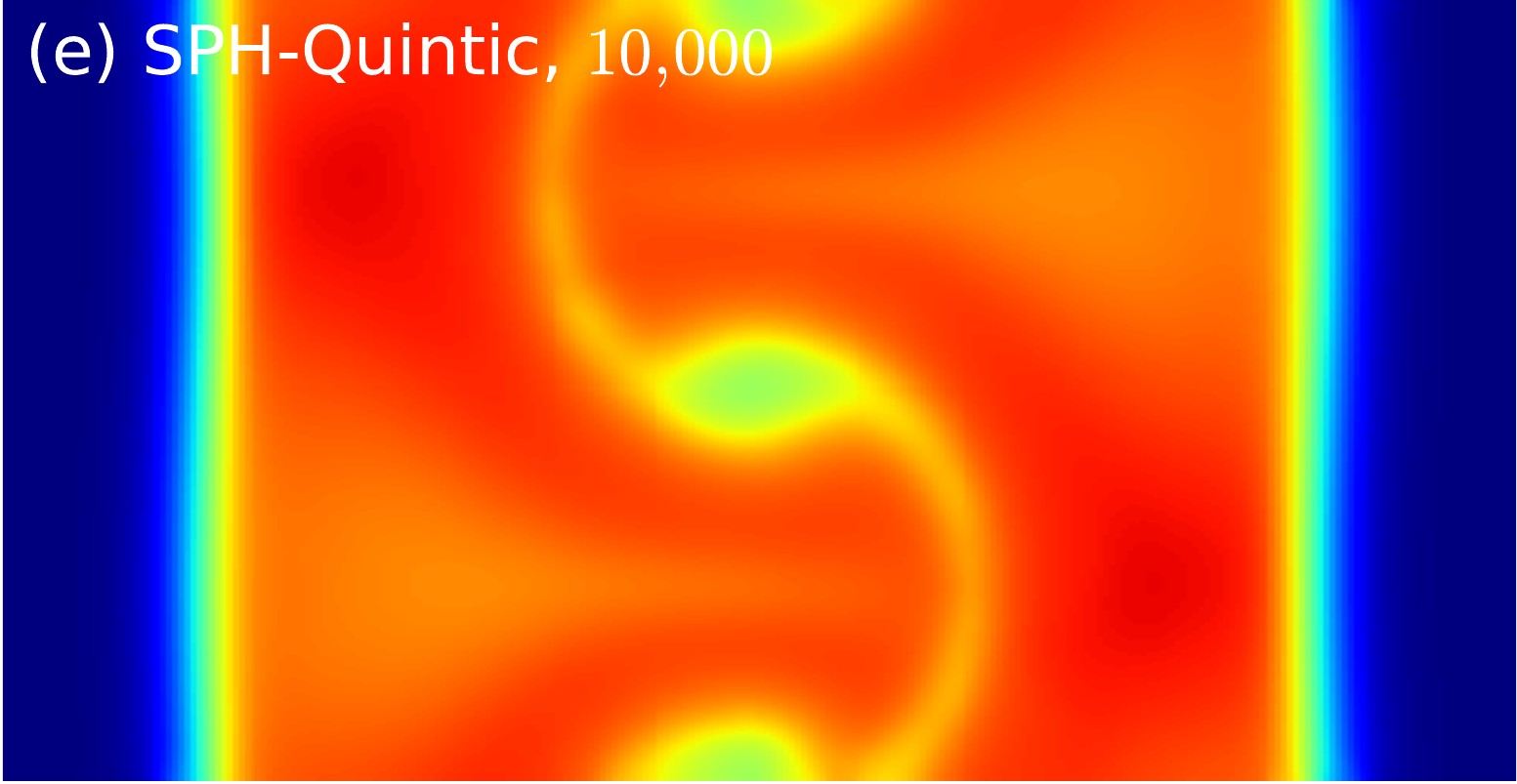}
\includegraphics[width=4.35cm]{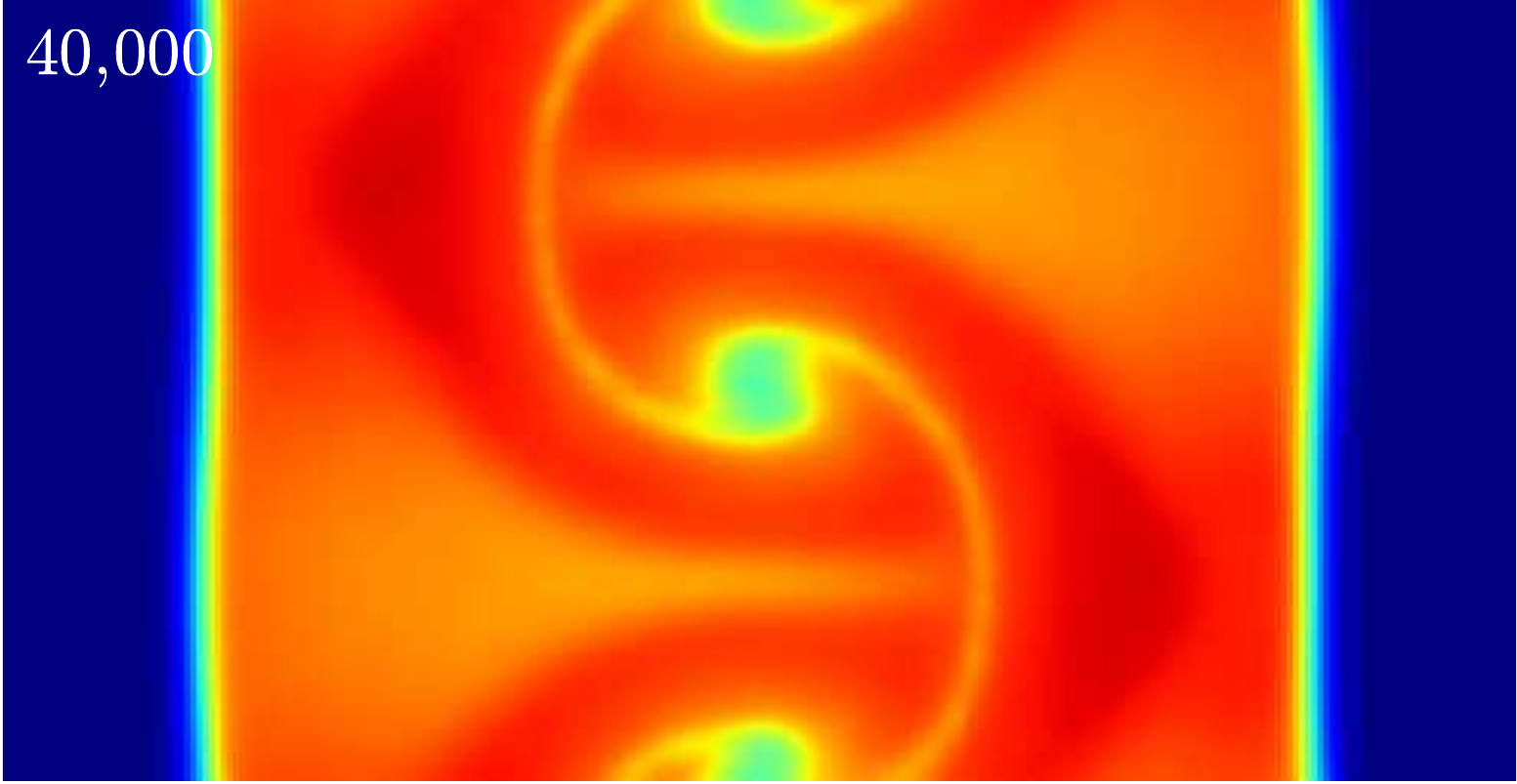}
\includegraphics[width=4.35cm]{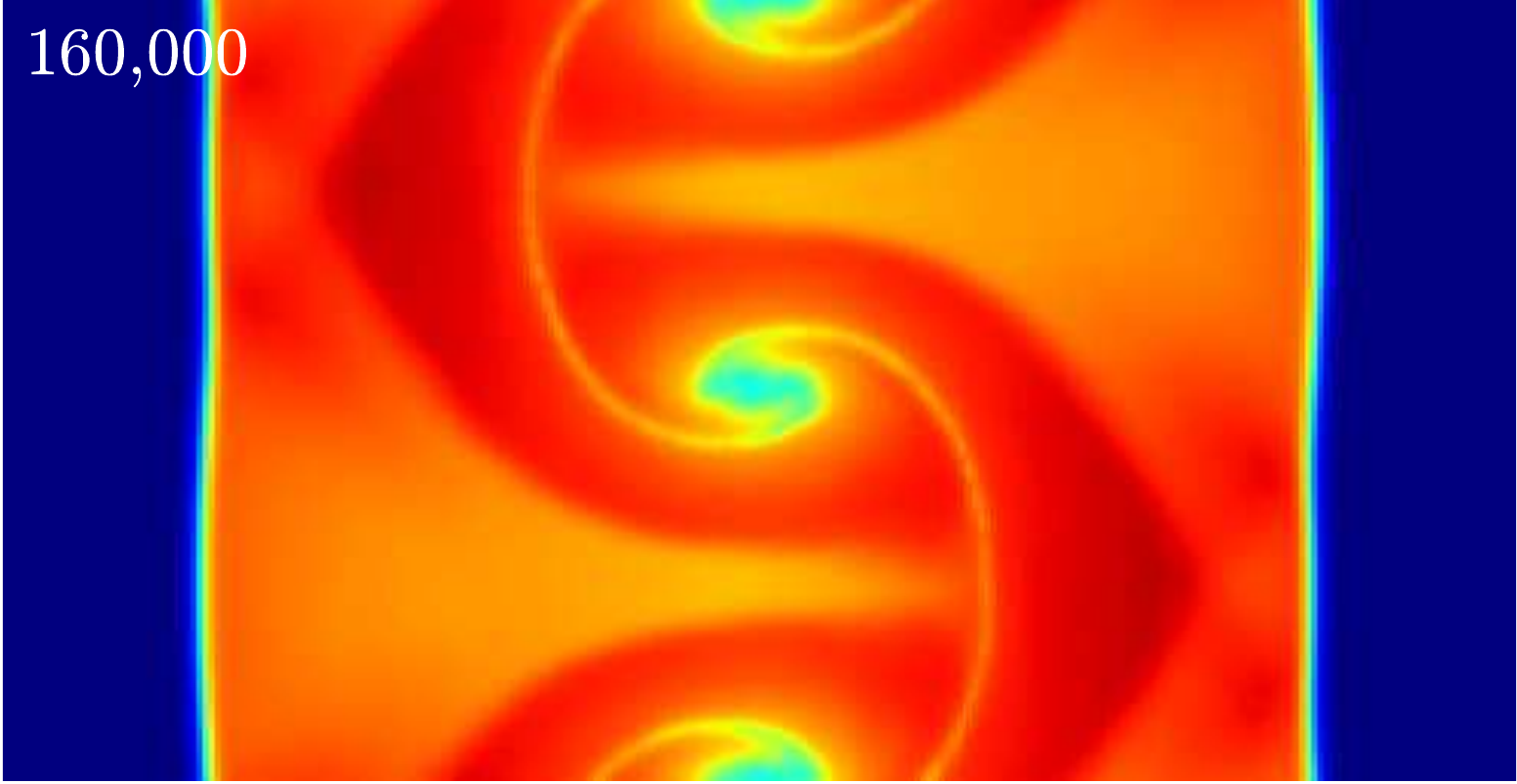}
\includegraphics[width=4.35cm]{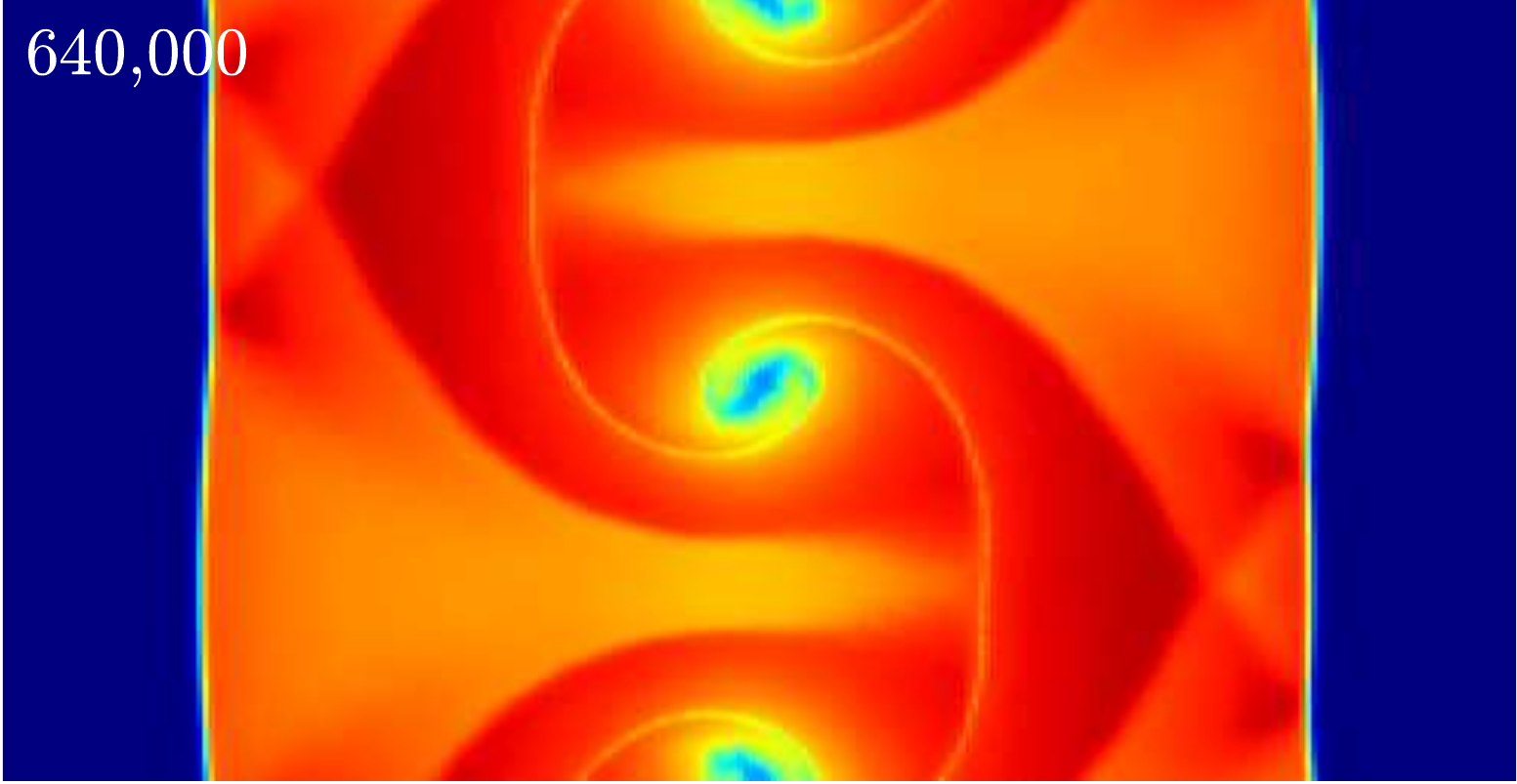}}
\vspace{0.04cm}
\centerline{
\includegraphics[width=4.35cm]{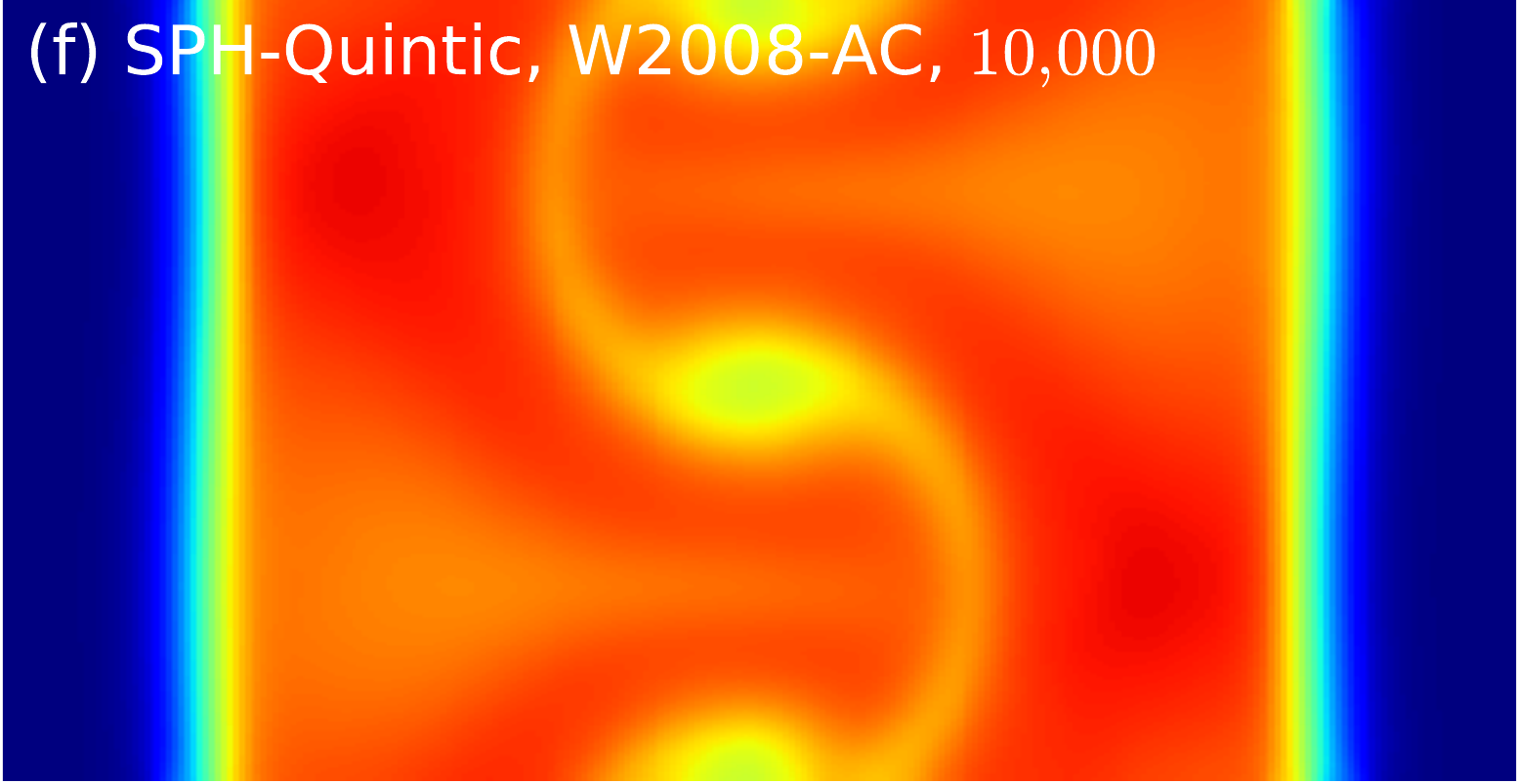}
\includegraphics[width=4.35cm]{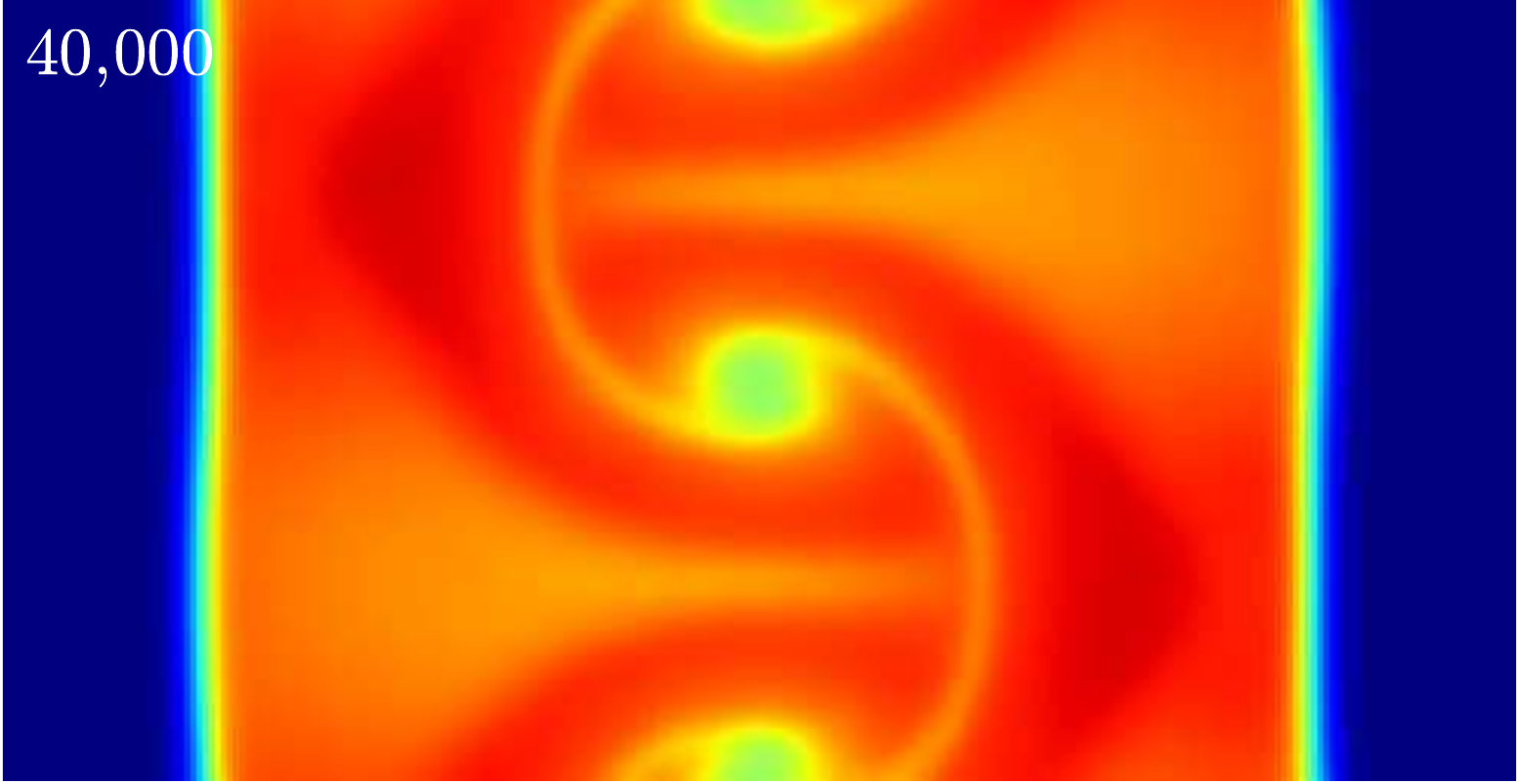}
\includegraphics[width=4.35cm]{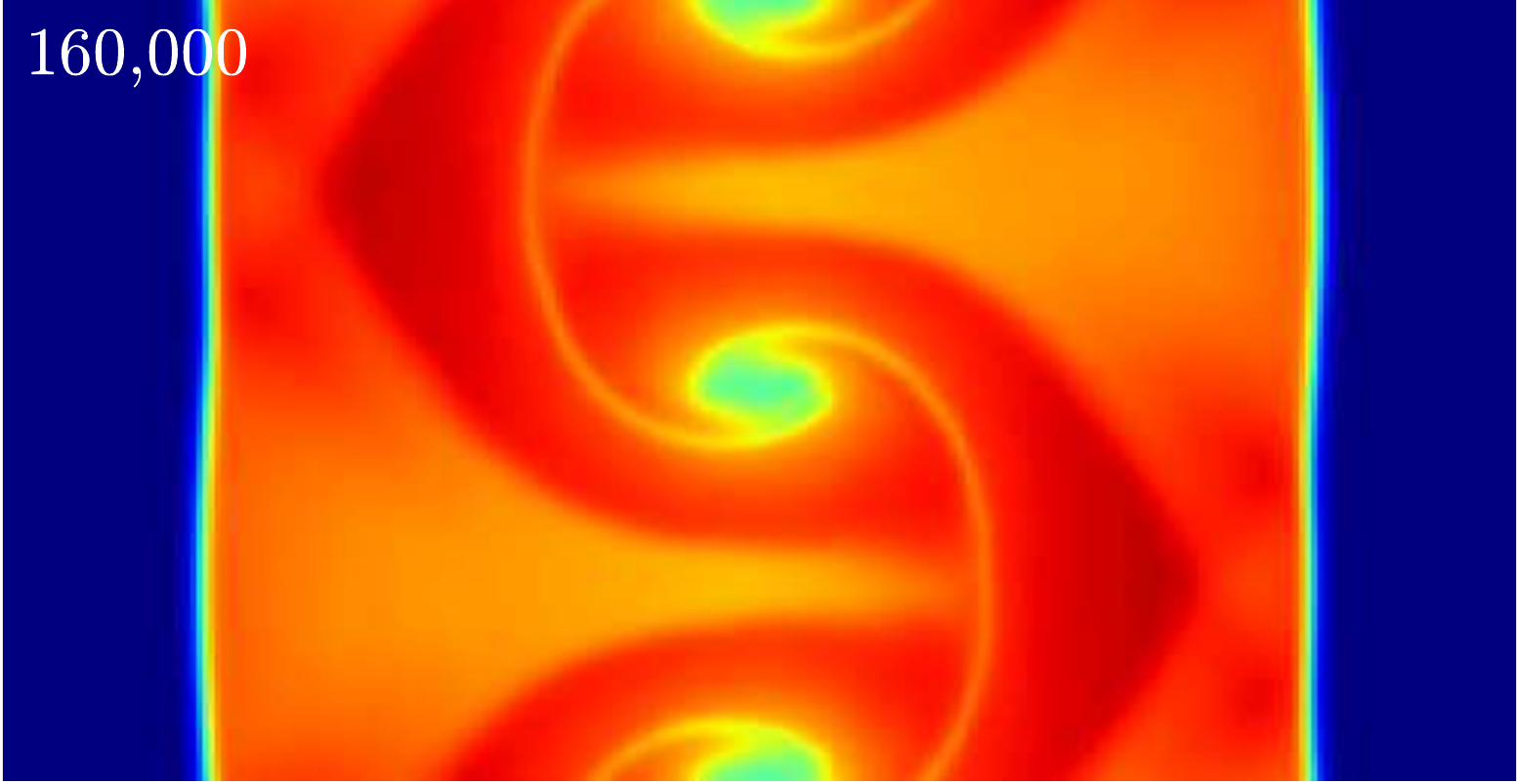}
\includegraphics[width=4.35cm]{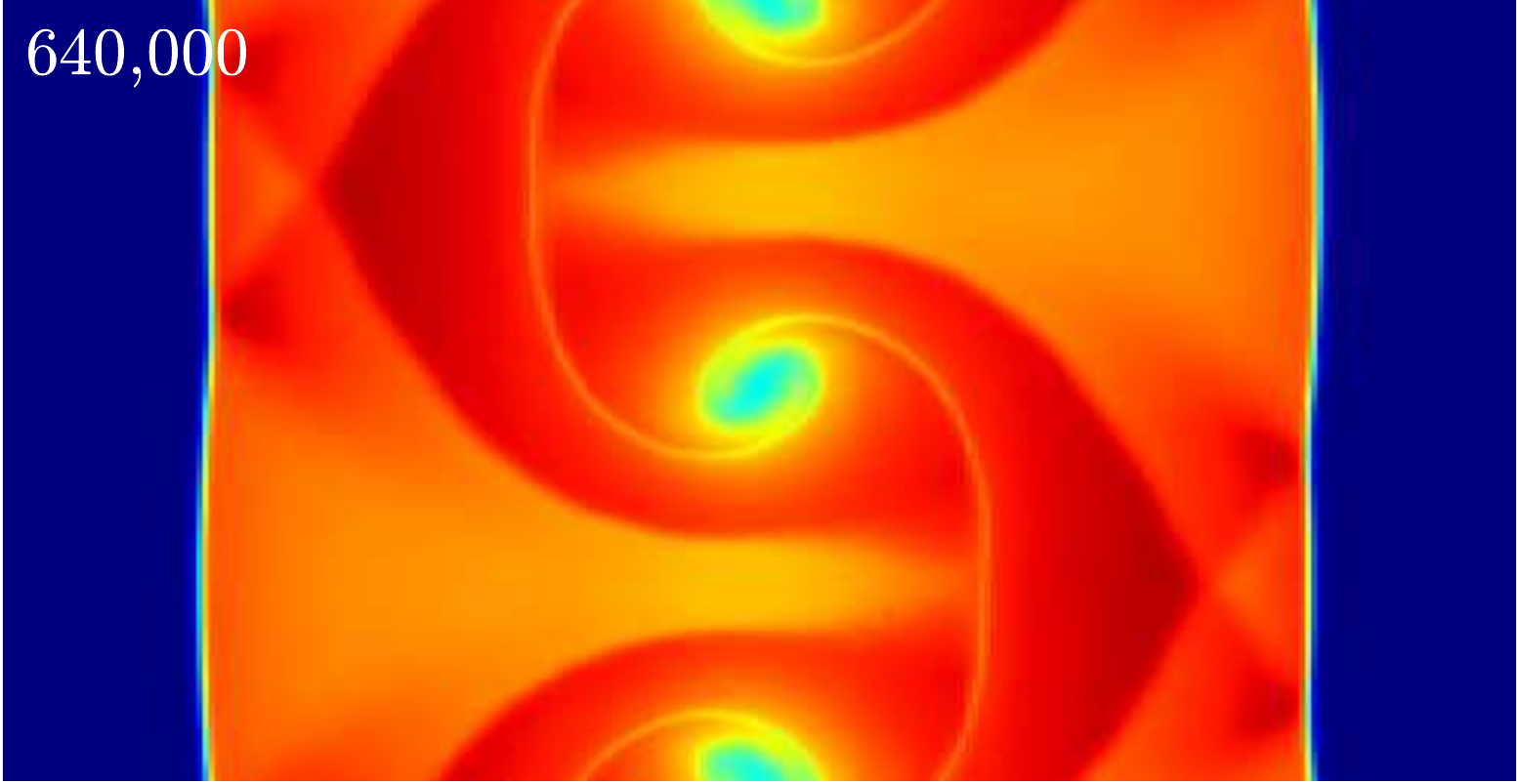}}
\vspace{0.04cm}

\caption{The density structure of the ${\cal M}' = 2$ non-linear thin
  shell instability in the range $-1 < x < +1$, $0 < y < 1$ at a time
  $t = 1.2$ modelled with (a) (a) {\small MG} using uniform grid, (b) {\small MG} using
  AMR with $1$,$2$,$3$ \& $4$ refinement levels,
  (c) {\small SEREN} using the M4 kernel,  (d) {\small SEREN} using the M4 kernel and
  the \citet{Wadsley2008} conductivity, (e) {\small SEREN} using the quintic
  kernel,  (f) {\small SEREN} using the quintic kernel and the \citet{Wadsley2008}
  conductivity.  The left-hand column shows the NTSI using
  the smallest resolution ($160 \times 16$ cells for the finite volume code and
  $N = 10,000$ for the SPH code) with increasing resolution moving
  right to the highest resolution ($1280 \times 128$ for the finite volume code
  and $N = 640,000$ particles for the SPH code).  The AMR simulations 
  have the same equivalent resolution as the corresponding uniform grid 
  simulation in the above row. Each sub-figure shows
  the density field (blue : low density - red : high density).} 
\label{FIG:NTSI-MACH2}
\end{figure*}

\subsubsection{Initial conditions} \label{SSS:NTSI-IC}
We model the NTSI with two uniform density gas flows with the same
initial density ($\rho = 1$), pressure ($P = 1$) and ratio of specific
heats ($\gamma = 5/3$).  The initial velocity profile is  
\begin{equation}
v_x(x,y) = 
\begin{cases}
+{\cal M}'\,c_0 & \;\;\;\; x < A\,\sin{\left(k\,y\right)} \\
-{\cal M}'\,c_0 & \;\;\;\; x > A\,\sin{\left(k\,y\right)}
\end{cases}
\end{equation}
where $A = 0.1$ is the amplitude of the sinusoidal boundary
perturbation, $k = 2\,\pi/\lambda$ is the wave number of the
perturbation, $\lambda = 1$ is the perturbation wavelength, 
$c_0$ is the sound speed of the unshocked gas and ${\cal M}' = 2$.  
We set the y-velocity, $v_y = 0$ everywhere initially.  
The initial velocities are then smoothed in the same manner as 
the shock tube tests (Section \ref{SSS:SHOCK-IC}, Eqn. \ref{EQN:SMOOTHV}).  
One caveat is that we model the gas adiabatically, not isothermally as originally considered by \citet{NTSI}.  Although this will lead to the instability growing on a slightly different timescale, we are principally concerned with comparing the two numerical methods than comparing to theory.

The computational domain extends between the limits $-5 < x < 5$ and
$0 < y < 1$ with open boundaries in the x-dimension and periodic
boundaries in the y-dimension.  Both codes use the standard algorithms 
and parameters described in Section \ref{S:NUMERICALMETHODS} for this test.  
For the finite volume code,
we use $160 \times 16$, $320 \time 32$, $640 \times 64$ and $1,280
\time 128$ uniform grid cells.  With the AMR simulations, we use
initially $160 \times 16$ with up to four refinement levels.  For
the SPH simulations, we initially set-up particles by relaxing a glass
from $10,000$, $40,000$, $160,000$ and $640,000$ particles.

\subsubsection{Simulations} \label{SSS:NTSI-SIMS}

We model the NTSI using both finite volume and SPH with different
code options and resolutions in order to study the development of the
instability under different conditions and to compare the convergence
with resolution of the two codes.  For the AMR code, we perform
simulations using both a uniform grid, and with $5$ levels of
refinement.  For the SPH code, we model the NTSI with both the M4 and
quintic kernels, and also with and without an artificial conductivity
term \citep{Wadsley2008}.  We model the growth of the instability 
and subsequent complex gas-flow until a time of $t = 1.2$.

Figure \ref{FIG:NTSI-MACH2-TIME} shows the time evolution of the density 
of the gas flow as the NTSI develops, saturates and evolves into a 
complex density structure for the highest resolution AMR 
(Fig \ref{FIG:NTSI-MACH2-TIME}(a)) and SPH 
(Fig \ref{FIG:NTSI-MACH2-TIME}(b)) simulations.  
While there is at first no shock due to the initial uniform density,
this quickly forms from the above initial conditions.  The NTSI 
develops rapidly since the sinusoidal boundary amplitude is
comparable to the shock thickness.  At $t = 0.3$ (Fig
\ref{FIG:NTSI-MACH2-TIME}; column 1), we can see that the instability
has already developed creating two density enhancements near the
concave sections of the boundary, where material is funneled to from
the inflowing gas.  By $t = 0.6$ (column 2), the instability
has already saturated such that the initial sinusoidal interface has 
been enhanced by bending modes to the point where the amplitude is 
comparable to the wavelength of the interface.
A complex sinusoidal density pattern containing a lower density cavity
at the centre, along with a lower density filament which defines the
original interface of the shock (this is likely a wall-heating effect
which is retained in the latter evolution).  We also note that as the
instability saturates, the contact layer between the low-density
inflowing material and the shocked-region becomes more planar as the
`feedback' of material from the shocked region fills out this cavity
and effectively dampens the generation of any future instability. 
The AMR and SPH results are nearly identical with only small 
noticeable deviations which will be discussed later.

Figure \ref{FIG:NTSI-MACH2} shows the density structure of the 
${\cal M}' = 2$ NTSI at a time $t = 1.2$ using both AMR and SPH with
various different options and resolutions. For the very lowest
resolution finite volume simulations on a uniform with $160 \times 16$
cells\footnote{We note that this is the lowest resolution possible
  since any smaller resolution would mean the sinusoidal boundary
  would not be resolved and therefore there would be no instability}
(Fig \ref{FIG:NTSI-MACH2}(a), column 1), the main density enhancements
due to the focusing from the sinusoidal boundary are apparent.  There
is not enough resolution to adequately represent the more complex
density structures.  As the resolution is increased (columns 2--4),
the simulation converges towards the complex density structure
described earlier.  When using AMR (Fig \ref{FIG:NTSI-MACH2}(b)), the
overall density structure is the same as the finite volume code with a
uniform mesh.  The
only noticeable difference is that some of the fine structure is a
little more diffuse due to numerical diffusion across different 
refinement levels.

For SPH simulations of the NTSI using the M4 and quintic kernels with
and without artificial conductivity (Figure \ref{FIG:NTSI-MACH2}
(c)-(f)), the lowest resolution simulations (column 1) clearly show the
generation of the principle large-scale density structure, including the
two density enhancements at the top-left and bottom-right of each
panel.  However, the density enhancements are not as strong as the
finite volume codes for the lowest resolution.  As the resolution is increased
for each set of options, the simulations clearly converge with each
other and with the uniform and AMR grid solutions.  The principle
differences lie within the small low-density filament that lies at the
original contact point between the two flows due to wall-heating, and
the low-density cavity in the middle of the domain.  The filament is much
more prominent in the SPH case; although both finite volume and SPH experience
wall-heating problems, artificial diffusion causes this feature to be
smeared out in finite volume codes as the simulation progresses.  
SPH on the other hand, has
no such in-built diffusion and source of dissipation must explicitly
added.  The artificial conductivity reduces this effect a little, but
it is still extremely prominent, particularly for the lower
resolutions.  The other noticeable difference is the simulations with
the M4 kernel have more noise in the otherwise smooth density fields,
and some of the sharp features prominent in the finite volume code are more
diffuse.  The simulations with the quintic kernel, despite having
formally lower resolution, are more sharp than the corresponding M4
simulations.  However, the fundamental features of the evolution of
the NTSI are the same in all simulations regardless of the details of
the SPH implementation.  This is because the NTSI is
principally a large-scale instability generated by large-inflows.
Therefore, noise and low accuracy do not affect the bulk evolution.

\begin{figure*}
\centerline{
\includegraphics[width=3.5cm]{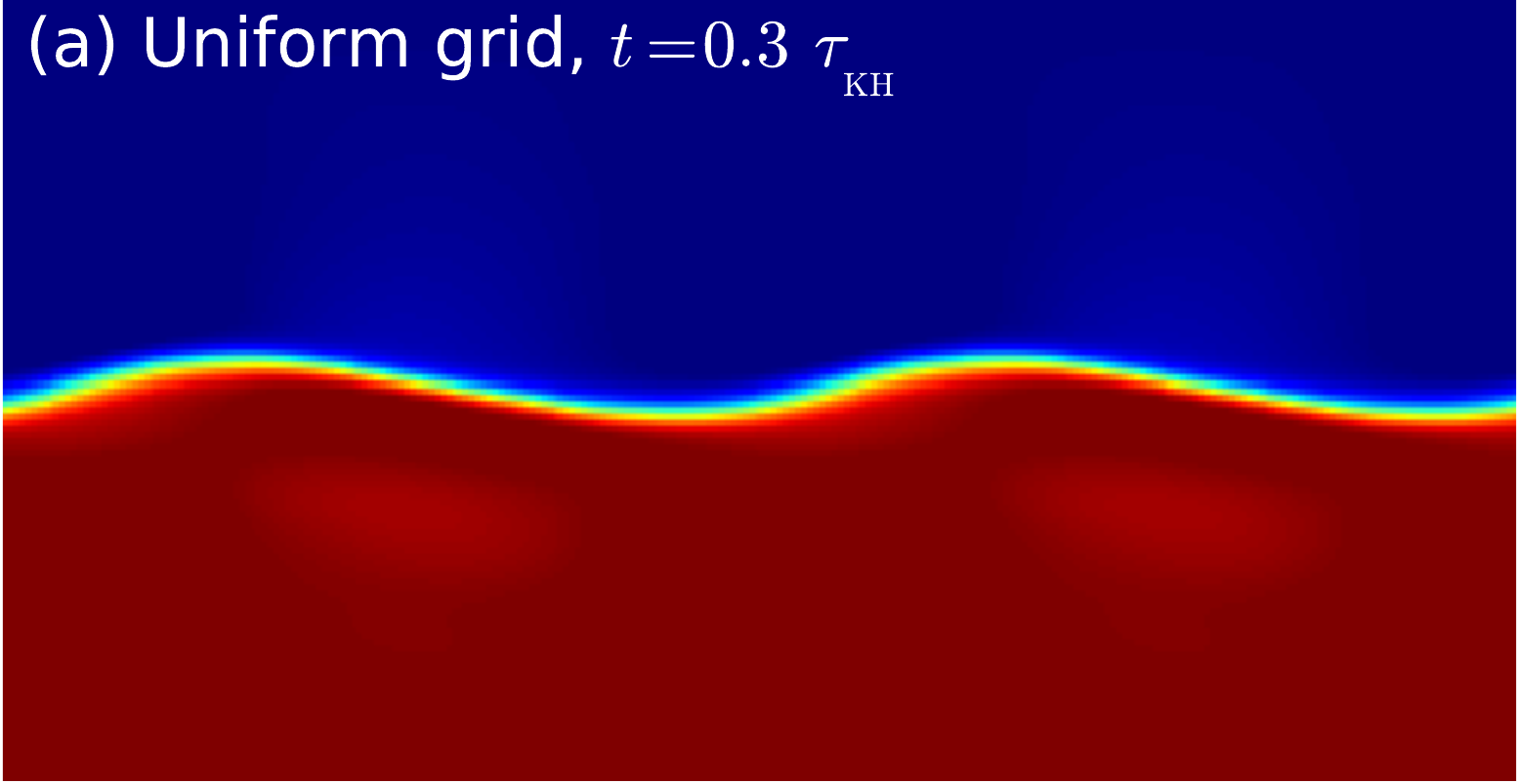}
\includegraphics[width=3.5cm]{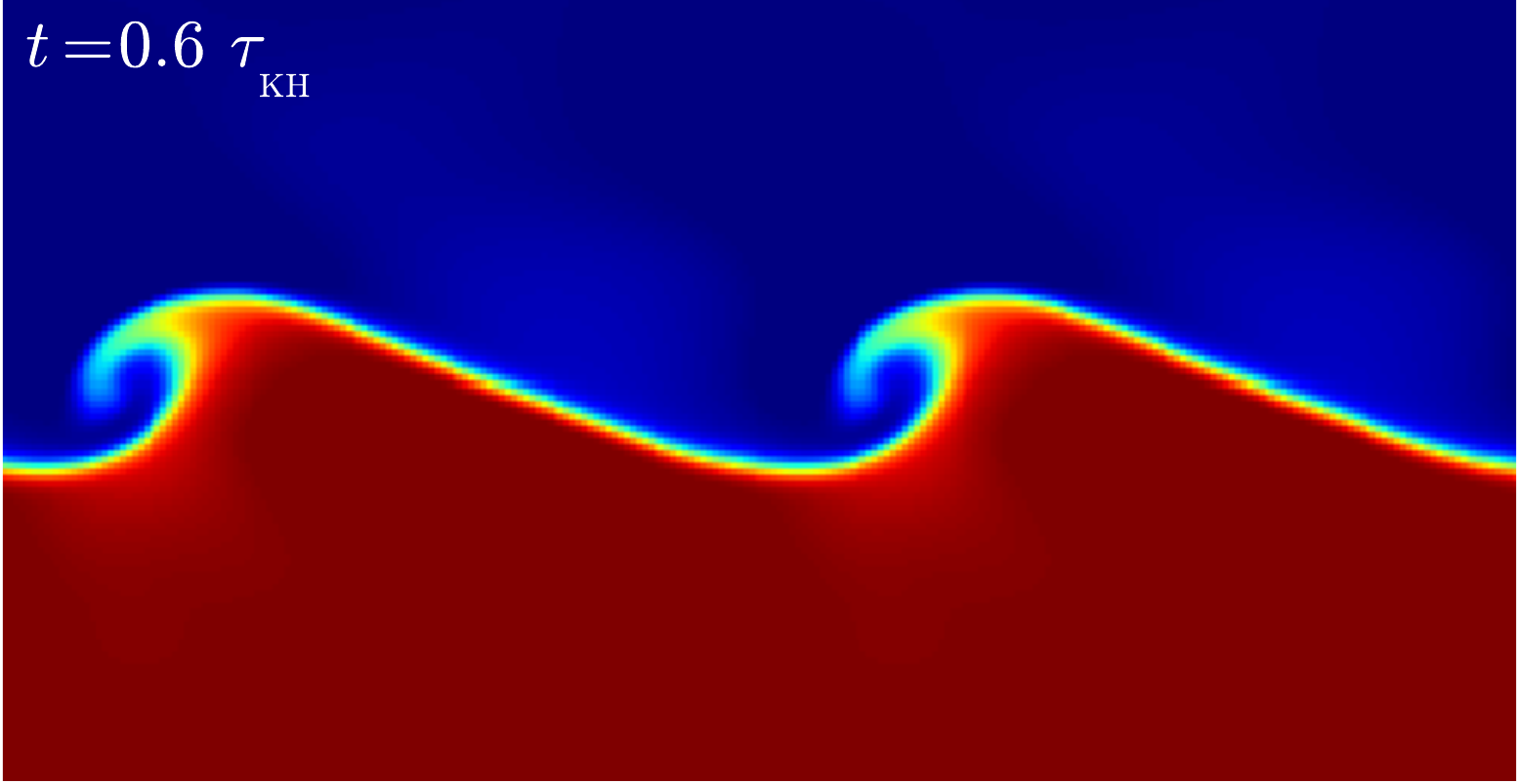}
\includegraphics[width=3.5cm]{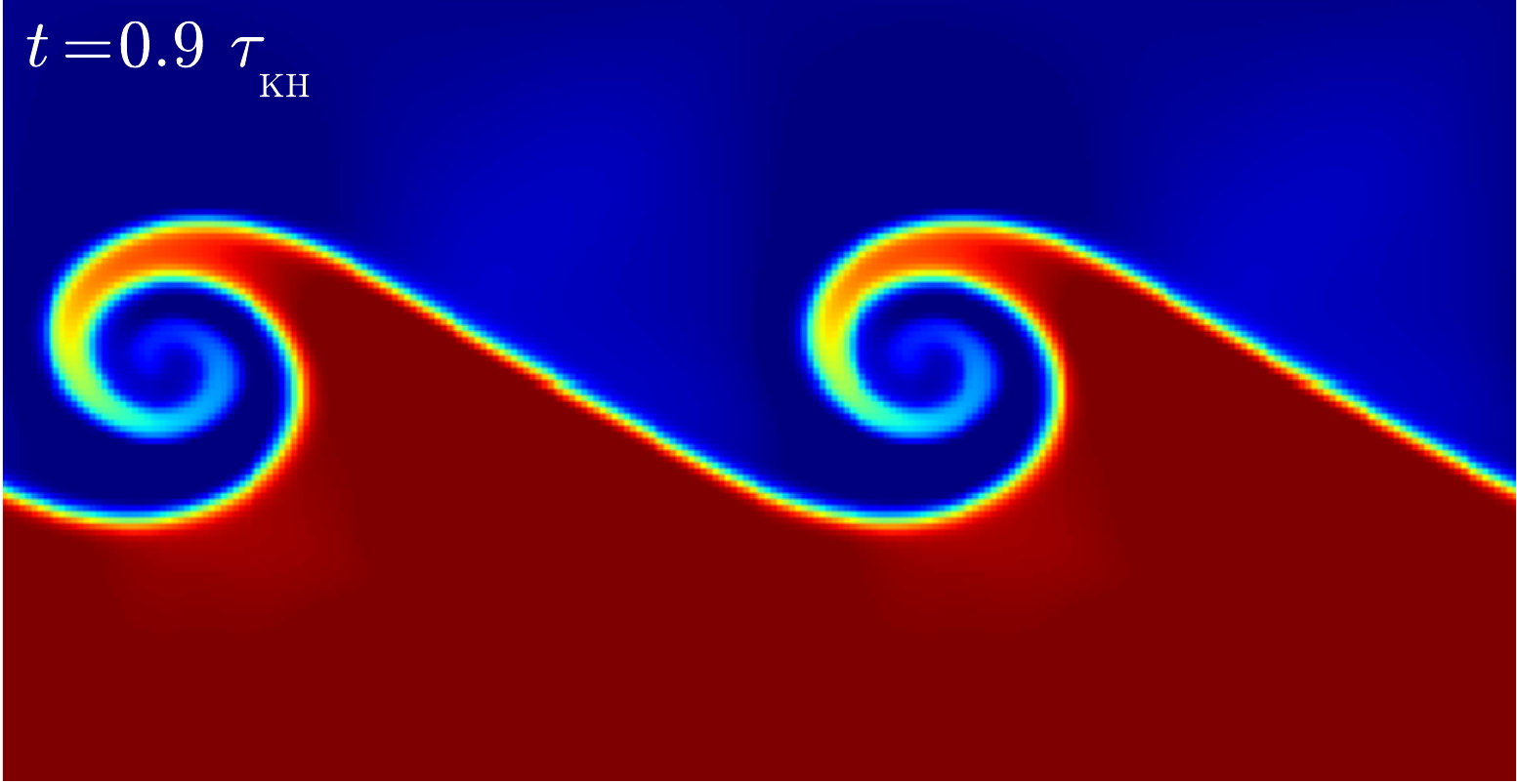}
\includegraphics[width=3.5cm]{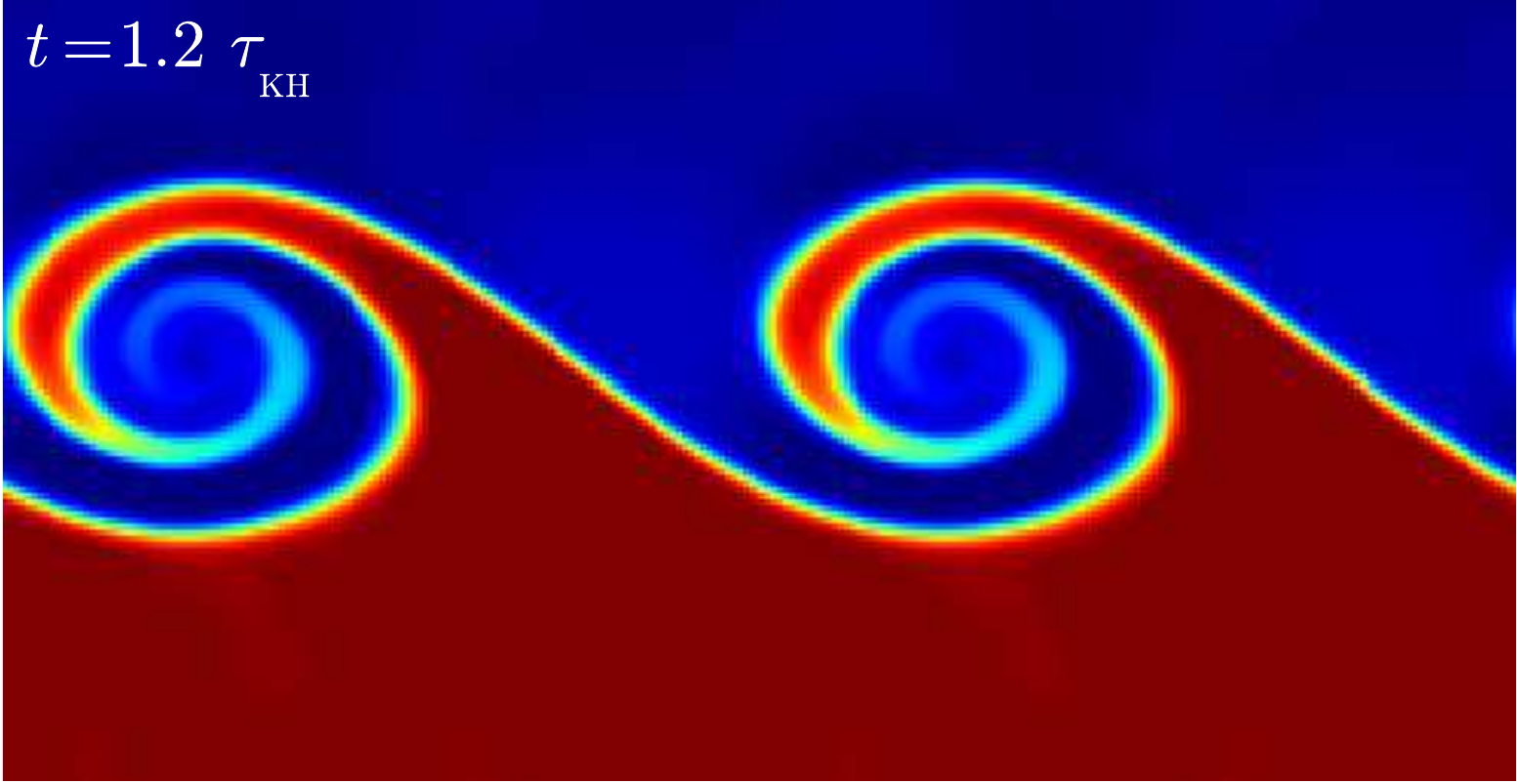}
\includegraphics[width=3.5cm]{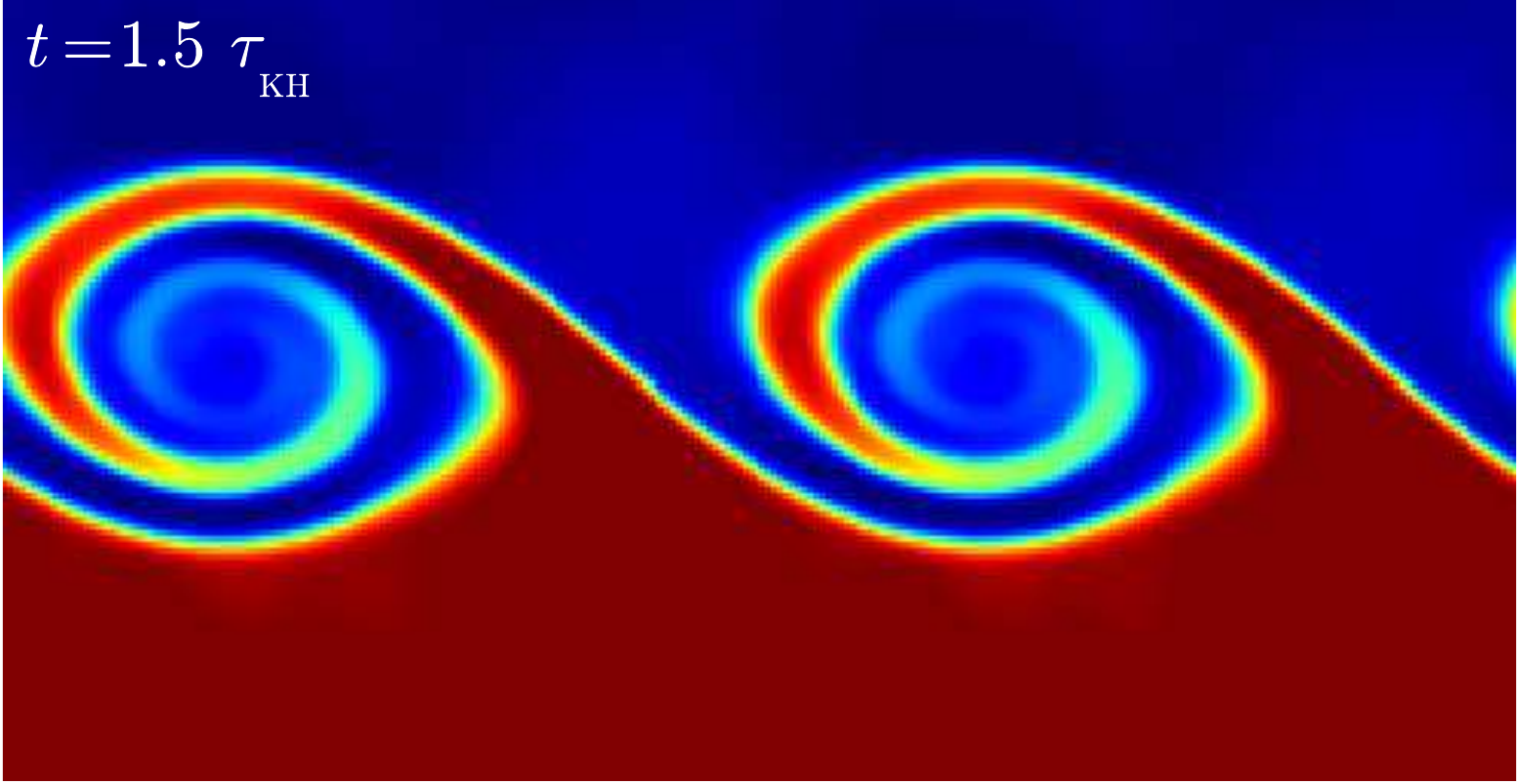}}
\centerline{
\includegraphics[width=3.5cm]{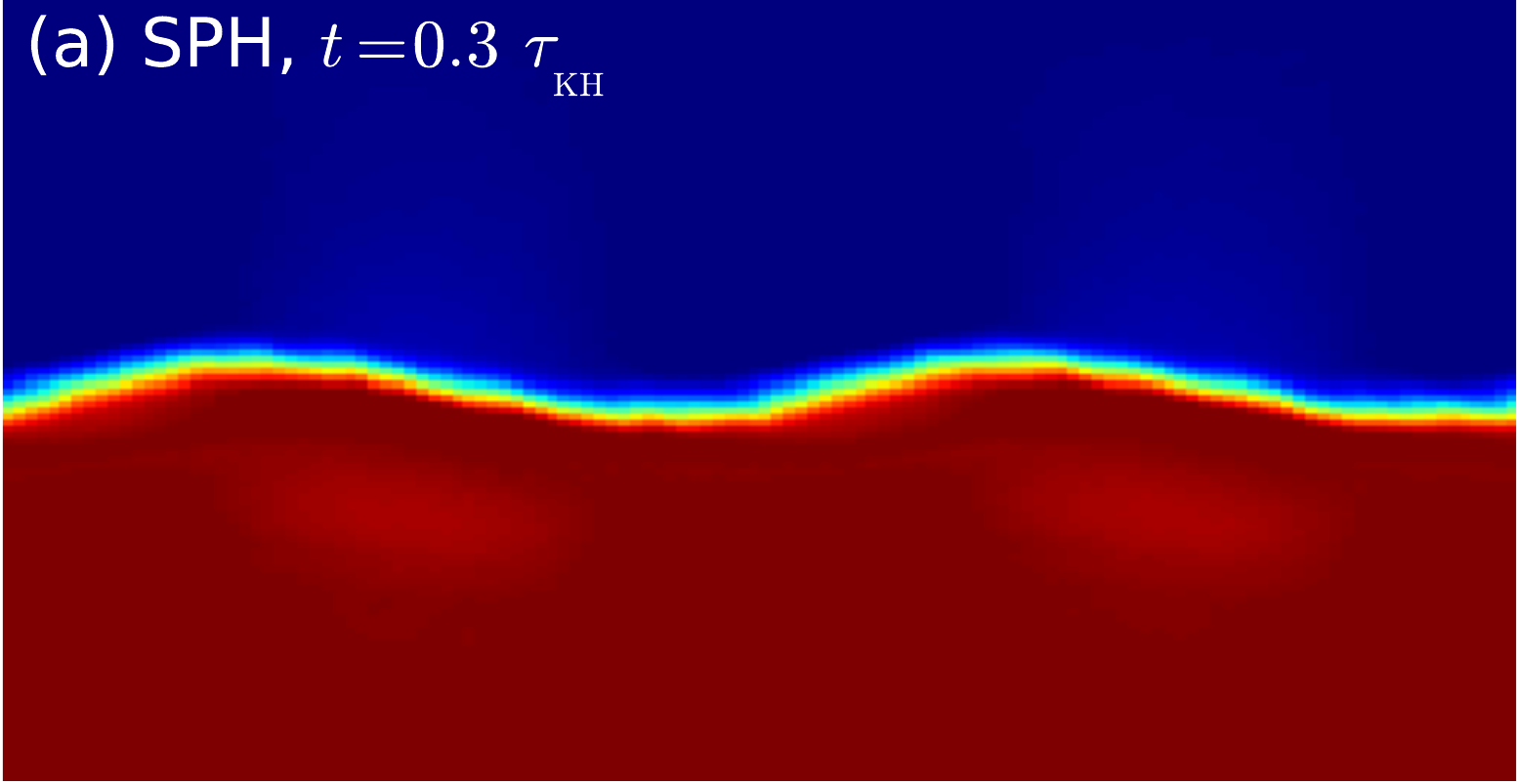}
\includegraphics[width=3.5cm]{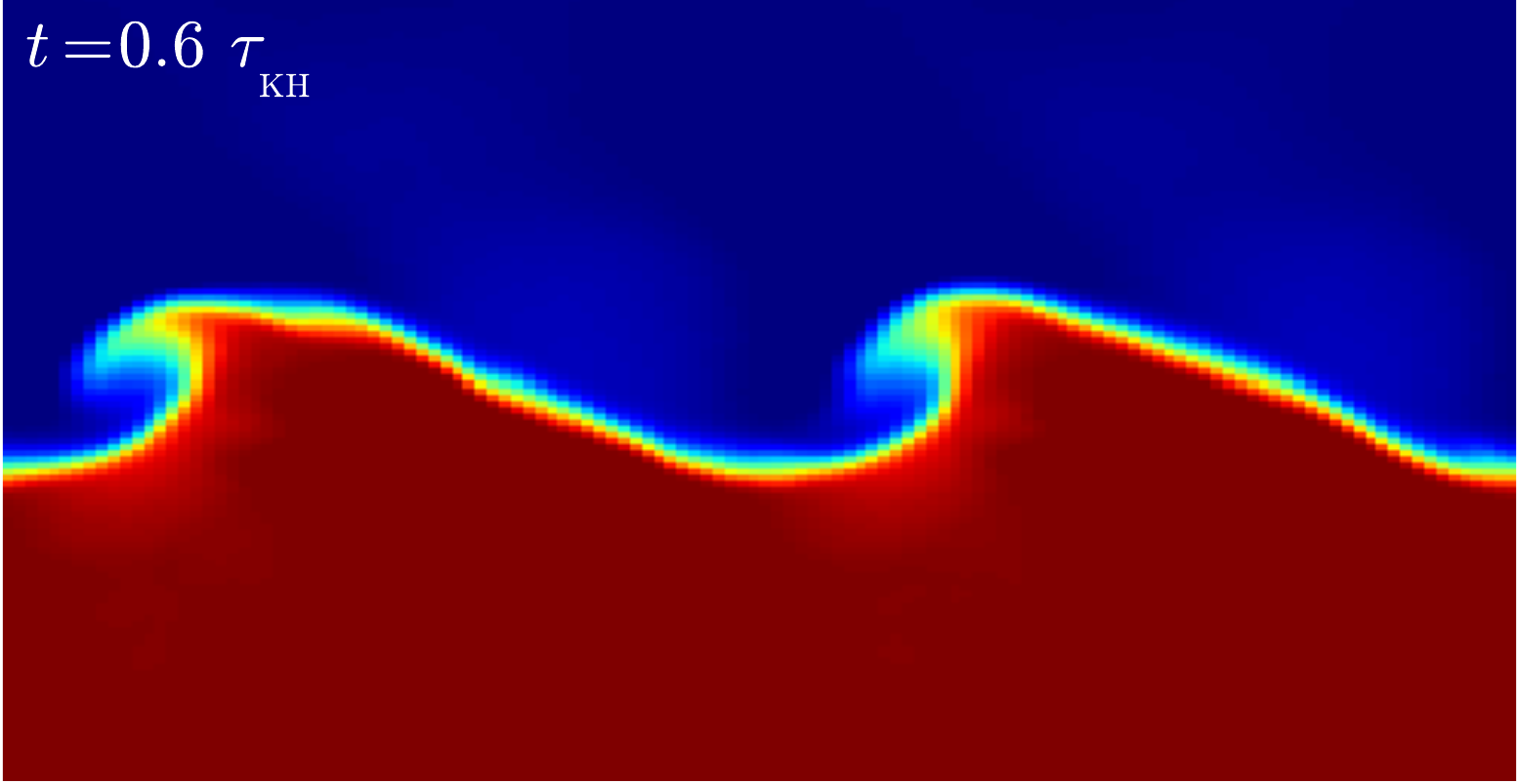}
\includegraphics[width=3.5cm]{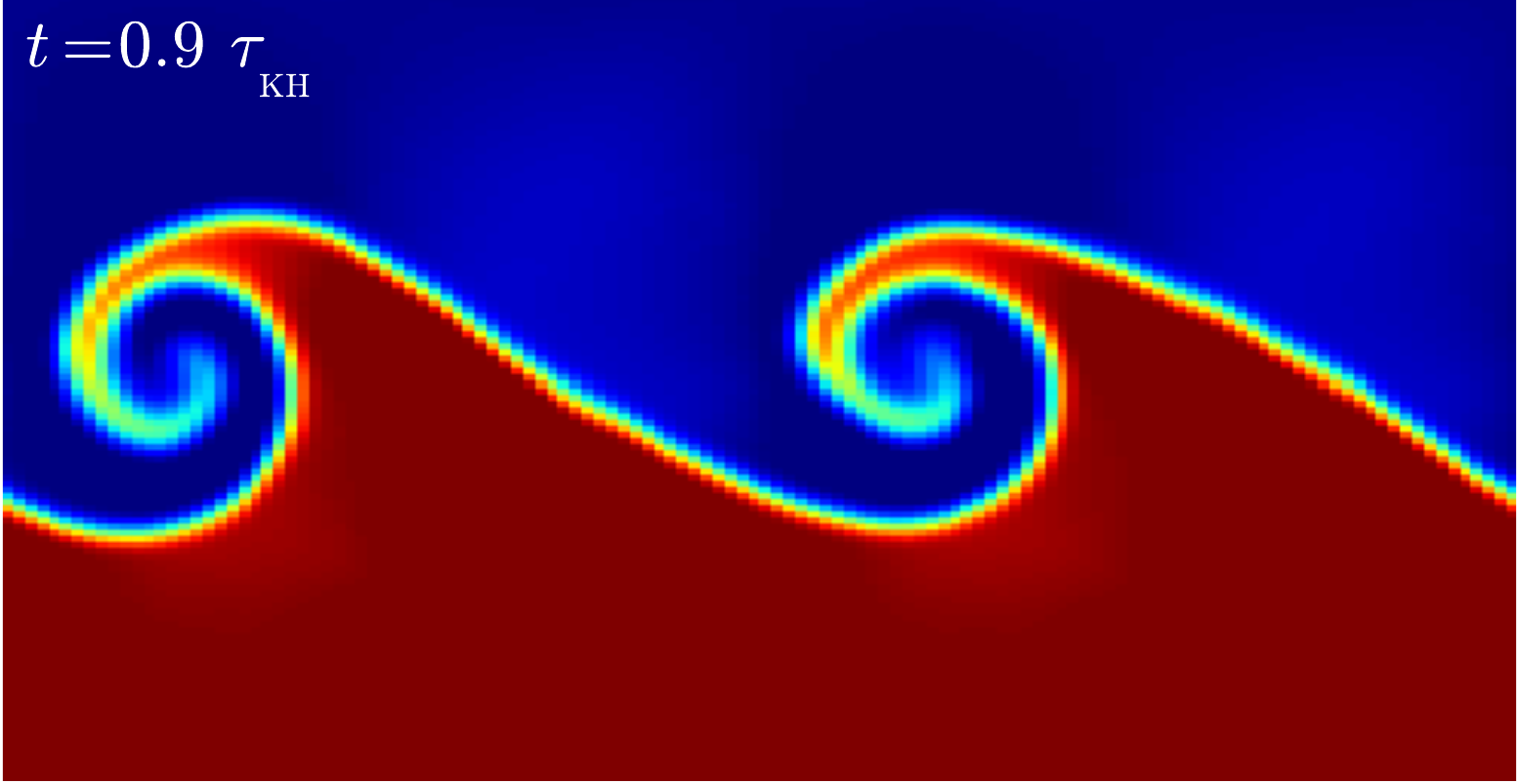}
\includegraphics[width=3.5cm]{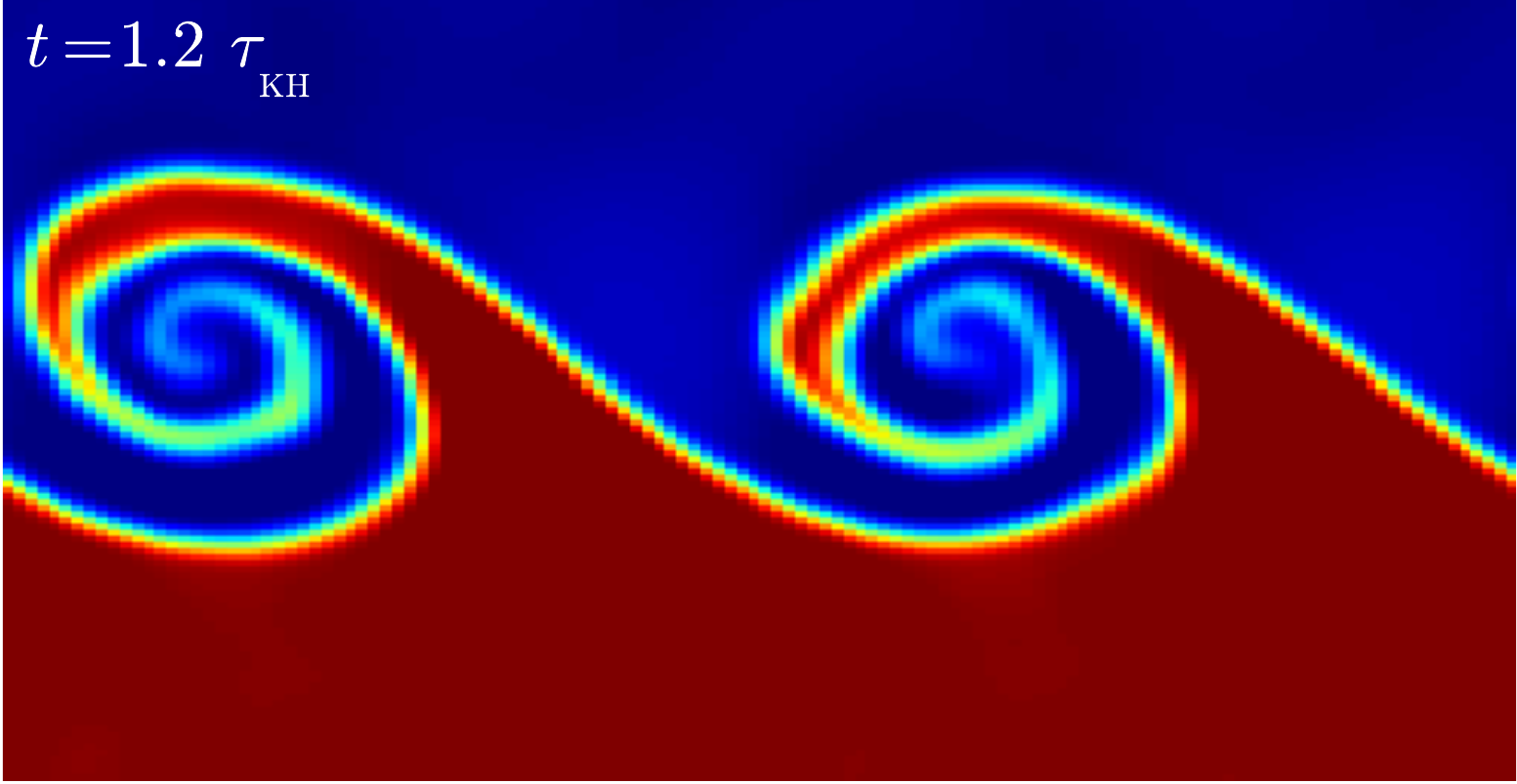}
\includegraphics[width=3.5cm]{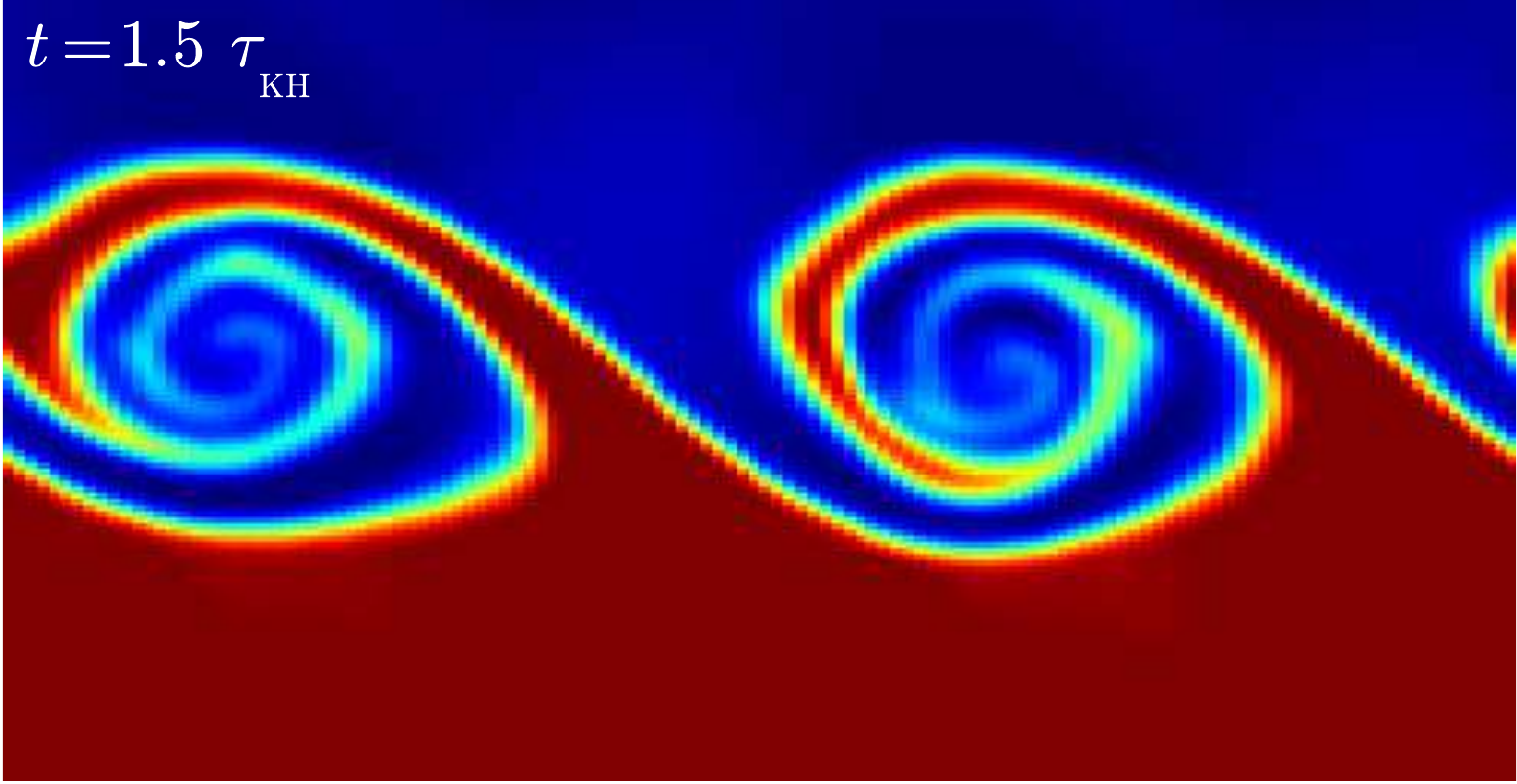}}
\vspace{0.04cm}
\caption{Development of the Kelvin-Helmholtz instability with a $2:1$ density contrast using (a) {\small MG} with a $256 \times 256$ uniform grid, and (b) {\small SEREN} {\refone with $195,872$ particles using} conservative SPH with the quintic kernel and \citet{Price2008} artificial conductivity.  The columns from left to right show the development of the instability at times $t = 0.3\tau_{_{\rm KH}}$, $t = 0.6\tau_{_{\rm KH}}$, $t = 0.9\tau_{_{\rm KH}}$, $t = 1.2\tau_{_{\rm KH}}$ and $t = 1.5\tau_{_{\rm KH}}$.}
\label{FIG:KHI-2:1-TIME}
\end{figure*}

\begin{figure*}
\centerline{
\includegraphics[width=4.35cm]{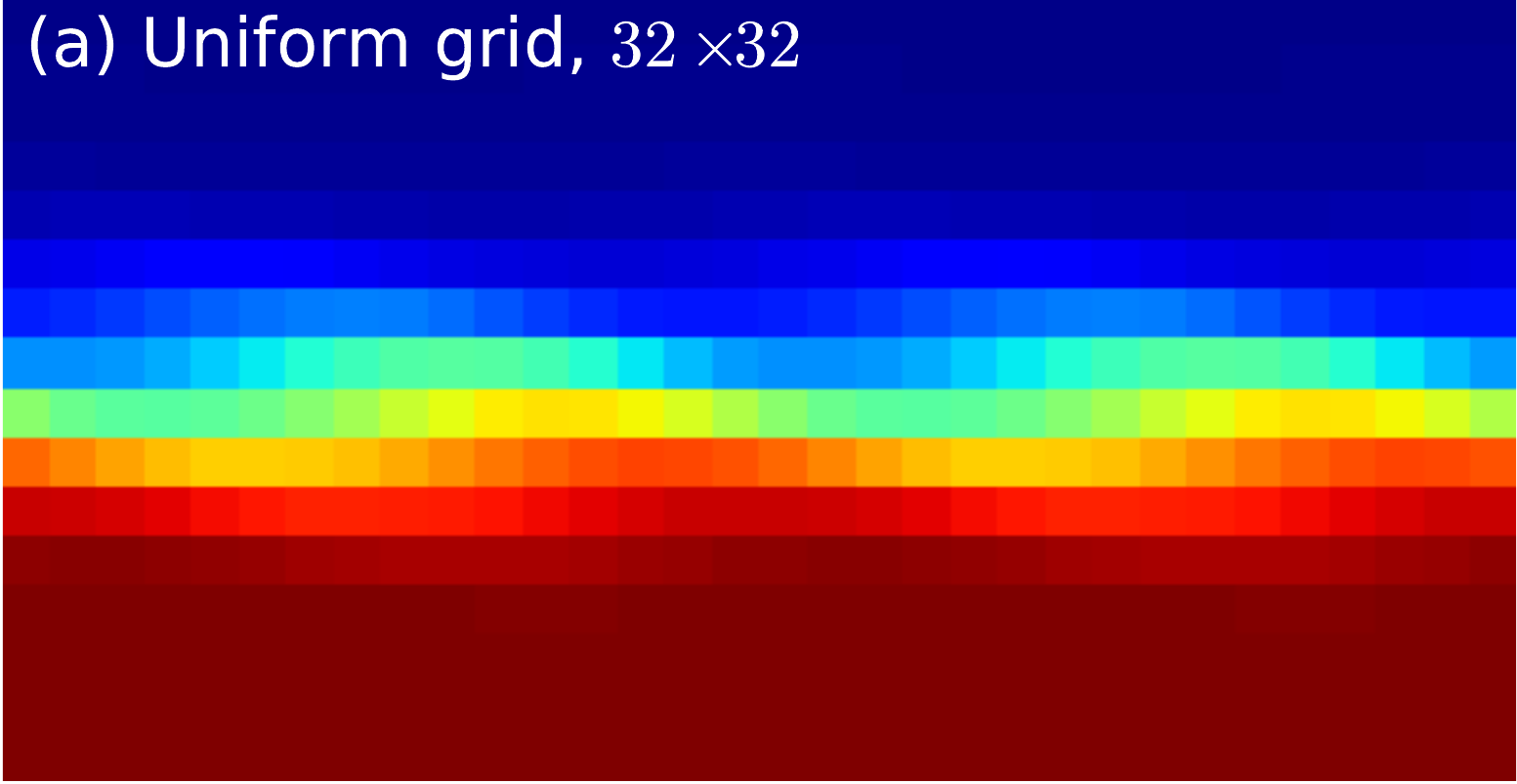}
\includegraphics[width=4.35cm]{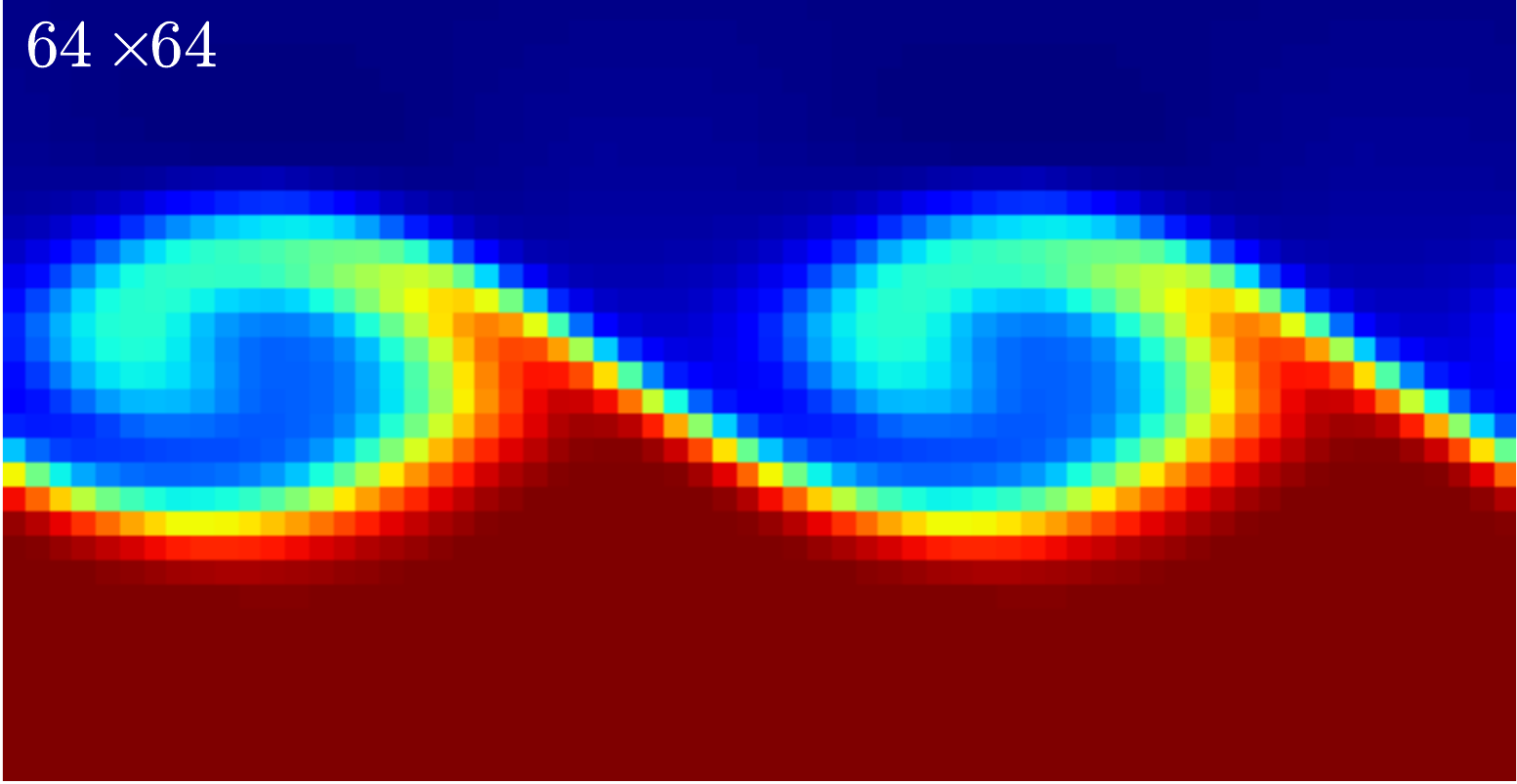}
\includegraphics[width=4.35cm]{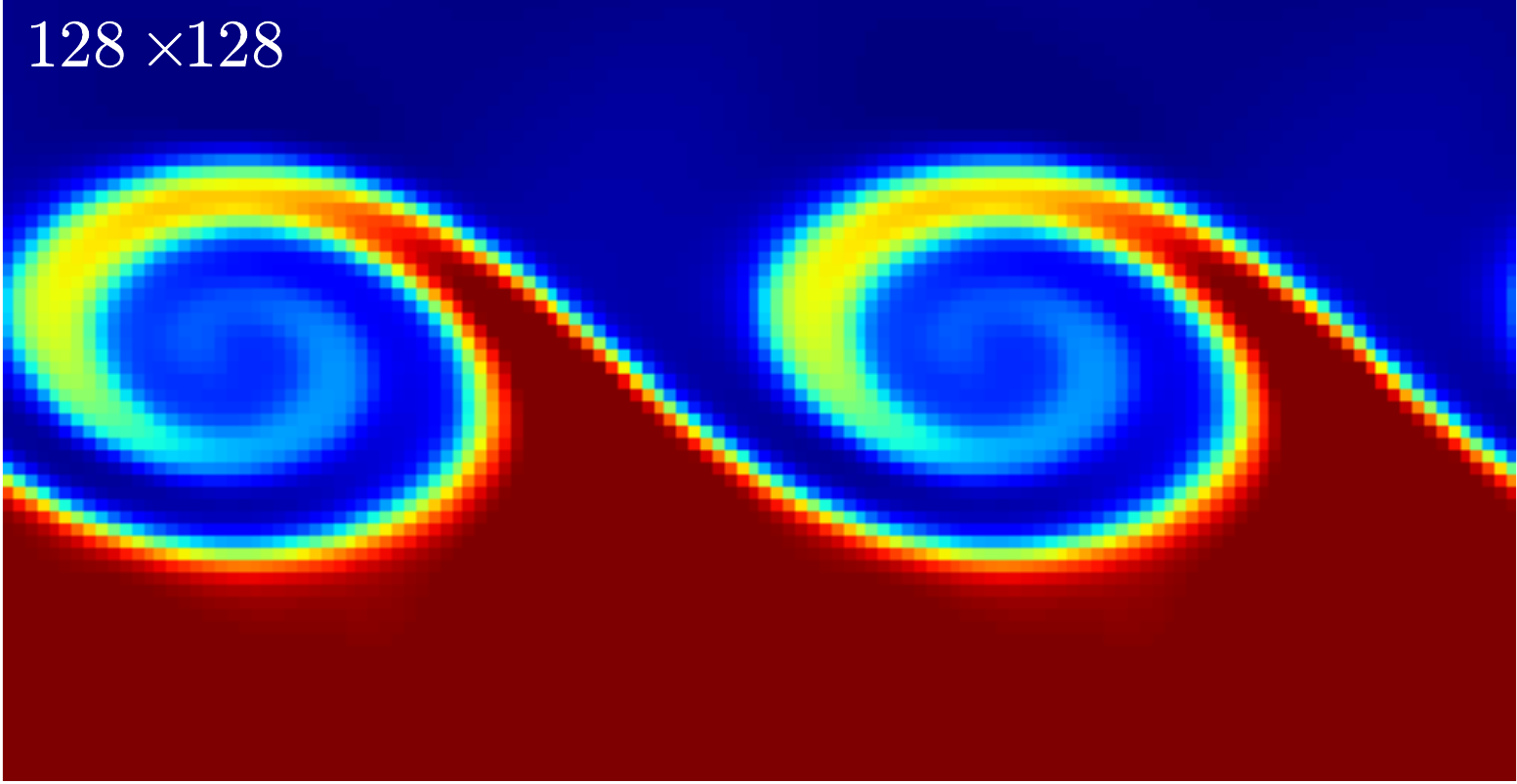}
\includegraphics[width=4.35cm]{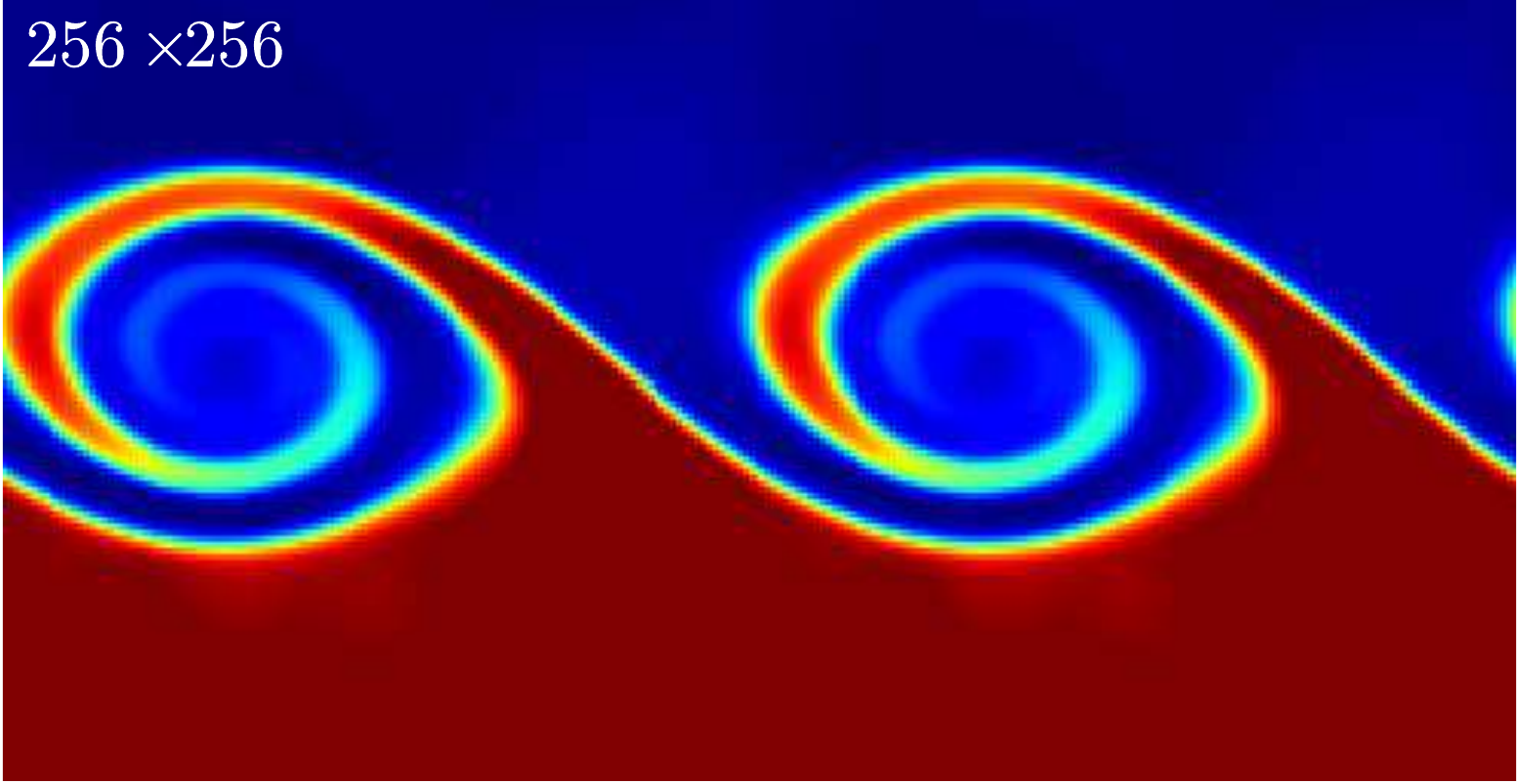}}
\vspace{0.04cm}
\centerline{
\includegraphics[width=4.35cm]{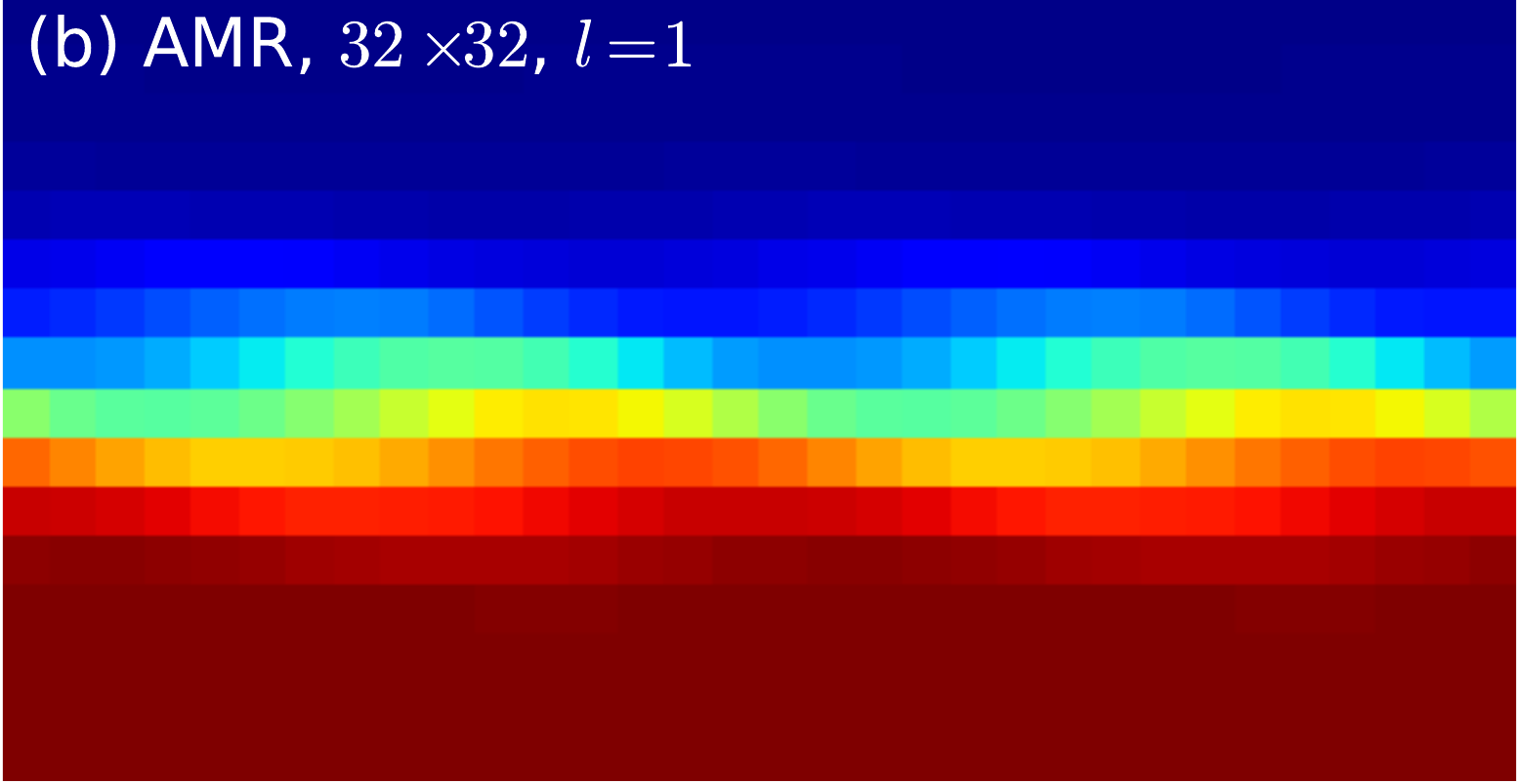}
\includegraphics[width=4.35cm]{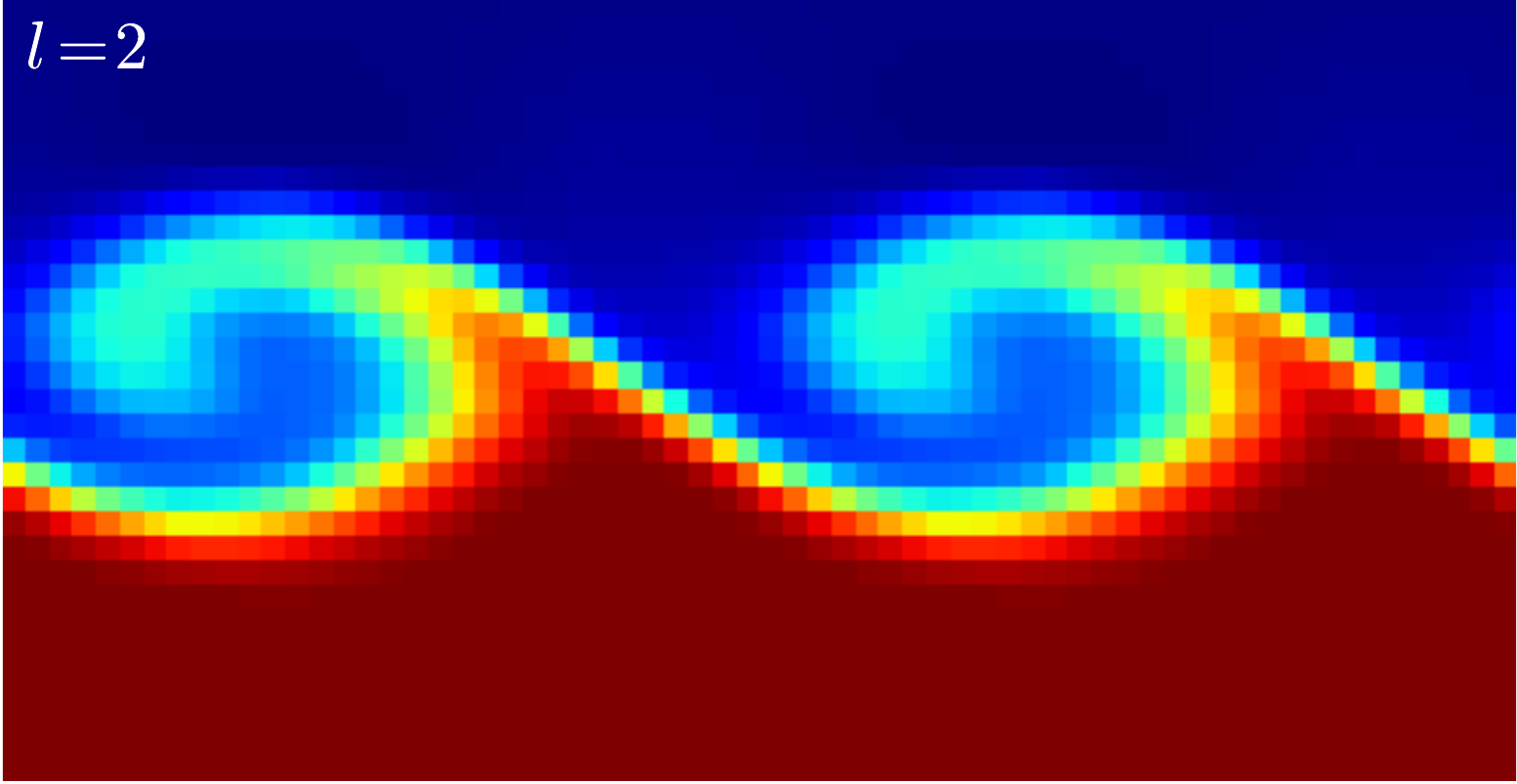}
\includegraphics[width=4.35cm]{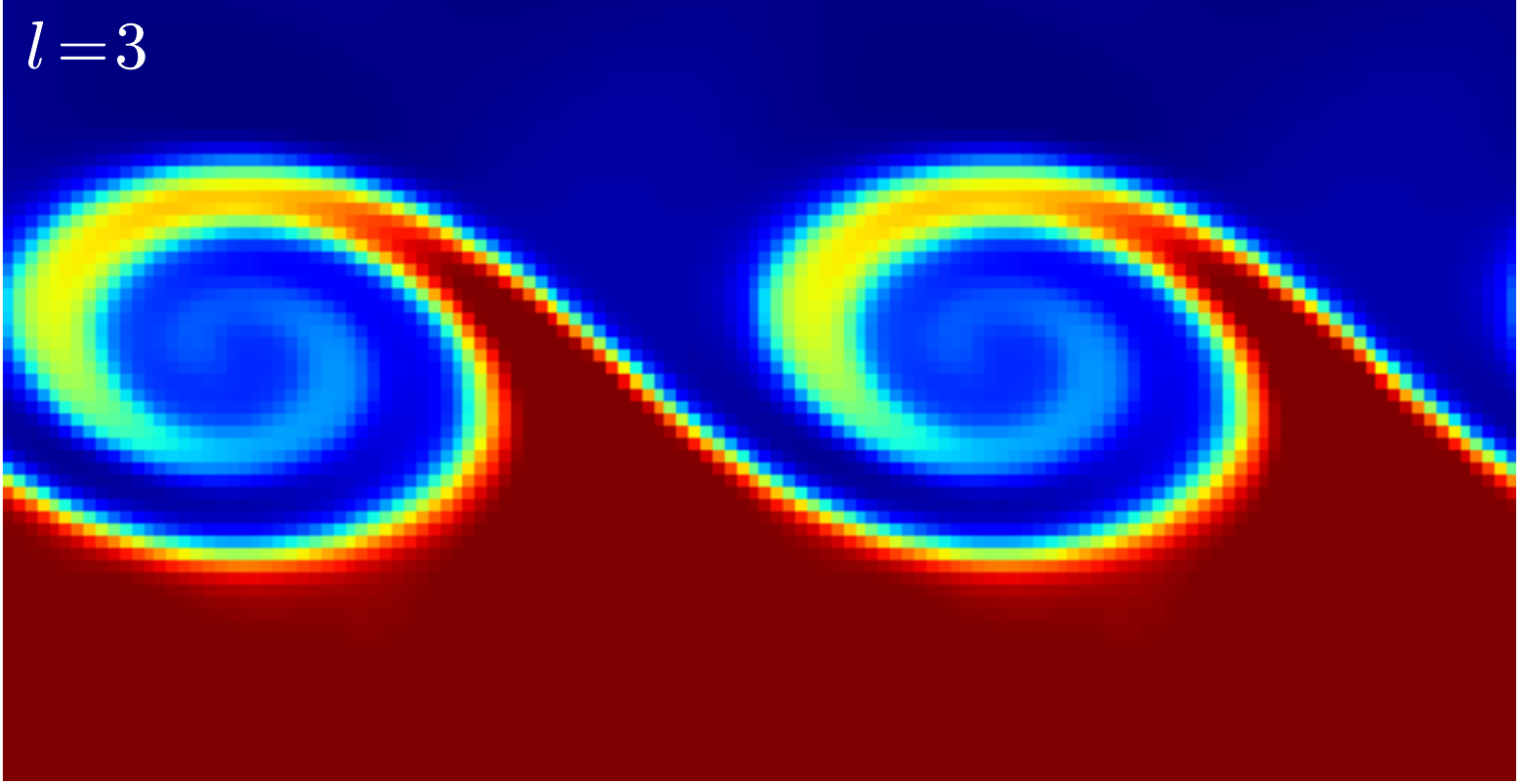}
\includegraphics[width=4.35cm]{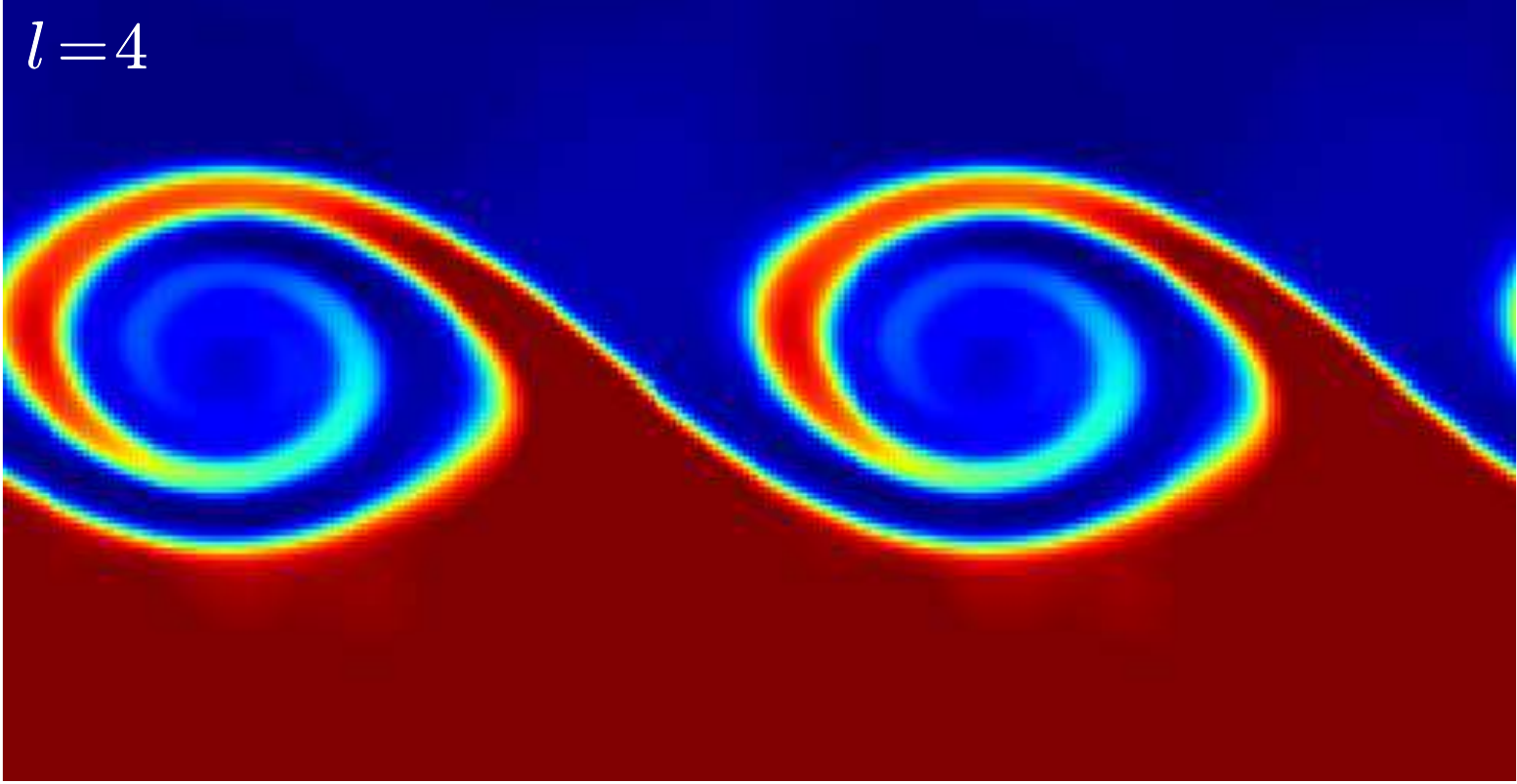}}
\vspace{0.04cm}
\centerline{
\includegraphics[width=4.35cm]{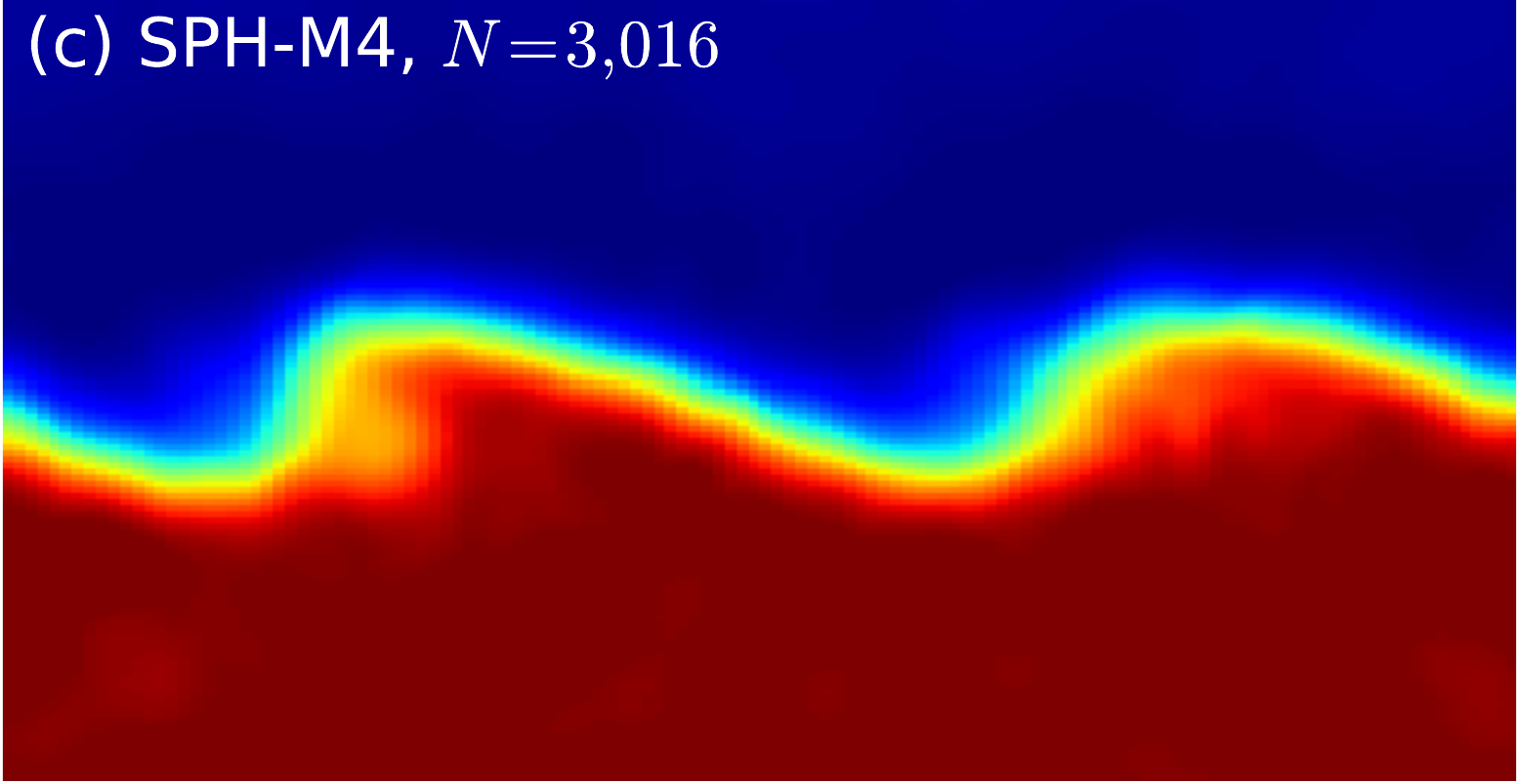}
\includegraphics[width=4.35cm]{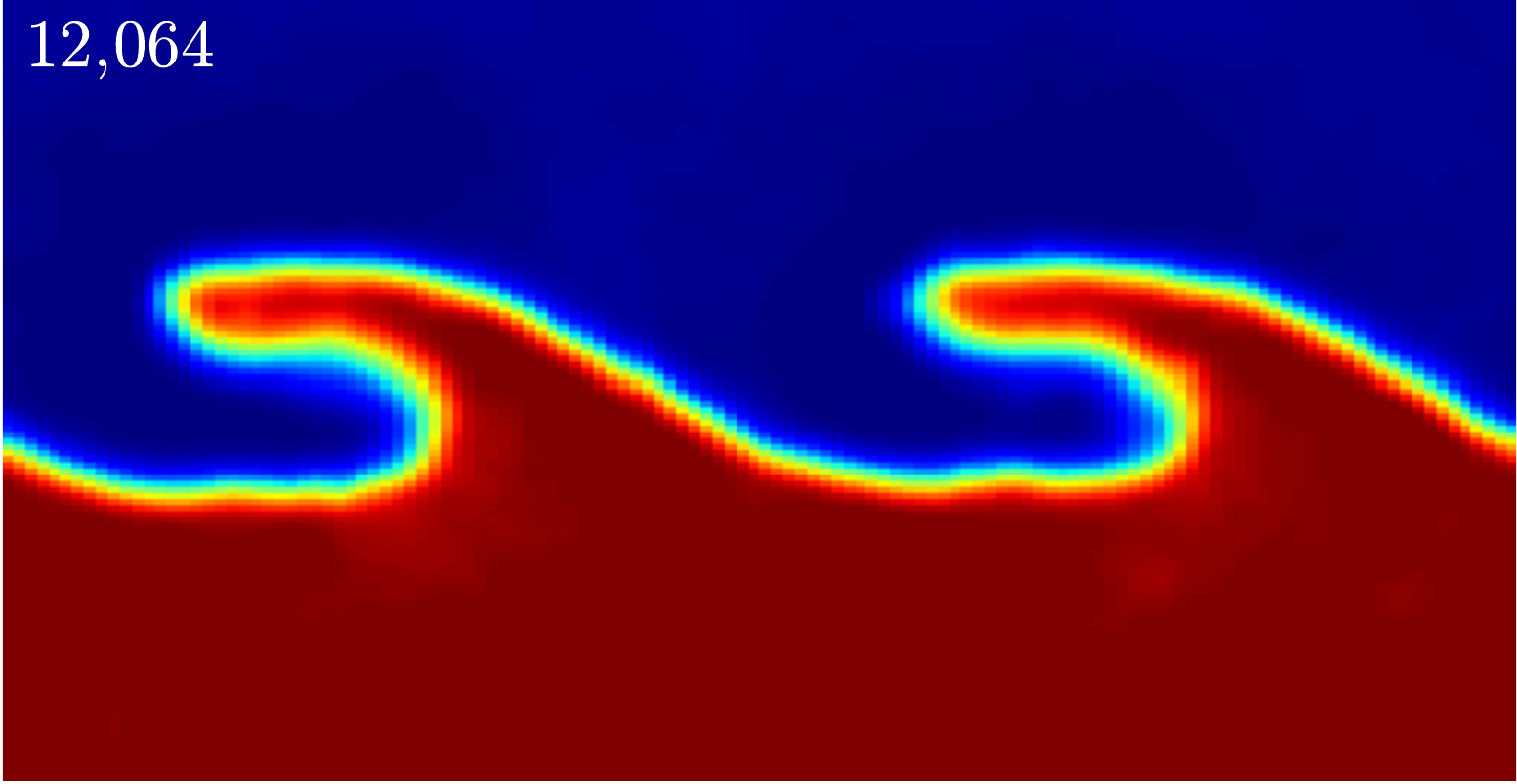}
\includegraphics[width=4.35cm]{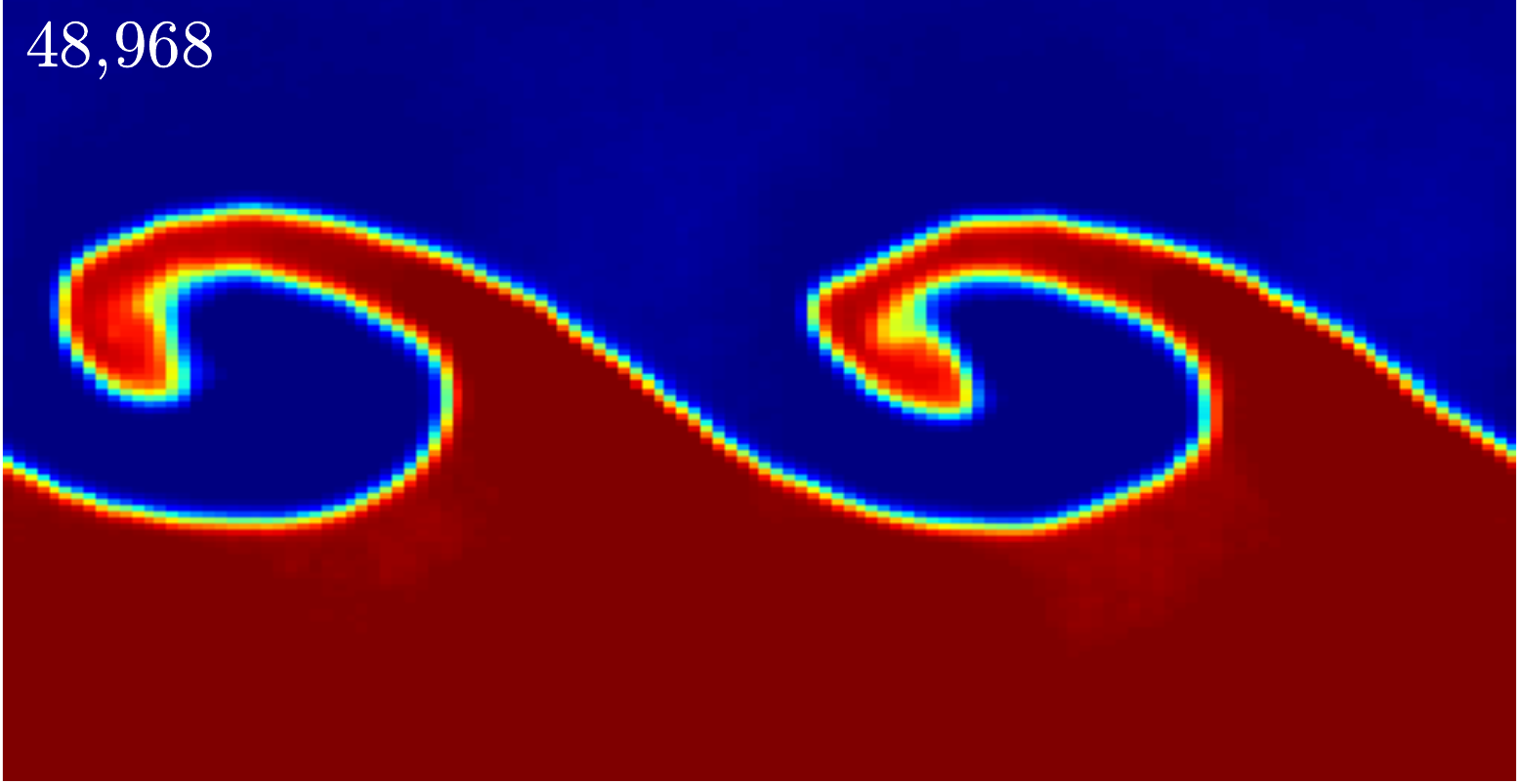}
\includegraphics[width=4.35cm]{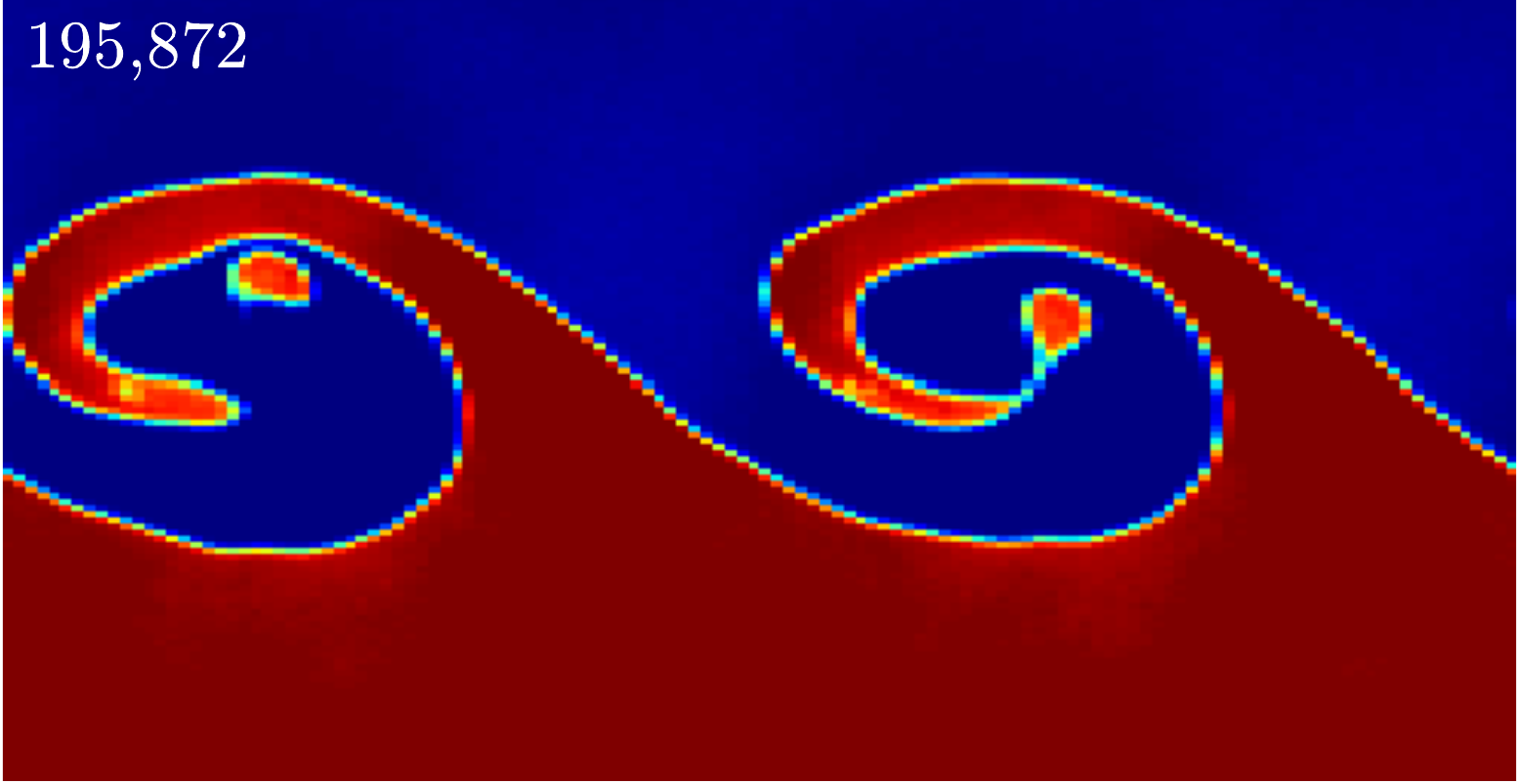}}
\vspace{0.04cm}
\centerline{
\includegraphics[width=4.35cm]{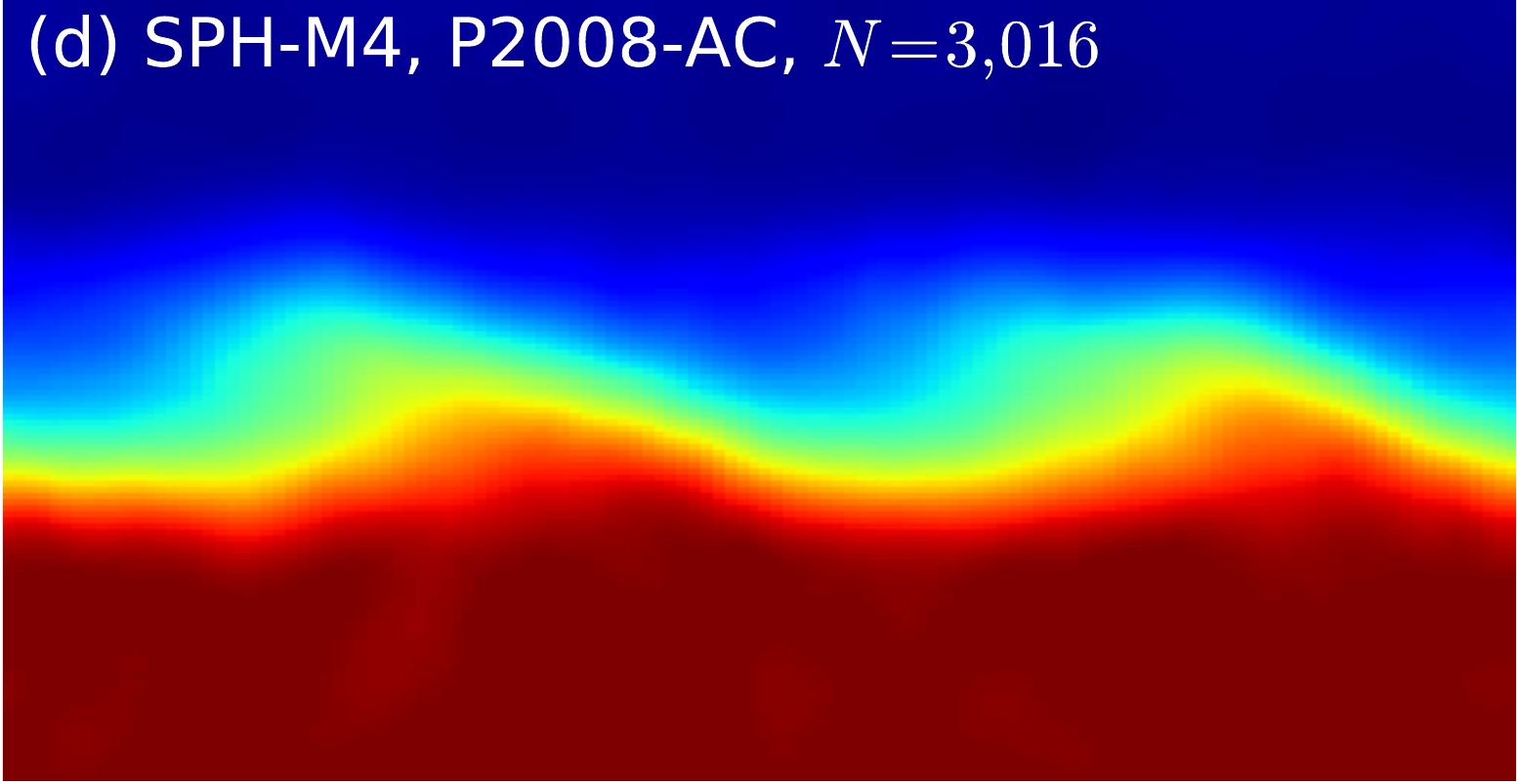}
\includegraphics[width=4.35cm]{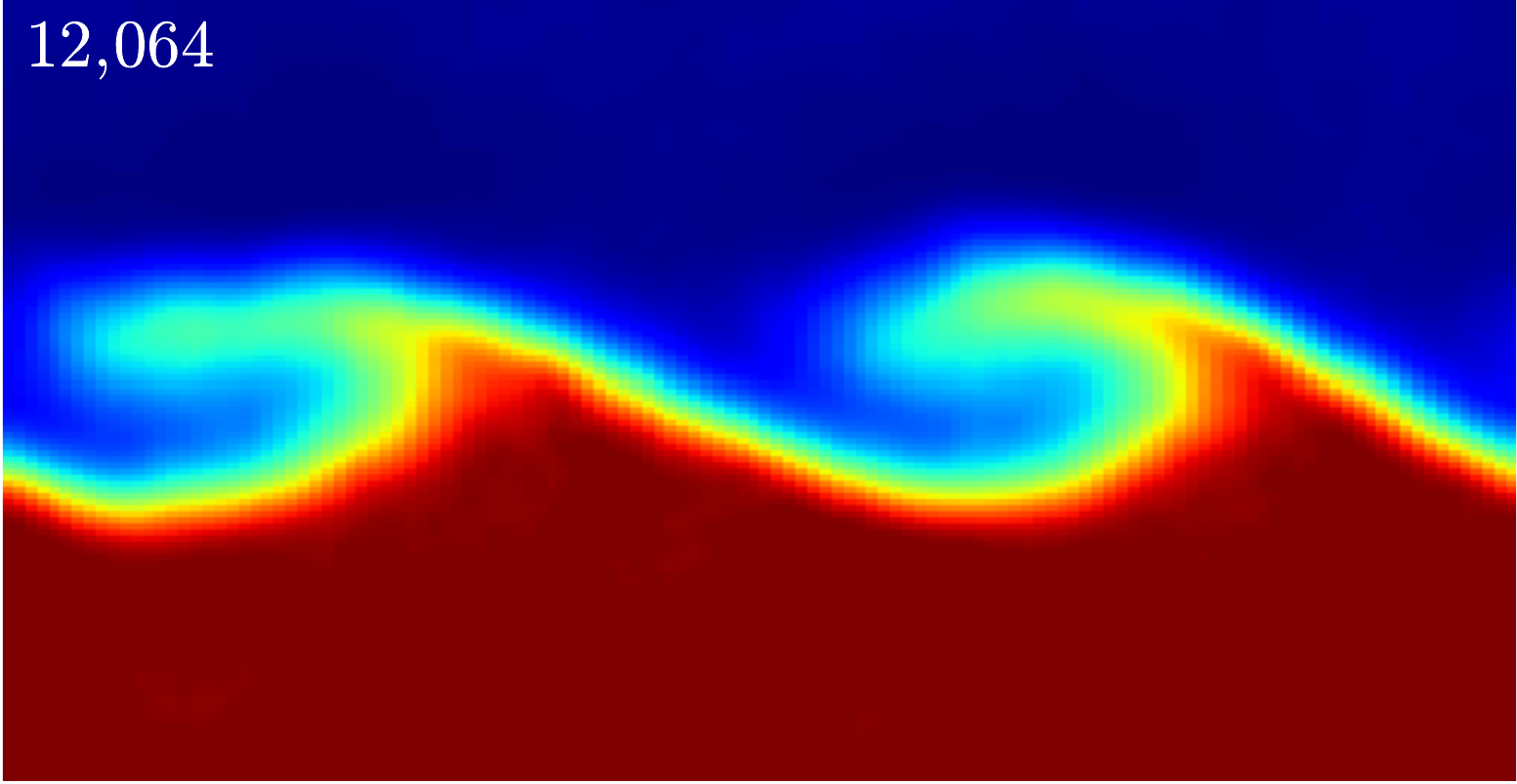}
\includegraphics[width=4.35cm]{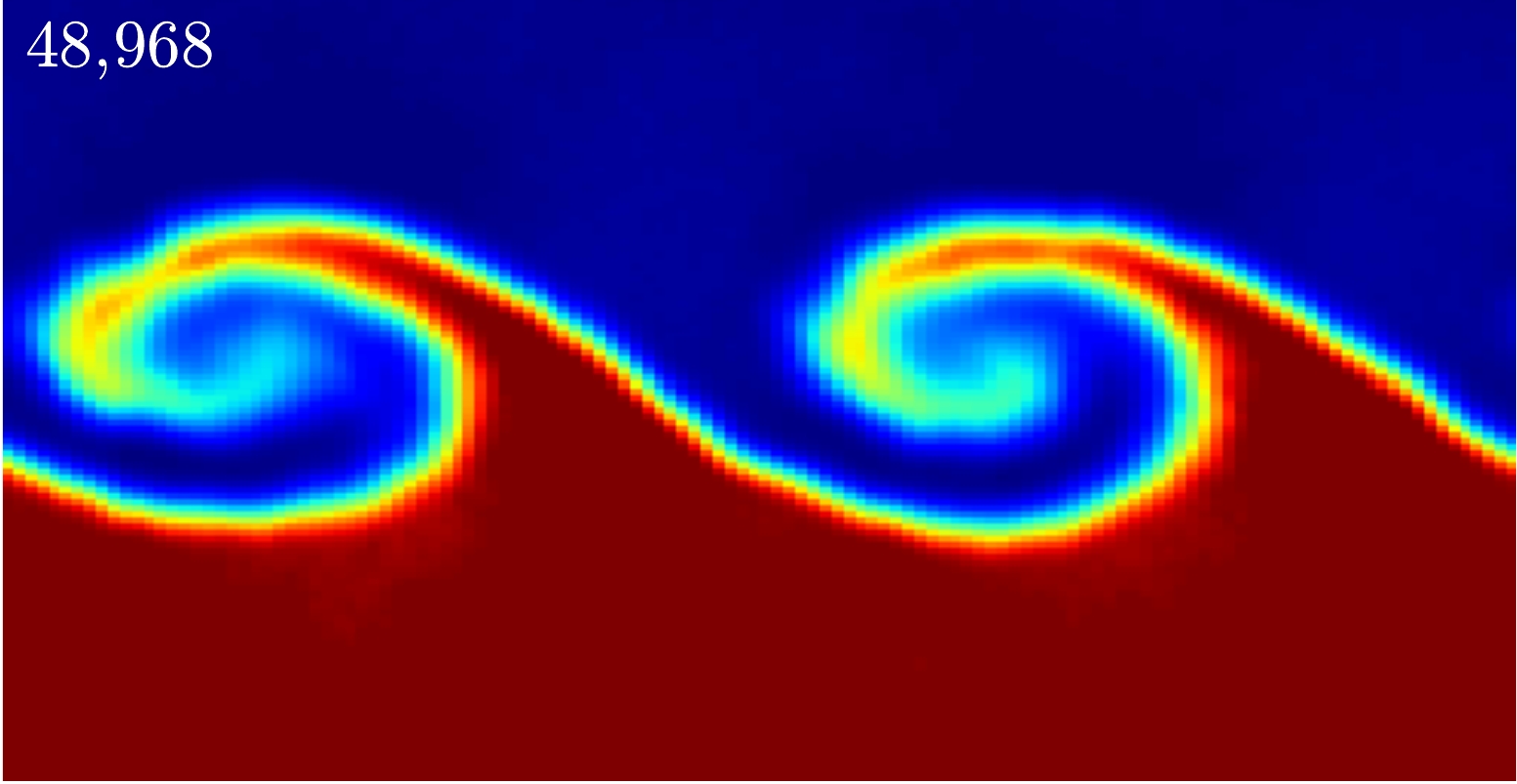}
\includegraphics[width=4.35cm]{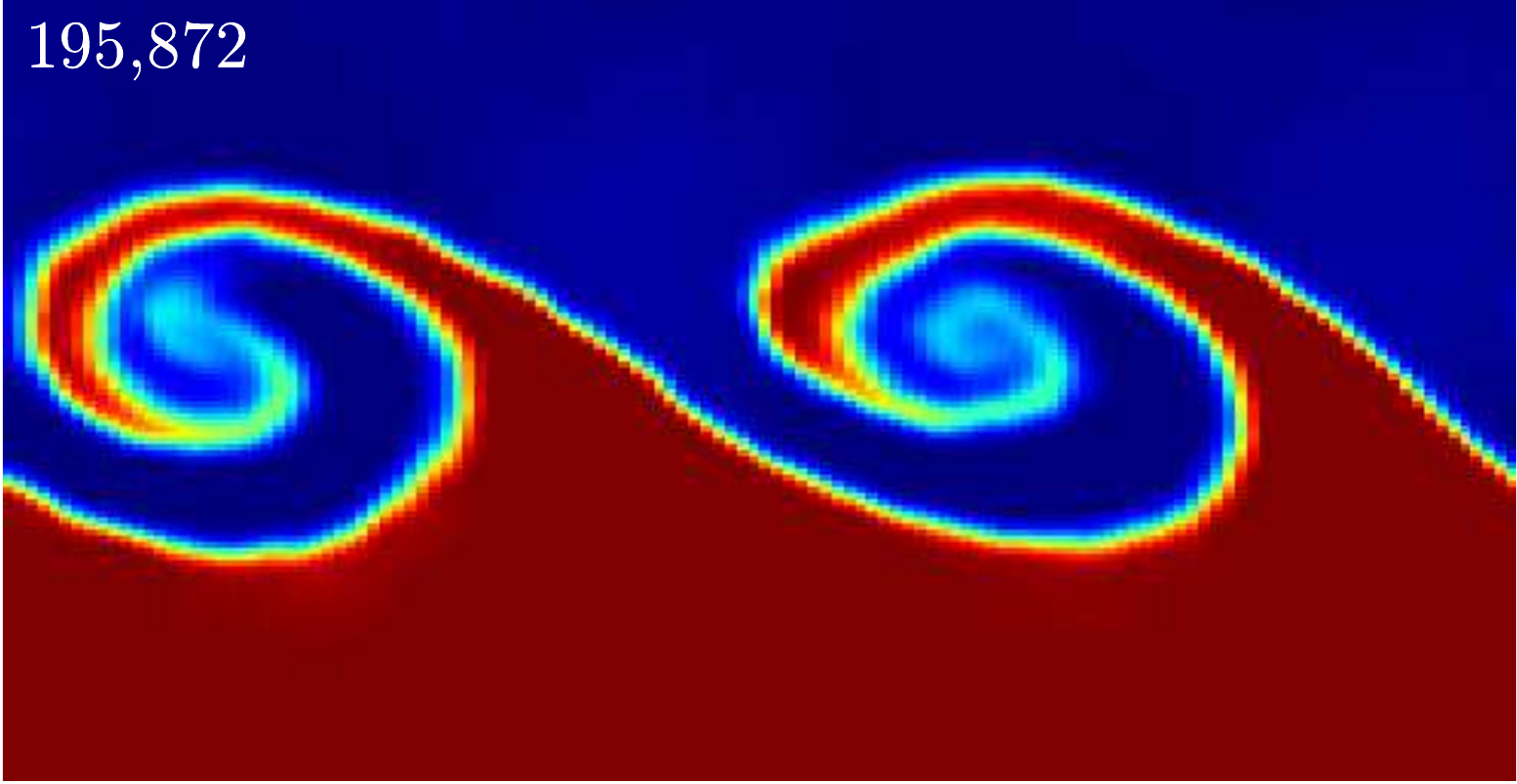}}
\vspace{0.04cm}
\centerline{
\includegraphics[width=4.35cm]{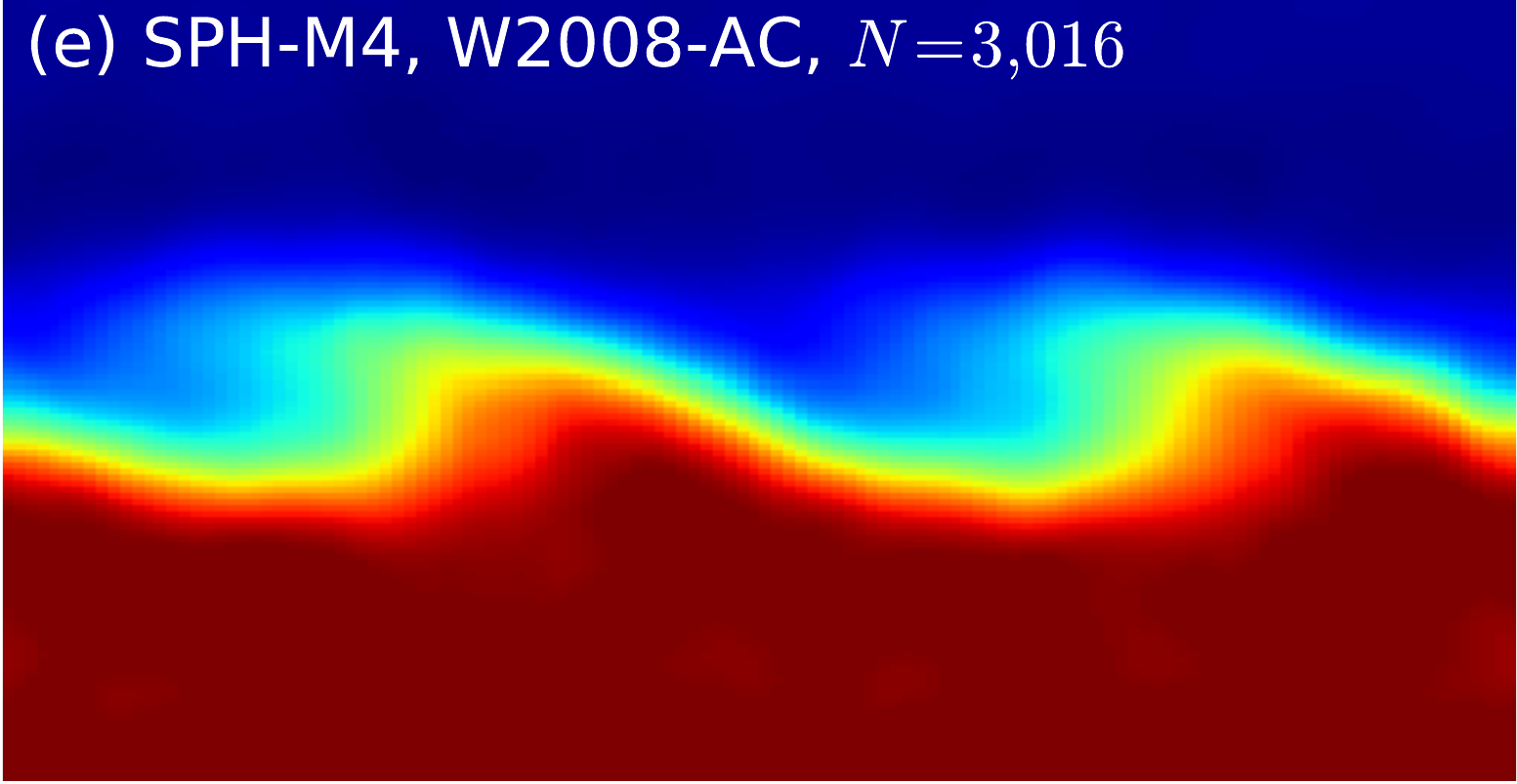}
\includegraphics[width=4.35cm]{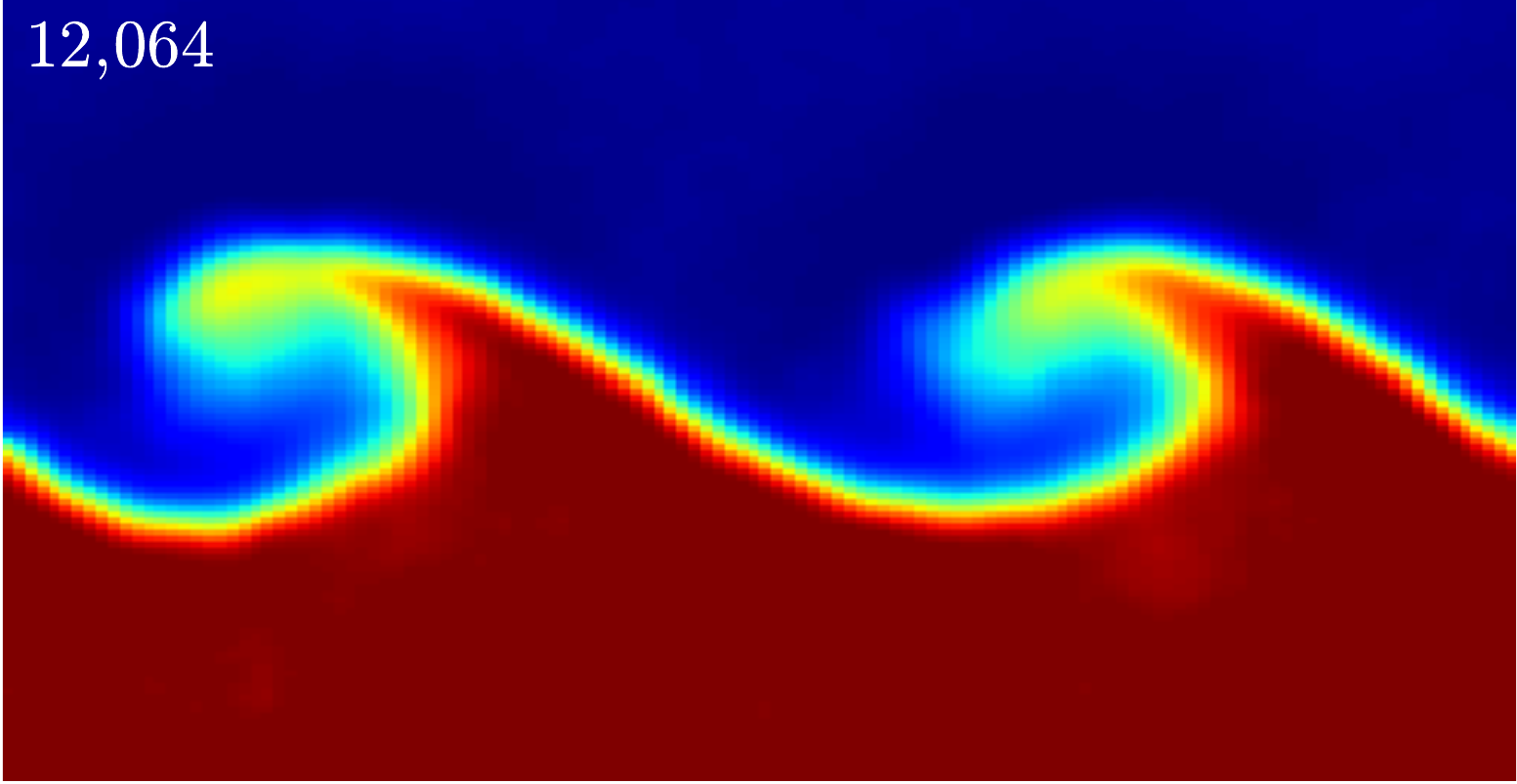}
\includegraphics[width=4.35cm]{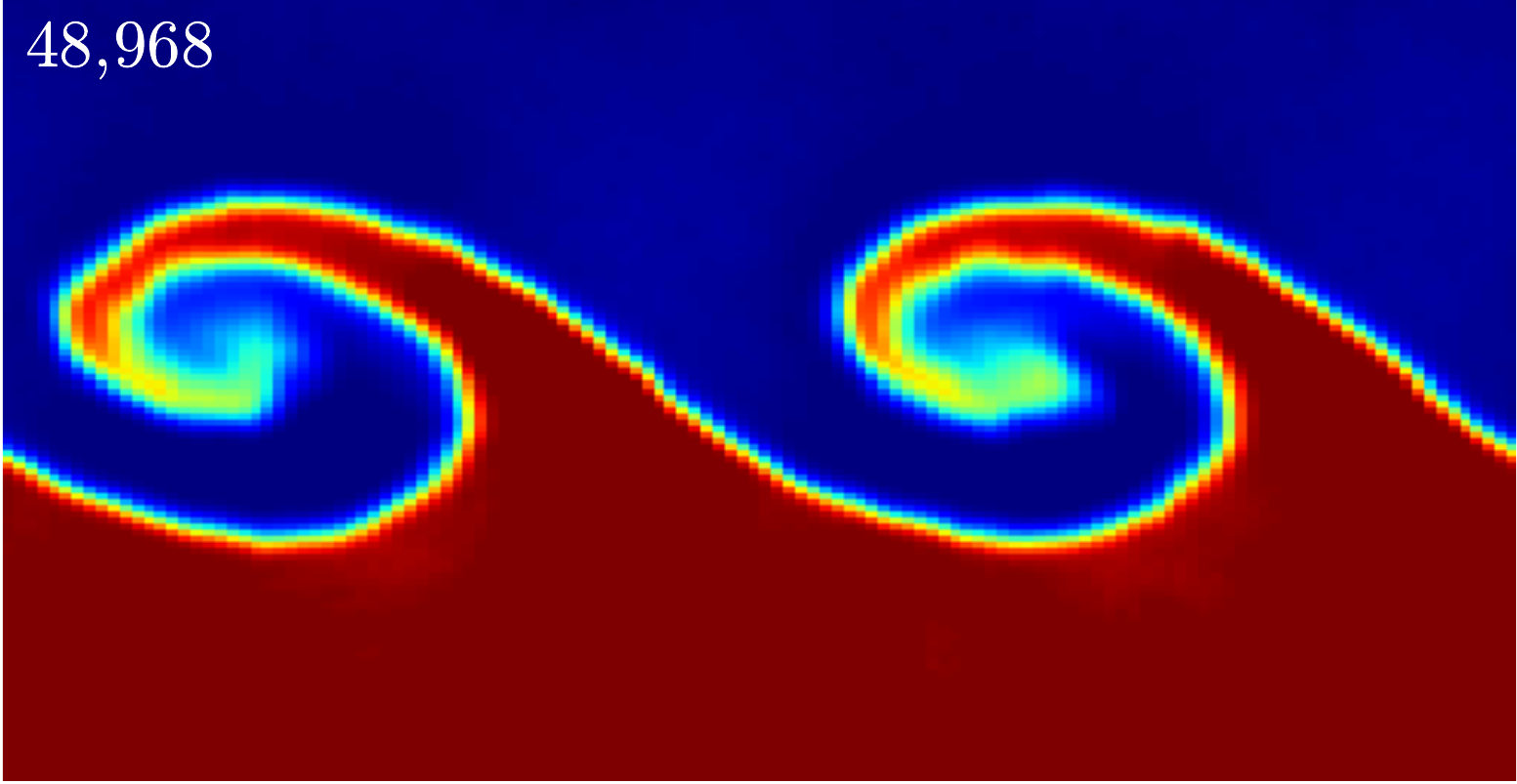}
\includegraphics[width=4.35cm]{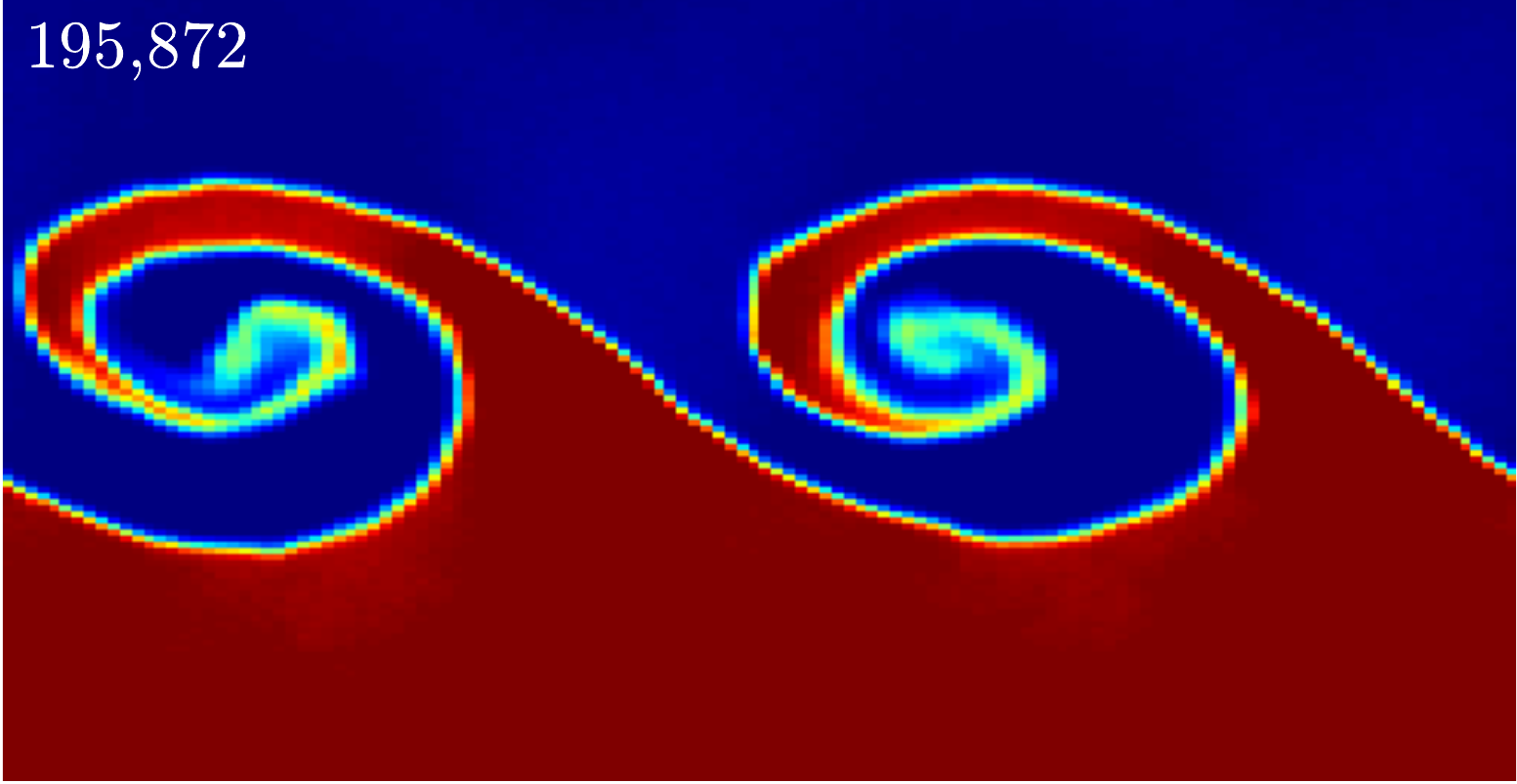}}
\vspace{0.04cm}
\centerline{
\includegraphics[width=4.35cm]{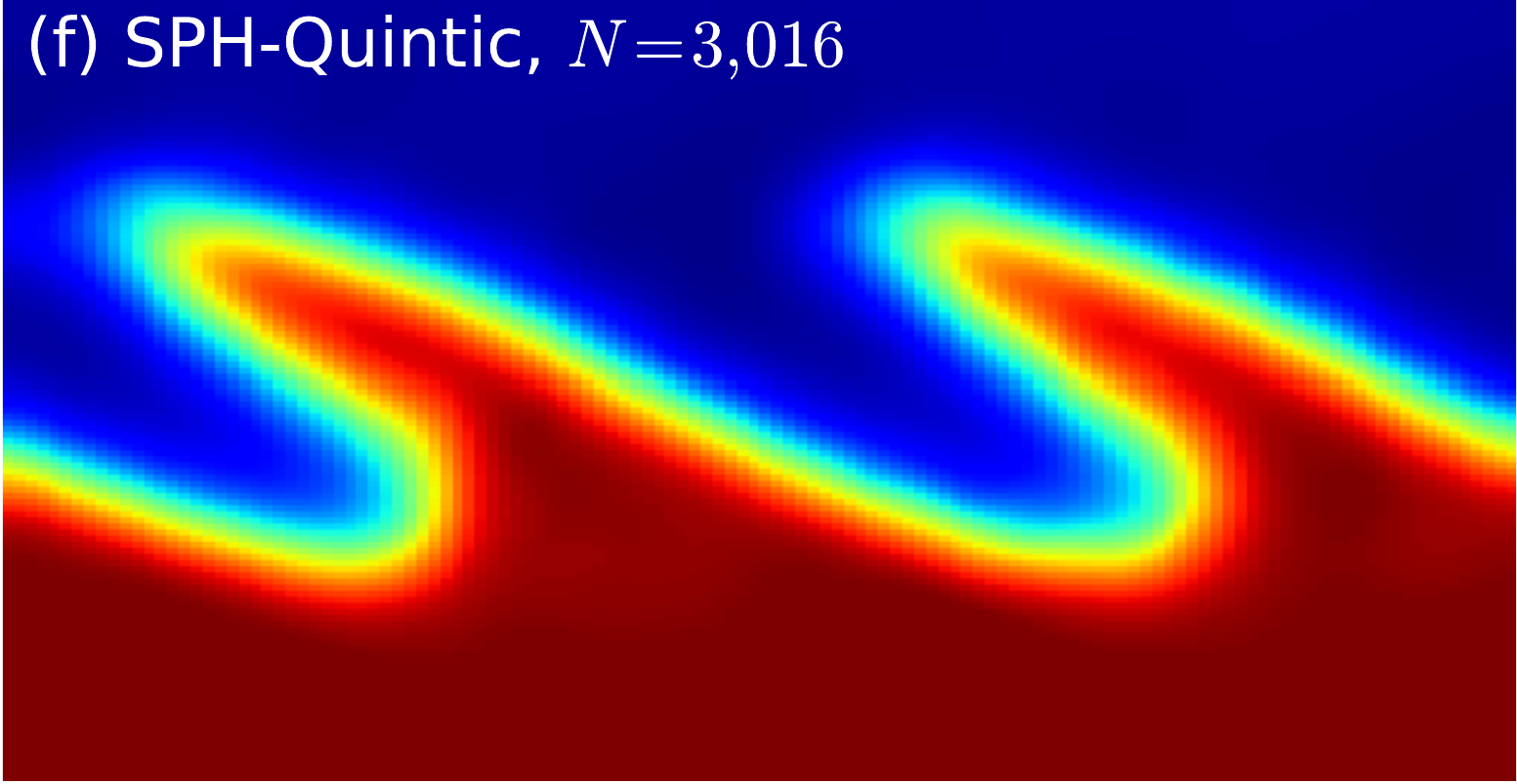}
\includegraphics[width=4.35cm]{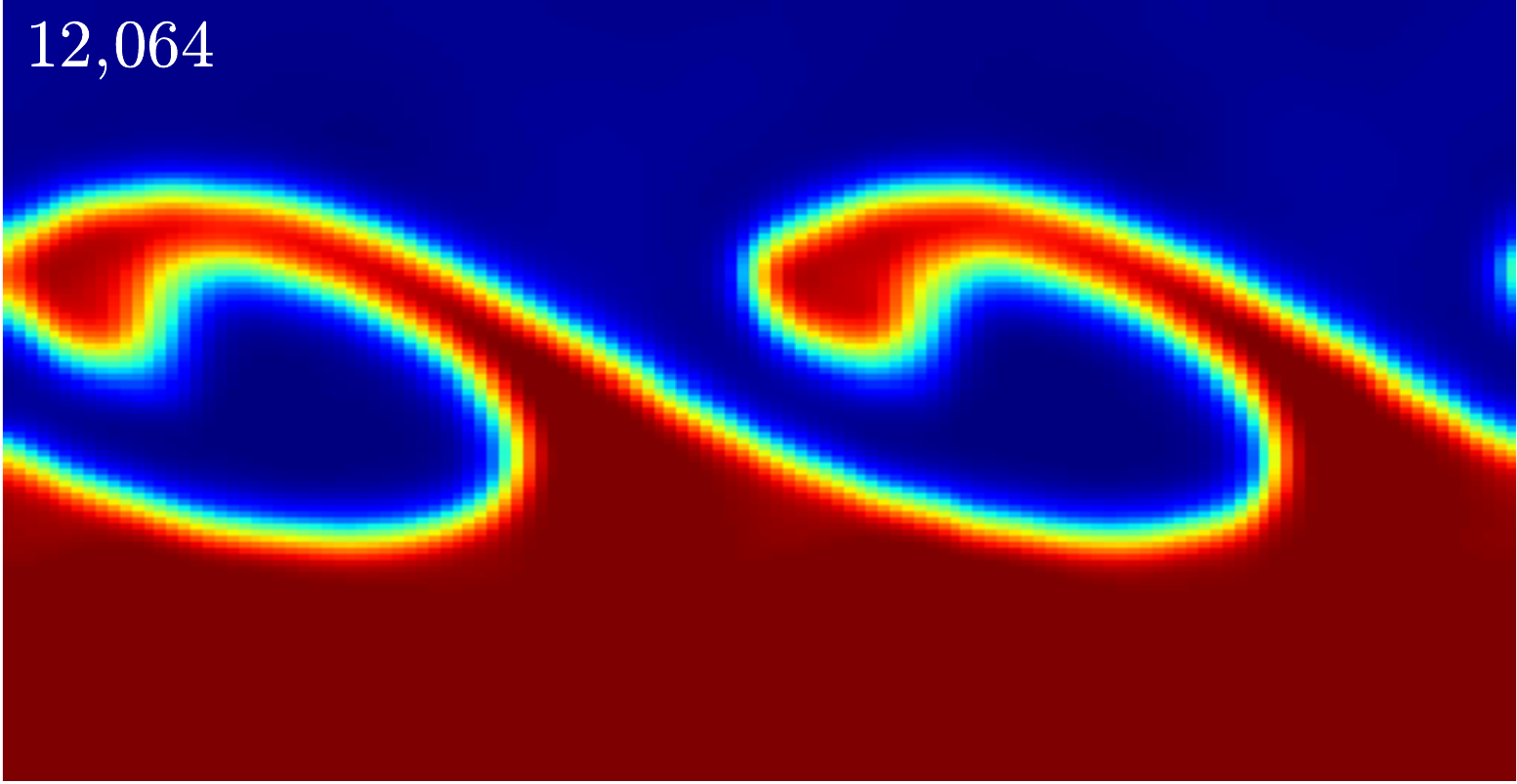}
\includegraphics[width=4.35cm]{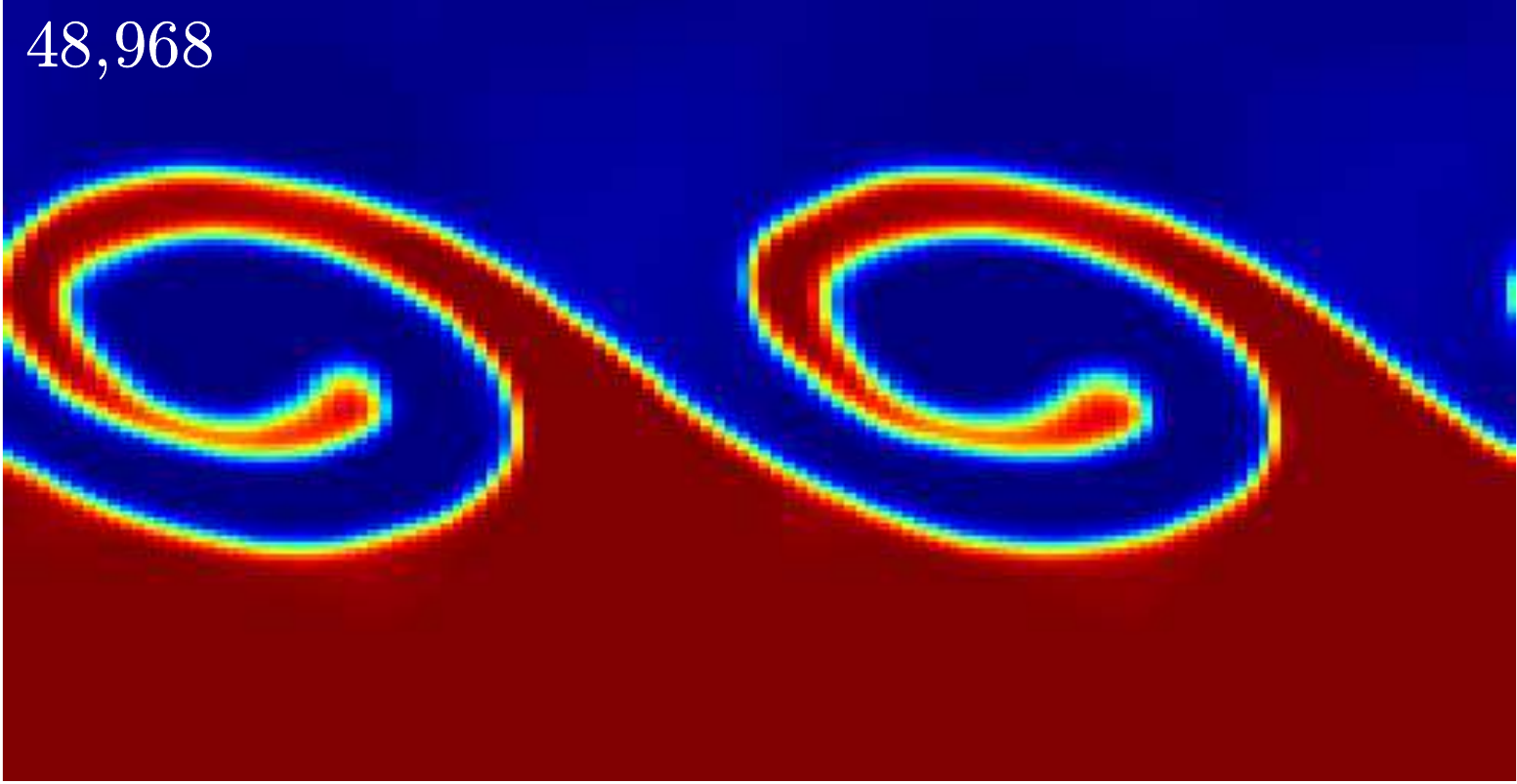}
\includegraphics[width=4.35cm]{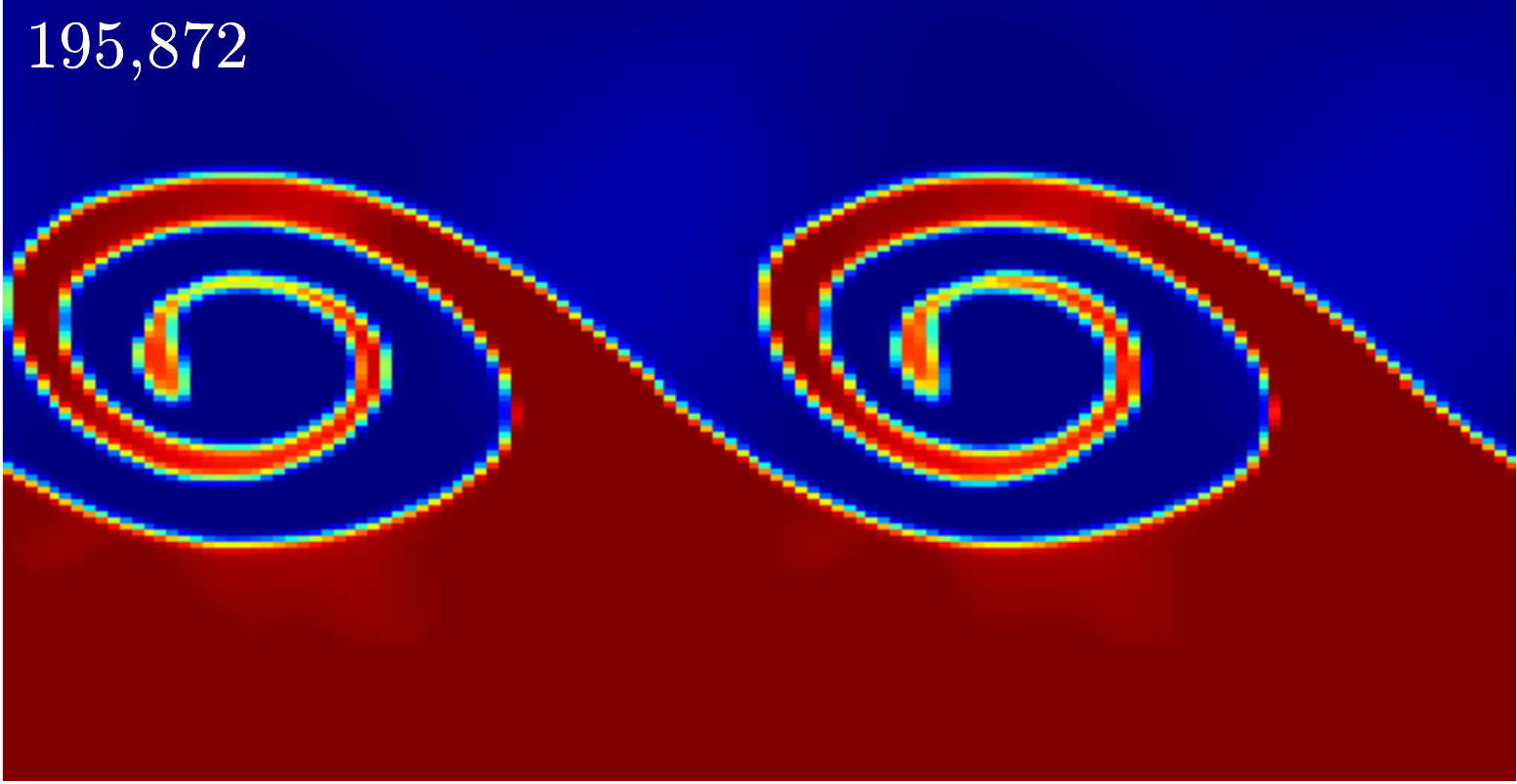}}
\vspace{0.04cm}
\centerline{
\includegraphics[width=4.35cm]{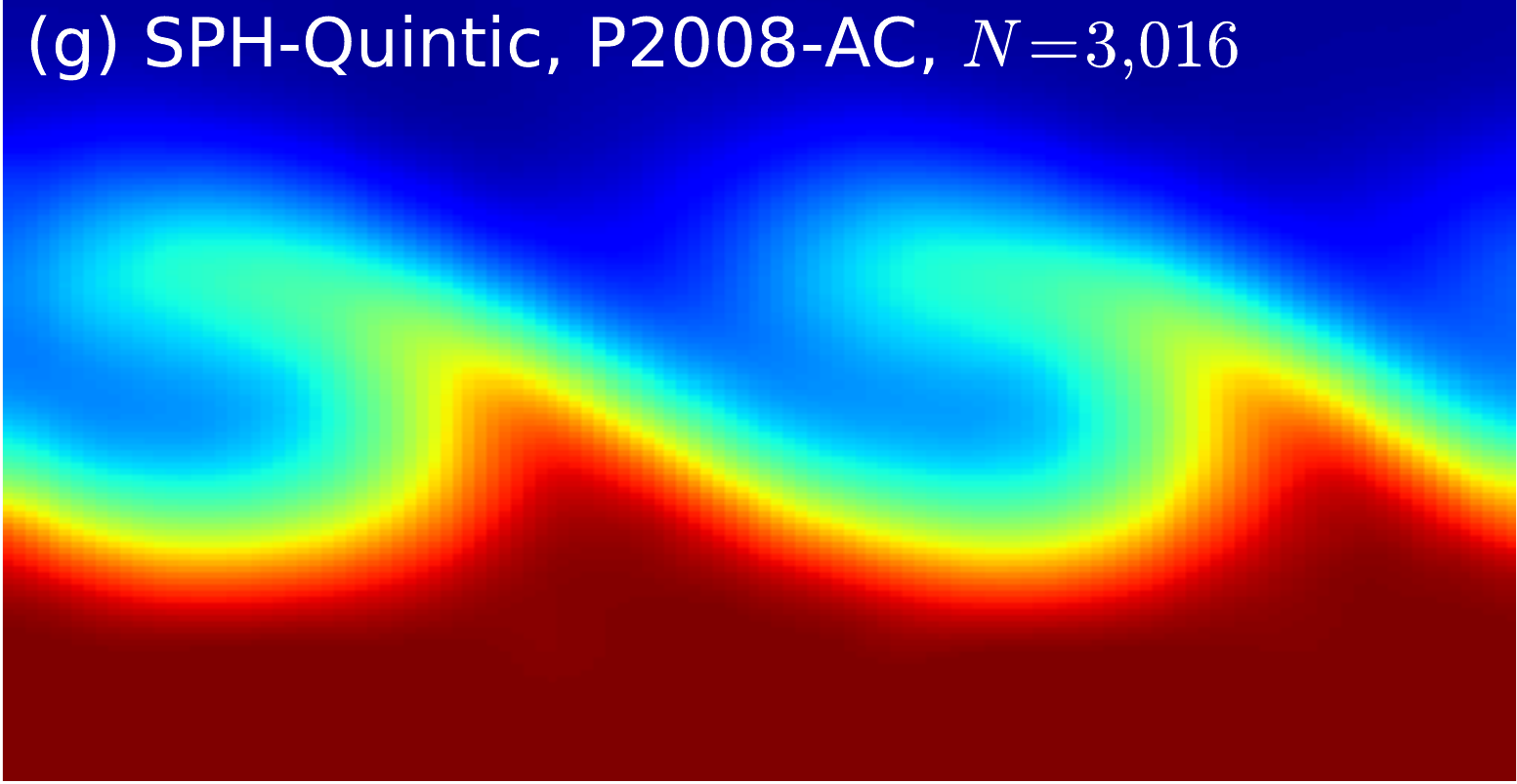}
\includegraphics[width=4.35cm]{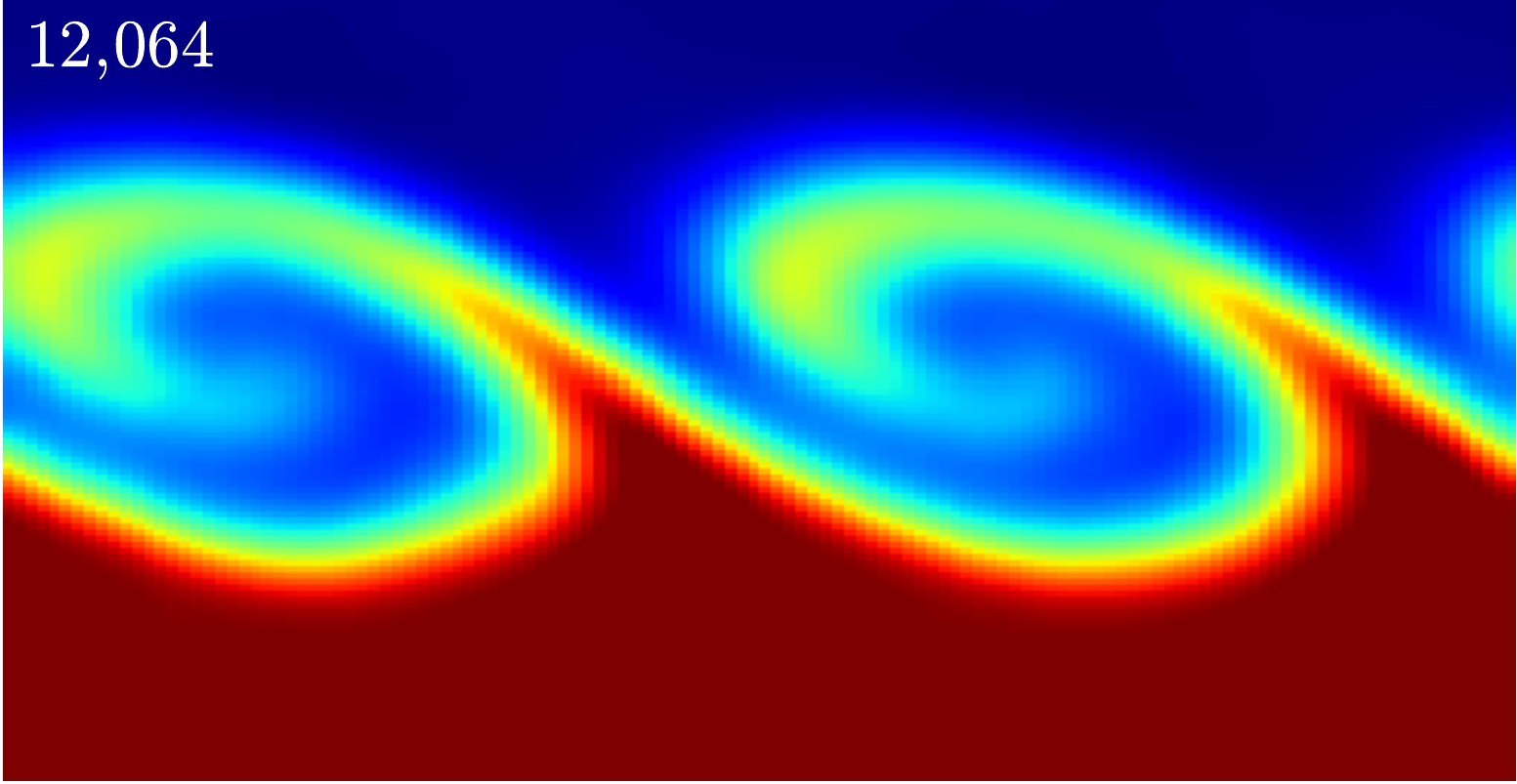}
\includegraphics[width=4.35cm]{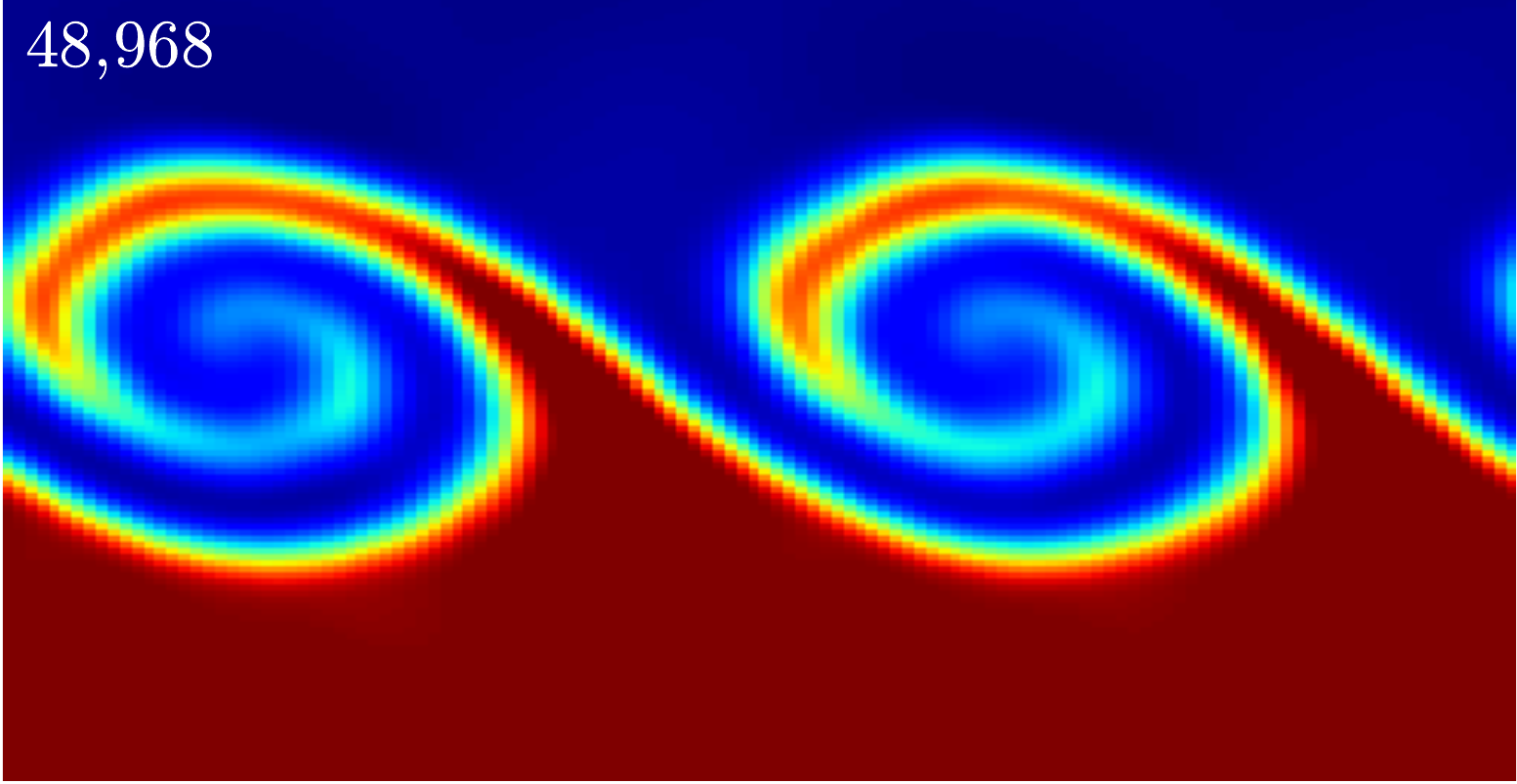}
\includegraphics[width=4.35cm]{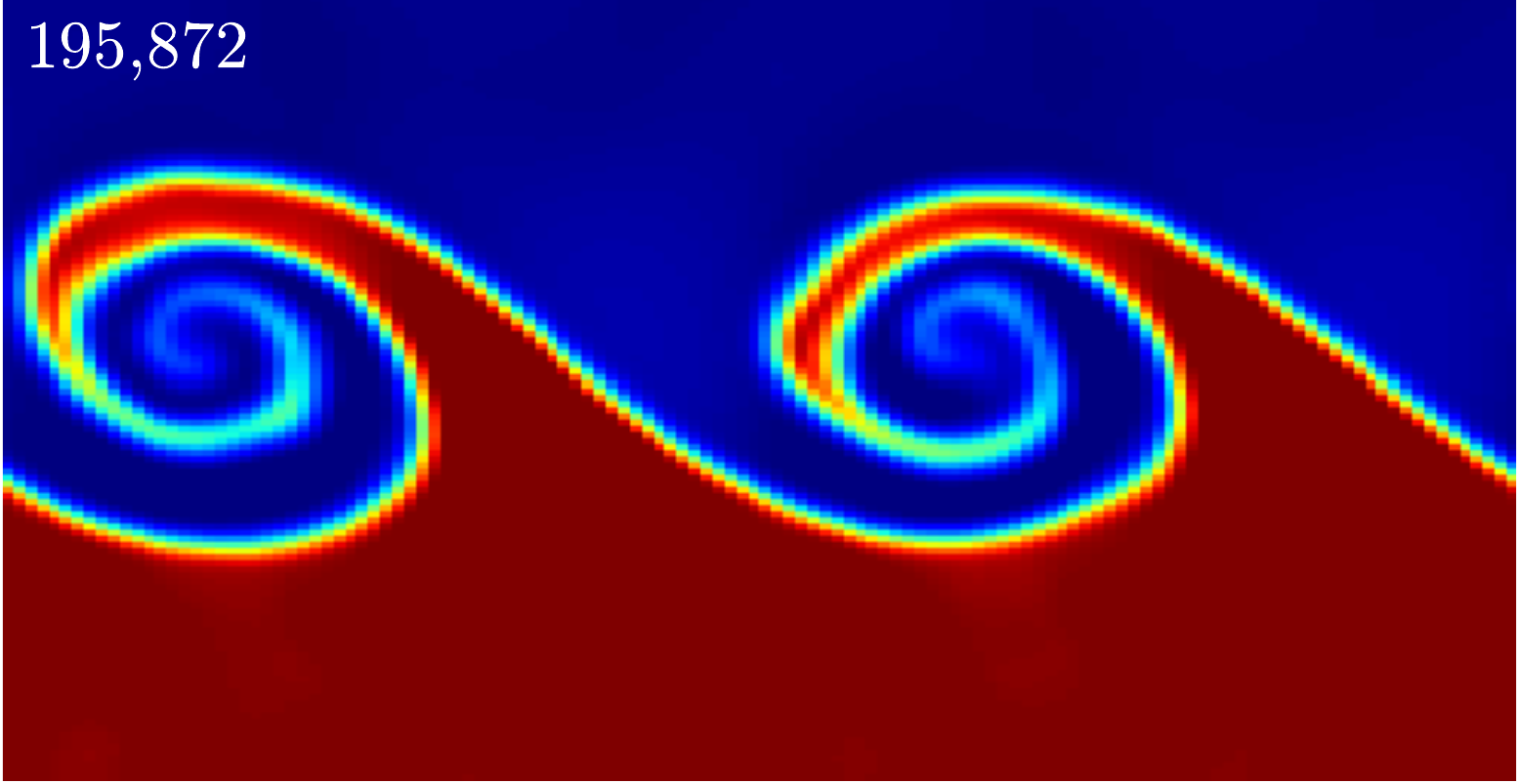}}
\vspace{0.04cm}
\centerline{
\includegraphics[width=4.35cm]{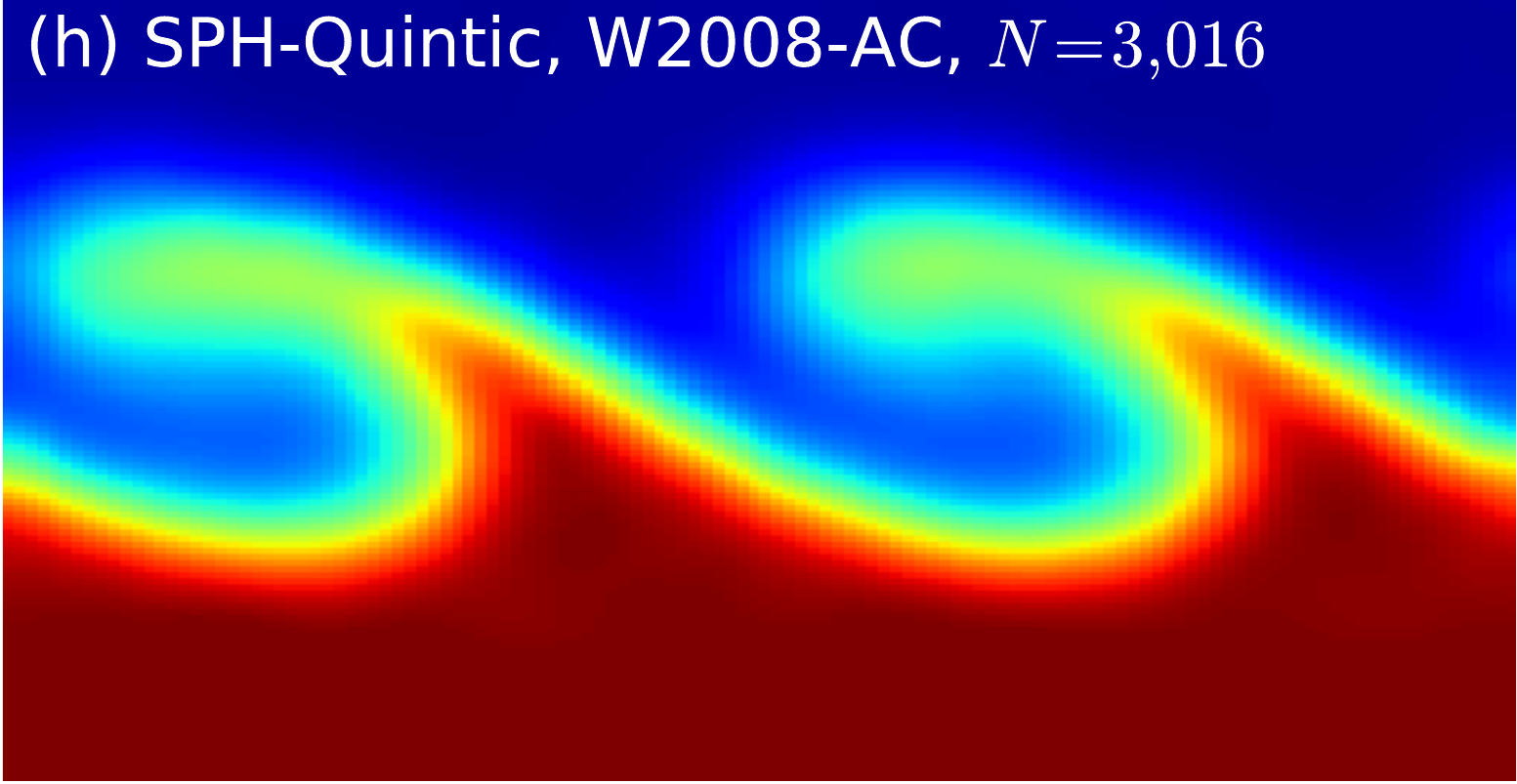}
\includegraphics[width=4.35cm]{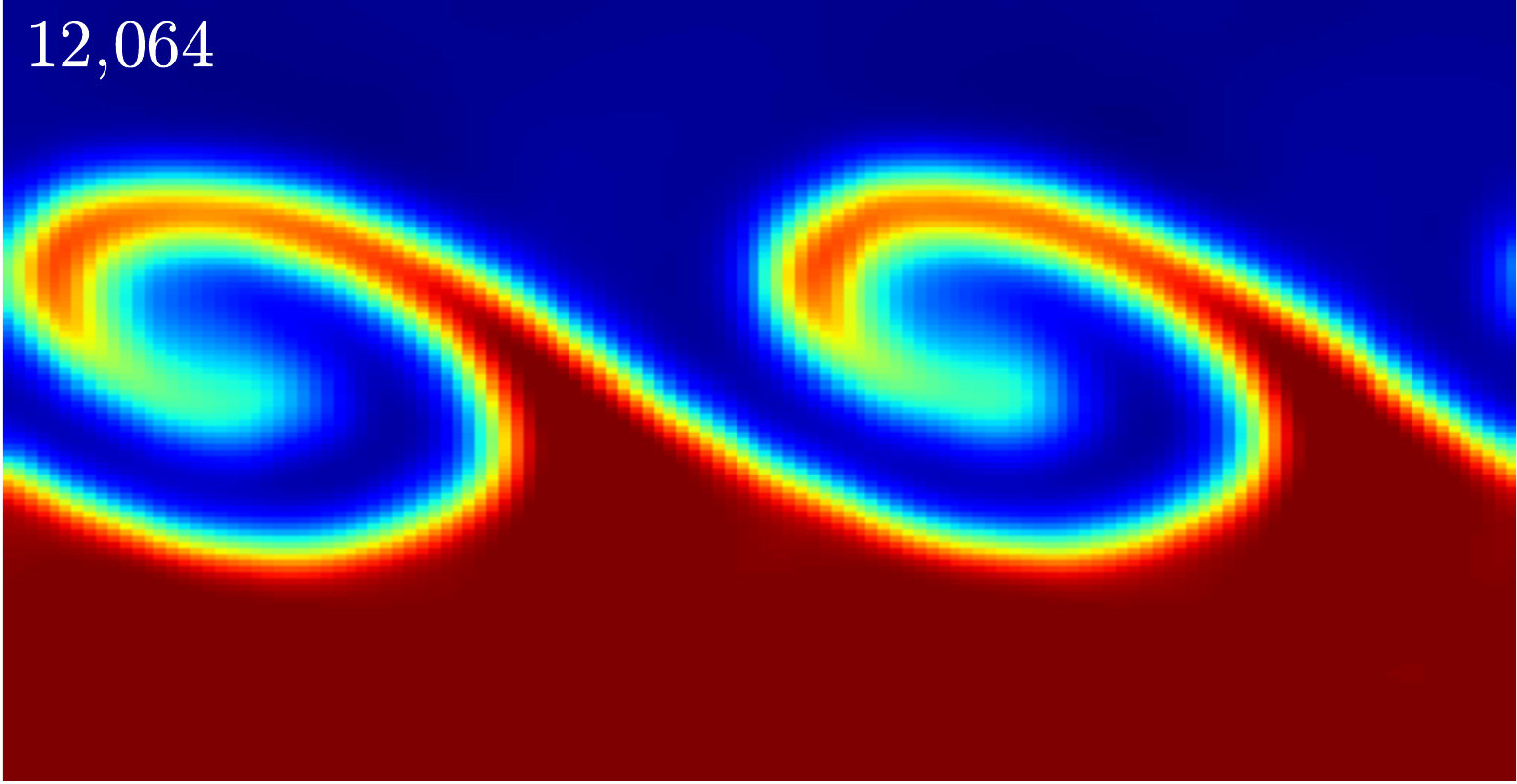}
\includegraphics[width=4.35cm]{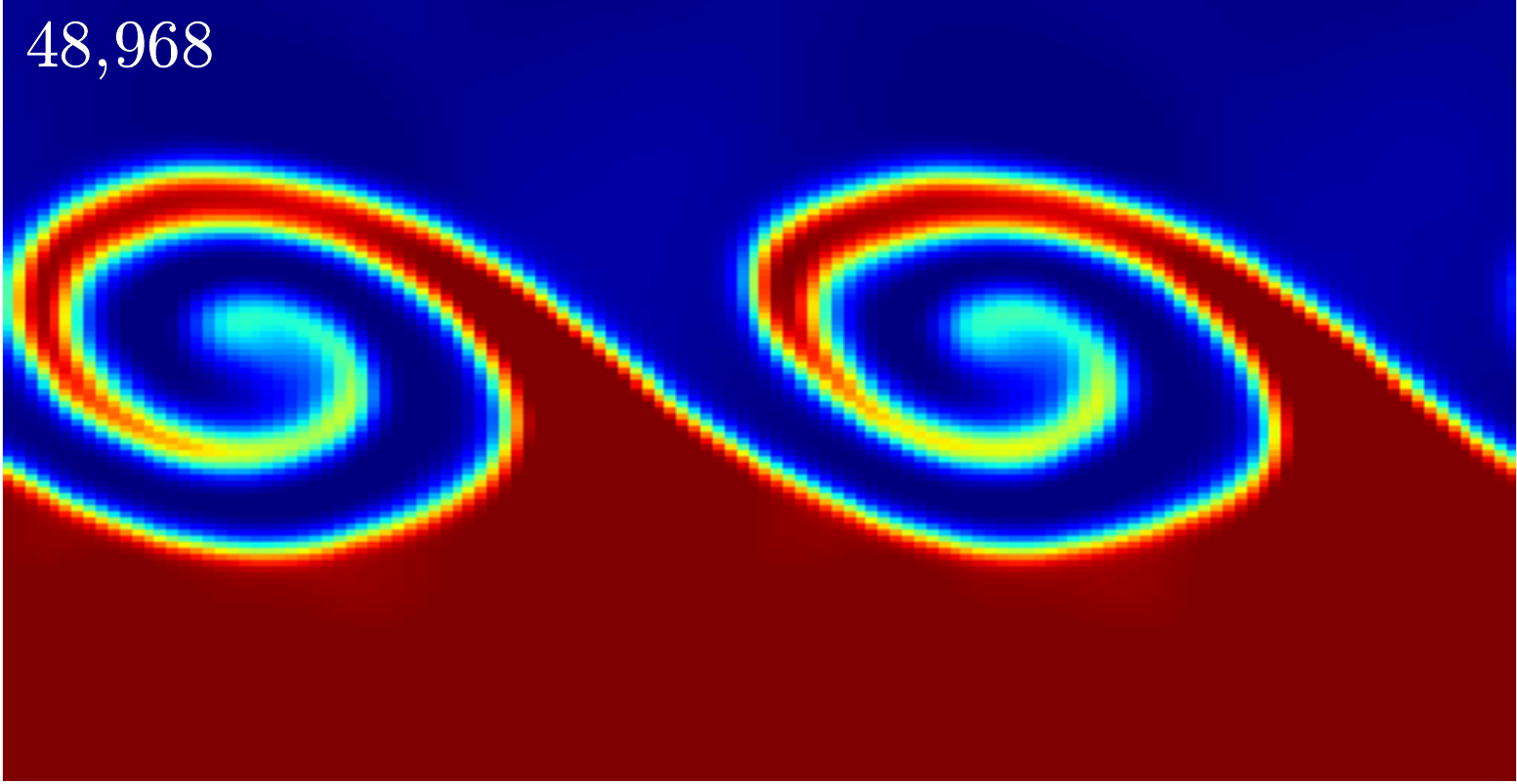}
\includegraphics[width=4.35cm]{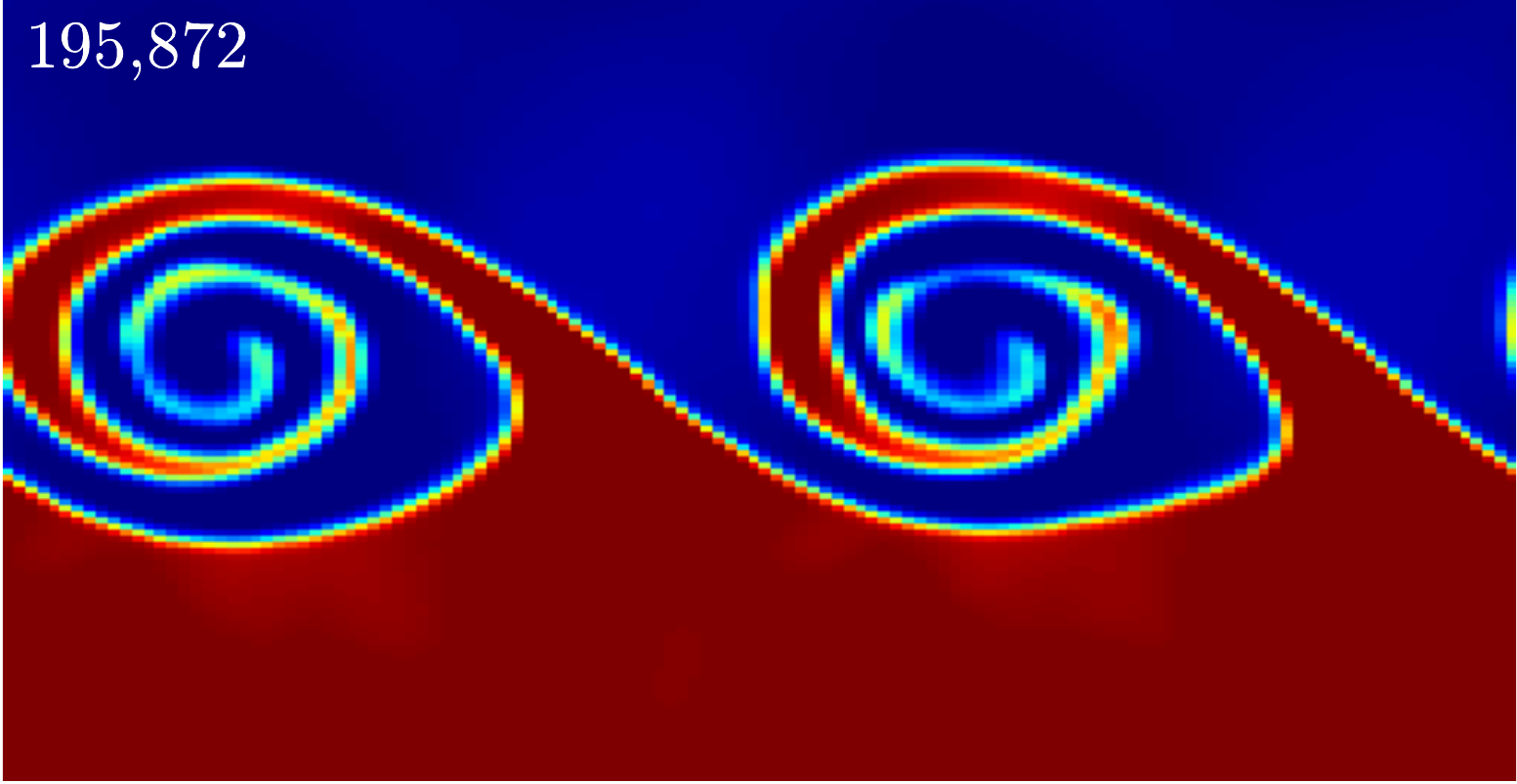}}
\vspace{0.04cm}

\caption{The   density  structure   of   the  $2:1$   Kelvin-Helmholtz
  instability at  a time $t  = 1.5\,\tau_{_{\rm KH}} =  1.59$ modeled
  with (a)  {\small MG} using uniform grid, (b)  {\small MG} using AMR with  5 levels of
  refinement, (c)  {\small SEREN} using the M4  kernel, (d) {\small SEREN}  using the M4
  kernel and the Price (2008) artificial conductivity, (e) {\small SEREN} using
  the M4 kernel and the  Wadsley et al. (2008) conductivity, (f) {\small SEREN}
  using the quintic kernel, (d) {\small SEREN} using the quintic kernel and the
  Price (2008)  artificial conductivity,  (e) {\small SEREN} using  the quintic
  kernel and  the Wadsley et  al. (2008) conductivity.   The left-hand
  column shows the  KHI using the smallest resolution  ($32 \times 32$
  cells for the  finite volume code and $N = 3,016$  for the SPH code)
  with increasing  resolution moving  right to the  highest resolution
  ($256  \times 256$  for the  finite volume  code and  $N  = 195,872$
  particles for the SPH code).  Note that we only show the top half of
  the  computational domain  ($y  > 0$)  due  to the  symmetry of  the
  initial conditions.  Each sub-figure shows
  the density field (blue : low density - red : high density).} 
\label{FIG:KHI-2:1}
\end{figure*}

\subsection{Kelvin-Helmholtz instability} \label{SS:KHI}

The  Kelvin-Helmholtz  instability  (hereafter  KHI)  is  a  classical
hydrodynamical  instability  generated  at  the boundary  between  two
shearing  fluids  which  can  lead  to vorticity  and  mixing  at  the
interface.  It has been modeled extensively in recent years by various
authors \citep[e.g.][]{Agertz2007,Price2008,Junk2010,AREPO,Valcke2010}
to compare  the ability of finite volume codes, and in particular SPH,  
to model such
mixing   processes.   It   was   first  used   in   this  context   by
\cite{Agertz2007} to highlight the  inability of standard SPH codes to
model mixing  between shearing layers  when there is  a discontinuity.
\cite{Agertz2007}   demonstrate  that in standard SPH implementations,
 the   two  fluids   exhibit  an artificial  repulsion  force on  each  
other,  even  when in  pressure equilibrium, which inhibits the two 
fluids from interacting and thus preventing the KHI from developing.

\cite{Price2008} explained that the specific internal energy
discontinuity at the density interface was responsible for a spurious
surface-tension effect that `repulsed' the two fluids.  
He suggests that this is
because of the inability of SPH to correctly model discontinuities
due to errors in the particle approximation and that all
quantities in SPH should include explicit dissipation/diffusion terms
in order to be `smear out' the discontinuity over several
smoothing lengths.  This is demonstrated by including an additional
artificial conductivity term, often ignored in most SPH
implementations, which allows the KHI to develop.  
\cite{Price2008} also discusses that due to SPH's Lagrangian nature,
the specific entropy (measured by the entropic function $A \equiv P /
\rho^{\gamma}$) of a fluid is conserved in adiabatic expansion or
contraction.  Therefore explicit dissipation or diffusion terms are
also required to allow entropy mixing.  Otherwise, the two fluids form
an oily `lava-lamp' effect with no true mixing or exchange between the
two.  

There are also
alternative derivations of SPH that can help solve the discontinuity problem. 
\cite{OSPH} suggested a new set of SPH fluid equations, a new smoothing
kernel function and the use of more neighbours.  
Their `Optimised SPH' uses a smoothed-pressure term that effectively 
smoothes out the specific internal energy discontinuity and therefore 
reduces the effective repulsive force. 
\cite{GSPH2010} also showed that Godunov SPH, a Godunov-type SPH scheme
using Riemann solvers, intrinsically smoothes specific internal energy 
discontinuities in the momentum and energy equations, and can model 
the KHI without any additional dissipation terms.

While  finite volume  codes can  model  the KHI  without any  explicit
dissipation terms,  numerical diffusion  due to advection  can provide
some unavoidable mixing at  the grid-scale.  \cite{AREPO} used the KHI
test amongst others to demonstrate that static finite volume codes can
have  problems dealing  with  some hydrodynamical  processes when  the
fluid is moving with a large supersonic advection velocity relative to
the grid.  He demonstrated that if the advection velocity was set high
enough, the  KHI would  not form in  the fluid and  instead, excessive
diffusion would  dominate the evolution  of the fluid  (preventing the
generation  of almost  all  fluid instabilities,  not  just the  KHI).
However,  \citet{Advection2010} have  argued that  the problem  can be
prevented  by  including sufficiently  high  resolution for  high-Mach
number advection velocities.  This  problem can therefore in principle
be greatly reduced  in AMR codes that use  appropriate mesh refinement
criteria.  We do not consider this problem with the finite volume code
further in this paper.

\begin{figure*}
\centerline{\includegraphics[width=4.35cm]{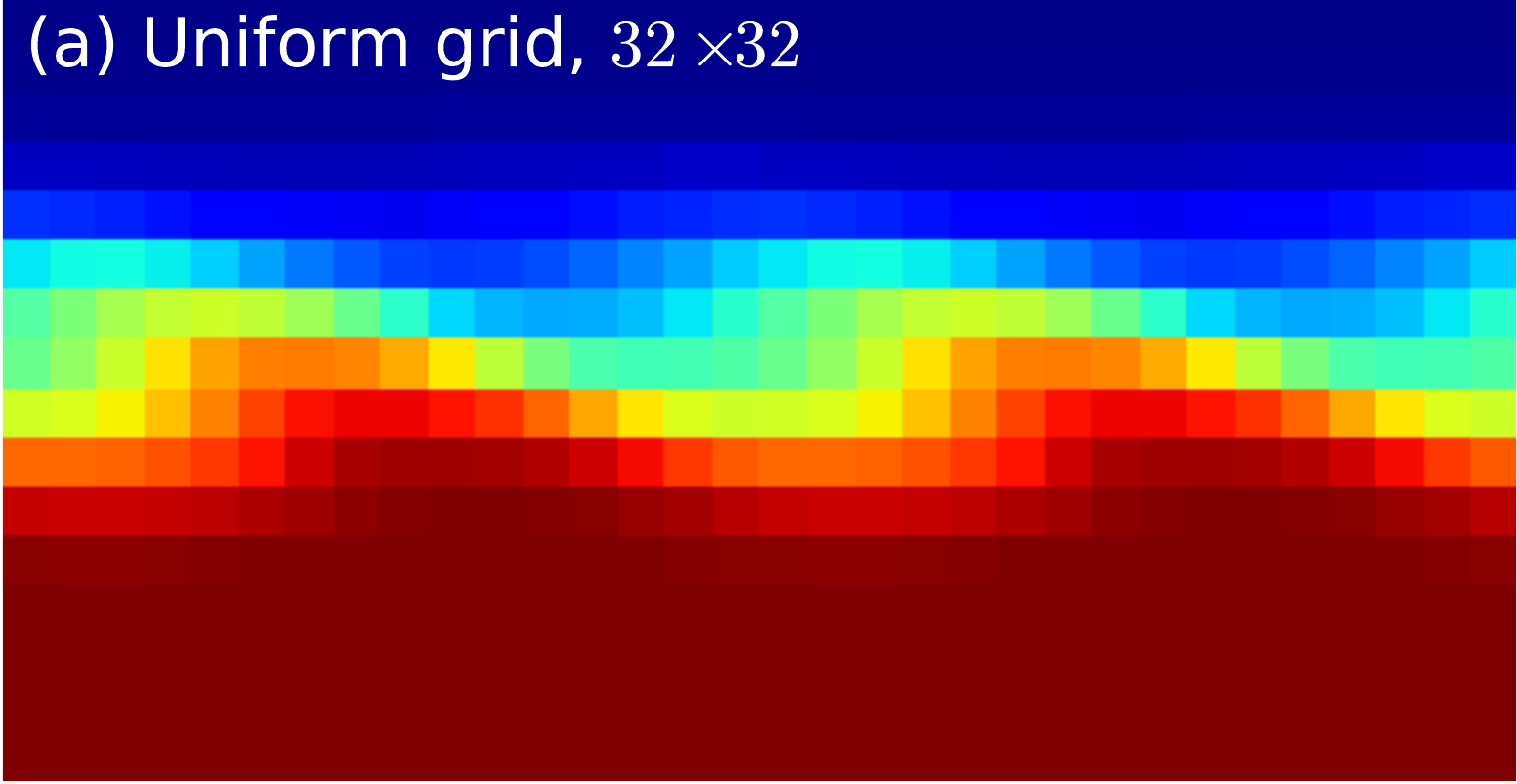}
\includegraphics[width=4.35cm]{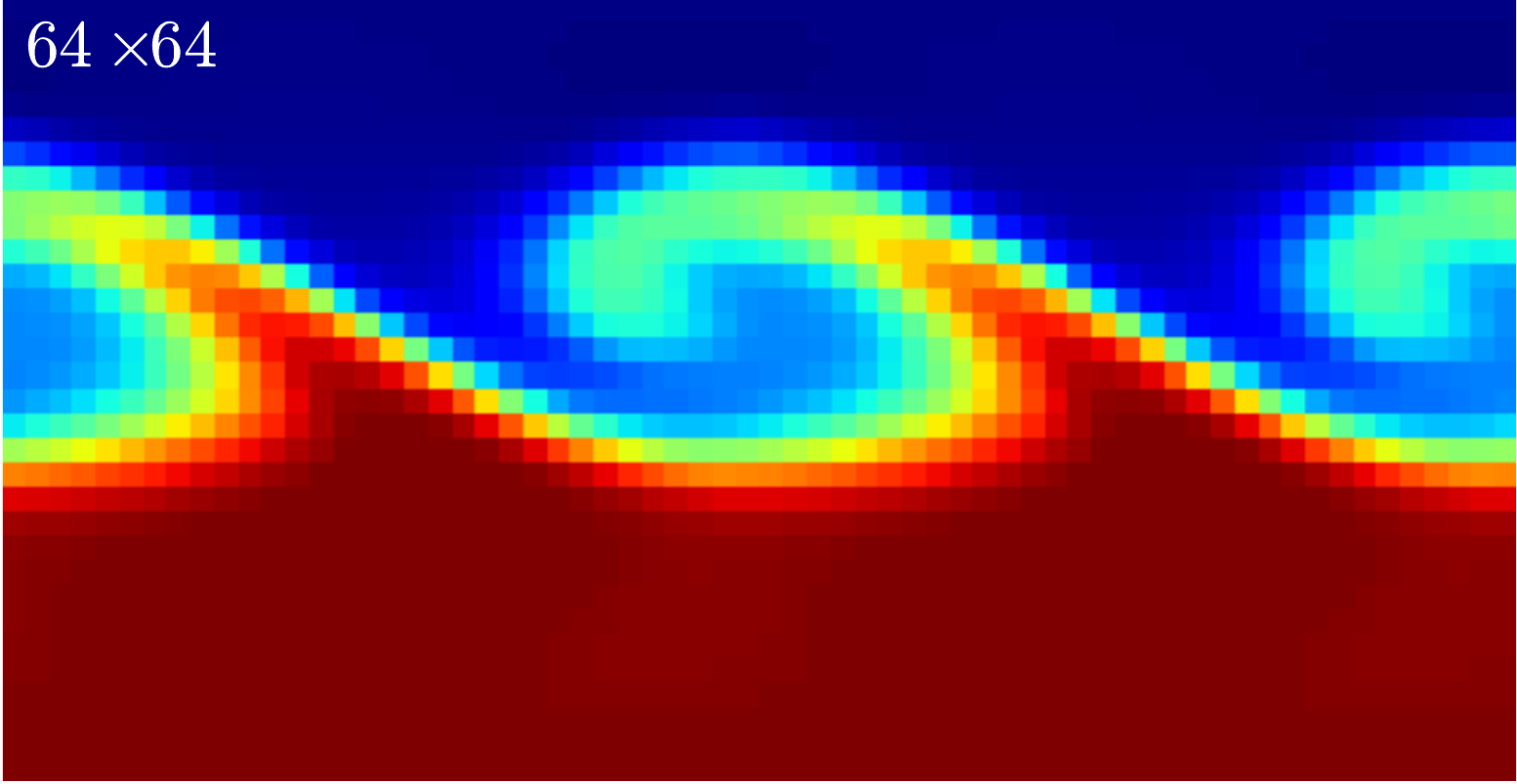}
\includegraphics[width=4.35cm]{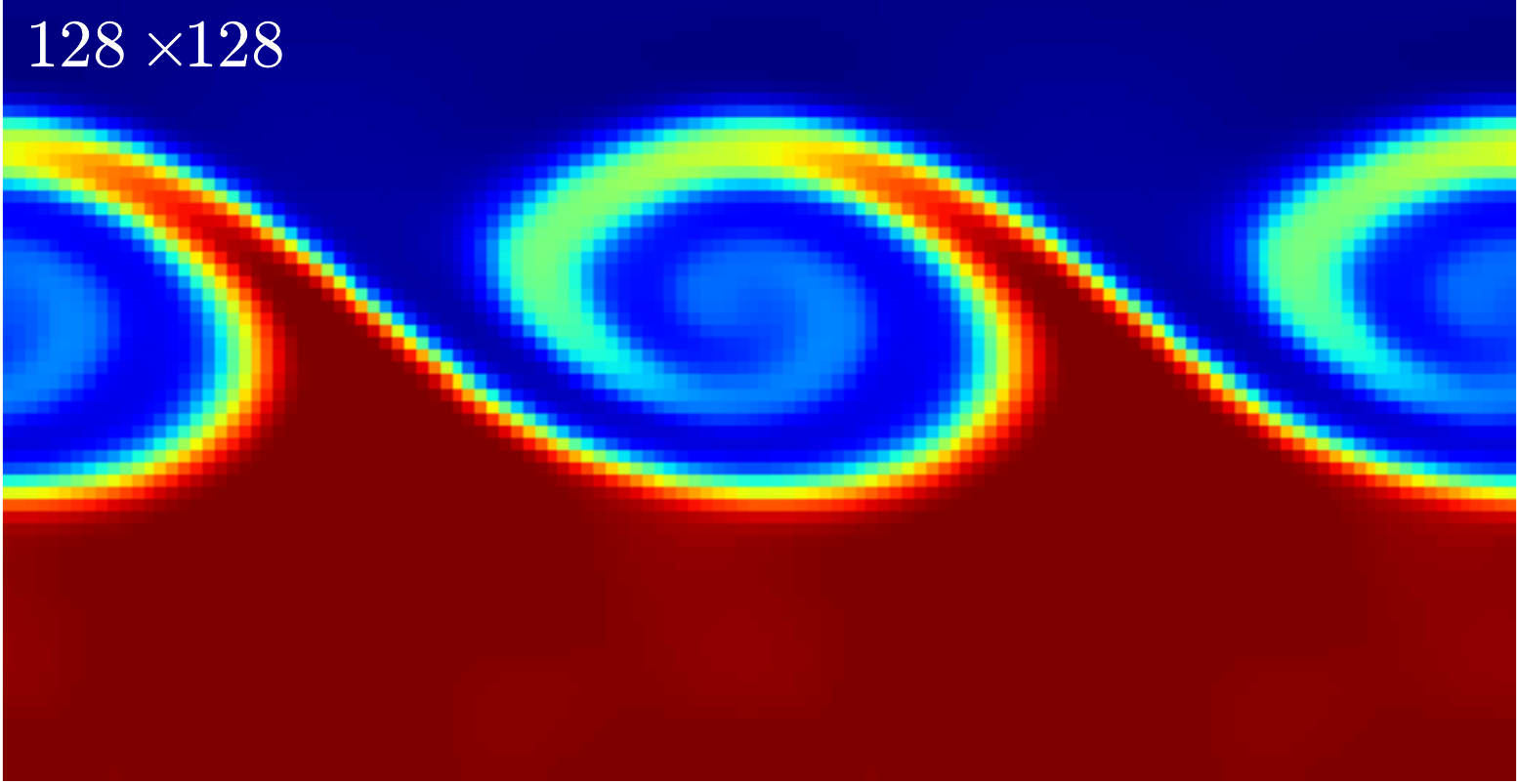}
\includegraphics[width=4.35cm]{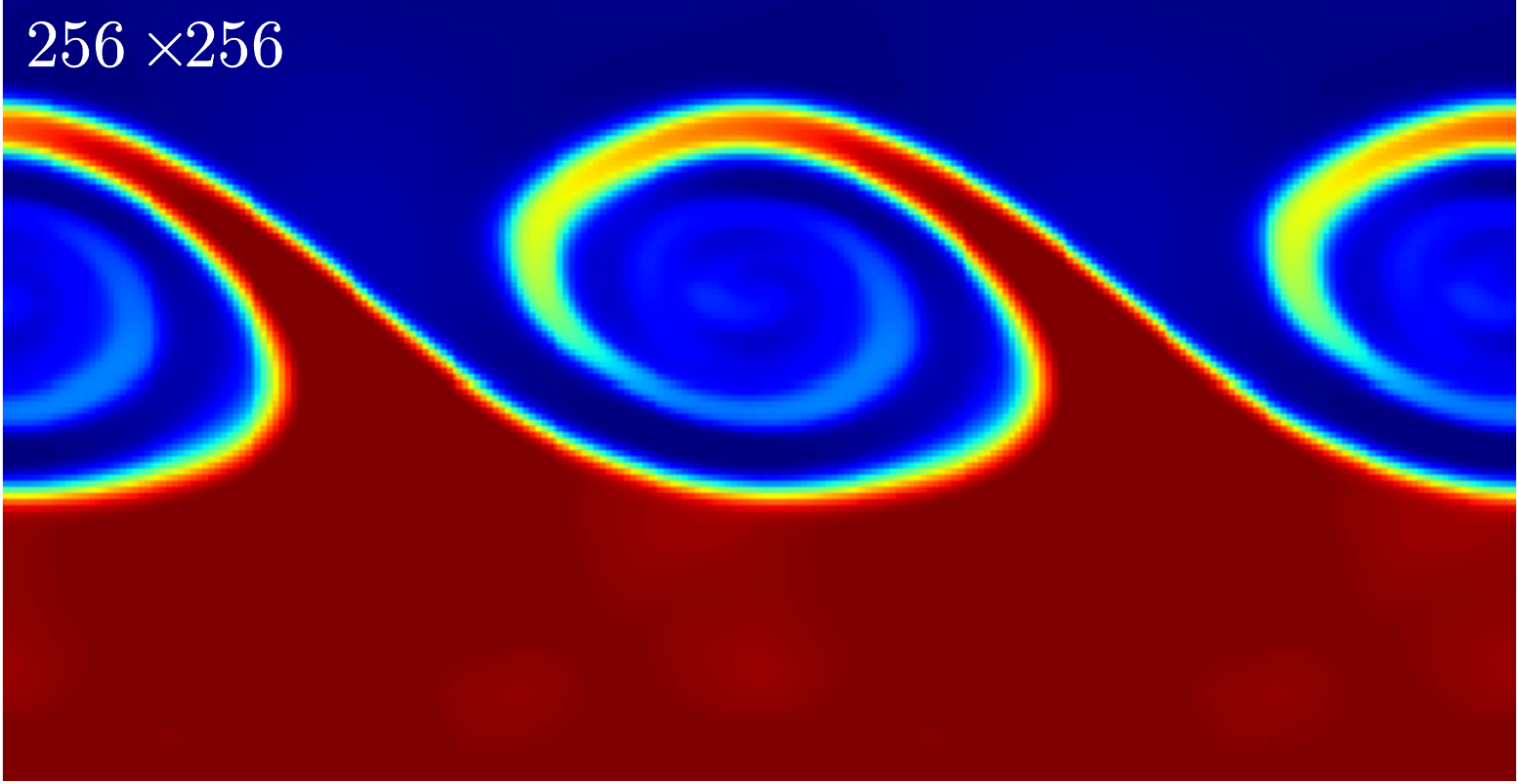}}
\vspace{0.04cm}
\centerline{\includegraphics[width=4.35cm]{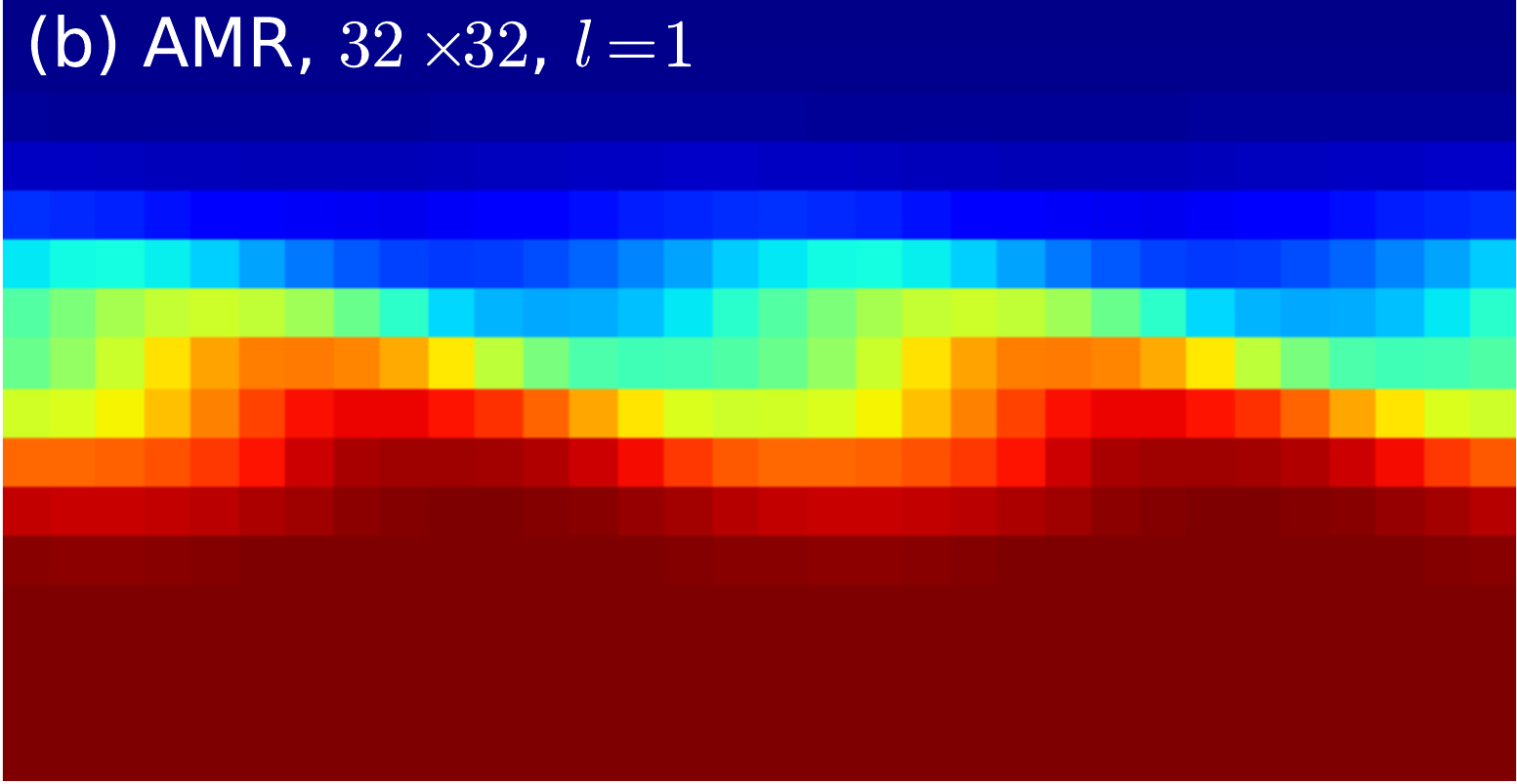}
\includegraphics[width=4.35cm]{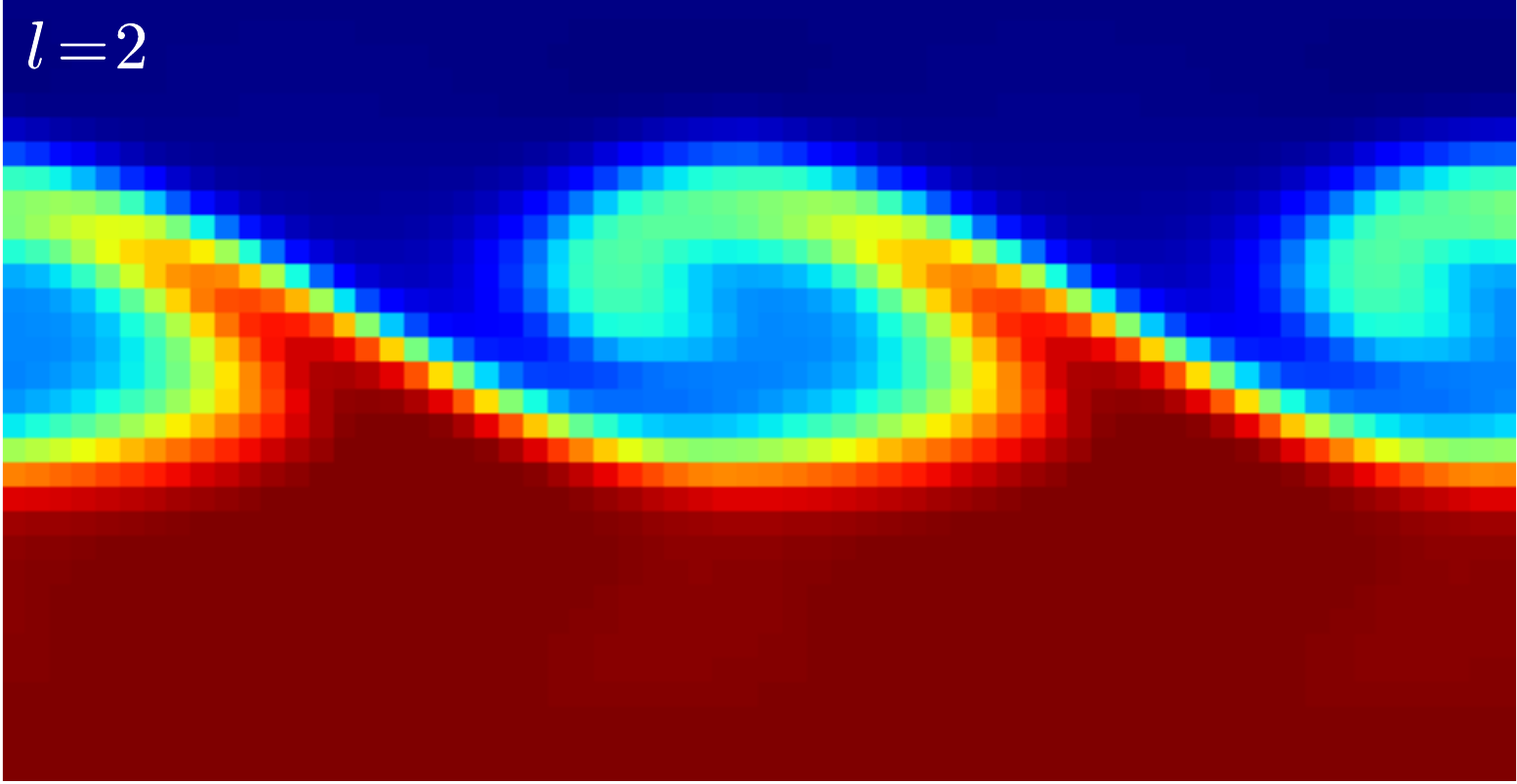}
\includegraphics[width=4.35cm]{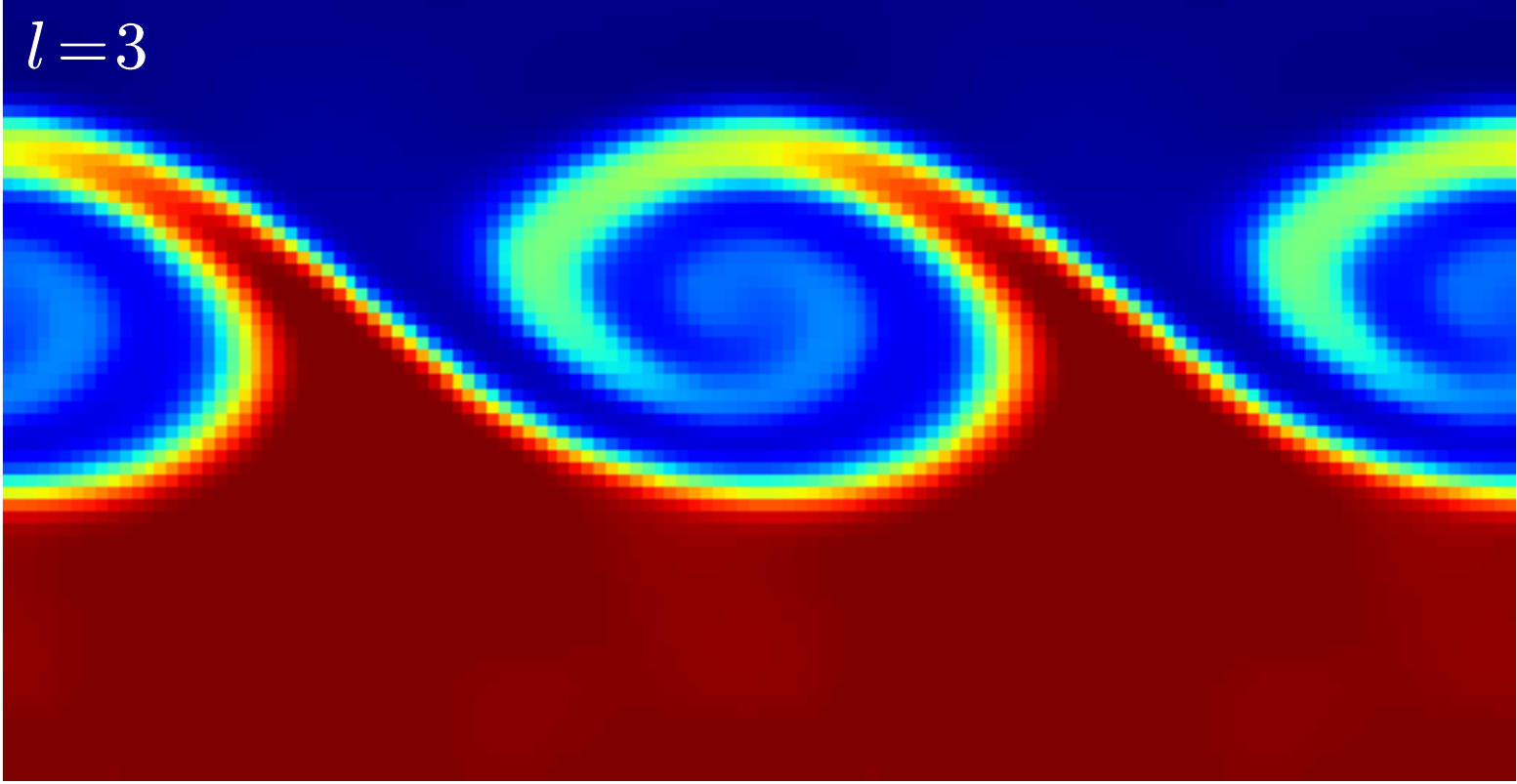}
\includegraphics[width=4.35cm]{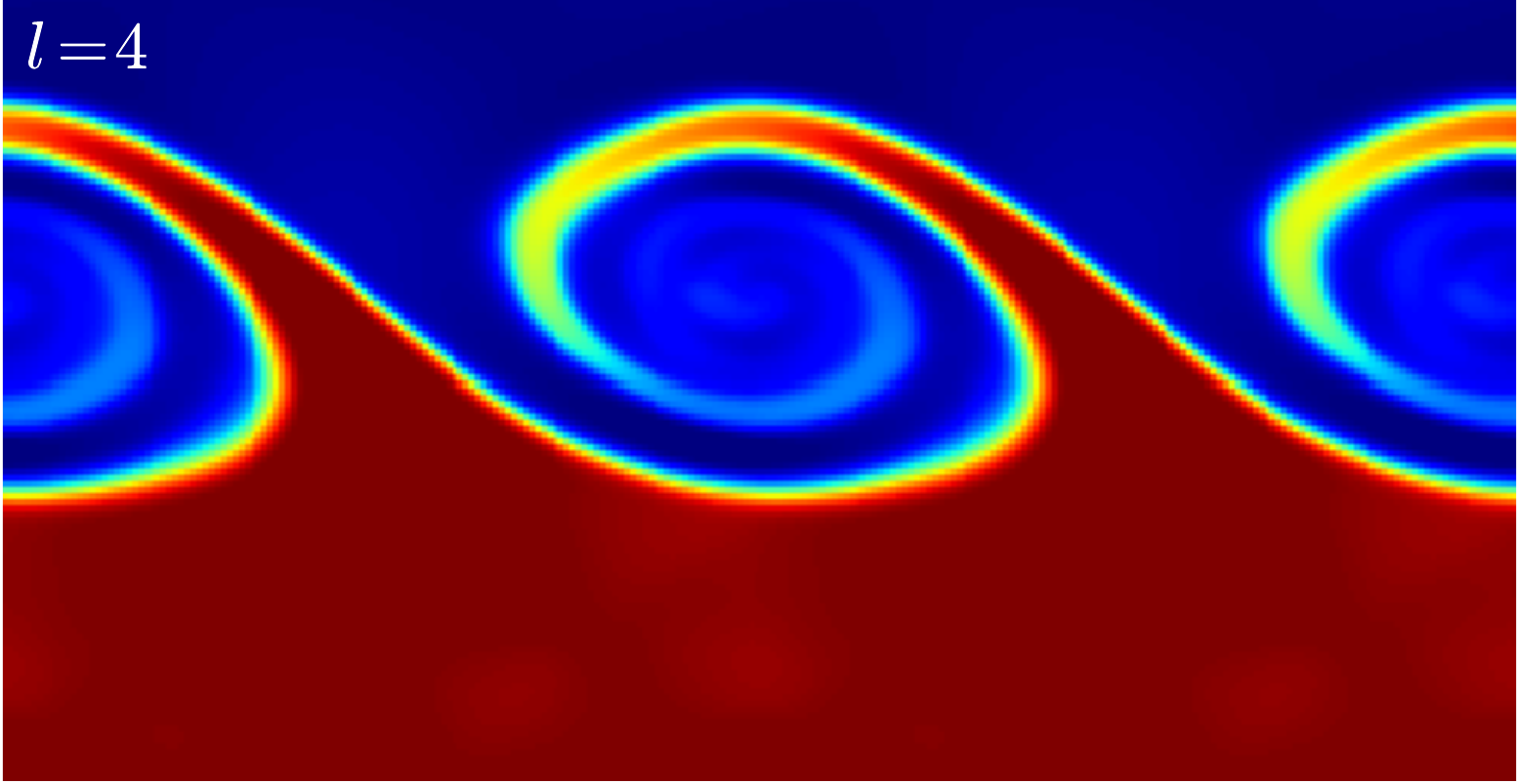}}
\vspace{0.04cm}
\centerline{\includegraphics[width=4.35cm]{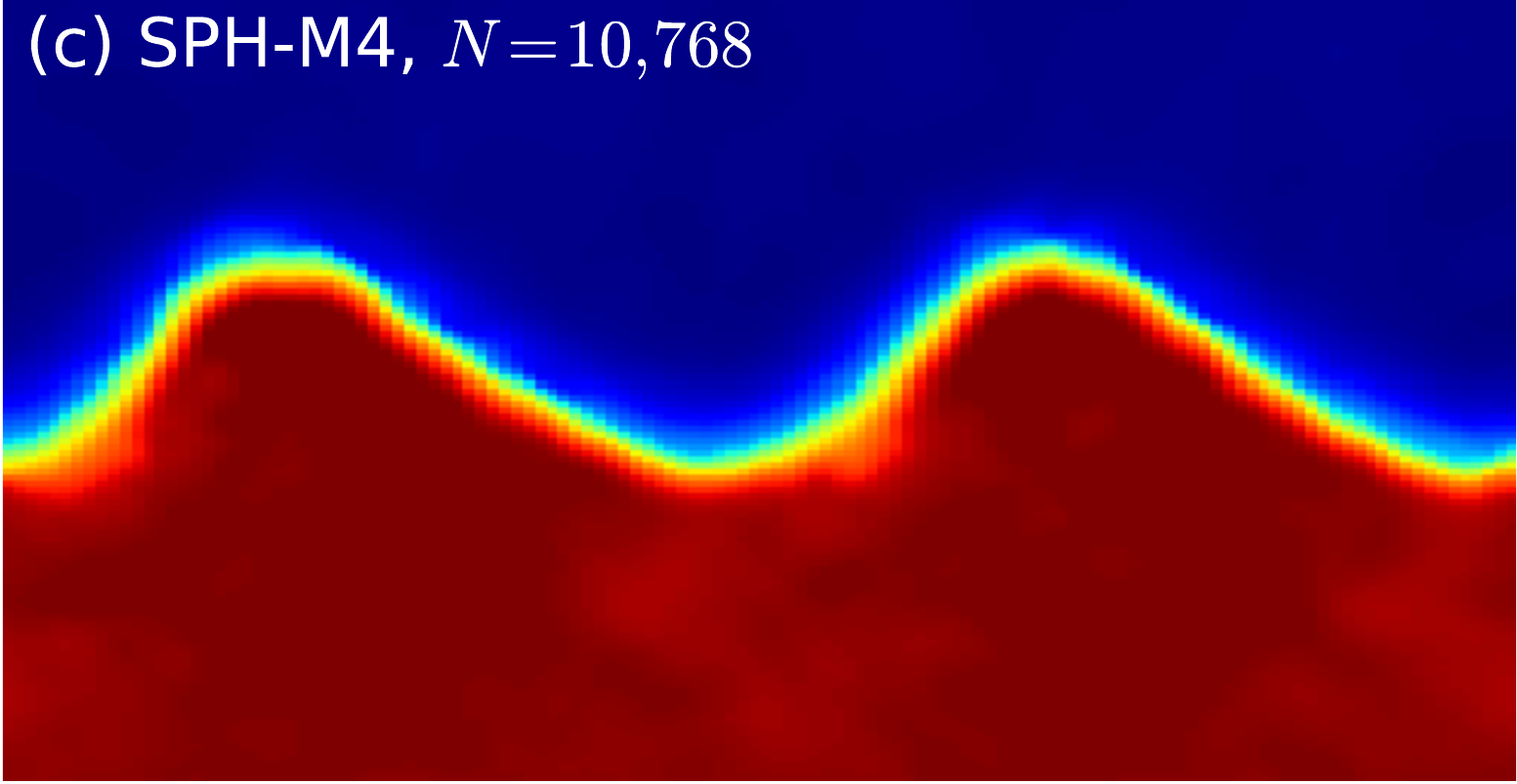}
\includegraphics[width=4.35cm]{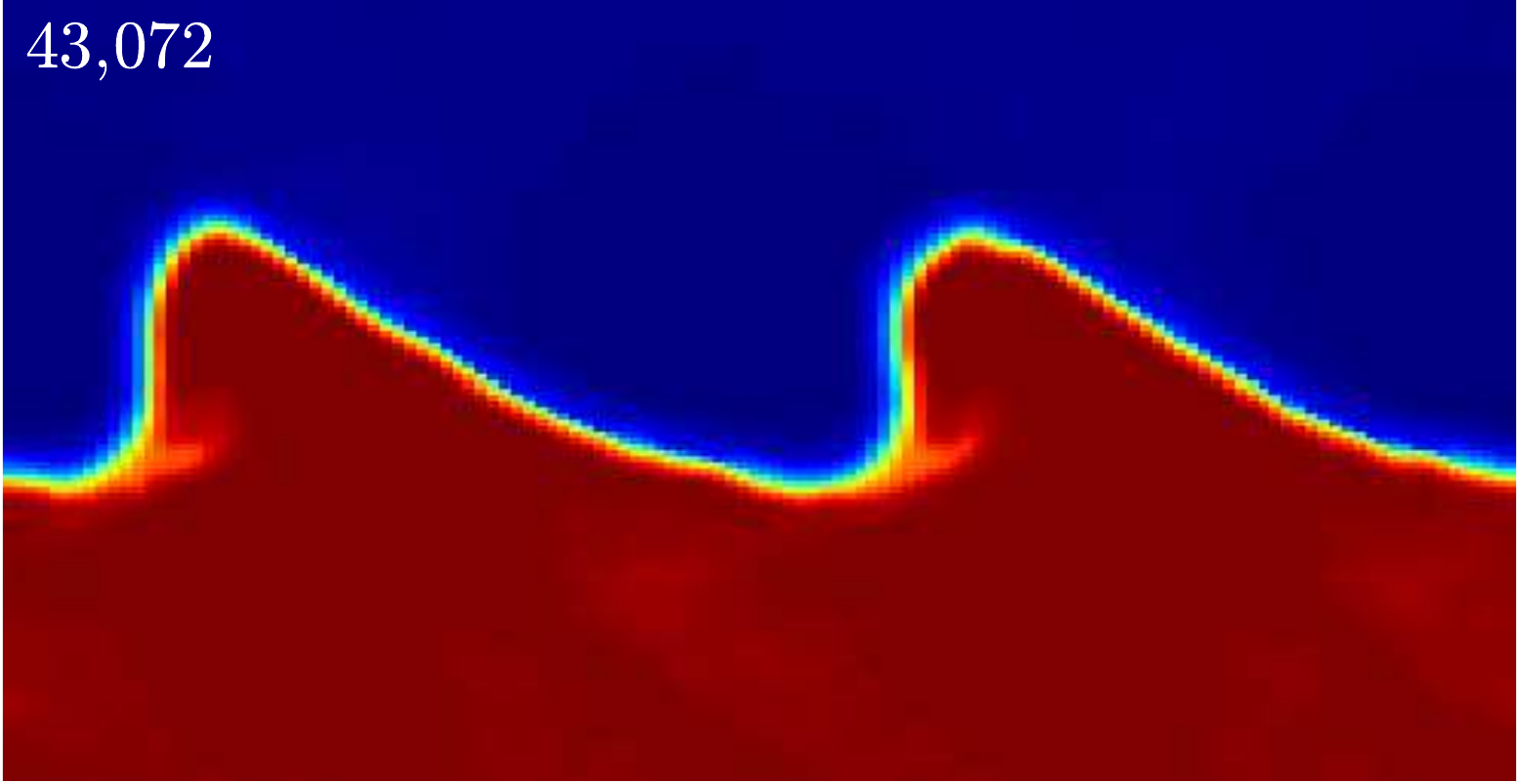}
\includegraphics[width=4.35cm]{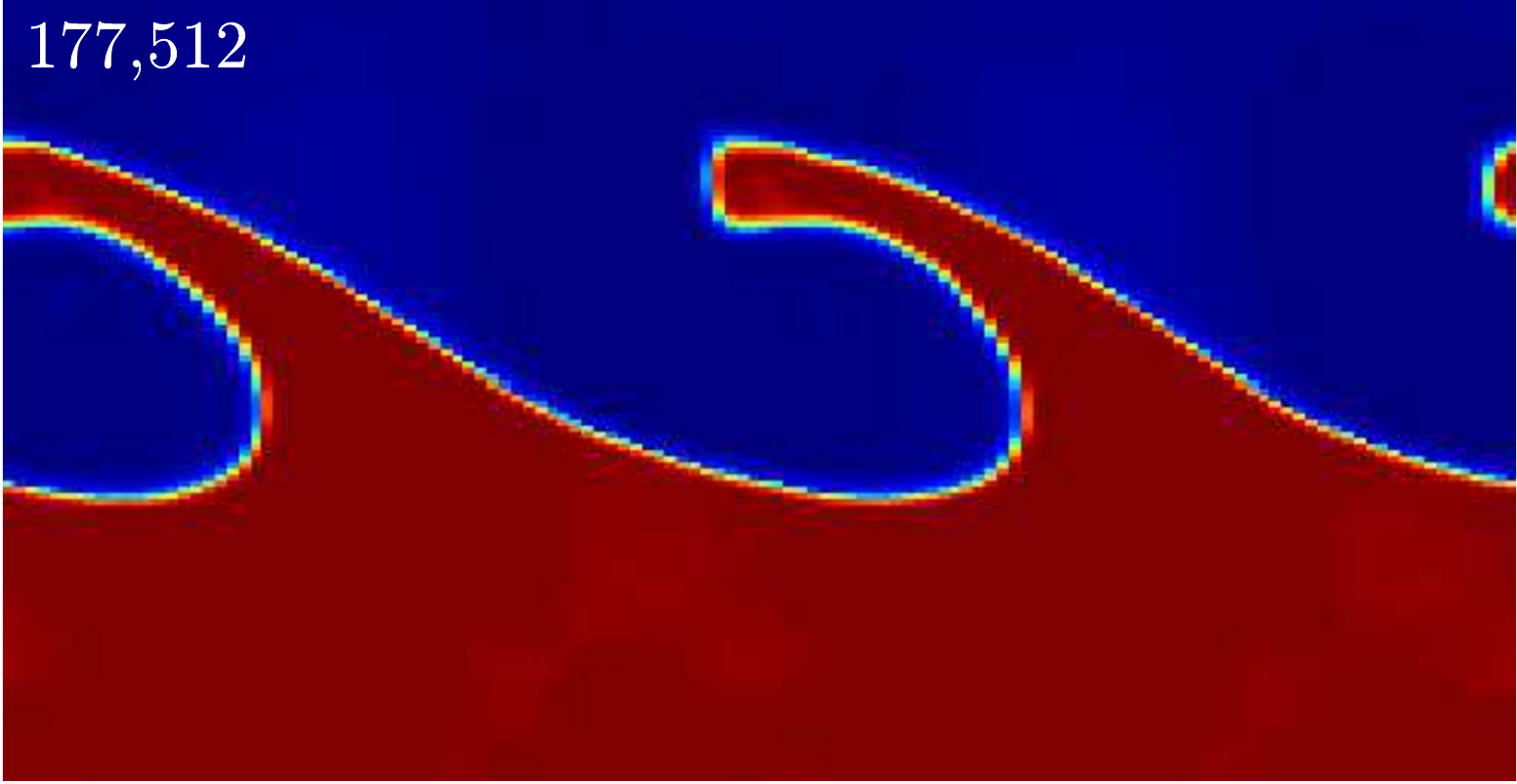}
\includegraphics[width=4.35cm]{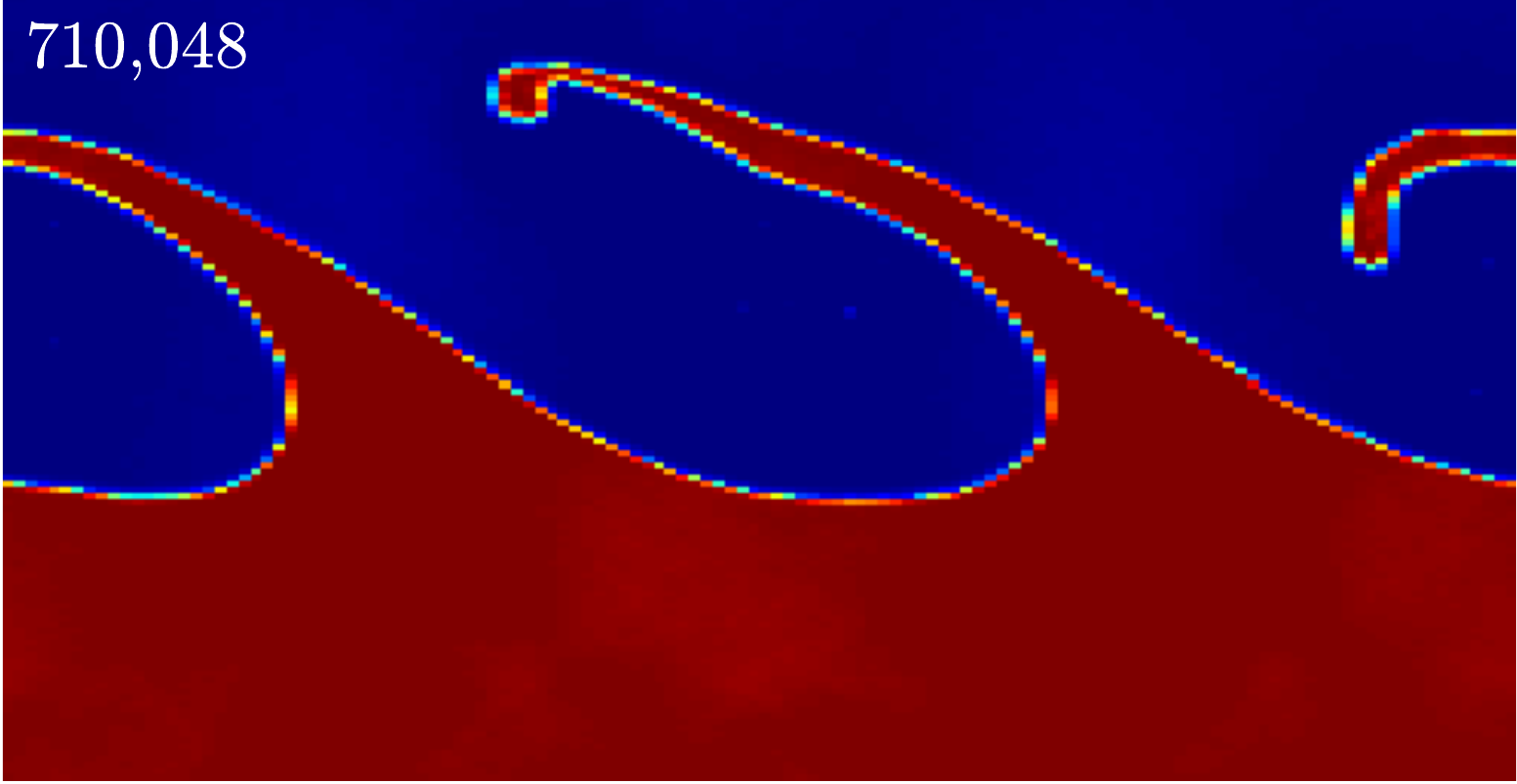}}
\vspace{0.04cm}
\centerline{\includegraphics[width=4.35cm]{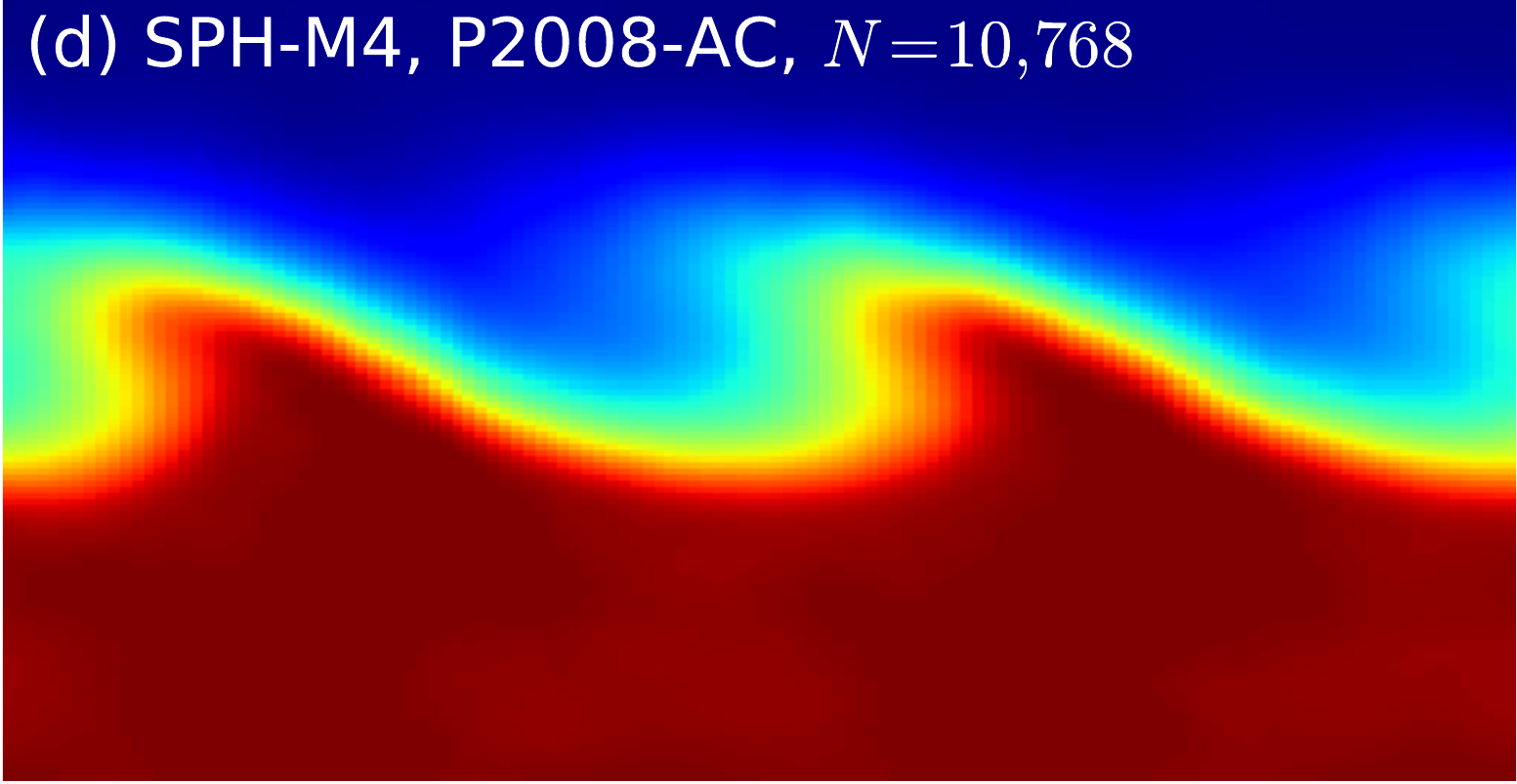}
\includegraphics[width=4.35cm]{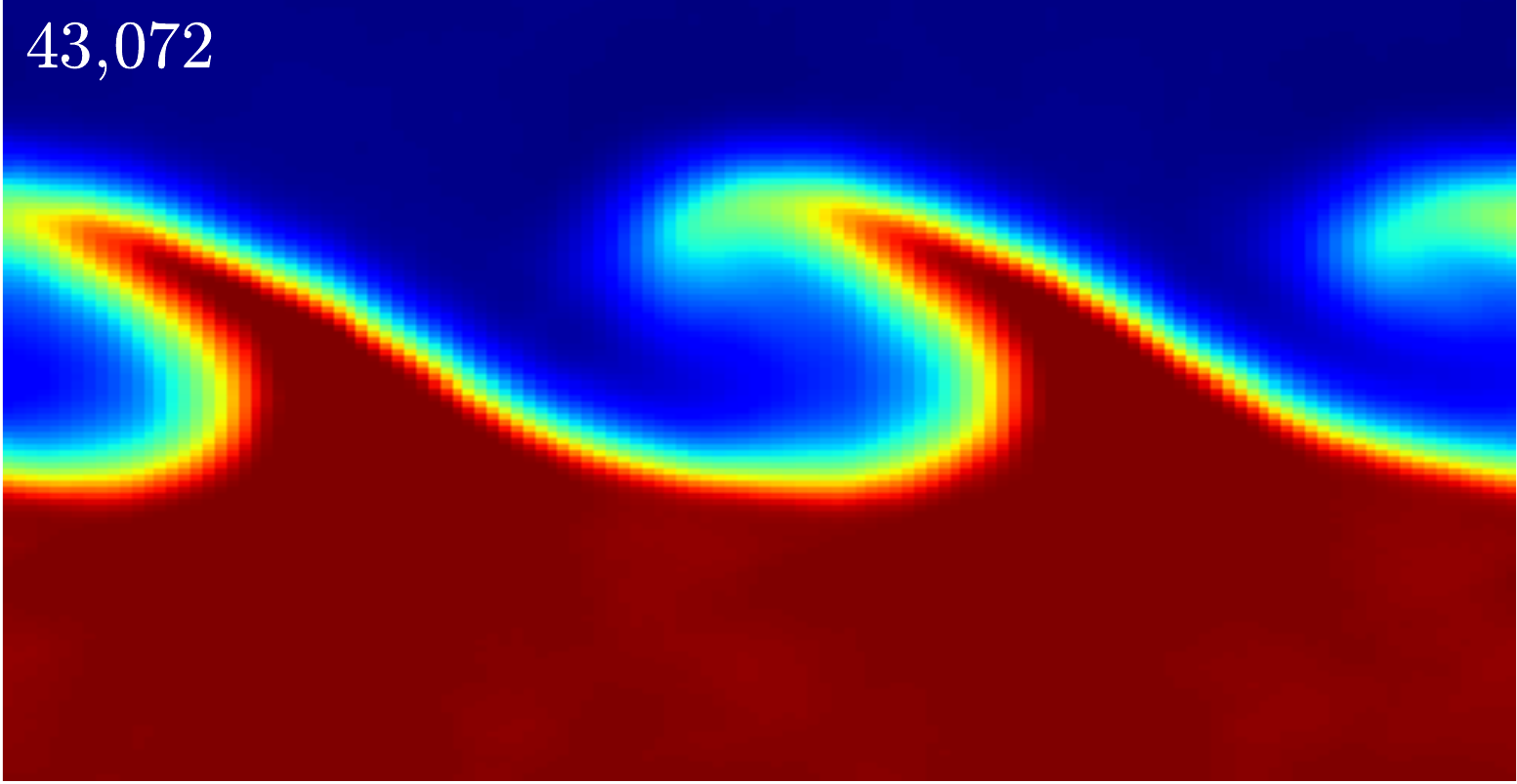}
\includegraphics[width=4.35cm]{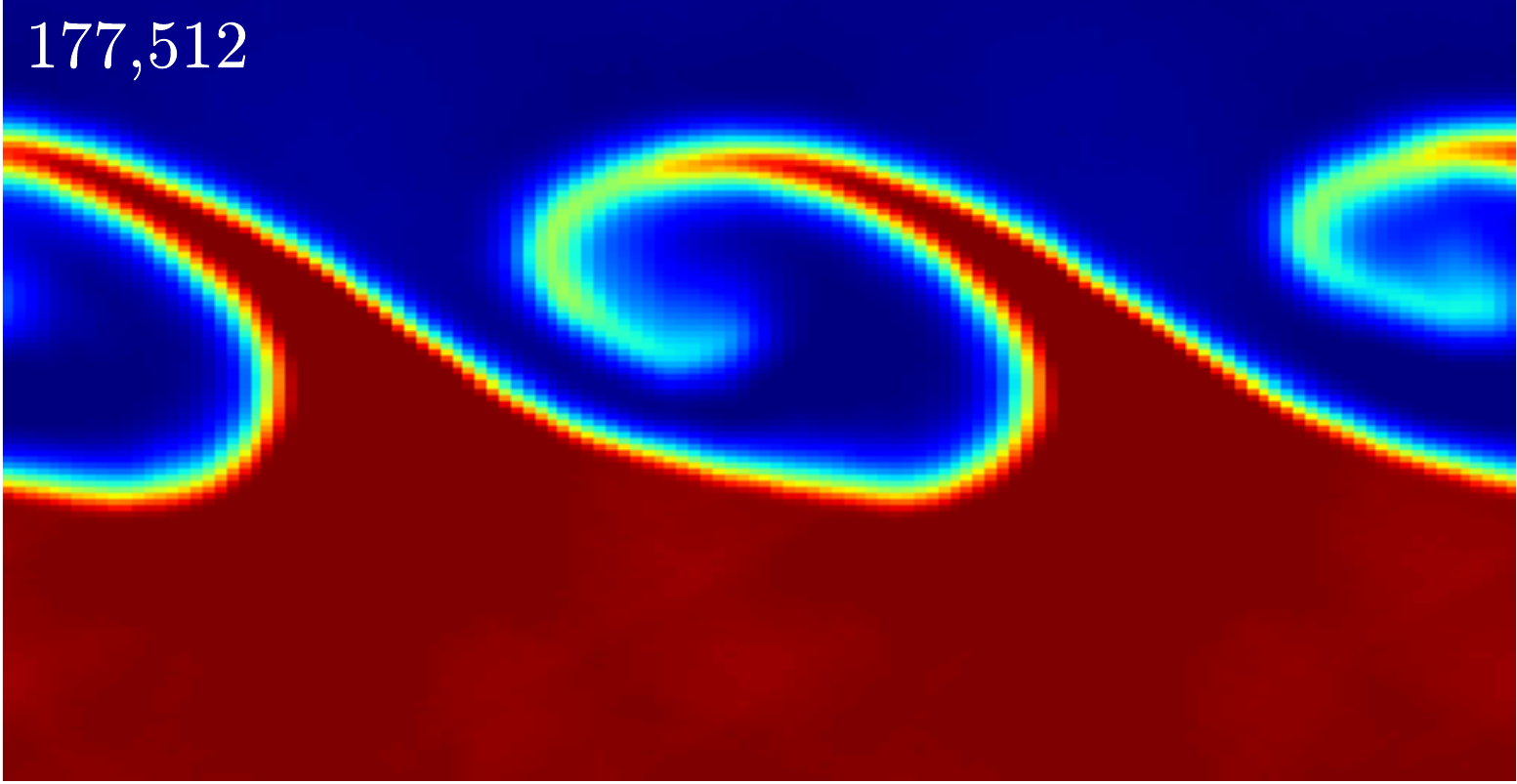}
\includegraphics[width=4.35cm]{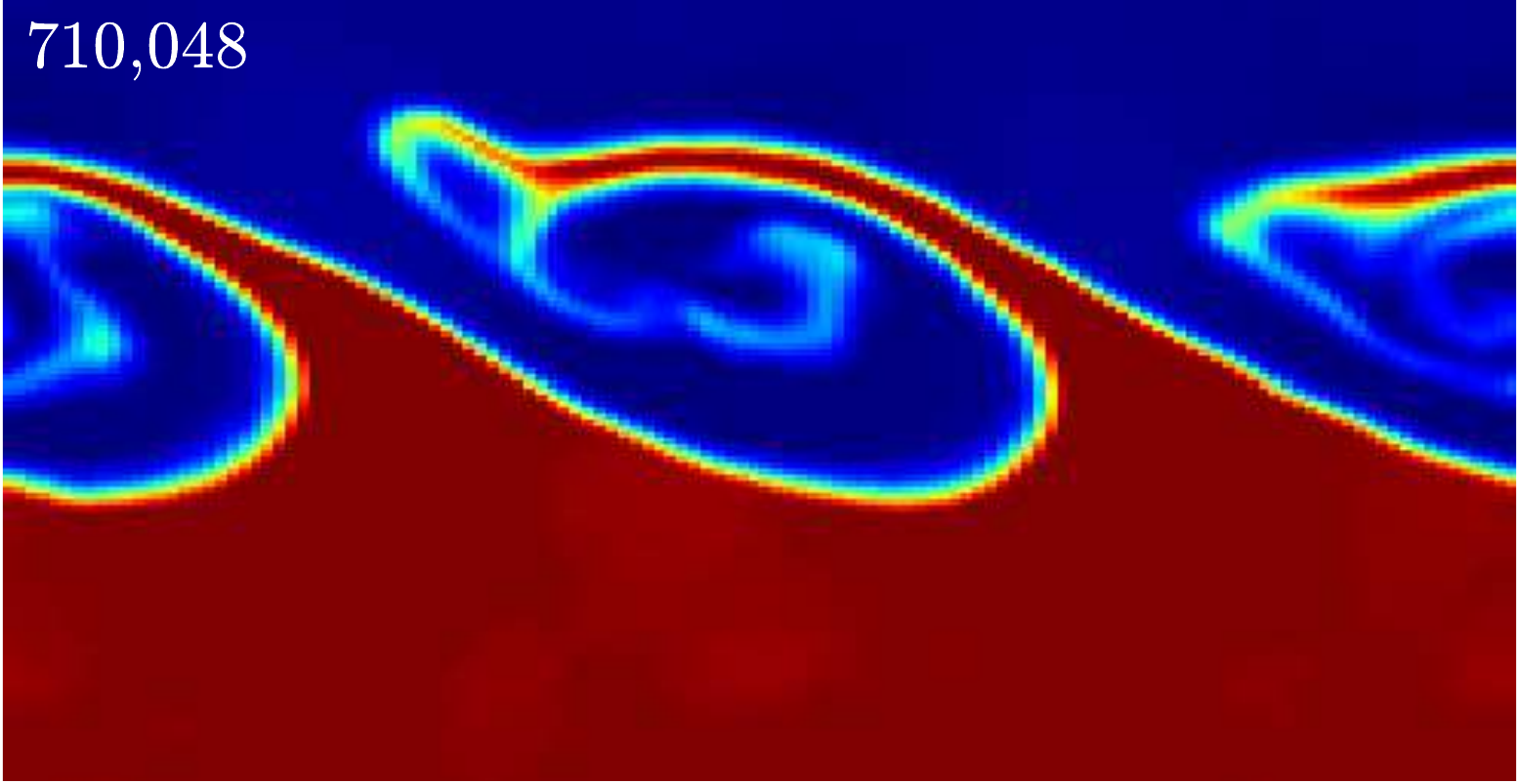}}
\vspace{0.04cm}
\centerline{\includegraphics[width=4.35cm]{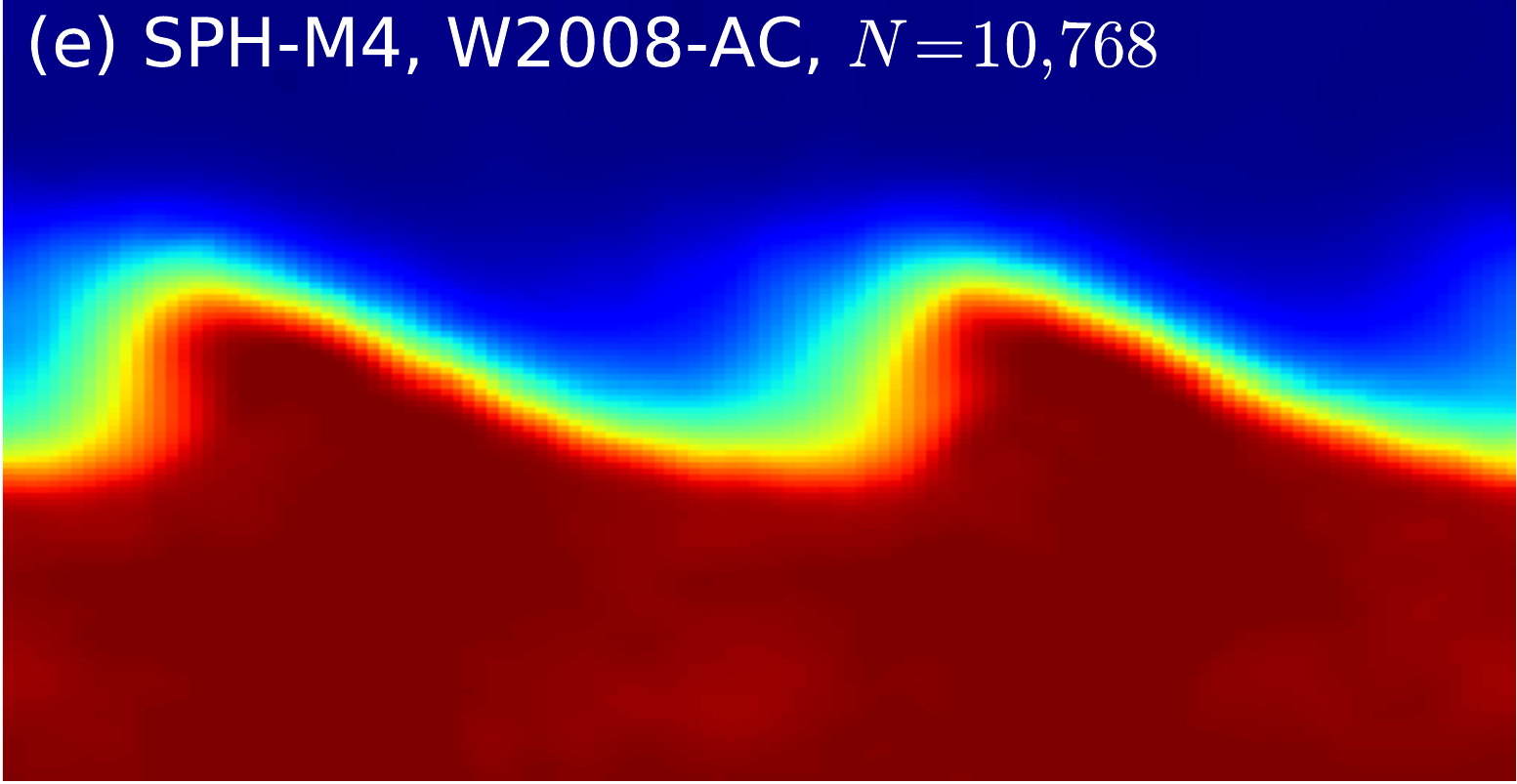}
\includegraphics[width=4.35cm]{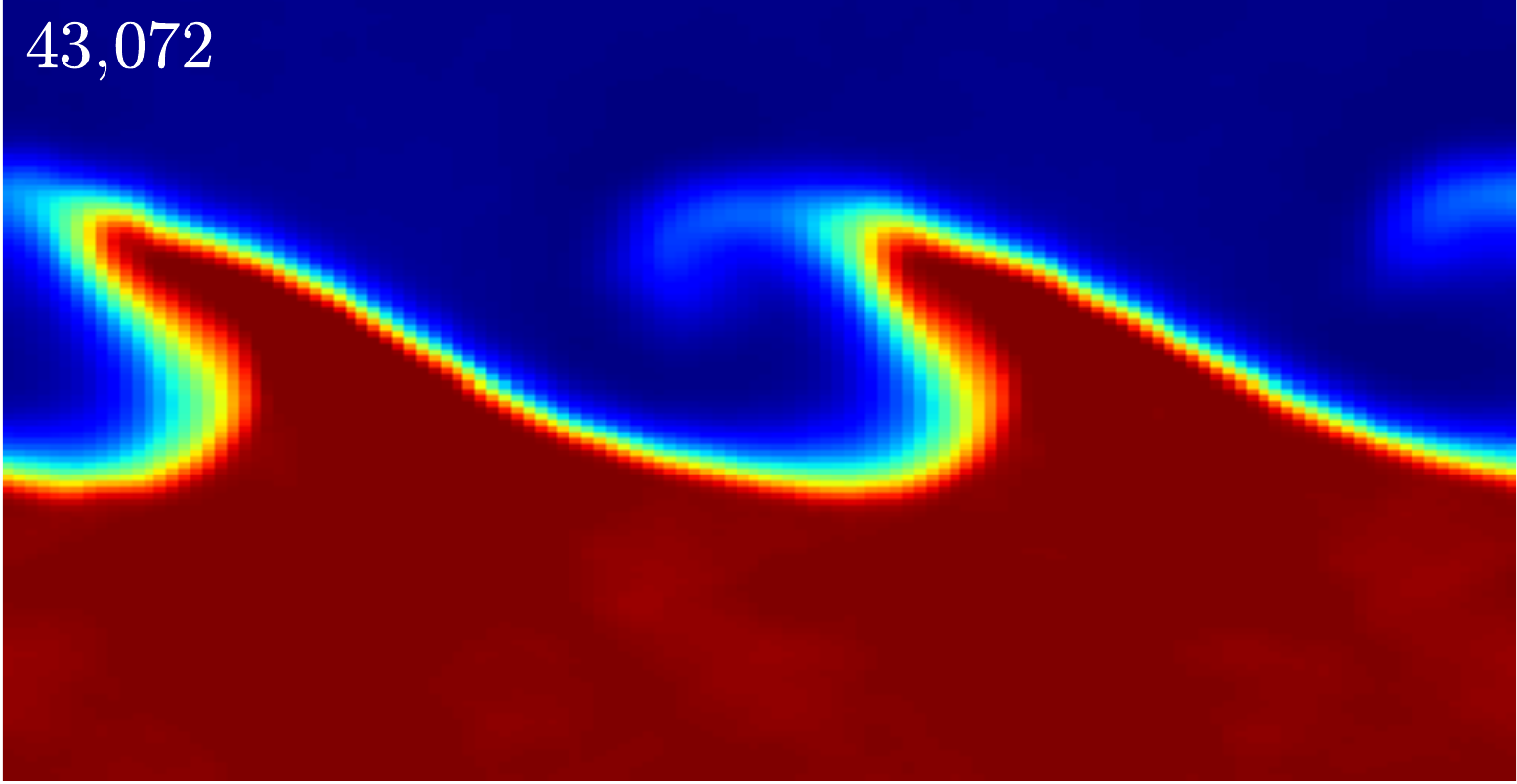}
\includegraphics[width=4.35cm]{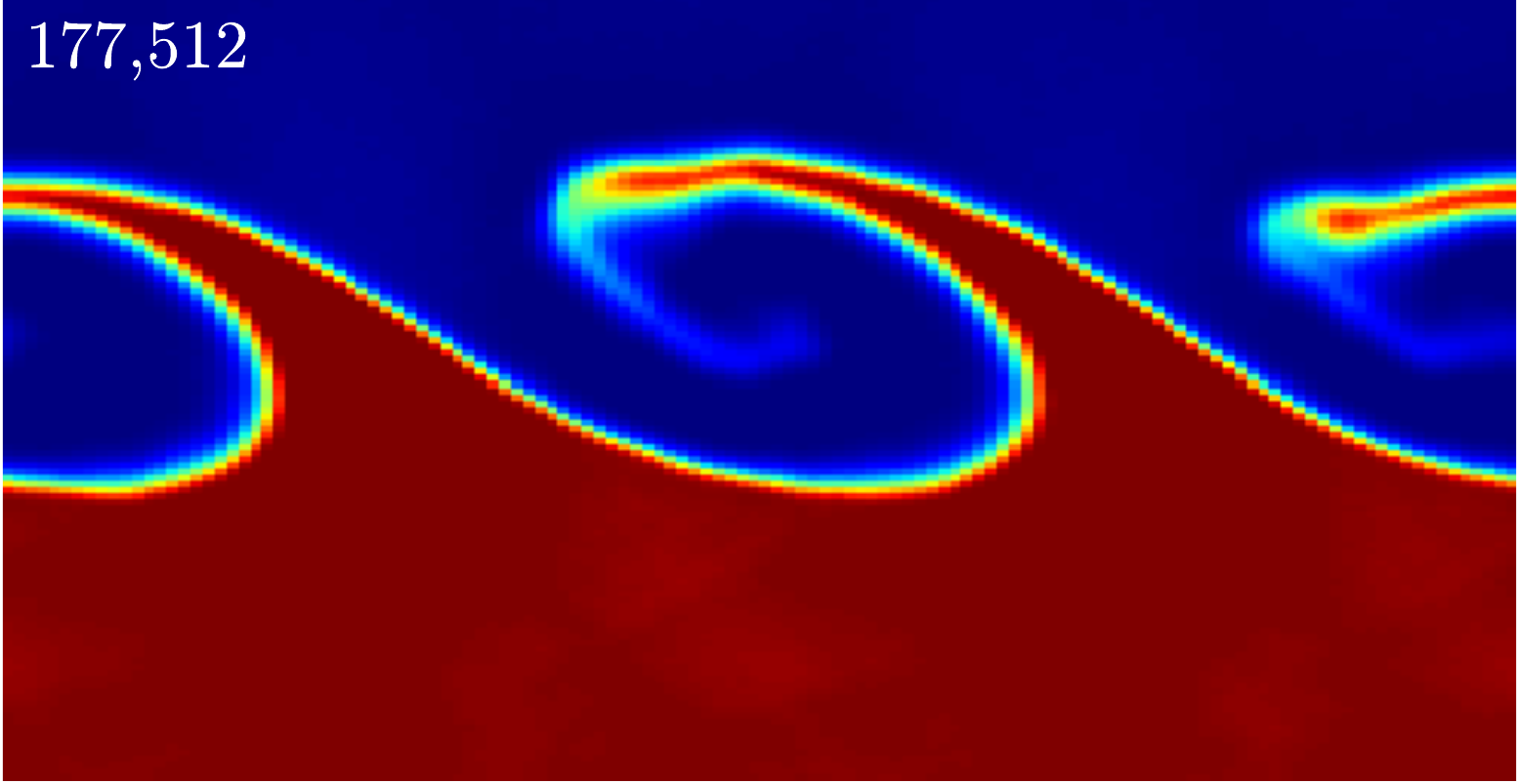}
\includegraphics[width=4.35cm]{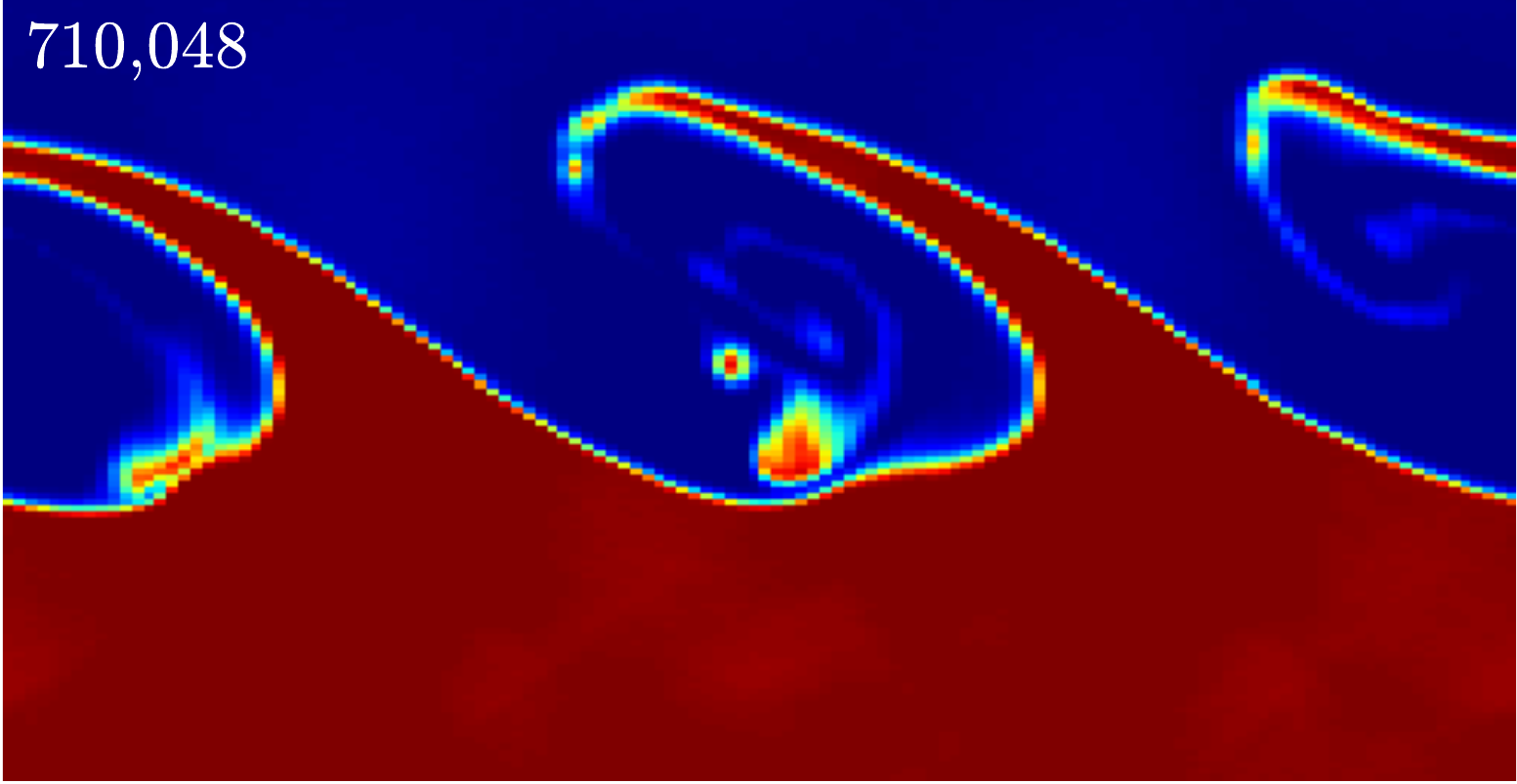}}
\vspace{0.04cm}
\centerline{\includegraphics[width=4.35cm]{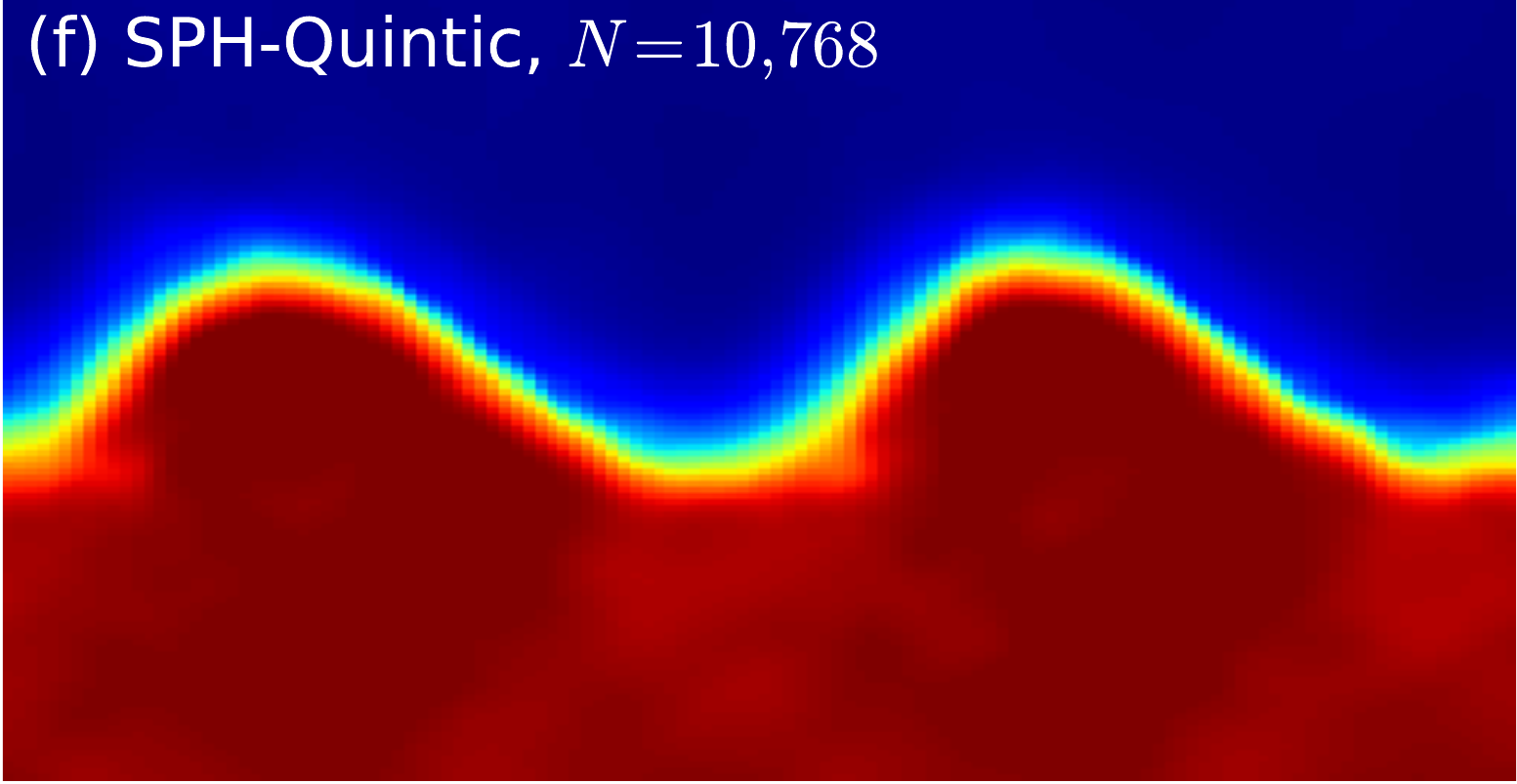}
\includegraphics[width=4.35cm]{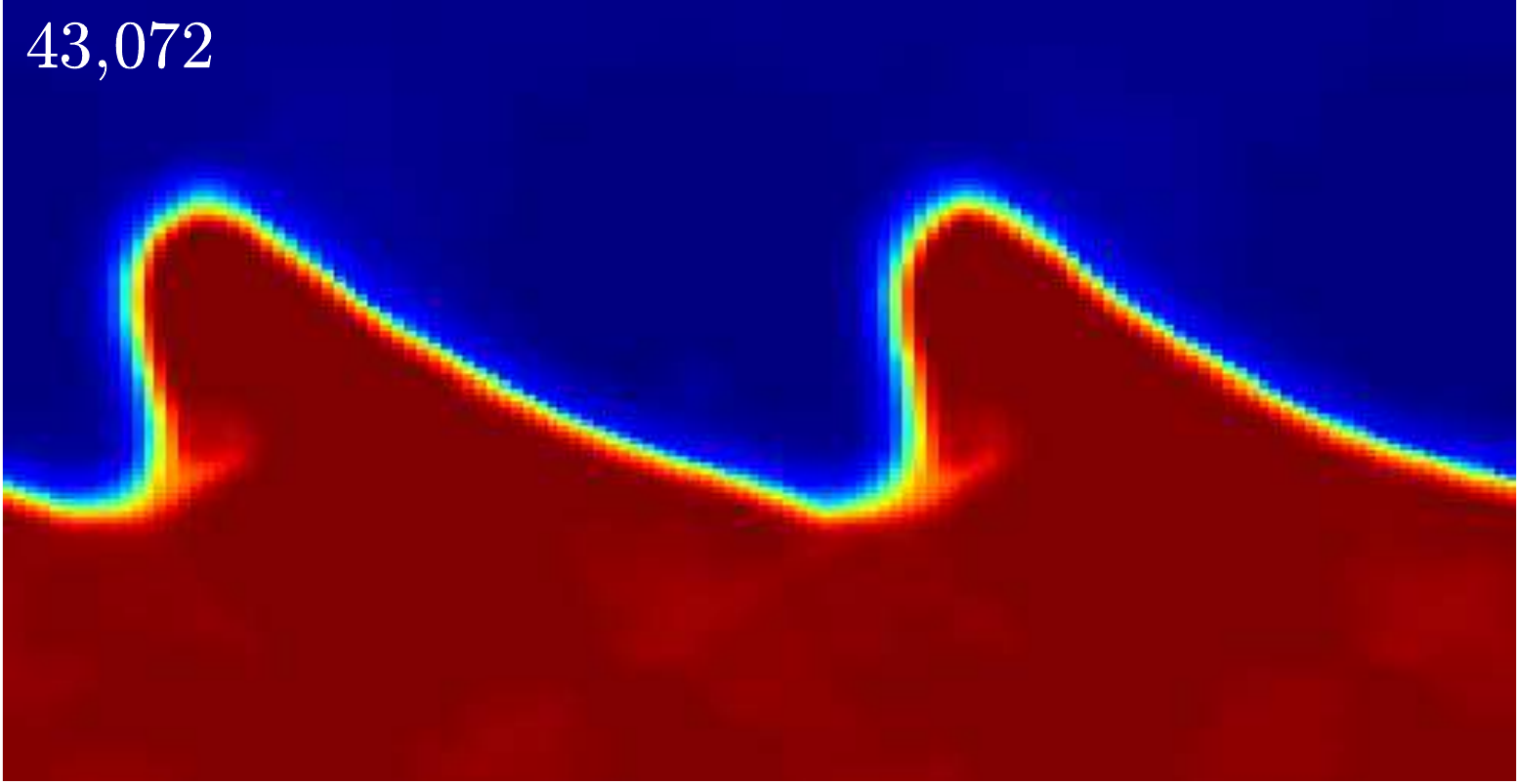}
\includegraphics[width=4.35cm]{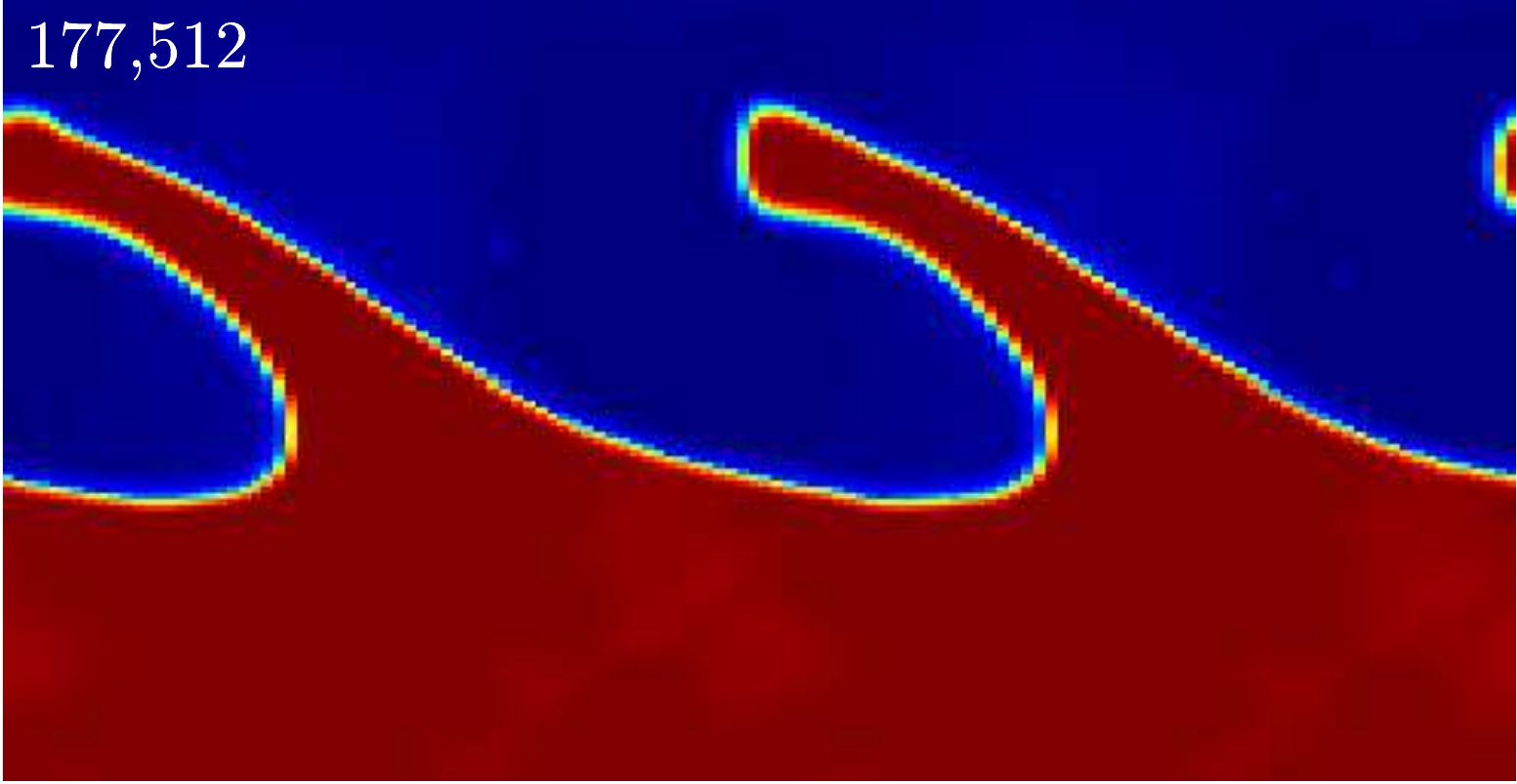}
\includegraphics[width=4.35cm]{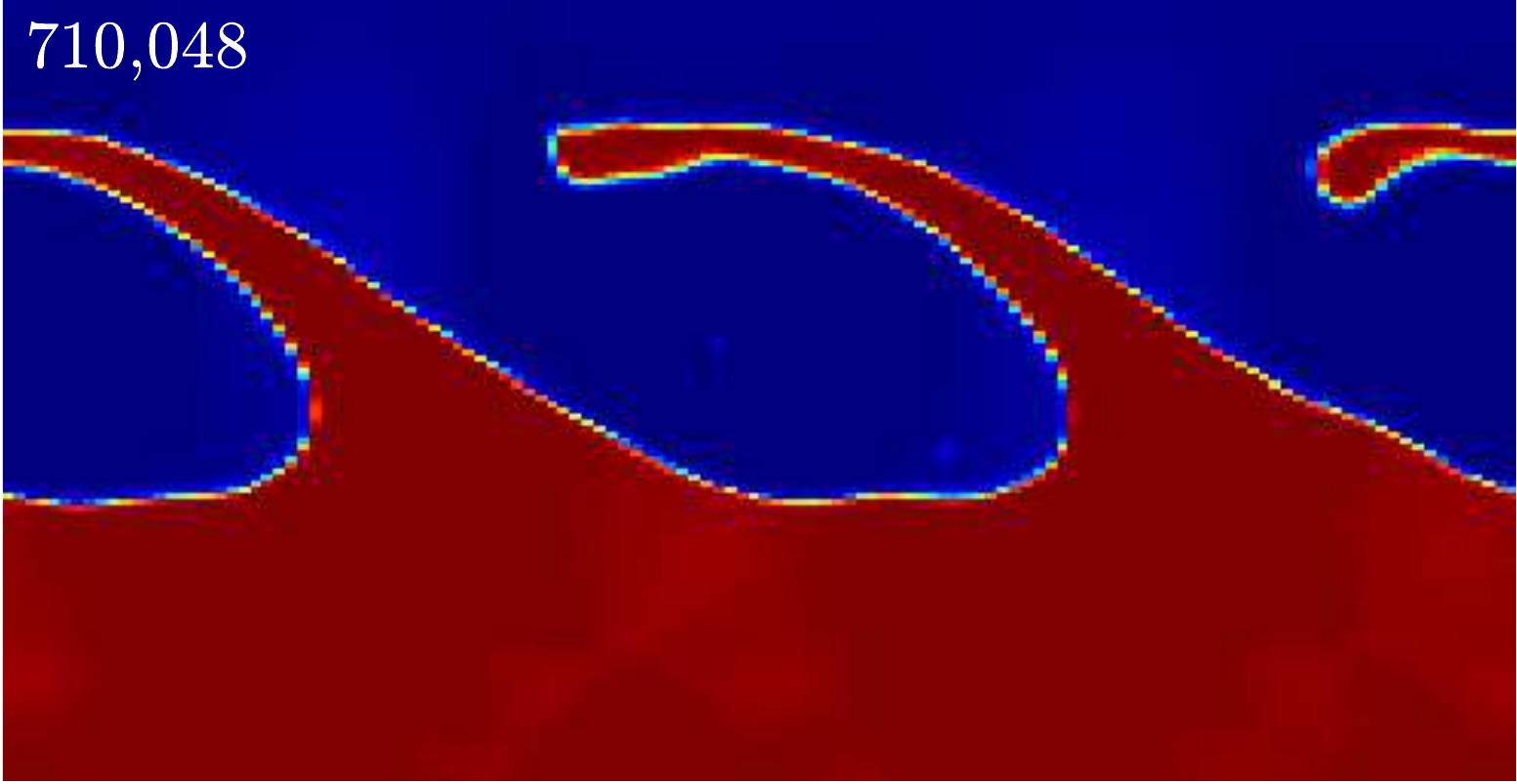}}
\vspace{0.04cm}
\centerline{\includegraphics[width=4.35cm]{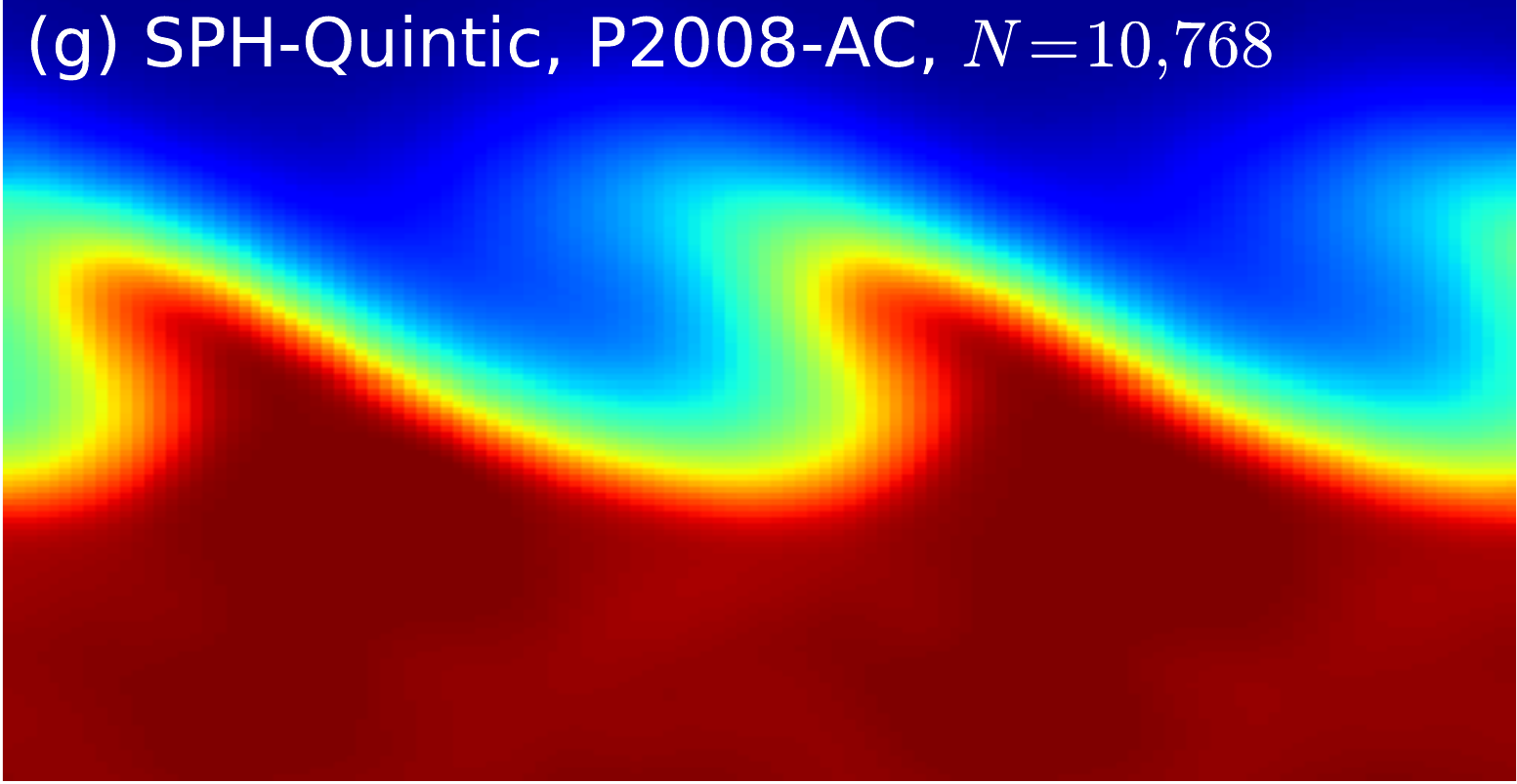}
\includegraphics[width=4.35cm]{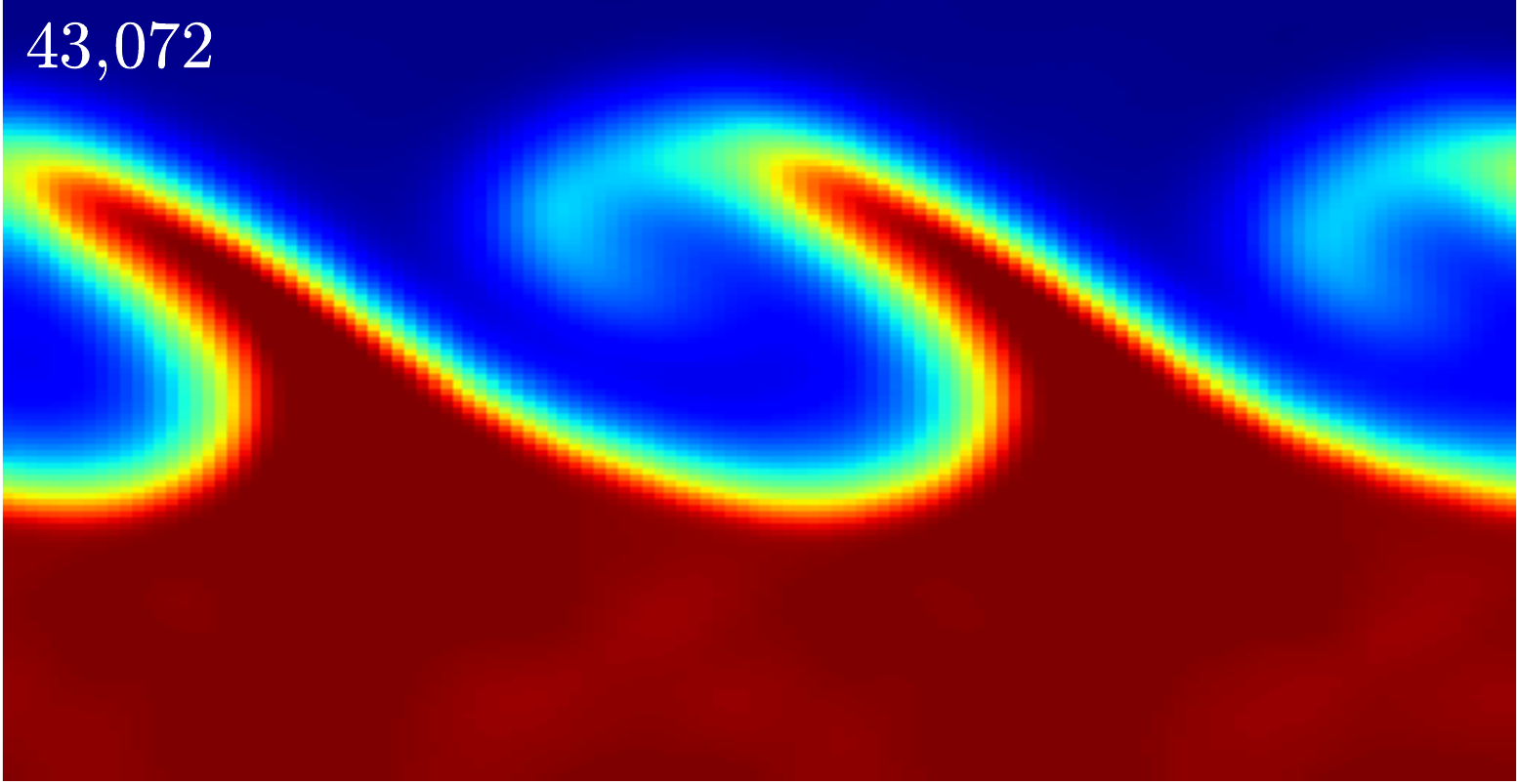}
\includegraphics[width=4.35cm]{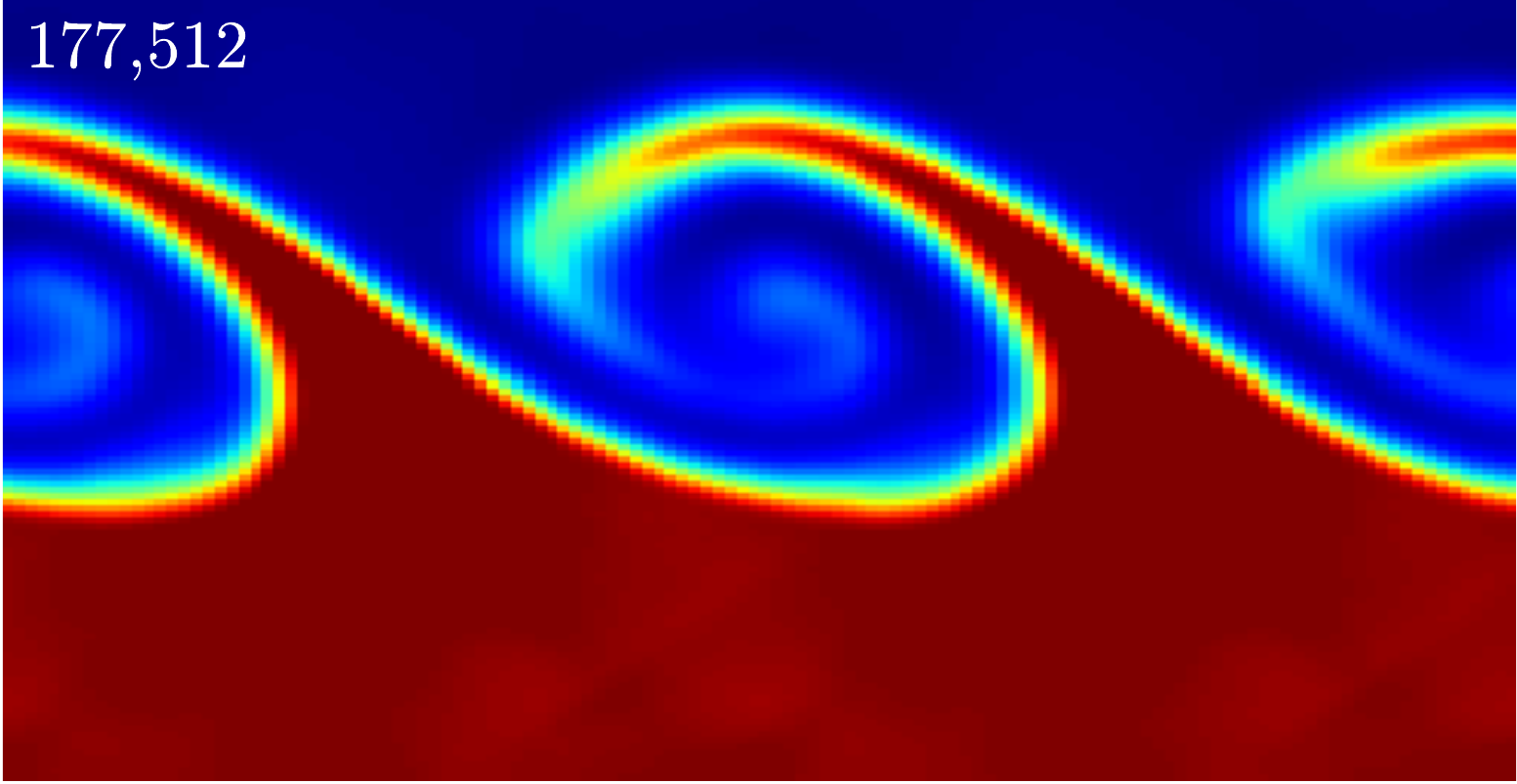}
\includegraphics[width=4.35cm]{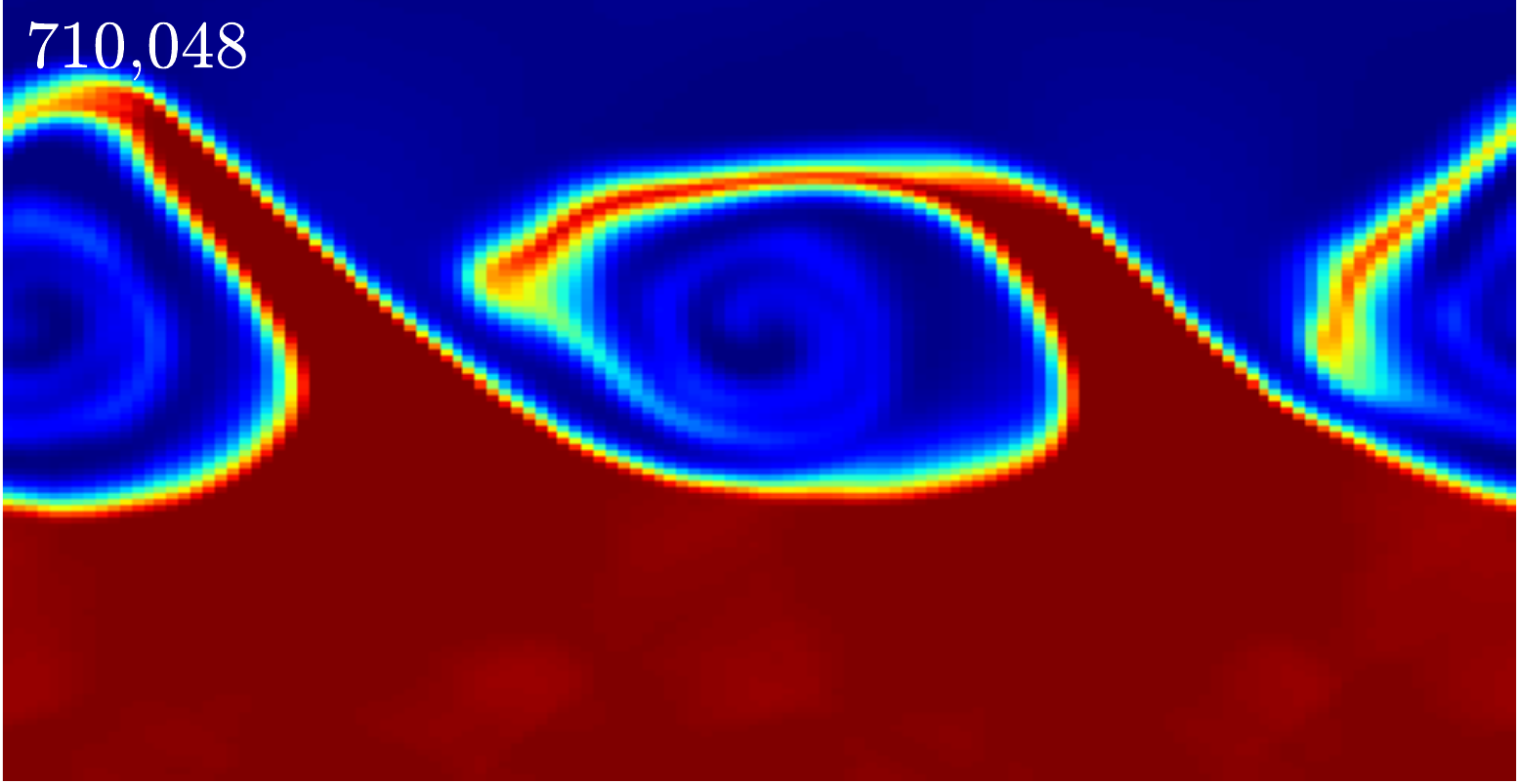}}
\vspace{0.04cm}
\centerline{\includegraphics[width=4.35cm]{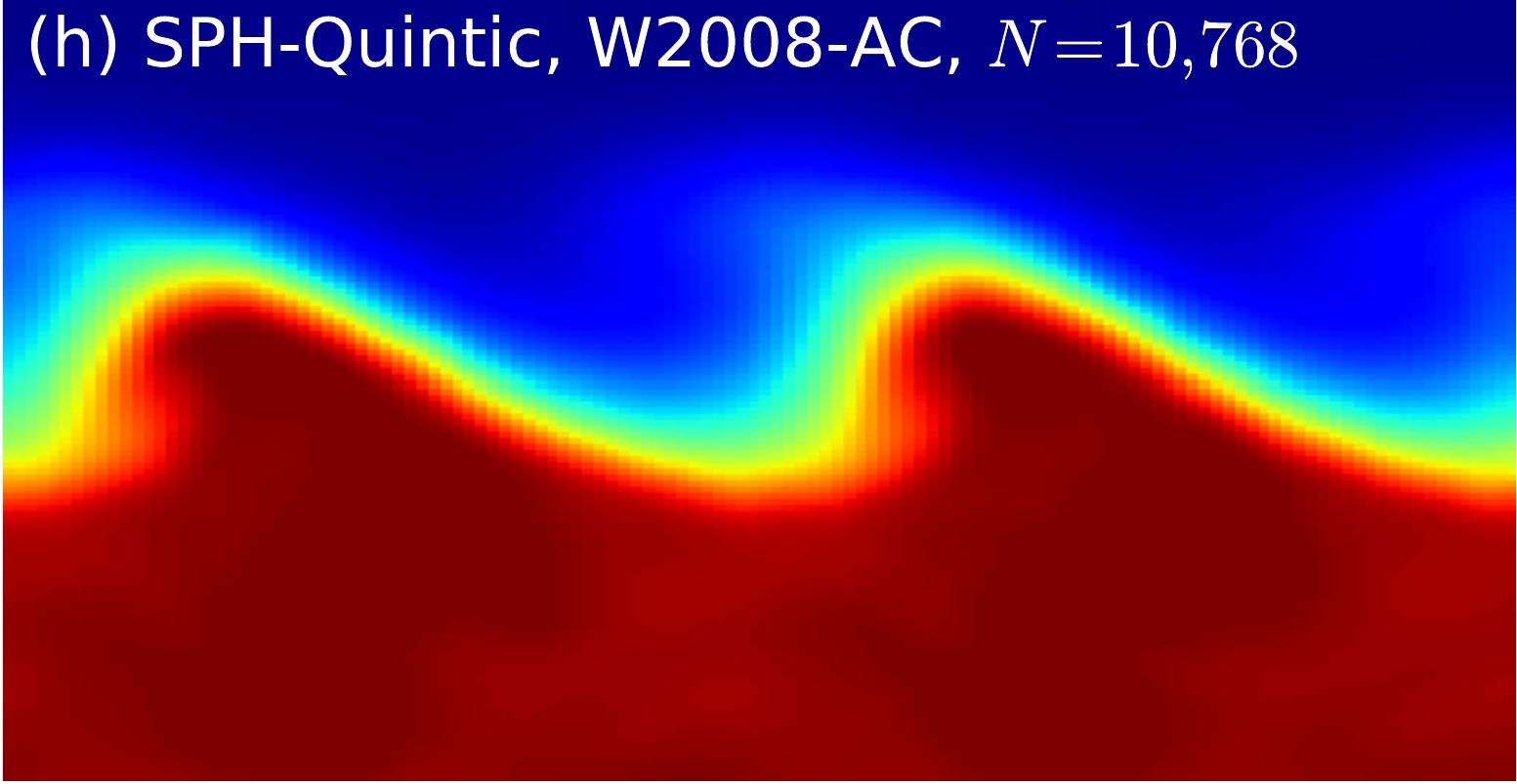}
\includegraphics[width=4.35cm]{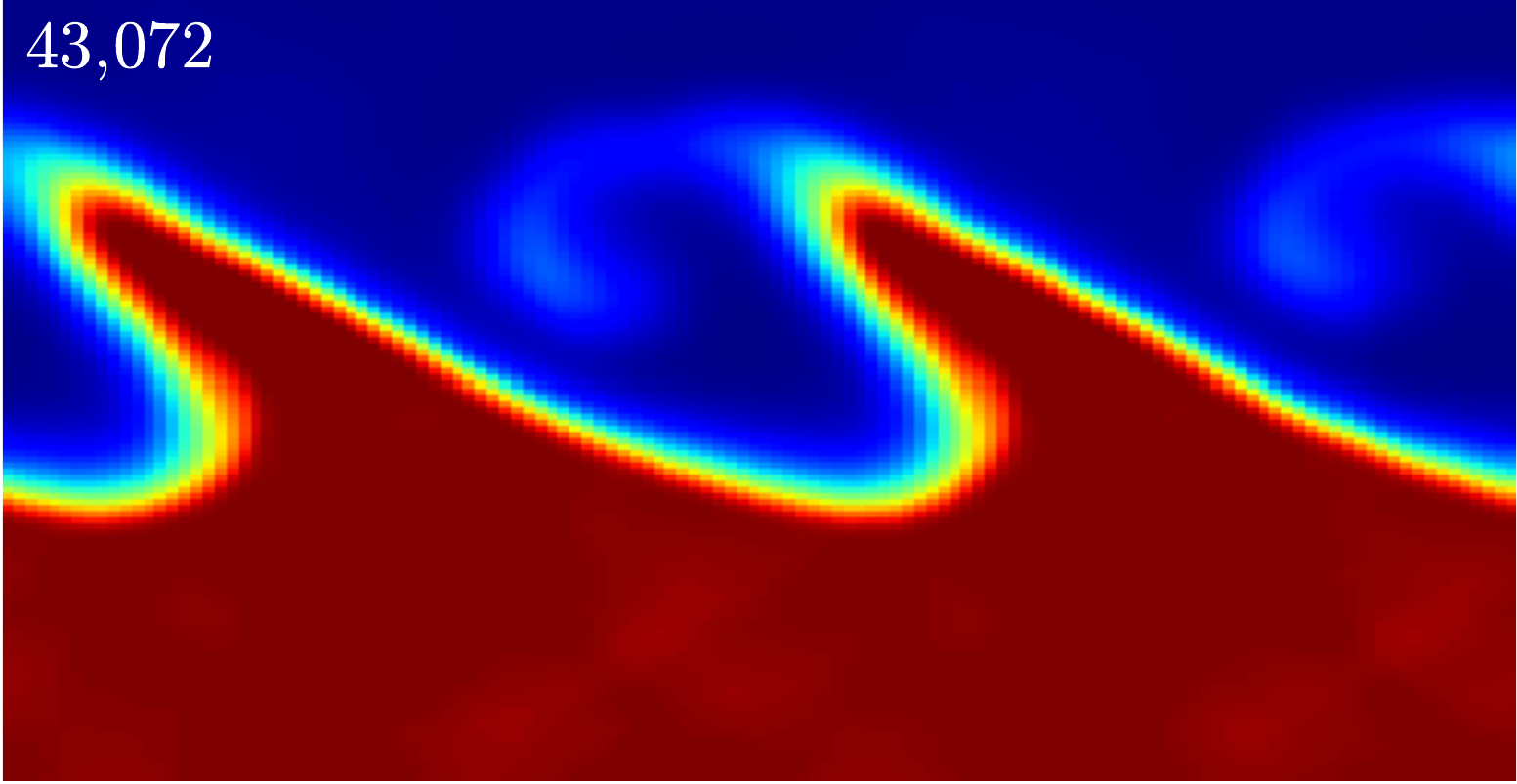}
\includegraphics[width=4.35cm]{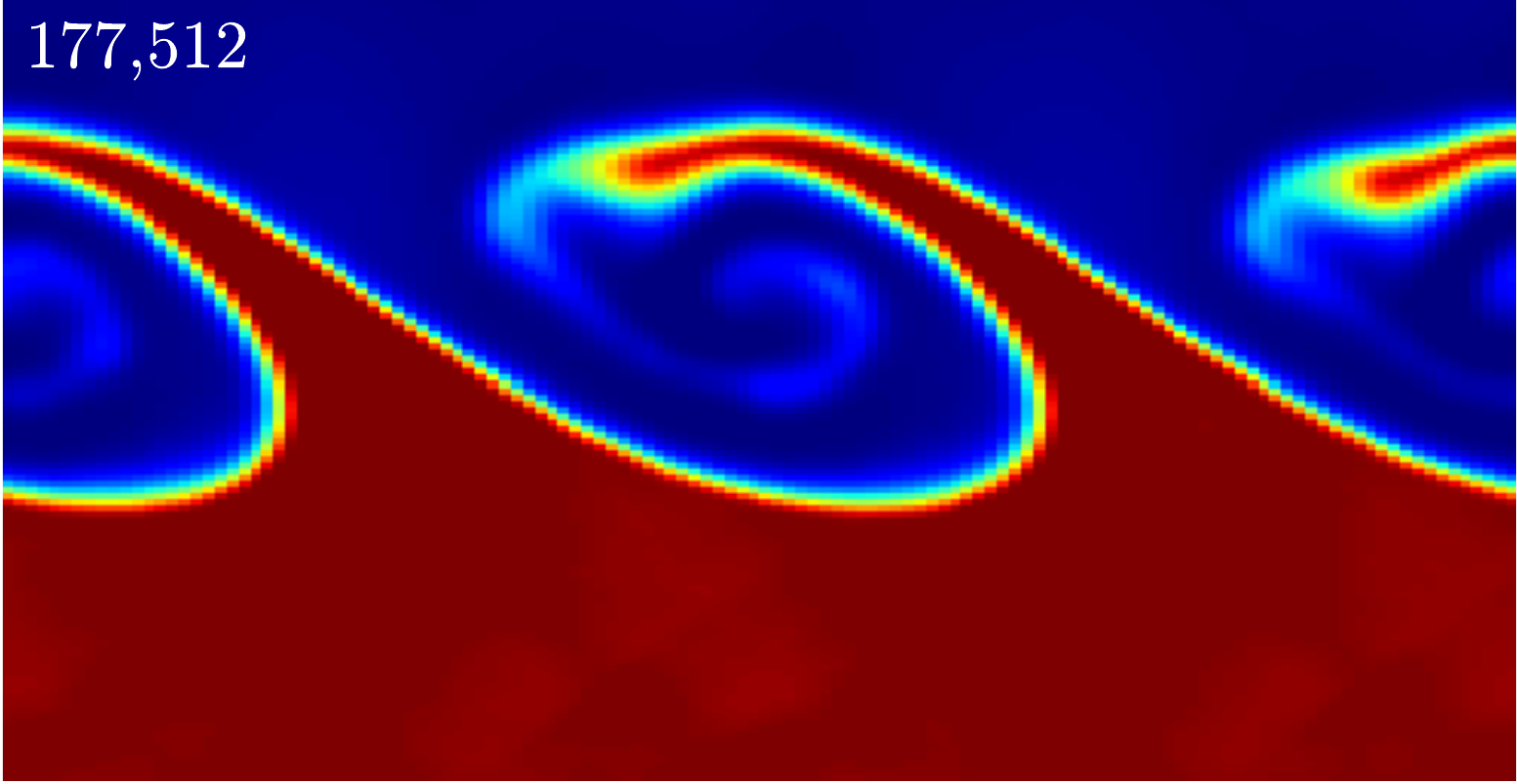}
\includegraphics[width=4.35cm]{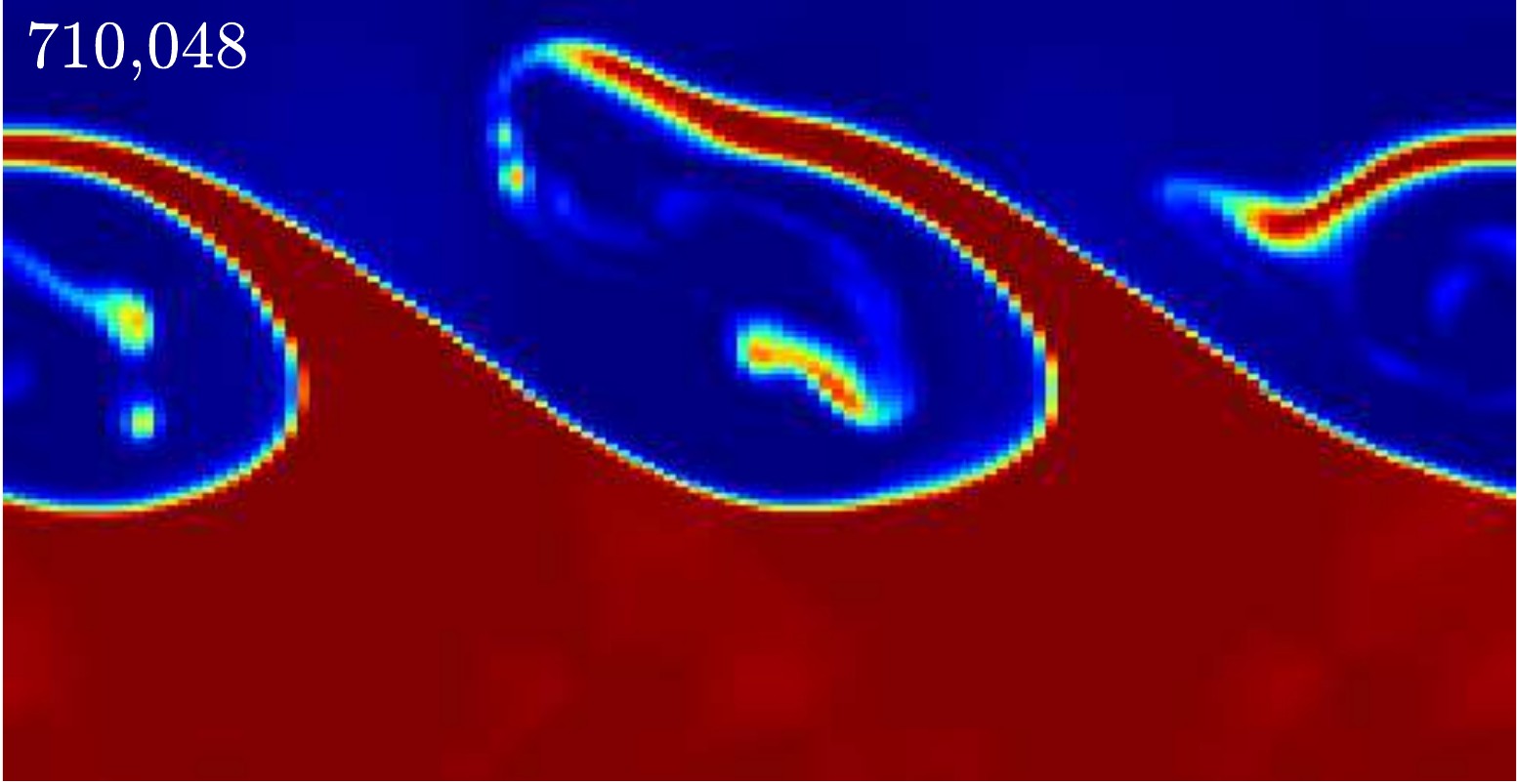}}
\vspace{0.04cm}

\vspace{0.04cm}

\caption{The   density  structure   of  the   $10:1$  Kelvin-Helmholtz
  instability at  a time $t  = \tau_{_{\rm KH}} =  1.74$ modeled
  with (a) {\small MG} using uniform grid, (b) {\small MG} using AMR, 
  (c) {\small SEREN} using the M4
  kernel,  (d)  {\small SEREN}  using  the  M4  kernel  and  the  Price  (2008)
  artificial  conductivity, (e)  {\small SEREN}  using the  M4  kernel and  the
  Wadsley  et al.  (2008)  conductivity, (f)  {\small SEREN} using  the quintic
  kernel,  (d) {\small SEREN}  using the  quintic kernel  and the  Price (2008)
  artificial conductivity, (e) {\small SEREN}  using the quintic kernel and the
  Wadsley et al. (2008)  conductivity.  The left-hand column shows the
  KHI  using the  smallest resolution  ($32 \times  32$ cells  for the
  finite volume code and $N = 10,768$ for the SPH code) with increasing
  resolution moving right to  the highest resolution ($256 \times 256$
  for the finite  volume code and $N = 710,048$  particles for the SPH
  code).  Note  that we  only show the  top half of  the computational
  domain ($y > 0$) due to the symmetry of the initial conditions.  
  Each sub-figure shows
  the density field (blue : low density - red : high density).} 
\label{FIG:KHI-10:1}
\end{figure*}

\subsubsection{Initial conditions} \label{SSS:KHI-ICS}

Analysis of the linear growth is given in many classical textbooks
and papers (e.g \citealt{Chandra1961}; \citealt{Junk2010}).
Following \cite{Price2008}, we model both a $2:1$ density contrast
and a $10:1$ density contrast.  The two fluids are separated along
the x-axis and have a x-velocity shear, but are in pressure balance
with $P = 2.5$.  The ratio of specific heats is $\gamma = 5/3$.
Fluid 1 ($y > 0$) has a density $\rho_1 = 1$ and x-velocity $v_1 =
0.5$.  Fluid 2 ($y < 0$) has a density $\rho_2$ ($=2$ or $10$) and
x-velocity $v_2 = -0.5$.  The velocity perturbation in the
y-direction is given by  
\begin{eqnarray} 
v_y &=& w_0\,\sin{ \left( \frac{2\,\pi\,x}{\lambda} \right)}\, \nonumber \\
&&\left\{ \exp{ \left[ -\frac{(y - y_{1})^2}{2\sigma^2} \right]} + 
\exp{ \left[ -\frac{(y - y_{2})^2}{2\sigma^2} \right]} \right\} \,, 
\label{EQN:KHIPERT}
\end{eqnarray}
where $\lambda = 0.5$ and $y_{1} = 0.25$ and $y_{2} = -0.25$ are the
locations of the shearing layers between the two fluids.  The
computational domain is $-0.5 < x < 0.5$ and $-0.5 < y < 0.5$ with
periodic boundaries in both the x-dimension and y-dimension.  The
growth timescale, $\tau_{\rm KH}$, of the Kelvin-Helmholtz instability
in the linear regime is  
\begin{eqnarray} \label{EQN:KHITIME}
\tau_{\rm KH} &=& \frac{(\rho_1 + \rho_2)}
{\sqrt{\rho_1\,\rho_2}}\frac{\lambda}{|{\bf v}_2 - {\bf v}_1|}\,.
\end{eqnarray}
For the $2:1$ density contrast, $\tau_{\rm KH} = 1.06$, and for the $10:1$ density contrast, $\tau_{\rm KH} = 1.74$.  We follow the evolution of the KHI until a time of $t = 2\,\tau_{\rm KH}$ using both {\small MG} and {\small SEREN}, beyond the linear growth of the instability and into the non-linear regime where vorticity develops.

We model each both KHI  at various resolutions.  For the finite volume
code, we model both the $2:1$ and $10:1$ instabilities with $32 \times
32$, $64  \times 64$,  $128 \times 128$  and $256 \times  256$ uniform
grid  cells.   When  using   AMR,  these  are  the  maximum  effective
resolutions of  the simulations.  For  the SPH simulations,  we set-up
each  part of the  fluid as  a separate  cubic lattice  arrangement of
particles.  For the  $\rho = 1$ fluid, we set-up  the particles on $44
\times 22$, $88 \times 44$, $180 \times 90$ and $360 \times 180$ grids
for  the different resolution  tests.  For  the $\rho  = 2$  fluid, we
set-up the particles  on $64 \times 32$, $128  \times 64$, $256 \times
128$ and $512 \times 256$ grids.  For the $\rho = 10$ fluid, we set-up
the particles on  $140 \times 70$, $280 \times  140$, $568 \times 284$
and $1136  \times 568$  grids.  The masses  for the particles  in each
density  fluid are  selected  to give  the  required average  density.
Therefore, the masses of the SPH  particles in the two regions are not
necessarily the same (but are as close as possible while maintaining a
uniform  grid of  particles  on  each side).   We  set-up the  thermal
properties  of  the  gas  to  give  pressure  equilibrium  across  the
interface.   We   first  calculate  the  SPH   density  from  Equation
\ref{EQN:RHO-SPH},  and then calculate  the specific  internal energy,
$u\ssi =  P\,/\rho\ssi/(\gamma - 1)$.  An  initially smoother internal
energy discontinuity  helps to minimise  the repulsive effects  at the
boundary between the two fluids \citep{GSPH2010}.

\subsubsection{Simulations} \label{SSS:KHI-SIMS}

We model both the $2:1$ and $10:1$ density-contrast KHI using both AMR
and SPH with a variety of different options and resolutions to assess
the effect of both on the development of the instability.  For the AMR
simulations, we perform simulations with both a uniform grid, and with
4 levels of refinement.   For the SPH simulations, we model the KHI
with both the M4 and Quintic kernels, and also with and without the
artificial conductivity terms advocated by both \cite{Price2008} and
\cite{Wadsley2008}.  We follow the growth of the instability until a
time $t = 1.5\,\tau_{_{\rm KH}} = 1.59$ for the $2:1$ instability and
$t = 1\,\tau_{_{\rm KH}}$ for the $10:1$ instability.  Figure
\ref{FIG:KHI-2:1} shows the development of the $2:1$ instability at
five successive time snapshots for the highest resolution AMR and SPH
simulations.  Figures \ref{FIG:KHI-2:1} \& \ref{FIG:KHI-10:1} shows
the development of the KHI for four different resolutions (columns 1 -
4, increasing resolution to the right) with these various combinations
of options for both the finite volume and SPH codes.   

For the very  lowest resolution using the finite  volume code with $32
\times  32$ grid  cells (Figure  \ref{FIG:KHI-2:1}(a), column  1), the
instability grows in the  linear regime with approximately the correct
timescale, but  there is insufficient resolution  to model small-scale
vorticity and  therefore, the instability stalls and  does not proceed
into  the  non-linear regime.   Using  $64  \times  64$ cells  (Figure
\ref{FIG:KHI-2:1}(a),  column 2),  there is  now enough  resolution to
model  vorticity, and  the  instability proceeds  into the  non-linear
regime generating a partial  spiral vortex at the shearing interfaces.
{\refone We note that this agrees with the previous result by 
\citet{Federrath2011} who find that mesh codes cannot adequately 
resolve vorticies with less than $\sim 30$ grid cells.}
As we increase  the resolution further to $128  \times 128$ grid cells
(Figure \ref{FIG:KHI-2:1}(d)) and $256  \times 256$ grid cells (Figure
\ref{FIG:KHI-2:1}(e)), the general  effect of increasing resolution is
to allow more  highly detailed spiral structure to  be resolved in the
simulation.  The general evolution of the instability (e.g. the growth
timescale, the size of the  spiral vortex) is converged by this point.
Increasing the resolution further  can lead to secondary instabilities
which  are  seeded  by   the  grid.   In  principle,  these  secondary
instabilities can  be suppressed by  using a physical  viscosity which
has a dissipation length scale independent of resolution.

For   the    SPH   simulations    using   the   M4    kernel   (Figure
\ref{FIG:KHI-2:1}(c,d,e)),  the  lowest  resolution  simulations  show
little  evidence  for  the   generation  of  vorticity.   SPH  without
conductivity demonstrates  some growth  of the seeded  perturbation in
the  distortion of  the interface,  similar to  the  lowest resolution
finite volume simulations.  When  using either of the two conductivity
options,  the  instability  appears  to  be  dominated  by  the  extra
dissipation  and  noise at  the  interface.   For the  no-conductivity
simulations, increasing  the resolution increases the  degree that the
instability grows into generating a spiral vortex.  Due to the lack of
any explicit entropy mixing source (except the small contribution from
artificial viscosity),  the instability grows  into longer finger-like
structures  that pertrude  into the  adjacent fluid.   At  the highest
resolution, the  finger forms one  complete spiral but still  does not
mix with the second fluid.  If we include artificial conductivity, the
fluid can readily  mix and generate vorticity similar  in structure to
the finite volume simulations.

Figure \ref{FIG:KHI-10:1} shows the density snapshot at a time $t =
1\,\tau_{_{\rm KH}} = 1.74$ for the $10:1$ KHI for the same
combination of options as the $2:1$ case.  In principle, the $10:1$
KHI is a sterner test for SPH since there is a much larger particle
number gradient at the fluid interface which leads to a larger
potential summation noise due to the assymetry in the kernel sampling.
As can be seen in Figure \ref{FIG:KHI-10:1}(c,f), the SPH simulations,
using both the M4 and quintic kernel but without artificial
conductivity, reproduce the growth of the perturbation on roughly the
correct timescale, but do not generate vorticity or mixing to an even
lesser degree than the $2:1$ KHI.  This is primarily due to the
surface tension effects at the interface being even stronger than the
$2:1$ case and therefore  suppressing any vorticity.  As we add either
kinds of artificial conductivity to the SPH simulations, then as with
the $2:1$ case, we generate vorticity and mixing following a similar
morphology to the finite volume code evolution.  Including artificial 
conductivity allows SPH to model the instability correctly including 
mixing.  We note that at higher resolutions ($710,048$ particles), 
small-scale wavelenghts seeded by SPH noise begin to corrupt the 
principle instability mode.

\subsubsection{Mixing in SPH} \label{SSS:KHI-SPHMIXING}
Our convergence test of the KHI reveals several important conclusions
regarding comparisons between SPH and AMR codes.   

Firstly, as is already known, there are clearly significant numerical 
effects in
SPH (namely in this case the artificial repulsion force between fluids
with different specific entropies) which can inhibit the growth of
hydrodynamical instabilities.  These can be mitigated to an extent by
increasing the number of neighbour (via using a larger kernel such as
the quintic kernel) and to a lesser degree by increasing the resolution.
However, as the $10:1$ KHI demonstrates, the degree to which
increasing resolution and neighbour number helps is dependent on the
size of the discontinuity and can not guarantee any
degree of convergence for the general case.  Therefore special
treatment (such as artificial conductivity) is required to suppress
unwanted numerical effects. 

Secondly, once the spurious numerical effects have been addressed, our
convergence study shows  that both the finite volume  code and the SPH
code  can  agree  very  well  and demonstrate  similar  evolution  and
convincing convergence with  increased resolution, although eventually
both codes will diverge due principally to noise-seeded asymmetries in
the SPH simulations leading to  the growth of other small scale modes.
Although  the sources of  diffusion/dissipation are  different (finite
volume:  advection   errors;  SPH  :   artificial  conductivity),  the
instabilities  in the  two codes  agree  in almost  every sense  (i.e.
growth  timescale, physical  size  of spirals,  number  of spirals  in
vortex).

Thirdly, regarding the SPH simulations, comparisons between the SPH
simulations with conductivity using either the M4 kernel or the
quintic kernel demonstrate that in some cases, accuracy (in the form
of reduced noise using more neighbours) can be more important than
resolution.  Formally, the resolution of the quintic kernel
simulations is lower than that of the corresponding M4 kernel
simulations since it contains fewer kernel volumes (approximately
half).  Despite having less resolution, the quintic kernel simulations
appear well converged with the finite volume simulations.  Although we do not
advocate using the quintic kernel based solely on these results, this
demonstrates the need for users of SPH to also consider using larger
kernels when testing new algorithms in SPH, as well as 
resolution-convergence studies.

\section{Discussion} \label{S:DISCUSSION}
The aim of our suite of  comparison simulations is to examine how well
finite volume and SPH methods converge with each other, what numerical
issues affect convergence, and what is their relative performance when
converged.  Firstly we will discuss some of the known issues with both
methods in  the context  of our simulations.   Then we will  examine a
number of  issues on the  relative accuracies and resolutions  of both
methods.  We note that there is an emphasis on SPH in this paper since
SPH is  expected to perform more  poorly than finite  volume in purely
hydrodynamical tests such as those in this paper.

\subsection{Accuracy} \label{SS:ACCURACY}
The accuracy of a numerical hydrodynamics scheme is the precision to which 
the solution of the original fluid equations can be determined.  
This is affected by various factors, such as how the fluid is discretised, 
how the gradients or fluxes of fluid quantities are calculated, and how those 
quantities are numerically integrated in time.  The accuracy is often 
parameterised by the {\it order} of the scheme.  If the scheme uses the 
first spatial gradient to construct quantities, it is said to be 
  {\it spatially 1st order} and the errors in spatial quantities are of 
order ${\cal O}(\Delta x^2)$, where $\Delta x$ is the spacing between 
fluid elements.  If the scheme uses also the second spatial 
gradient, then it is said to be {\it spatially 2nd order}, and the errors 
are of order ${\cal O}(\Delta x^3)$.  Another important aspect that 
determines the accuracy is the {\it consistency}.  If a scheme can 
calculate a linear gradient exactly as $\Delta x \rightarrow 0$, then 
the scheme is said to have {\it 1st order consistency}.  
If it can calculate a second-order gradient exactly, then it has 
{\it 2nd order consistency}.  For example, a numerical 
scheme may use linear gradients to calculate terms, and 
therefore be spatially 1st-order, but may not correctly calculate these 
gradients and so therefore does not have 1st-order consistency.

\subsubsection{Finite volume code accuracy} \label{SSS:GRIDACCURACY}

In  finite   volume  codes,  the   domain  is  usually   divided  into
equal-volume cells, at least in Cartesian coordinates, but this is not
strictly  necessary.  However, variable  mesh  spacing  makes it  more
expensive to  ensure high order accuracy  and also leads  to errors in
the shock conditions when shock-capturing is used.  AMR codes overcome
this  problem by using  a mesh  that is  locally uniform  and refining
where necessary, such as in  the neighbourhood of a shock.  This means
that  shocks always  propagate through  a  uniform grid.   It is  also
relatively easy to ensure high  order at boundaries between coarse and
fine grids.

Note that  although most modern upwind  codes are 2nd  order in smooth
regions,  Godunov's theorem  \citep{Godunov1959} 
tells  us  that a  code  that is  second  order  everywhere cannot  be
monotonic in regions  where there are sharp changes  in the gradients,
such as shocks.  It is therefore necessary to  use a non-linear switch
that reduces the scheme to 1st order in such regions. In any case, all
shock-capturing  codes are  1st  order if  the  flow contains  shocks,
however, it  is still  worth using 2nd  order in smooth  regions since
this leads to faster convergence.

\begin{figure*}
\centerline{
\includegraphics[height=7.7cm,angle=0]{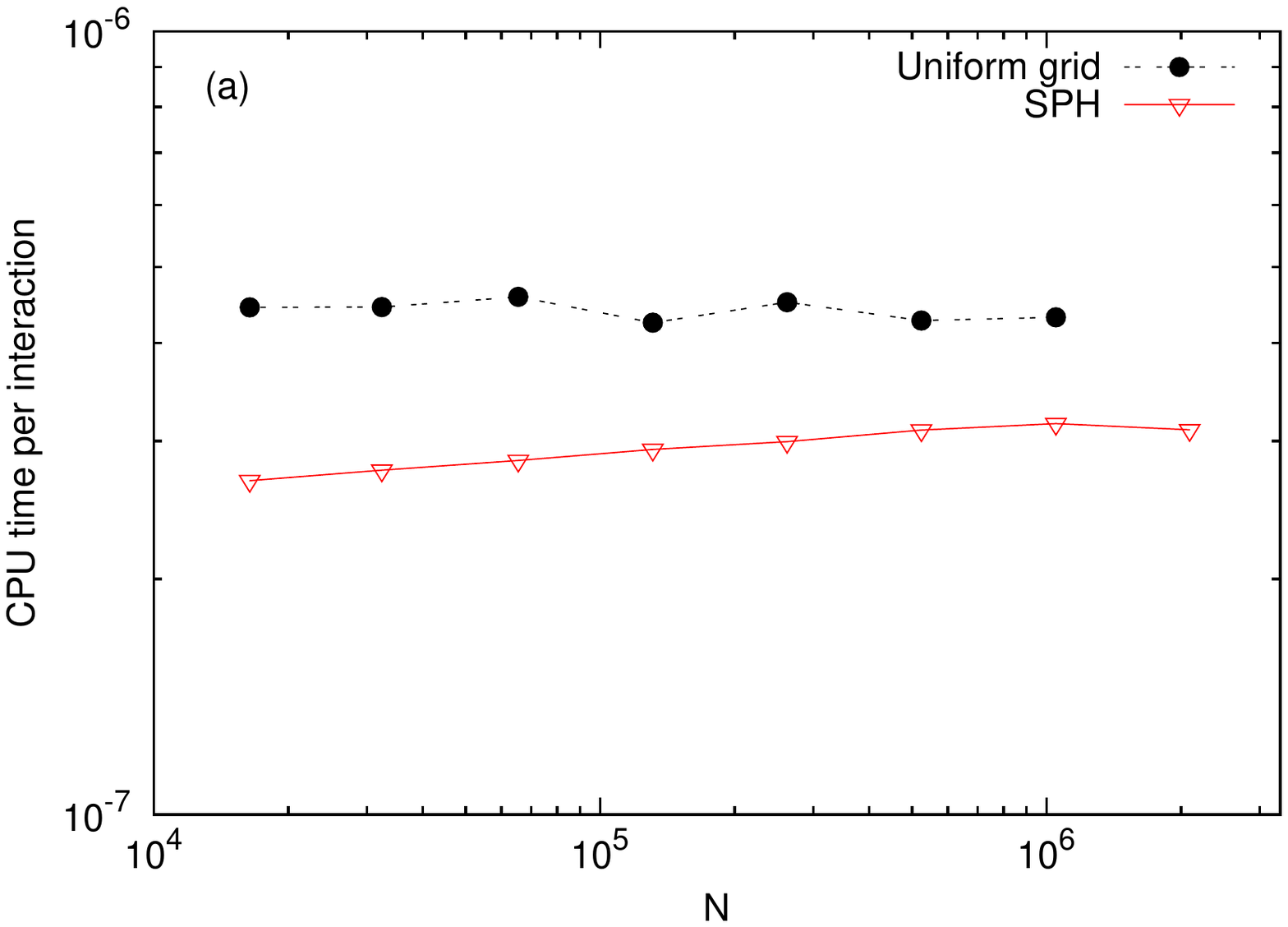}\,
\includegraphics[height=7.7cm,angle=0]{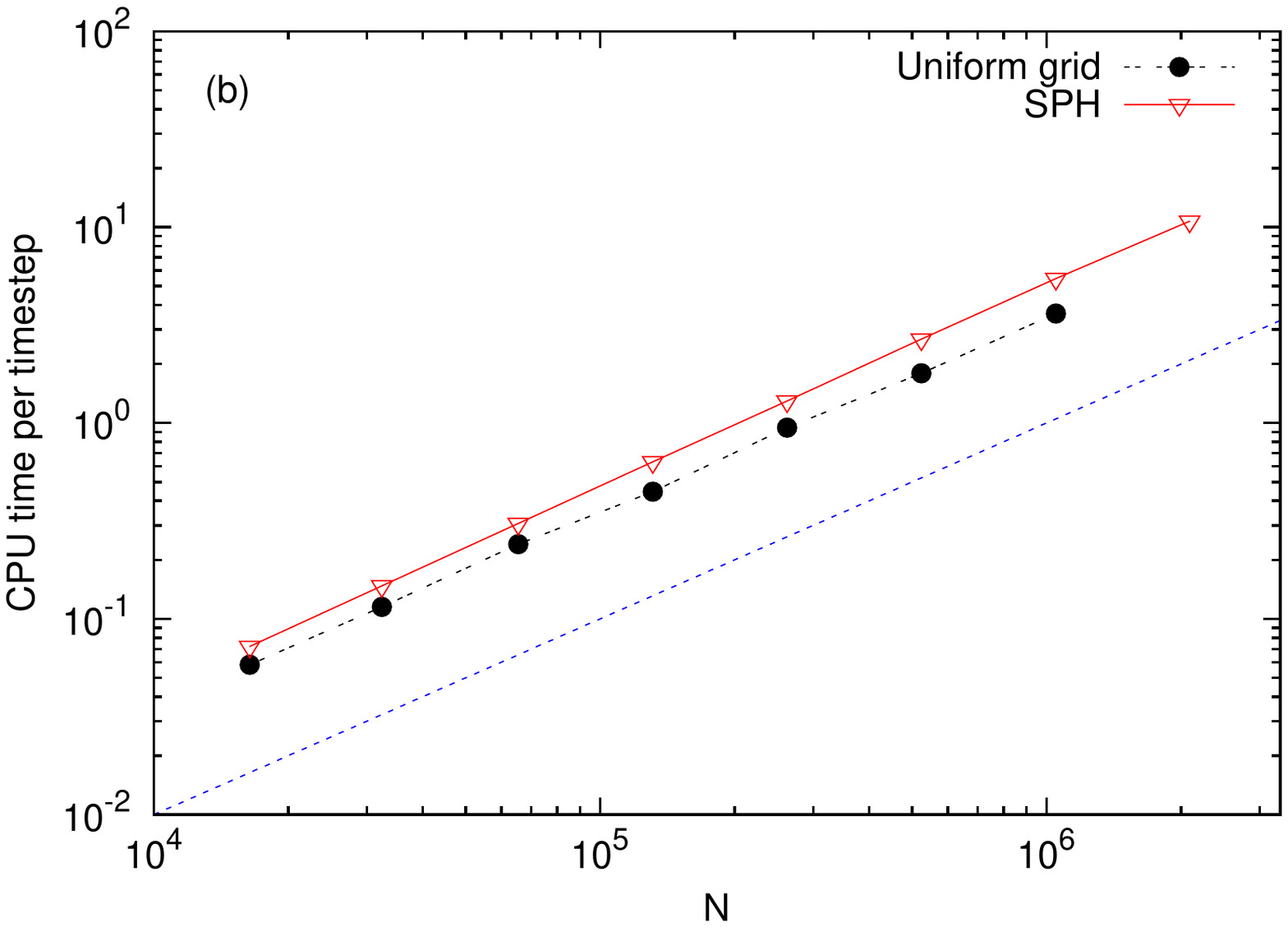}}
\caption{(a) The CPU time per cell-cell, or particle-particle,
  interaction in the finite volume and SPH codes. (b) The total CPU
  time of all cells or particles per timestep for the finite volume and SPH codes.  The trend expected for constant no. of operations per cell/particle ($t_{\rm CPU} \propto N$) is shown for reference.}
\label{FIG:TIMING}
\end{figure*}

\subsubsection{SPH code accuracy} \label{SSS:SPHACCURACY}

The accuracy of SPH is less well-defined than with finite volume codes.  SPH represents a solution variable, $A$, by computing the kernel-weighted volume integral
\begin{equation} \label{EQN:INTEGRALSPH}
\langle A({\bf r}) \rangle = \int { A({\bf r}) W(|{\bf r} - {\bf r}'|, h) \, dV}\,.
\end{equation}
One can show that, for reasonable kernels, the convergence is $O(h^2)$ 
\citep{GM1982}, so that $h$ is equivalent to the mesh spacing in a 
second order finite volume code.  Also, kernels are expected to have 
the property that $W(|{\bf r} - {\bf r}'|, h) \rightarrow \delta(|{\bf r} - {\bf r}'|)$ as $h \rightarrow 0$.  Therefore, Equation \ref{EQN:INTEGRALSPH} has at least first-order consistency. 
However, SPH discretises the integral by splitting the fluid into discrete mass elements of volume $dV = m/\rho$ into a summation of the form 
\begin{equation}
\langle A({\bf r}) \rangle = \sum_{j=1}^{N_n} 
{ m_j \frac{A_j}{\rho_j} W(|{\bf r} - {\bf r_j}|, h)},
\label{EQN:SUMSPH}
\end{equation}
\noindent where $N_n$ is the number of neighbouring particles, $A_j$ is the value of $A$ of particle $j$.  This approximation introduces a discretisation error into every SPH sum which is dependent on the number of neighbouring particles and the distribution of particles inside the smoothing kernel, but independent of the underlying fluid quantities that we are trying to solve.  Even for a constant function with no spatial gradients, i.e. $A({\bf r}) = const$, Equation \ref{EQN:SUMSPH} will not return this constant value unless $\sum_i \left\{ m_i\,W_i / \rho_i \right\} = 1$ exactly, which is not guaranteed in general.  Therefore, standard SPH does not even have zeroth-order consistency \citep[See][for a more detailed discussion]{GSPH2010}.

For a random/disordered distribution of particles, the discretisation error is Poissonian and scales as $1/\sqrt{N_n}$.  However, SPH tends to evolve the particles into a minimum-energy, glass-like lattice in sub-sonic flows.  \citet{Niederreiter1978} has shown that the error in such lattice configurations scales as $1/N_n\, \log N_n$.  
Since the particle positions are determined by the integration scheme, 
we do not have direct control unless we employ a particle 
re-mapping scheme, such as in `Regularised SPH' \citep{RSPH2001}.  
In principle, we could obtain more control over the discretisation error by fine-tuning the number of neighbours (via the smoothing length) where required.  For small $N_n$, the discretisation error will dominate the total error.  For much larger $N_n$, the smoothing error will dominate at which point increasing the neighbour number has no further effect on reducing the total error.  Therefore, one optimal approach is to attempt to constrain the discretisation error such that it is the same order as the smoothing error by fine-tuning the smoothing length rather than using Equation \ref{EQN:RHO-SPH}.

One further practical limitation on the accuracy of SPH codes is the particle 
clumping or tensile instability \citep{Swegle1995} which is an unwanted numerical effect where close-approaching particles artificially clump together due to the mathematical form of the SPH equations of motion.  The clumping instability is activated when the inter-particle distance becomes less than some fraction of the smoothing length.  This therefore limits the maximum possible number of neighbours allowed inside the smoothing kernel and subsequently the maximum obtainable accuracy using the summation approximation \citep{Price2012}.  This can partly be solved using a higher-order kernel \citep[e.g. quartic, Quintic, see][]{PriceChp3}, but this does not provide a general solution to this problem.

\subsection{Relative resolution requirements of finite volume and SPH codes}

In finite volume codes, the spatial resolution is defined by several local mesh spacings, $\Delta x$.  In SPH, the spatial resolution is also well-defined, this time by several particle smoothing lengths, $h$.  
Since SPH fluid elements are divided by mass, it is more common to consider mass resolution.  However, for consistency we will refer to the spatial resolution of SPH.

For the shock tube tests (Section \ref{SS:SHOCKTUBE}), it is
problematic to compare both methods since the finite volume code uses a
uniform-mesh spacing, whereas the SPH uses an adaptive smoothing
length (Eqn. \ref{EQN:HRHO-SPH}).  Of the three shock tube tests, only
the cooling shocks have an intrinsic length-scale that must be
resolved.  For finite volume methods, we find that at least four grid cells are required to resolve the cooling region using the standard options.  For SPH methods, for the simulation that just captures the cooling region without signs of post-shock oscillations, the initial resolution is $\Delta x = 1/4$ ($\lambda_{_{\rm COOL}} \sim \frac{1}{3}\,h$), rising to $\Delta x \sim 1/16$ ($\lambda_{_{\rm COOL}} \sim \frac{4}{3}\,h$) near the location of peak-shock temperature, finally peaking at $\Delta x = 1/4000$ once the gas has passed through the cooling region.  This suggests that the key diagnostic of resolution is the smoothing length of the initial adiabatic shock (before cooling takes place) compared to the size of the cooling region.  If this is not resolved, then the shock is broadened significantly before the peak temperature has been attained (so-called pre-shock heating) and significant cooling will have already reduced the peak temperature.

Our simulations of the NTSI and the KHI have shown that finite volume and SPH,
given enough resolution and accuracy, can show very good agreement in
hydrodynamical problems with complex flow patterns.  
Although it is difficult to know exactly when 
two simulations using two different hydrodynamical methods are producing 
the same results (due to their own individual errors), we inspect and 
compare the results visually, i.e. observing when the same features 
are present in both simulations.
It is noticeable that the NTSI simulations converge very well with
standard options and parameters, whereas the SPH KHI required
additional algorithms or modifications (e.g. artificial conductivity)
plus more neighbours.  One principle difference between the two cases
is the NTSI is seeded by a large scale, super-sonic perturbation where
small-scale particle noise and errors are not important, whereas the 
KHI is the growth of a seeded, low-amplitude velocity perturbation where 
noise and errors can corrupt the instability before it can grow.
The accuracy of the SPH method (controlled somewhat by the number of
neighbours) required to converge on the same results as the finite volume code
is therefore dependent somewhat on the problem studied.  This is clear
from the KHI convergence tests where the M4 kernel does not appear to
completely converge no matter how high the resolution.  
An important consequence of this is that future SPH convergence studies,
particularly testing new physics implementations, should consider varying 
both the total particle number and neighbour 
number (via using larger kernels).

One notion often assumed in comparison studies 
\citep[e.g.][]{Tasker2008} is that the resolution of SPH and finite volume codes 
are the same when the number of cells equals the number of SPH particles. 
This ignores the fact that SPH requires several dozen neighbours to be able 
to calculate hydrodynamical quantities, whereas finite volume  codes require far 
less neighbouring cell information to calculate interaction terms 
(two interactions per dimension for a second-order scheme).  
It is therefore better to equate 'kernel volumes' with 
`neighbouring-cell volumes' when determining the comparative resolution 
of SPH and finite volume codes.  
For a finite-difference finite volume code that is spatially second-order, the number 
of cell-cell interactions per cell is $N_{\rm int} = 2\,D$ where $D$ is the 
dimensionality.  For SPH, the number of neighbours is 
$N_n = 2\,{\cal R}\,\eta$, $\pi\,\left({\cal R}\,\eta \right)^2$ and 
$\left(4\,\pi/3\right)\,\left({\cal R}\,\eta \right)^3$ for 
one, two and three dimensions respectively where ${\cal R}$ is the compact 
support of the kernel (i.e. the extent of the kernel in multiples of $h$) {{\refone and 
$\eta$ is the dimensionless constant (default value $1.2$) as defined in Section \ref{SS:SPH}}. 
When using the M4-kernel for SPH (${\cal R} = 2$), the ratio of particle-to-cell interactions is $\sim 2$, $\sim 4$, and $\sim 9$ for one, two and three dimensions respectively.
Alternatively for the quintic kernel, the ratios are $\sim 3$, $\sim 6$ and $\sim 15$.
{\refone Therefore using the quintic kernel will incur a performance penalty of up to $~60\%$ longer run times compared to using the M4 kernel for the same number of particles.}

\subsection{Relative performance of AMR and SPH} \label{SS:PERFORMANCE}

We first compare the performance of {\small SEREN} and {\small MG} by simulating the simplest possible fluid simulation, a static, uniform density fluid.  We evolve each fluid for some set time at various resolutions, and then record the total wall-clock time and the number of timesteps required to complete the simulation.  
Figure \ref{FIG:TIMING}(a) shows the time per cell-cell, or
particle-particle, interaction for both finite volume and SPH
simulations.  We see that the time per particle interaction for SPH is
shorter than the corresponding finite volume interaction time by
approximately a factor of two.  This is understood in that most finite
volume codes use Riemann solvers for every cell-cell interaction.
This requires more arithmetic operations than particle-particle
interactions, even for a single iteration, whereas most SPH codes use
simpler hydrodynamical sums.  However, this is off-set by the fact
that mesh cells require fewer total interactions per cell than SPH
interactions per particle as discussed above.  Figure
\ref{FIG:TIMING}(a) shows the total run-time per cell/particle per
timestep for both finite volume and SPH codes.   We can therefore see
that the total computational time is shorter for finite volume codes for a simulation of the same effective resolution.  We should note that these timing statistics only apply for the two code implementations used and will likely differ to some extent in other similar Finite-Volume and SPH codes.

Second, we consider simple scaling arguments regarding the relative 
performances.  Since the CPU time per interaction is constant for both
finite volume and SPH codes, the total CPU time is then dependent on
the total number of interactions and timesteps.  For finite volume codes, the number of interactions per step scales as $N_{int} \propto D\,\Delta x^{-D}$, and for explicit methods, the timestep obeying the CFL conditions scales as $\Delta t \propto \Delta x$.  Therefore the total CPU work, $\mathbb{W}_{_{\rm MESH}}$, scales as 
\begin{equation}
\mathbb{W}_{_{\rm MESH}} = N_{int} \times \frac{1}{\Delta t} \propto \frac{1}{\Delta x^{D+1}}
\end{equation}
In smooth flow (i.e. in the absence of shocks), the error scales as $\Delta E_{_{\rm MESH}} \propto \Delta x^2$.  Therefore, the relationship between work and error is 
\begin{equation}
\Delta E_{_{\rm MESH}} \propto \frac{1}{\mathbb{W}_{_{\rm MESH}}^{2/(D + 1)}}
\end{equation}

For SPH codes using a constant number of neighbours, $N_n$, the number
of interactions per step scales as $N_{int} \propto h^{-D}$ and the
timestep scales as $\Delta t \propto h$.  The total CPU work,
$\mathbb{W}_{_{\rm SPH}}$, scales similarly to finite volume codes, i.e.
\begin{equation} \label{EQN:SPHERROR1}
\mathbb{W}_{_{\rm SPH}} = N_n^2 \times N_{kernel} \times \frac{1}{\Delta t} \propto \frac{N_n^2}{\Delta h^{D+1}}
\end{equation}
For smooth flow, where we assume the smoothing kernel error dominates
over the particle discretisation error (i.e. error $\propto h^2$),
then the relationship between the error and the work is similar to
finite volume codes, i.e.
\begin{equation} \label{EQN:SPHERROR2}
\Delta E_{_{\rm SPH}} \propto \frac{1}{\mathbb{W}_{_{\rm SPH}}^{2/(D + 1)}}
\end{equation}
However, if the discretisation error dominates over the smoothing error, then the error is unbounded and results in much worse scaling performance.  One hypothetical approach is to follow the error analysis discussed in Section \ref{SSS:SPHACCURACY} and attempt to limit the error by controlling the number of neighbours.  Since the smallest error possible is the smoothing kernel error, then the optimal approach would be to set the number of neighbours locally so that the discretisation error equals the smoothing error.  For smooth flows, the particles settle into a glass-like lattice whose error scales as $\propto 1 / N_n\,\log N_n$.  In order for the discretisation error to match the smoothing error, we require that 
\begin{equation} \label{EQN:SPHERROR3}
N_n\,\log N_n \propto \frac{1}{h^2}
\end{equation}
Ignoring the log-term and substituting into Equation \ref{EQN:SPHERROR1}, we obtain 
\begin{equation}
\Delta E_{_{\rm SPH}} \propto \frac{1}{\mathbb{W}_{_{\rm SPH}}^{2/(D + 5)}}
\end{equation}
In 3D, the error-work relation  for finite volume and SPH codes scales
as  $\Delta E_{_{\rm  MESH}} \propto  \mathbb{W}_{_{\rm MESH}}^{-1/2}$
and  $\Delta E_{_{\rm  SPH}}  \propto \mathbb{W}_{_{\rm  SPH}}^{-1/4}$
respectively.    Therefore,  SPH  codes   are  not   competitive  when
high-accuracy is required for hydrodynamical phenomenon.  However, SPH
is most  often used  in astrophysics where  such high-accuracy  is not
necessarily  required.   The  accuracy  and  CPU  cost  of  additional
algorithms, such as self-gravity and radiative transport, must also be
considered.  {\refone For  example,  \citet{Federrath2010} found  that
FLASH was  significantly slower than  SPH for problems  involving sink
particles. However,  this may not be  relevant to our  two codes since
there is considerable variation in the performance of both AMR and SPH 
codes. These matters will be discussed further in subsequent papers.}

\section{Conclusions} \label{S:CONCLUSIONS}
We have performed a suite of standard hydrodynamical tests comparing the 
convergence between simulations using the AMR finite volume code {\small MG} \citep{VanLoo2006} and the SPH code {\small SEREN} \citep{Hubber2011}.  We have tested how well the two methods compare, how well they converge with each other, in what ways they do not 
converge, and what these simulations inform us about resolution of the 
two methods.  

{\raw 
\begin{enumerate}
\item We find that in most cases, both methods converge with each other 
given enough resolution, and for SPH enough neighbours to reduce the 
discretisation error.  For some cases, improved accuracy in the SPH 
approximation is gained by using a larger kernel (e.g. the quintic kernel) 
to increase the number of neighbouring particles. 
For roughly uniform density problems, SPH codes require approximately 
$3$ times as many particles than grid cells to produce the same results 
as finite volume  codes.
\item For finite volume codes, adiabatic shocks or cooling shocks where the cooling length 
is resolved are correctly modeled using a second-order Riemann solver 
without the need for artificial viscosity.  For strictly isothermal 
shocks, we must use a first order Riemann solver, or artificial viscosity, 
to correctly capture the shock.
\item For SPH codes, adiabatic shocks, or cooling shocks where the cooling length 
is resolved, the standard artificial viscosity parameters ($\alpha_{_{\rm AV}} = 1$, 
$\beta_{_{\rm AV}} = 2$) suffice to allow shock capturing for all Mach numbers explored 
($1 < {\cal M}' < 64$).  For isothermal shocks, or cooling shocks where the 
cooling region is not resolved, higher values of $\alpha$ and $\beta$ may 
be required.  Alternatively, we find that using the harmonic mean of the 
density in SPH dissipation terms, instead of the arithmetic mean, performs 
better in preventing post-shock oscillations in strong shock problems.
\item In mixing problems (e.g. the Kelvin-Helmholtz instability), increasing 
the number of neighbours (by way of using kernels with larger compact support)
can partly resolve the energy discontinuity problem in SPH that leads to gap 
formation between the two fluids.  The reduced effect is sufficient to allow 
vorticity to be generated between the two fluids.  However, since there is 
no intrinsic mixing in SPH, an artificial conductivity term must still be 
added to allow convergence with finite volume methods (which contain intrinsic mixing 
through advection).  For larger discontinuities (e.g. the 10:1 KHI), the 
artificial repulsive force is too great for even the larger quintic kernel 
to amend the problem.  Therefore, artificial dissipation is required in 
this case to allow vorticity and mixing.
\item For roughly uniform density problems, finite volume codes out-perform SPH 
codes by an order of magnitude in wall-clock time, assuming the effective 
resolution of both codes is the same.  The CPU time for particle-particle 
interactions is less than the corresponding cell-cell interaction time in 
finite volume codes, but the larger number of interactions required per particle, 
plus the larger number of particles required to achieve similar results 
in an overall much longer total CPU time. 
\end{enumerate}
}

\section*{Acknowledgements}
DAH is funded by a Leverhulme Trust Research Project Grant (F/00 118/BJ) and an STFC post-doc.  Part of this work was done as part of an International Team at the International Space Science Institute in Bern.  We thank the anonymous referee for useful comments that improved an earlier draft of this paper.


\appendix

\section{Derivation of cooling shock solution} \label{A:COOLSHOCK}

Consider a steady flow in the frame of the shock. Then the continuity
and momentum equations give
\be
\rho v = Q = \rho_0 s,
\label{1}
\ee
\be
p + \rho v^2 = \Pi = p_0 + \rho_0 s^2,
\label{2}
\ee
where $s$  is the speed of  the shock, $\rho_0$  the pre-shock density
and $p_0$ the pre-shock pressure.
The energy equation is 
\be
\frac{\rd }{\rd x} \left[ {v \left( { \frac{\gamma}{\gamma - 1} p +
    \frac{1}{2} \rho v^2 } \right) } \right] = A \rho (T_0 - T).
\label{3}
\ee
The temperature is given by
\be
T = \frac{p}{\rho}.
\label{4}
\ee
We therefore have from (\ref{1}) and (\ref{2})
\be
T = \left( {\frac{\Pi}{\rho} - \frac{Q^2}{\rho^2}} \right)
\label{5}
\ee

Using (\ref{1}) and (\ref{4}), the energy equation can be written
\[
Q\frac{\rd }{\rd  x} \left(  { \frac{\gamma}{\gamma -  1} T +
    \frac{1}{2} \frac{Q^2}{\rho^2} } \right) = A \rho (T_0 -
    T).
\]
Using (\ref{5}) to eliminate $T$ gives
\be
\frac{Q}{(\gamma   -    1)}   \frac{\rd   }{\rd    x}   \left(   {\gamma
    \frac{\Pi}{\rho}  -  \frac{\gamma  +  1}{2}  \frac{Q^2}{\rho^2}  }
    \right) = A \rho \left[ {T_0 - \left( {\frac{\Pi}{\rho} -
	  \frac{Q^2}{\rho^2}} \right)} \right].
\label{6}
\ee
Define a new variable
\be
y = \frac{1}{\rho}.
\label{7}
\ee
Then (\ref{6}) becomes
\[
y   \frac{\rd   }{\rd    x}   \left(   {\gamma
    \Pi y  -  \frac{\gamma  +  1}{2} Q^2 y^2  }
    \right) = (\gamma - 1) \frac{A}{Q} (T_0 - \Pi y + Q^2 y^2),
\]
which is
\be
(\gamma \Pi y - (\gamma + 1)  Q^2 y^2) \frac{\rd y}{\rd x} = (\gamma -
1) \frac{A}{Q} (T_0 - \Pi y + Q^2 y^2).
\label{8}
\ee

Integrating this gives
\be
f(y) = (\gamma - 1) \frac{A}{Q} x + C,
\label{9}
\ee
where
\be
\begin{array}{l}
\ds{f(y) =  \frac{[\Pi^2/Q^2 - 2 (\gamma  + 1) T_0]}{\surd(\Pi^2  - 4 T_0
    Q^2)} \tanh^{-1} \left[ \frac{(2 Q^2 y - \Pi)}{\surd(\Pi^2 - 4 T_0
    Q^2)} \right]}\\
\\
~~~~~~~~~\ds{ - (\gamma + 1) y -  \frac{\Pi}{2 Q^2} \ln (\Pi y - T_0 -
Q^2 y^2)}.
\end{array}
\label{10}
\ee
Imposing the strong shock condition
\[
y = \frac{1}{\rho} = \frac{\gamma - 1}{\gamma + 1},
\]
at $x = 0$, gives
\be
f(y) - f  \left( \frac{\gamma - 1}{\gamma + 1} \right)  = (\gamma - 1)
\frac{A}{Q} x.
\label{11}
\ee


\begin{thebibliography}{49}
\expandafter\ifx\csname natexlab\endcsname\relax\def\natexlab#1{#1}\fi

\bibitem[{{Agertz} {et~al}\mbox{.}(2007){Agertz}, {Moore}, {Stadel}, {Potter},
  {Miniati}, {Read}, {Mayer}, {Gawryszczak}, {Kravtsov}, {Nordlund}, {Pearce},
  {Quilis}, {Rudd}, {Springel}, {Stone}, {Tasker}, {Teyssier}, {Wadsley}, \&
  {Walder}}]{Agertz2007}
{Agertz} O. {et~al.}, 2007, \mnras, 380, 963

\bibitem[{{Barnes} \& {Hut}(1986)}]{BH1986}
{Barnes} J., {Hut} P., 1986, \nat, 324, 446

\bibitem[{{Berger} \& {Colella}(1989)}]{AMR1989}
{Berger} M.~J., {Colella} P., 1989, Journal of Computational Physics, 82, 64

\bibitem[{{Berger} \& {Oliger}(1984)}]{AMR1984}
{Berger} M.~J., {Oliger} J., 1984, Journal of Computational Physics, 53, 484

\bibitem[{{Blondin} \& {Marks}(1996)}]{Blondin1996}
{Blondin} J.~M., {Marks} B.~S., 1996, \na, 1, 235

\bibitem[{{B{\o}rve} {et~al}\mbox{.}(2001){B{\o}rve}, {Omang}, \&
  {Trulsen}}]{RSPH2001}
{B{\o}rve} S., {Omang} M., {Trulsen} J., 2001, \apj, 561, 82

\bibitem[{{Boss} \& {Bodenheimer}(1979)}]{BBSIT1979}
{Boss} A.~P., {Bodenheimer} P., 1979, \apj, 234, 289

\bibitem[{{Cha} {et~al}\mbox{.}(2010){Cha}, {Inutsuka}, \&
  {Nayakshin}}]{GSPH2010}
{Cha} S.-H., {Inutsuka} S.-I., {Nayakshin} S., 2010, \mnras, 403, 1165

\bibitem[{{Chandrasekhar}(1961)}]{Chandra1961}
{Chandrasekhar} S., 1961, {Hydrodynamic and hydromagnetic stability},
  {Chandrasekhar, S.}, ed.

\bibitem[{{Commer{\c c}on} {et~al}\mbox{.}(2008){Commer{\c c}on}, {Hennebelle},
  {Audit}, {Chabrier}, \& {Teyssier}}]{Commercon2008}
{Commer{\c c}on} B., {Hennebelle} P., {Audit} E., {Chabrier} G., {Teyssier} R.,
  2008, \aap, 482, 371

\bibitem[{{Creasey} {et~al}\mbox{.}(2011){Creasey}, {Theuns}, {Bower}, \&
  {Lacey}}]{Creasey2011}
{Creasey} P., {Theuns} T., {Bower} R.~G., {Lacey} C.~G., 2011, \mnras, 415,
  3706

\bibitem[{{Dawson}(1983)}]{PIC1983}
{Dawson} J.~M., 1983, Reviews of Modern Physics, 55, 403

\bibitem[{{Falle}(1991)}]{Falle1991}
{Falle} S.~A.~E.~G., 1991, \mnras, 250, 581

\bibitem[{{Falle} {et~al}\mbox{.}(1998){Falle}, {Komissarov}, \&
  {Joarder}}]{Falle1998}
{Falle} S.~A.~E.~G., {Komissarov} S.~S., {Joarder} P., 1998, \mnras, 297, 265

\bibitem[{{Federrath} {et~al}\mbox{.}(2010){Federrath}, {Banerjee}, {Clark}, \&
  {Klessen}}]{Federrath2010}
{Federrath} C., {Banerjee} R., {Clark} P.~C., {Klessen} R.~S., 2010, \apj, 713,
  269

\bibitem[{{Federrath} {et~al}\mbox{.}(2011){Federrath}, {Sur}, {Schleicher},
  {Banerjee}, \& {Klessen}}]{Federrath2011}
{Federrath} C., {Sur} S., {Schleicher} D.~R.~G., {Banerjee} R., {Klessen}
  R.~S., 2011, \apj, 731, 62

\bibitem[{{Frenk} {et~al}\mbox{.}(1999){Frenk}, {White}, {Bode}, {Bond},
  {Bryan}, {Cen}, {Couchman}, {Evrard}, {Gnedin}, {Jenkins}, {Khokhlov},
  {Klypin}, {Navarro}, {Norman}, {Ostriker}, {Owen}, {Pearce}, {Pen},
  {Steinmetz}, {Thomas}, {Villumsen}, {Wadsley}, {Warren}, {Xu}, \&
  {Yepes}}]{Frenk1999}
{Frenk} C.~S. {et~al.}, 1999, \apj, 525, 554

\bibitem[{{Gingold} \& {Monaghan}(1977)}]{GM1977}
{Gingold} R.~A., {Monaghan} J.~J., 1977, \mnras, 181, 375

\bibitem[{{Gingold} \& {Monaghan}(1982)}]{GM1982}
{Gingold} R.~A., {Monaghan} J.~J., 1982, Journal of Computational Physics, 46,
  429

\bibitem[{{Godunov}(1959)}]{Godunov1959}
{Godunov} S.~K., 1959, Math. Sbornik, 47, 271

\bibitem[{{Heitsch} {et~al}\mbox{.}(2007){Heitsch}, {Slyz}, {Devriendt},
  {Hartmann}, \& {Burkert}}]{Heitsch2007}
{Heitsch} F., {Slyz} A.~D., {Devriendt} J.~E.~G., {Hartmann} L.~W., {Burkert}
  A., 2007, \apj, 665, 445

\bibitem[{{Hubber} {et~al}\mbox{.}(2011){Hubber}, {Batty}, {McLeod}, \&
  {Whitworth}}]{Hubber2011}
{Hubber} D.~A., {Batty} C.~P., {McLeod} A., {Whitworth} A.~P., 2011, \aap, 529,
  A27+

\bibitem[{{Junk} {et~al}\mbox{.}(2010){Junk}, {Walch}, {Heitsch}, {Burkert},
  {Wetzstein}, {Schartmann}, \& {Price}}]{Junk2010}
{Junk} V., {Walch} S., {Heitsch} F., {Burkert} A., {Wetzstein} M., {Schartmann}
  M., {Price} D., 2010, \mnras, 407, 1933

\bibitem[{{Kitsionas} {et~al}\mbox{.}(2009){Kitsionas}, {Federrath}, {Klessen},
  {Schmidt}, {Price}, {Dursi}, {Gritschneder}, {Walch}, {Piontek}, {Kim},
  {Jappsen}, {Ciecielag}, \& {Mac Low}}]{Kitsionas2009}
{Kitsionas} S. {et~al.}, 2009, \aap, 508, 541

\bibitem[{{Klein} \& {Woods}(1998)}]{Klein1998}
{Klein} R.~I., {Woods} D.~T., 1998, \apj, 497, 777

\bibitem[{{Lucy}(1977)}]{Lucy1977}
{Lucy} L.~B., 1977, \aj, 82, 1013

\bibitem[{{Mitchell} {et~al}\mbox{.}(2009){Mitchell}, {McCarthy}, {Bower},
  {Theuns}, \& {Crain}}]{Mitchell2009}
{Mitchell} N.~L., {McCarthy} I.~G., {Bower} R.~G., {Theuns} T., {Crain} R.~A.,
  2009, \mnras, 395, 180

\bibitem[{{Monaghan}(1997)}]{Monaghan1997}
{Monaghan} J.~J., 1997, Journal of Computational Physics, 136, 298

\bibitem[{Niederreiter(1978)}]{Niederreiter1978}
Niederreiter H., 1978, Bulletin of the American Mathematical Society, 84, 957

\bibitem[{{Price}(2005)}]{PriceChp3}
{Price} D., 2005, ArXiv Astrophysics e-prints

\bibitem[{{Price}(2008)}]{Price2008}
{Price} D.~J., 2008, Journal of Computational Physics, 227, 10040

\bibitem[{{Price}(2012)}]{Price2012}
{Price} D.~J., 2012, Journal of Computational Physics, 231, 759

\bibitem[{{Price} \& {Federrath}(2010)}]{PF2010}
{Price} D.~J., {Federrath} C., 2010, \mnras, 406, 1659

\bibitem[{{Price} \& {Monaghan}(2007)}]{PM2007}
{Price} D.~J., {Monaghan} J.~J., 2007, \mnras, 374, 1347

\bibitem[{{Quirk}(1994)}]{Quirk1994}
{Quirk} J.~J., 1994, International Journal for Numerical Methods in Fluids, 18,
  555

\bibitem[{{Read} {et~al}\mbox{.}(2010){Read}, {Hayfield}, \& {Agertz}}]{OSPH}
{Read} J.~I., {Hayfield} T., {Agertz} O., 2010, \mnras, 405, 1513

\bibitem[{{Robertson} {et~al}\mbox{.}(2010){Robertson}, {Kravtsov}, {Gnedin},
  {Abel}, \& {Rudd}}]{Advection2010}
{Robertson} B.~E., {Kravtsov} A.~V., {Gnedin} N.~Y., {Abel} T., {Rudd} D.~H.,
  2010, \mnras, 401, 2463

\bibitem[{{Saitoh} \& {Makino}(2009)}]{SM2009}
{Saitoh} T.~R., {Makino} J., 2009, \apjl, 697, L99

\bibitem[{{Shore}(2007)}]{Shore2007}
{Shore} S.~N., 2007, {Astrophysical Hydrodynamics: An Introduction}

\bibitem[{{Springel}(2005)}]{Gadget2}
{Springel} V., 2005, \mnras, 364, 1105

\bibitem[{{Springel}(2010)}]{AREPO}
{Springel} V., 2010, \mnras, 401, 791

\bibitem[{{Springel} \& {Hernquist}(2002)}]{SH2002}
{Springel} V., {Hernquist} L., 2002, \mnras, 333, 649

\bibitem[{{Swegle}(1995)}]{Swegle1995}
{Swegle} J., 1995, Journal of Computational Physics, 116, 123

\bibitem[{{Tasker} {et~al}\mbox{.}(2008){Tasker}, {Brunino}, {Mitchell},
  {Michielsen}, {Hopton}, {Pearce}, {Bryan}, \& {Theuns}}]{Tasker2008}
{Tasker} E.~J., {Brunino} R., {Mitchell} N.~L., {Michielsen} D., {Hopton} S.,
  {Pearce} F.~R., {Bryan} G.~L., {Theuns} T., 2008, \mnras, 390, 1267

\bibitem[{{Valcke} {et~al}\mbox{.}(2010){Valcke}, {de Rijcke}, {R{\"o}diger},
  \& {Dejonghe}}]{Valcke2010}
{Valcke} S., {de Rijcke} S., {R{\"o}diger} E., {Dejonghe} H., 2010, \mnras,
  408, 71

\bibitem[{{van Leer}(1977)}]{VanLeer1977}
{van Leer} B., 1977, Journal of Computational Physics, 23, 276

\bibitem[{{Van Loo} {et~al}\mbox{.}(2006){Van Loo}, {Falle}, \&
  {Hartquist}}]{VanLoo2006}
{Van Loo} S., {Falle} S.~A.~E.~G., {Hartquist} T.~W., 2006, \mnras, 370, 975

\bibitem[{{Vishniac}(1994)}]{NTSI}
{Vishniac} E.~T., 1994, \apj, 428, 186

\bibitem[{{Wadsley} {et~al}\mbox{.}(2008){Wadsley}, {Veeravalli}, \&
  {Couchman}}]{Wadsley2008}
{Wadsley} J.~W., {Veeravalli} G., {Couchman} H.~M.~P., 2008, \mnras, 387, 427

\end{thebibliography}
\end{document}